\documentclass[letterpaper,12pt,titlepage,openright,twoside,final]{report}

\newcommand{\href}[1]{#1} 

\usepackage{ifthen}
\newboolean{ElectronicVersion}
\setboolean{ElectronicVersion}{false} 

\usepackage{amsmath,amssymb,amstext} 
\usepackage[pdftex]{graphicx} 
\usepackage{subfigure}
\usepackage{dcolumn}
\usepackage{bm}
\usepackage{afterpage}

\ifthenelse{\boolean{ElectronicVersion}}{
    \usepackage[letterpaper=true,pdftex,bookmarks,pagebackref,
	plainpages=false, 
        pdfpagelabels=true, 
        pdftitle=Title\ goes\ here, 
        pdfauthor=Sean\ Stotyn	 
        ]{hyperref}}{}

\let\origdoublepage\cleardoublepage
\newcommand{\clearemptydoublepage}{\clearpage{\pagestyle{empty}\origdoublepage}}
\let\cleardoublepage\clearemptydoublepage

\begin{document}

\pagestyle{empty}
\pagenumbering{roman}

\begin{titlepage}
        \begin{center}
        \vspace*{1.0cm}

        \Huge
        {\bf Exact Solutions and Black Hole Stability in Higher Dimensional Supergravity Theories}

        \vspace*{1.0cm}

        \normalsize
        by \\

        \vspace*{1.0cm}

        \Large
        Sean Michael Anton Stotyn \\

        \vspace*{3.0cm}

        \normalsize
        A thesis \\
        presented to the University of Waterloo \\ 
        in fulfillment of the \\
        thesis requirement for the degree of \\
        Doctor of Philosophy \\
        in \\
        Physics \\

        \vspace*{2.0cm}

        Waterloo, Ontario, Canada, 2012 \\

        \vspace*{1.0cm}

        \copyright\ Sean Michael Anton Stotyn 2012 \\
        \end{center}
\end{titlepage}

\pagestyle{plain}
\setcounter{page}{2}

\cleardoublepage

  \noindent
I hereby declare that I am the sole author of this thesis. This is a true copy of the thesis, including any required final revisions, as accepted by my examiners.

  \bigskip
  
  \noindent
I understand that my thesis may be made electronically available to the public.

\cleardoublepage


\begin{center}\textbf{Abstract}\end{center}

This thesis examines exact solutions to gauged and ungauged supergravity theories in space-time dimensions $D\ge5$ as well as various instabilities of such solutions.

I begin by using two solution generating techniques for five dimensional minimal ungauged supergravity, the first of which exploits the existence of a Killing spinor to generate supersymmetric solutions, which are time-like fibrations over four dimensional hyper-K\"ahler base spaces.  I use this technique to construct a supersymmetric solution with the Atiyah-Hitchin metric as the base space.  This solution has three independent parameters and possesses mass, angular momentum, electric charge and magnetic charge.  Via an analysis of a congruence of null geodesics, I determine that the solution contains a region with naked closed time-like curves.  The centre of the space-time is a conically singular pseudo-horizon that repels geodesics, otherwise known as a repulson.  The region exterior to the closed time-like curves is outwardly geodesically complete and possesses an asymptotic region free of pathologies provided the angular momentum is chosen appropriately.  

The second solution generating technique exploits a hidden $G_2$ symmetry in five dimensional minimal supergravity.  I use this hidden symmetry approach to construct the most general black string solution in five dimensions, which is endowed with mass, angular momentum, linear momentum, electric charge and magnetic charge.  This general black string satisfies the first law of thermodynamics, with the Bekenstein-Hawking entropy being reproduced via a microstate counting in terms of free M-branes in the decoupling limit.  Furthermore it reduces to all previously known black string solutions in its various limits.  A phase diagram for extremal black strings is produced to draw conclusions about extremal black rings, in particular why supersymmetric black rings exhibit a lower bound on the electric charge.  The same phase diagram further suggests the existence of a new class of supersymmetric black rings, which are completely disconnected from the previously known class.

A particular limit of this general black string is the magnetically charged black string, whose thermodynamic phase behaviour and perturbative stability were previously studied but not very well understood.  I construct magnetically charged topological solitons, which I then show play an important role in the phase structure of these black strings.  Topological solitons in Einstein-Maxwell gravity, however, were previously believed to generically correspond to unstable ``bubbles of nothing" which expand to destroy the space-time.  I show that the addition of a topological magnetic charge changes the stability properties of these Kaluza-Klein bubbles and that there exist perturbatively stable, static, magnetically charged bubbles which are the local vacuum and the end-point of Hawking evaporation of magnetic black strings.  

In gauged supergravity theories, bubbles of nothing are stabilised by the positive energy theorem for asymptotically anti-de Sitter space-times.  For orbifold anti-de Sitter space-times in odd dimensions, a local vacuum state of the theory is just such a bubble, known as the Eguchi-Hanson soliton.  I study the phase behaviour of orbifold Schwarzschild-anti-de Sitter black holes, thermal orbifold anti-de Sitter space-times, and thermal Eguchi-Hanson solitons from a gravitational perspective; general agreement is found between this analysis and the previous analysis from the gauge theory perspective via the AdS/CFT correspondence.  I show that the usual Hawking-Page phase structure is recovered and that the main effect of the soliton in the phase space is to widen the range of large black holes that are unstable to decay despite the positivity of their specific heat.  Furthermore, using topological arguments I show that the soliton and orbifold AdS geometry correspond to a confinement phase in the boundary gauge theory while the black hole corresponds to a deconfinement phase.

An important instability for rotating asymptotically anti-de Sitter black holes is the superradiant instability.  Motivated by arguments that the physical end point of this instability should describe a clump of scalar field co-rotating with the black hole, I construct asymptotically anti-de Sitter black hole solutions with scalar hair.  Perturbative results, \emph{i.e.} low amplitude boson stars and small radius black holes with low amplitude scalar hair, are presented in odd dimensions relevant to gauged supergravity theories, namely $D=5,7$.  These solutions are neither stationary nor axisymmetric, allowing them to evade the rigidity theorem; instead the space-time plus matter fields are invariant under only a single helical Killing vector.  These hairy black holes are argued to be stable within their class of scalar field perturbations but are ultimately unstable to higher order perturbative modes.

\cleardoublepage


\begin{center}\textbf{Acknowledgements}\end{center}

It is with great delight that I acknowledge the numerous people throughout my academic career that helped make the production of this thesis possible.  First and foremost, I would like to thank my Advisor, Robert Mann, who has selflessly given much time and devotion to training me to become an independent researcher and future colleague.  Without his continued guidance, support, and friendship, I would likely not have had the patience and courage to persist in the academe.  Similarly I would like to express sincere gratitude toward my undergraduate mentors, Kristin Schleich and Don Witt, for their continued support and friendship, especially during periods of hardship and uncertainty.  Thanks are due to my first academic supervisor, Jasper Wall, whom I worked for as a summer student at the end of my third year of undergraduate studies and who taught me important lessons about the nature and frustration of academic research.  Furthermore my gratitude extends to Don Marolf, who supervised my six month visit to the University of California at Santa Barbara during the first half of 2010.  In addition to those who have directly advised me in my research, Robert Myers, Raphael Sorkin, and Eric Poisson deserve special thanks for their service as members of my Ph.D. Advisory Committee.

The work presented in Chapter 3 of this thesis would not have been possible without the collaboration of Geoffrey Comp\`ere, Sophie de Buyl, and Amitabh Virmani; I worked very closely with Geoffrey and Sophie during my stay at the University of California at Santa Barbara, while Amitabh was collaborating with us from l'Universit\'e Libre de Bruxelles in Belgium.  I am especially thankful to Geoffrey and Sophie for their guidance and support as well as their friendship.  Similarly, the work presented in Chapter 6 would not have been possible without the devoted collaborative efforts of Paul McGrath and Miok Park.  Although I very much spear-headed the project, their hard work and insights at various stages were indispensable to the timely outcome of the results.

There are numerous peers in a context other than collaborative work who deserve named recognition for their inspiration, guidance and friendship throughout my graduate studies: Chanda Prescod-Weinstein and Jonathan Hackett, for their support and advice in academic and non-academic related matters respectively, and Paul McGrath, Miok Park, Eric Brown, Razieh Pourhassan, Danielle Leonard, Wilson Brenna, Uzair Hussein, and Aida Ahmadzadegan for their friendship, which has not only fostered my passion for physics by allowing an outlet to talk about it with people who will actually understand but has also provided necessary distraction from physics when needed.  An apology is also due to Eleanor Kohler, Stephanie Crawford, Gregory Lyons, and any other non-physics person who has been the victim of seemingly boring and abstract physics talk during social outings.

I certainly would not have gotten this far without the love and support of family and friends.  For their continual support in the face of not understanding a single thing I do, I would like to thank my mother, Angela, my father, Dwayne, my two sisters, Jennifer and Michelle, my grandparents, Anne, Virginia, and Richard, my aunts, Pamela, Brenda, Candy, and Barb, and my uncles, Brad, Brent, Brad, and Terry.  I am furthermore grateful to Gregory Lyons for the love, comfort, and steady support that only a partner can provide, Kevin Nixon for being a best friend and all around support structure, Andrew Zurell for being essentially the only non-physics friend who has shown a genuine interest in and a sincere attempt to understand the work that I do, Ricardo Fonseca for being an instant, inseparable, and true friend, Audrey Lopez and Jack Loehr for being needed silly distractions from reality, Simon Reid for being an incredible roommate and having an ability to match my sense of humour perfectly, Dwayne Appleby for mutual academic inspiration in vastly disparate fields, and Michael Kushnir for simply understanding me through and through.  

Finally, I am grateful to my love affair with art for keeping me sane and balanced; it has added another dimension of colour and beauty to the world that I simply could not live without.

\cleardoublepage


\begin{center}\textbf{Dedication}\end{center}

This is dedicated to the minority groups who are underrepresented in the physics community.

\cleardoublepage

\renewcommand\contentsname{Table of Contents}
\tableofcontents
\cleardoublepage


\listoffigures
\addcontentsline{toc}{chapter}{List of Figures}
\cleardoublepage


\pagenumbering{arabic}


\chapter{Introduction}

\section{A Brief Review of Quantising Gravity}

Two of the greatest intellectual advancements in our understanding of the nature of the universe in the twentieth century were the discoveries of the general theory of relativity and the theory of quantum mechanics.  Although impressive in its predictive power of the utterly bizarre laws governing the world of the very small, it was quickly realised that the initial development of quantum mechanics, which treats particles as fundamental, was inconsistent with the theory of special relativity.  Thus the subject of quantum field theory was born, in which quantum fields are fundamental and interact on a static space-time background with global Lorentz symmetry explicitly built in.  However, general relativity states that space-time itself is dynamic and hence promotes global Lorentz symmetry to a local symmetry.  This implies that gravity must have propagating degrees of freedom and a more fundamental theory of gravity would have these quantised.  Such quanta are massless spin-2 particles called gravitons.  It would appear that a natural way to build a theory that includes all the known forces of nature would be to incorporate this graviton field into the Standard Model of Particle Physics.  This approach fails abysmally, however, because such a theory is not renormalisable: interactions involving gravitons yield divergences that cannot be rendered finite.  On the one hand there is the Standard Model of Particle Physics, which is the most successful physical theory ever developed, while on the other hand there is a beautiful classical theory of gravity, which ought to be quantised but resists the standard attempts to do so.  How can these two beautifully successful theories be reconciled?

A breakthrough in understanding this paradox came about by a rather simple but profound consideration: what if particles are not point-like but instead the fundamental objects are strings whose vibrations give rise to the physical spectrum of observed particles?  It turns out that in positing this hypothesis, massless spin-2 excitations of the string are guaranteed; that is, string theory \emph{necessarily} includes gravity.  String theory is currently the most promising candidate for a theory that unifies gravity with all of the other fundamental interactions in nature.  A peculiar feature, however, is that ten dimensions of space-time are required in order for the theory to be consistent, whereas we only observe four space-time dimensions.  Nevertheless, it was hoped that there would be a unique theory that would give rise to our observable universe.  This dream was short lived as it was soon realised that not only are there seemingly distinct string theories but also a huge number of vacua of the theories, possibly with our observable universe amidst this vast landscape.

The non-uniqueness of string theory is now believed to be solved by eleven dimensional M-theory, which \emph{is} unique.  Although a complete picture of exactly what M-theory is currently is not understood, its low energy limits are.  It is known that the fundamental objects of the theory are M2-branes and M5-branes, which are respectively electrically and magnetically charged under a 3-form gauge potential.  Furthermore, dimensional reduction of M-theory on a circle leads to type IIA string theory with its whole spectrum of D-branes having interpretations in terms of M2- and M5-branes\cite{Polchinski}: the fundamental IIA string is an M2-brane wrapping the compact dimension, for instance.  From type IIA string theory, the other string theories and their full spectra of objects can be obtained by various dualities: type IIB is related to type IIA by T-duality, type I is obtained as an orientifold of type IIB, the SO(32) heterotic theory is related to type I by S-duality, and the $E_8\times E_8$ heterotic theory is related to the SO(32) heterotic theory by T-duality.  Thus, the different string theories are unified into a single framework starting from a unique theory, which is certainly an elegant feature of a fundamental theory.

\section{Supersymmetry and Supergravity}

A common ingredient in M-theory and all of the string theories is supersymmetry, without which the theories are inconsistent.  Supersymmetry is an extension of the Poincar\'e symmetry and relates bosonic and fermionic degrees of freedom through transformations involving a spinor parameter, $\epsilon$.  As a simple example, in a theory with a free scalar field $\phi$ and fermion field $\psi$, the fields transform under global supersymmetry according to
\begin{equation}
\begin{split}
\delta(\epsilon)\phi ={}& \bar\epsilon\psi, \\
\delta(\epsilon)\psi ={}& \gamma^\mu\epsilon\partial_\mu\phi,
\end{split}
\end{equation}
where $\gamma^\mu$ is the $D$ dimensional representation of the gamma matrices.  Two successive supersymmetry transformations on $\phi$ then leads to general coordinate transformations $\delta(\epsilon_2)\delta(\epsilon_1)\phi=-\xi^\mu\partial_\mu\phi$, where the infinitesimal displacement vector is given by
\begin{equation}
\xi^\mu=\bar\epsilon_2\gamma^\mu\epsilon_1.
\end{equation}
Thus the supersymmetry algebra contains the Poincar\'e algebra.  In fact, the classical supersymmetry superalgebra is precisely the Poincar\'e algebra with additional fermionic supercharges $Q^i$ satisfying (anti-)commutation relations
\begin{align}
&\big\{Q^i,\bar Q_j\big\}=-\frac12\delta^i_j\gamma_\mu P^\mu, \nonumber\\
&\big[M_{\mu\nu},Q^i\big]=-\frac12\gamma_{\mu\nu}Q^i, \label{eq:superalgebra}\\
&\big[P_\mu,Q^i\big]=0, \nonumber
\end{align}
where $P^\mu$ are the generators of space-time translations, $M_{\mu\nu}$ are the generators of Lorentz boosts, and $\gamma_{\mu\nu}=\gamma_{[\mu}\gamma_{\nu]}$.  The supercharges can be seen to have a global internal symmetry: if there is a single supercharge the symmetry group is U(1) since the superalgebra is invariant under a global phase shift of $Q$.  The presence of multiple supercharges leads to a non-abelian symmetry group in general.  This internal symmetry is known as R-symmetry and can play an important role in supersymmetric theories.

Since global supersymmetry contains the Poincar\'e symmetry, promoting the global symmetry to a local symmetry necessarily induces a theory of gravity.  Such gravity theories are known as supergravities and are exactly the low energy limits of the string theories discussed above: eleven dimensional supergravity is the low energy limit of M-theory, type IIA supergravity is the low energy limit of type IIA string theory, and so on.  Gauging any global symmetry necessarily introduces a gauge field and in the context of supersymmetry this gauge field is a Majorana vector spinor field of spin-$\frac32$ known as the gravitino, $\Phi_\mu$.  In any space-time dimension, the number of allowed supercharges, $\cal N$, -- labeled by the indices $i,j$ in the superalgebra (\ref{eq:superalgebra}) -- is controlled by the number of gravitino fields.  This is tightly constrained by the supersymmetry algebra in four dimensions: for the four dimensional theory to have consistent particle interactions, the highest spin particles in the theory must have spin $s\le2$, which constrains the maximal supersymmetry representation to ${\cal N}=8$\cite{Polchinski}.  In eleven dimensional supergravity, a single gravitino reduces to eight gravitini when the theory is compactified to four dimensions and thus produces the full set of allowed supersymmetry generators.  

Eleven dimensions is the highest space-time dimension possible for a consistent theory of supergravity since any higher dimensions would require the four dimensional theory to have particles with spin $s>2$.  Furthermore, for the same reason the eleven dimensional theory has only a single gravitino, whose excitations transform in the vector spinor representation of SO(9), which gives 128 real fermionic states.  Supersymmetry then requires the number of fermionic and bosonic states to match.  The excitations of the gravitational field transform in the traceless symmetric tensor representation of SO(9), which gives 44 bosonic states.  This leaves 84 bosonic states, which are exactly accounted for by a gauge potential, whose excitations transform in the rank 3 antisymmetric tensor representation of SO(9).  This leads to a \emph{unique} supergravity theory in eleven dimensions, consisting of a single supermultiplet comprised of a graviton, gravitino, and rank 3 gauge potential; no further matter fields can be added, including a cosmological constant.  

At low energies, the fermionic degrees of freedom are not excited so the gravitino can be consistently set to zero, leaving the bosonic sector of the theory only.  In eleven dimensions, the bosonic action takes the form
\begin{equation}
{\cal S} =\frac{1}{16\pi G_{11}}\int{R_{11} \star_{11}  \mathbf{1} - \frac{1}{2}\star_{11}  \mathcal F \wedge \mathcal F +\frac{1}{6} \mathcal F \wedge \mathcal F \wedge \mathcal A}, \label{eq:11dsugra}
\end{equation}
where $G_{11}$ is the eleven dimensional Newton's constant, $R_{11}$ is the Ricci scalar, $\star_{11}$ is the Hodge dual operator, $\mathcal A$ is the 3-form gauge potential and $\mathcal F = {\bf d} \mathcal A$ is its field-strength.  The last term is a topological Chern-Simons term whose presence constrains the type of charge on the fundamental objects of the theory.  In the absence of the Chern-Simons term, the 3-form gauge potential could be dualised to an 6-form potential, interchanging the notions of electric and magnetic charge.  However, the form of the Chern-Simons term makes this dualisation impossible.  The result is that M2-branes carry the electric charge and M5-branes carry the magnetic charge.  

\subsection{Five Dimensional Minimal Supergravity}

From (\ref{eq:11dsugra}) all other supergravity theories can be derived via dimensional reduction over compact spatial directions.  A particularly interesting and simple example is five dimensional minimal (${\cal N}=1$) supergravity, which is the consistent truncation of (\ref{eq:11dsugra}) on a six-torus.  The five dimensional theory can be obtained by the ansatz
\begin{align}
ds^2_{11} ={}& ds^2_5 + \big(dx_1^2 + dx^2_2\big) + \big(dx_3^2 + dx^2_4\big) +\big(dx_5^2 + dx^2_6\big)\\
\mathcal A ={}& \frac{1}{\sqrt{3}} {\bf A} \wedge \big({\bf d} x_1 \wedge {\bf d} x_2+{\bf d}x_3 \wedge {\bf d}x_4 +{\bf d}x_5 \wedge {\bf d}x_6 \big) \label{eq:5dgauge}
\end{align}
where the moduli -- the volumes of the three two-tori that are parametrised by $(x_1,x_2)$, $(x_3,x_4),$ and $(x_5,x_6)$ -- are all the same and $\bf A$ is a 1-form gauge potential in five dimensions.  Five dimensional solutions carrying electric charge under $\bf A$ are necessarily point-like in five dimensions; from (\ref{eq:5dgauge}) these are interpreted as M2-branes wrapping the canonical (12), (34), and (56) two-cycles of the $T^6$.  Similarly, solutions carrying the corresponding magnetic charge of $\bf A$ are string-like in five dimensions; these are interpreted as M5-branes wrapping the four-cycles orthogonal to the canonical two-cycles of the $T^6$.  Furthermore, since the total electric and magnetic charges ought to be quantised, the number of branes of each type wrapping the various cycles are related to the moduli.  Thus in five dimensional minimal supergravity, in which the moduli are all equal, there are an equal number of M2-branes wrapping each two cycle and similarly there are an equal number of M5-branes wrapping each four-cycle.
Using the above ansatz for the dimensional reduction, the action of the bosonic sector of five dimensional minimal supergravity takes the form of Einstein-Maxwell theory with a Chern-Simons term:
\begin{equation}
{\cal S}=\frac{1}{16\pi G_5}\int{\left(R\star_5 1-\frac{1}{2}{\bf F\wedge}\star_5{\bf F}+\frac{1}{3\sqrt{3}}{\bf F}\wedge {\bf F}\wedge {\bf A}\right)},  \label{eq:UngaugedSUGRA}
\end{equation}
where $G_5=G_{11}/V_{T^6}$ is the five dimensional Newton's constant with $V_{T^6}$ being the volume of the $T^6$, ${\bf F}={\bf dA}$ is the 2-form field strength and $\star_5$ denotes the Hodge dual operator in five dimensions.  Although a consistent theory in its own right, solutions to (\ref{eq:UngaugedSUGRA}), denoted ${\cal M}_5$, are also solutions to eleven dimensional supergravity of the form ${\cal M}_5\times T^6$.

This five dimensional theory is rather simple but has a surprisingly rich spectrum and allows a simplified setting for studying supersymmetric solutions to the field equations, \emph{i.e.} solutions which satisfy the Bogomol'nyi-Prasad-Sommerfeld (BPS) bound.  The BPS bound is a lower bound on the energy of a solution in terms of its charges. Therefore solutions to the equations of motion that saturate the BPS bound represent local vacua of the theory.  A familiar example of such a BPS solution in four dimensions is the extremal Reissner-Nordstr\"om black hole, whose mass and electric charge satisfy $M^2=Q^2$.

BPS solutions describing extremal black hole configurations play a particularly important role in string theory: arguably one of the greatest successes of string theory is the statistical mechanical explanation of the Bekenstein-Hawking entropy of certain extremal and near-extremal black holes. The first example was the five dimensional extremal black hole studied in \cite{Strominger:1996sh}, followed by a number of other types of extremal black holes\cite{David:2002wn, Mathur:2005ai, Peet:2000hn}; the agreement between the statistical mechanical entropy and the Bekenstein-Hawking entropy has been shown to hold in each case.  Then, in \cite{Breckenridge:1996sn,Callan:1996dv, Horowitz:1996fn} the success of \cite{Strominger:1996sh} was extended to the five dimensional near-extremal setting, \emph{i.e.} excitations above the BPS bound. The black holes considered in \cite{Breckenridge:1996sn,Callan:1996dv, Horowitz:1996fn,Strominger:1996sh} all have topologically spherical horizons, however Elvang, Emparan, Mateos, and Reall \cite{Elvang:2004rt} presented the first example of a supersymmetric black ring solution with horizon topology $S^1\times S^2$. The Bekenstein-Hawking entropy of the supersymmetric black ring was then successfully reproduced from a statistical mechanical description in \cite{Bena:2004tk,Cyrier:2004hj}.  
This is rather remarkable because not only does string theory give us quantised gravity for free but it also yields a microscopic explanation of the origin of black hole entropy!   

\subsection{Gauged Supergravities}

The mechanism for reproducing the entropy from a microscopic perspective is that BPS supergravity solutions described by extremal black $p$-branes have near-horizon geometries given by AdS$_{p+2}\times S^n$, where $p+2+n=10~\mathrm{or}~11$.  Fluxes are typically carried on the $S^n$, which relates the radius of the $S^n$ to the curvature radius of AdS$_{p+2}$.  The anti-de Sitter conformal field theory (AdS/CFT) correspondence\cite{Maldacena:1997re} states that the bulk gravity theory in AdS$_{p+2}$ is dual to a conformal field theory without gravity living on the boundary of the space-time, ${\mathbb R}\times S^p$.  The statistical entropy of the boundary gauge theory is then found to be identical to the entropy of the bulk solution, \emph{i.e.} the Bekenstein-Hawking entropy.  

These near-horizon geometries of (near-)extremal black holes and black $p$-branes are themselves valid solutions in their own right.  In fact, they can be obtained by the reduction of ten or eleven dimensional supergravity theories on spheres.  The lower dimensional supergravity theories contain an effective negative cosmological constant and have non-abelian gauge fields in the supermultiplet of the graviton; such supergravity theories are known as \emph{gauged} supergravity.  The non-abelian gauge fields essentially come from the compactification on non-toroidal manifolds\footnote{There are other global symmetries that can be gauged, which lead to gauged supergravity theories, but these are not considered here.}: they gauge (a subgroup of) the symmetry group of the manifold on which the higher dimensional supergravity theory is compactified.  Furthermore, like five dimensional ungauged minimal supergravity above, some gauged supergravities are consistent truncations of higher dimensional supergravity theories.  In such cases, these gauged supergravities are self-consistent independent theories whose solutions also automatically satisfy the corresponding higher dimensional theory.  

An example of this is the consistent truncation of type IIB supergravity on $S^5$, yielding five dimensional maximal gauged supergravity\cite{Gunaydin:1984fk,Kim:1985ez}.  The isometry group of the five-sphere is SO(6) and indeed this is the symmetry group that is gauged in the five dimensional theory.  This is, in fact, the original theory that led Maldacena to his famous AdS/CFT conjecture in \cite{Maldacena:1997re}.  Solutions to five dimensional gauged supergravity are asymptotically AdS$_5$, hence the boundary of such solutions, ${\mathbb R}\times S^3$, possesses the conformal group as a symmetry group.  Maldacena conjectured that the supergravity solution in the bulk space-time is dual to a conformal field theory living on the boundary; in the case of five dimensional gauged supergravity the corresponding four dimensional boundary gauge theory is ${\cal N}=4$ super-Yang-Mills theory with a U($N$) gauge symmetry.  Another example of a consistent truncation is that of eleven dimensional supergravity reduced on $S^4$, yielding seven dimensional gauged supergravity\cite{Nastase:1999cb}.  Again, the SO(5) isometry of the $S^4$ becomes the gauge group in the seven dimensional theory and by the AdS/CFT conjecture the bulk supergravity theory is dual to a conformal field theory living on the boundary of the AdS$_7$.  

In these two gauged supergravity theories, setting the gauge fields to zero yields empty AdS$_{p+2}$ solutions as the supersymmetric ground state; this is similar to Minkowski space-time being the supersymmetric ground state of ungauged supergravity theories.  Excitations above the AdS vacuum, for instance black hole space-times, put the boundary gauge theory at finite temperature.  Such solutions have been proven useful for studying non-abelian gauge theories at finite temperatures.  For instance, such studies have led to insights into the nature of high temperature superconductors \cite{Hartnoll:2008vx} and the properties of quark-gluon plasmas \cite{Policastro:2001yc}.  The spectrum of solutions in these theories is rather rich and, given their connection to QCD-like gauge theories via the AdS/CFT correspondence, their stability properties are of particular interest.

\section{Outline of the thesis}

This thesis is essentially separable into two parts: chapters 2, 3, and 4 deal exclusively with five dimensional minimal ungauged supergravity while chapters 5 and 6 deal with odd dimensional gauged supergravities in the simplifying scenario that the gauge fields vanish.

In chapter 2, I construct a supersymmetric solution to five dimensional minimal supergravity by employing a solution generating technique outlined in \cite{Gauntlett:2002nw}, in which all supersymmetric solutions of the theory are classified according to whether the Killing vector constructed from the Killing spinor is time-like or null.  In the time-like case, the solution takes the form of a time-like fibration over a four dimensional hyper-K\"ahler base space.  I use the Atiyah-Hitchin base space to construct a solution and find a three-parameter family of solutions that possess mass, angular momentum, electric charge, and magnetic charge.  They are shown to describe conically singular, repulsive pseudo-horizons inside a region containing closed time-like curves.  Asymptotically, the space-times are free of pathologies and describe a twisted U(1) fibration over the conifold $M_4/{\mathbb Z}_2$ where $M_4$ is four dimensional Minkowski space-time.

I then broaden the focus to non-supersymmetric solutions in chapter 3 by constructing the most general black string in five dimensional minimal supergravity.  This is done by exploiting a hidden $G_2$ symmetry that becomes manifest when the theory is reduced to three dimensions.  This general black string is a five-parameter family of solutions which possess mass, angular momentum transverse to the string, linear momentum along the string, electric charge, and magnetic charge.  I show that it satisfies the first law of thermodynamics and that the Bekenstein-Hawking entropy is reproduced by a statistical microstate counting in the decoupling limit.  I then reproduce various known extremal and supersymmetric black string solutions by taking appropriate limits of the general string.  Since tensionless black strings correspond to infinite radius limits of black rings, I show that the extremal tensionless limits of the general black string predict a new class of supersymmetric black rings -- distinct from the one found by Elvang \emph{et al.} in \cite{Elvang:2004rt} -- and that there similarly appears to be two disjoint branches of general black rings of the theory.

In chapter 4, I study the thermodynamic stability of the black string with a topological magnetic charge, which is the limit of the general string in which the angular and linear momenta, and the electric charge vanish.  I show that if one considers boundary conditions that break supersymmetry, static topological solitons with the same magnetic charge as the string enter the spectrum and play a nontrivial role in the thermodynamic phase space.  Furthermore, these topological solitons are connected to a wider two-parameter family of dynamic bubbles.  For a given radius of the compact Kaluza-Klein direction, there exist both a small and a large soliton: I show that the small soliton is perturbatively stable while the large soliton is perturbatively unstable.  Under certain restrictions in parameter space, the small soliton then becomes the effective vacuum state and is the end point of Hawking evaporation of the string.  This is finally compared with the perturbative stability of the magnetic black string and it is shown that the Gubser-Mitra conjecture, which claims a correlation between thermodynamic and perturbative stability, is not satisfied for these objects.

In chapter 5, continuing in the spirit of thermodynamic stability, I analyse the thermodynamic phase structure of vacuum gauged supergravity theories on orbifold space-times AdS$/{\mathbb Z}_k$ in odd dimensions $D\ge5$.  If the boundary conditions break supersymmetry, there are three such vacuum space-times known: the Eguchi-Hanson soliton, the orbifold anti-de Sitter geometry and the orbifold Schwarzschild-anti-de Sitter black hole.  I show that the soliton is always thermodynamically favoured over the orbifold AdS geometry, indicating that both solutions are local vacua with the soliton as the stable ground state.  I then show that in addition to a normal Hawking-Page phase transition between the black hole and the orbifold AdS geometry, there is a further phase transition between the soliton and the black hole.  I find that the primary effect of the soliton in the phase space is to widen the range of large black holes that are thermodynamically unstable despite their positive specific heat; there is a range of phase space in which black holes are stable against Hawking evaporation yet are unstable toward decay to the soliton.  Such black holes thus undergo direct semiclassical tunneling to the soliton configuration.  In addition, these phase transitions are shown to be dual to a confinement-deconfinement phase transition in the dual gauge theory.

I continue studying instabilities in five and seven dimensional gauged supergravity in chapter 6 but with a shifted focus to the dynamical instability of superradiant scattering of scalar field perturbations off spinning black holes.  I construct semi-stable end states to this instability, which describe spinning black holes with co-rotating lumps of scalar hair.  The combined metric and scalar fields of such solutions have broken axisymmetry and are invariant under only a single Killing vector field, which is tangent to the generators of the horizon.  I construct these hairy black holes perturbatively as power expansions in the amplitude of the scalar field, $\epsilon$, and the horizon size relative to the AdS radius, $r_+/\ell$.  The equations of motion also admit boson star solutions, which are smooth horizonless solitonic solutions that describe a clump of rotating scalar field whose self gravitational attraction is balanced by centrifugal repulsion.  These boson star solutions can be obtained from the hairy black hole in the limit of vanishing horizon radius.  The hairy black holes are then argued to be stable within the class of scalar field perturbations governed by the scalar field ansatz.  However, they are argued to ultimately be unstable toward further superradiant instabilities from higher order perturbative modes.

\chapter{Supergravity on an Atiyah-Hitchin Base}

As discussed in the introduction, promoting global supersymmetry to a local symmetry leads directly to supergravity theories in ten and eleven dimensions, which are in turn the low energy limits of superstring theory and M-theory.  Exact solutions to these supergravity theories help to illuminate the theory's structure, for instance through brane dynamics and intersections, as well as the spectrum of possible objects contained therein.  Of particular importance are the supersymmetric solutions because they represent local ground states of the theory, hence they often have stability properties not shared by their non-supersymmetric counterparts.  Of practical interest is a complete classification of all supersymmetric solutions to ten and eleven dimensional supergravity, although this is a very involved task.  A much lighter and tractable undertaking is such classifications in various limits and consistent truncations of ten and eleven dimensional supergravity.  For example, a classification of supersymmetric solutions of ${\cal N}=2$ $D=4$ supergravity was provided in \cite{Tod:1983pm}, which classified solutions according to whether the Killing vector constructed from the Killing spinor is time-like or null.  Following in this vein, the authors of \cite{Gauntlett:2002nw} set out a prescription for generating supersymmetric solutions to ${\cal N}=1$ $D=5$ supergravity, \emph{i.e.} five dimensional minimal supergravity.  The solutions again fall into two classes depending on whether the Killing vector is time-like or null.  

There has been a large amount of work generating supersymmetric solutions to five dimensional minimal supergravity, both by the construction laid out in \cite{Gauntlett:2002nw} and by other means.  The Breckenridge-Myers-Peet-Vafa (BMPV) solution \cite{Breckenridge:1996is} can be constructed as a solution to five dimensional minimal supergravity and it describes an asymptotically flat supersymmetric black hole specified by an electric charge and two equal angular momenta.  Elvang \emph{et al.} constructed an asymptotically flat supersymmetric black ring solution specified by an electric charge and two independent angular momenta \cite{Elvang:2004rt}.  A supersymmetric G\"{o}del-like space-time was constructed in \cite{Gauntlett:2002nw}.  In \cite{Tomizawa:2007he}, the Eguchi-Hanson space was used to construct a supersymmetric black ring solution specified by an electric charge and two equal angular momenta.  In \cite{Gauntlett:2004wh,Gauntlett:2004qy} Gauntlett and Gutowski constructed supersymmetric analogues to the black saturn configuration; their solutions describe concentric black rings with an optional black hole at the common center.  These last solutions demonstrate conclusively that black hole uniqueness does not hold in dimensions higher than four, even for supersymmetric solutions.  
 
According to the prescription laid out in \cite{Gauntlett:2002nw}, all supersymmetric solutions to five dimensional minimal supergravity with a time-like Killing vector are described in terms of a time-like fibration over a four dimensional hyper-K\"ahler base manifold.  The examples of supersymmetric solutions above all have one of three base spaces: ${\mathbb R}^4$, Euclidean Taub-NUT, or Eguchi-Hanson.  In the present chapter, which is based on but slightly corrects and expands on some of the work published in \cite{Stotyn:2008fk}, I construct a supersymmetric solution using the Atiyah-Hitchin metric as the hyper-K\"{a}hler base space.  A similar but distinct solution using the ambi-polar generalisation of the Atiyah-Hitchin metric was constructed in \cite{Bena:2007ju}.  For simplicity I use a cohomogeneity-1 ansatz, \emph{i.e.} metric functions of the radius only, and the solution so constructed is generated by two first-order differential equations and one Poisson equation on the base.  By studying a congruence of null geodesics, it is shown that for most of the parameter space, the supersymmetric Atiyah-Hitchin solution describes singular pseudo-horizons in a region containing closed time-like curves; the region exterior to the closed time-like curves is outwardly geodesically complete and free of pathologies.  There is a set of parameter space of measure zero for which the geometry is completely smooth and pathology-free.

\section{Supersymmetric Solutions of 5D Ungauged Minimal Supergravity}
\label{sec:solutions}

In this section, I outline the salient features of \cite{Gauntlett:2002nw} with regard to constructing supersymmetric solutions to the supergravity equations.\footnote{In \cite{Gauntlett:2002nw}, the metric signature and field strength are $(+----)$ and ${\cal F}=2{\bf d}{\cal A}$ respectively, whereas here the metric signature is $(-++++)$ and the 1-form potential is taken to be ${\bf A}=-2{\cal A}$ such that the field strength is ${\bf F}={\bf dA}=-{\cal F}$.}  The action for the bosonic sector of five dimensional minimal supergravity is given by (\ref{eq:UngaugedSUGRA}), which result in the equations of motion
\begin{equation}
R_{\mu\nu}-\frac{1}{2}\left(F_{\mu\sigma}{F_{\nu}}^{\sigma}-\frac{1}{6}F^2g_{\mu\nu}\right)=0,
\end{equation}
\begin{equation}
{\bf d}\star_5{\bf F}-\frac{2}{\sqrt{3}}{\bf F}\wedge{\bf F}=0,    \label{eq:FEOM}
\end{equation}
where $\star_5$ is the five dimensional Hodge dual operator, $F^2\equiv F_{\mu\nu}F^{\mu\nu}$ and $F_{\mu\nu}$ is related to the 2-form field strength by ${\bf F}=\frac12F_{\mu\nu}{\bf d}x^\mu\wedge{\bf d}x^\nu$.  Solutions to the equations of motion are supersymmetric if they admit a supercovariantly constant spinor, also called a Killing spinor
\begin{equation}
\left(D_\mu-\frac{1}{8\sqrt{3}}\big({\gamma_{\mu}}^{\nu\rho}-4\delta_\mu^\nu\gamma^\rho\big)F_{\nu\rho}\right)\epsilon^a=0,
\end{equation}
where $\epsilon^a$ is a commuting symplectic Majorana spinor with symplectic index $a=1,2$ and $D_\mu$ is the Lorentz covariant derivative operator acting on spinors.  This Killing spinor leads to further important structures in the solutions, in particular a real scalar, $q$, a real 1-form, ${\bf V}$, and three complex 2-forms, ${\bf \Phi}^{11}$, ${\bf \Phi}^{12}$ and ${\bf \Phi}^{22}$ can be constructed by taking bilinears of $\epsilon^a$:
\begin{align}
q\epsilon^{ab} ={}& \bar{\epsilon}^a\epsilon^b, \\
V_{\mu}\epsilon^{ab}={}&\bar{\epsilon}^a\gamma_\mu\epsilon^b, \\
\Phi^{ab}_{\mu\nu}={}&\bar{\epsilon}^a\gamma_{\mu\nu}\epsilon^b,
\end{align}
where $\epsilon^{12}=1$.  The Killing spinor has 8 real degrees of freedom, but the objects created from the bilinears have $1+5+3\times10=36$ components.  To reconcile this mismatch, there are numerous algebraic and differential relations between the bilinears that reduce the independent degrees of freedom from 36 to 8; these relations are analysed in detail in \cite{Gauntlett:2002nw} and here I will simply quote the important results.

First, the 1-form is a Killing vector and it is related to the scalar via 
\begin{equation}
V_\mu V^\mu=-q^2,  \label{eq:V}
\end{equation}
which means that $V^\mu$ is either time-like or null; only the time-like case will be considered in what follows.  It is further assumed, without loss of generality, that $q>0$ is globally defined.  Since $V$ is a time-like Killing vector, coordinates can be introduced such that $V=\partial_t$.  Imposing the condition (\ref{eq:V}), the metric can then be written locally in the form
\begin{equation}
ds^2=-H^{-2}(dt+{\bf \omega})^2+Hh_{mn}dx^mdx^n,   \label{eq:5dmetric}
\end{equation}
where the scalar $H=q^{-1}$ is introduced for later convenience and $\omega$ is a 1-form connection.  The metric $Hh_{mn}$ is obtained by projecting the full five dimensional metric perpendicular to the orbits of $V$.  Furthermore $h_{mn}$ must be hyper-K\"{a}hler, which means it must possess 2-form structures ${\bf X}^{(i)}$ that satisfy the algebra of unit quaternions.  Furthermore, a positive orientation on $h_{mn}$ must be chosen so that these hyper-K\"ahler 2-forms are anti-self-dual.  The ${\bf X}^{(i)}$ are related to the ${\bf \Phi}^{ab}$ by ${\bf \Phi}^{11}={\bf X}^{(1)}+i{\bf X}^{(2)}$, ${\bf \Phi}^{22}={\bf X}^{(1)}-i{\bf X}^{(2)}$ and ${\bf \Phi}^{12}=-i{\bf X}^{(3)}$.  Hereafter $h_{mn}dx^mdx^n=ds_{\cal B}^2$ is denoted as the base space, $\cal B$, and is endowed with a hyper-K\"{a}hler metric.  

It is convenient to introduce a f\"unfbein, $\Sigma$, such that if the vierbein, $\sigma_{\cal B}$, defines the positive orientation volume 4-form on $\cal B$, then $\Sigma={\bf e^0}\wedge(H^2\sigma_{\cal B})$ defines the positive orientation volume 5-form on the full solution, where
\begin{equation}
{\bf e}^0={H^{-1}}({\bf d}t+\omega).
\end{equation}
Using this definition, the solution for ${\bf F}$ is
\begin{equation}
{\bf F}=-\frac{\sqrt{3}}{2}{\bf de}^0+\frac{1}{\sqrt{3}}{\bf G^+}, \label{eq:FSol}
\end{equation}
where ${\bf G^+}$ is a self-dual 2-form on $\cal B$ defined via
\begin{equation}
{\bf G}^+=\frac{1}{2H}({\bf d\omega}+\star_4 {\bf d\omega}). \label{eq:Gomega}
\end{equation}
Here $\star_4$ denotes the Hodge dual on the four dimensional base space, $\cal{B}$.  From equation (\ref{eq:FSol}), the Bianchi identity, ${\bf dF}=0$, gives
\begin{equation}
{\bf dG^+}=0,  \label{eq:Bianchi}
\end{equation}
meaning that ${\bf G}^+$ is closed and hence can be written locally as an exact form ${\bf G^+=\alpha d\Omega}$, for some constant $\alpha$ and 1-form ${\bf \Omega}$.  The equation of motion for $\bf F$, (\ref{eq:FEOM}), gives a differential equation for the scalar $H$: a somewhat tedious calculation leads to
\begin{equation}
\Delta H \star_41=\frac{4}{9}{\bf G^+}\wedge{\bf G^+}, \label{eq:Poisson}
\end{equation}
where $\star_41\equiv\sigma_{\cal B}$ and $\Delta$ is the Laplacian operator on the base space.

All of the ingredients are now present in order to construct supersymmetric solutions.  However it turns out that the analysis simplifies considerably if the base space admits a triholomorphic Killing vector field, $L$.  This is a Killing vector on the base which leaves the hyper-K\"ahler 2 forms invariant, \emph{i.e.} the Lie derivative vanishes ${\cal L}_L{\bf X}^{(i)}=0$.  Such base spaces can always be put into Gibbons-Hawking form
\begin{align}
&ds_{\cal B}^2=f^{-1}\big(d\chi+\vec a\cdot d\vec x\big)^2+fd\vec x\cdot d\vec x, \\
&\nabla \times {\vec a}=\nabla f, \label{eq:veca}
\end{align}
where $L=\partial_\chi$ and $\nabla$ is the flat connection on $\mathbb R^3$; notice that equation (\ref{eq:veca}) implies that $f$ is harmonic on $\mathbb R^3$.  If $L$ is a Killing vector of the full five dimensional solution, $\omega$ can be decomposed as
\begin{equation}
\omega=\omega_\chi\big({\bf d\chi}+\vec a\cdot {\bf d}\vec x\big)+\vec \omega\cdot {\bf d}\vec x.
\end{equation}
Because $\bf G^+$ and equation (\ref{eq:Poisson}) are tangent to the base space, the equations of motion of the full solution are solved completely in terms of the base.  In the case of Gibbons-Hawking bases, such an analysis reveals that the solution is given in terms of four functions, $f$, $K$, $L$ and $M$, all of which are harmonic on $\mathbb R^3$.  The results are
\begin{align}
&H^{-1}=K^2f^{-1}+L, \\
&\omega_\chi=K^3f^{-2}+\frac32KLf^{-1}+M, \\
&\nabla\times\vec\omega=f\nabla\omega_\chi-\omega_\chi\nabla f-3\big(K^2+Lf\big)\nabla\big(Kf^{-1}\big),
\end{align}
with the solution to the latter being unique up to an arbitrary gradient, which can be absorbed into the definition of $t$.  

This is a rather remarkable result because the full five dimensional solution is described in terms of four harmonic functions on $\mathbb R^3$, whereas according to equation (\ref{eq:Poisson}) the solution for $H$ is in general given by Poisson's equation on the full base space.

\section{The Atiyah-Hitchin Space}

In this section I briefly review the hyper-K\"ahler metric known as the Atiyah-Hitchin metric.  It is an example of a hyper-K\"ahler manifold without a tri-holomorphic Killing vector, hence the immediately preceding discussion does not apply to solutions with such base spaces.  I begin by discussing the physical interpretation of the Atiyah-Hitchin metric, followed by its explicit form and properties.

The dynamics of two non-relativistic BPS monopoles is described by a manifold $\cal{M}$ which has the product structure
\begin{equation}
{\cal{M}}=\mathbb{R}^3\times\frac{S^1\times M}{\mathbb{Z}_2}, \nonumber
\end{equation}
where a point in $\mathbb{R}^3\times S^1$ denotes the centre of mass of the system and a time-varying phase angle that determines the total electric charge, while a point in $M$ specifies the monopole separation and a relative phase angle.  The four dimensional manifold, $M$, is invariant under SO(3) and can be parametrised by a radial coordinate $r$, roughly giving the separation of the monopoles, and Euler angles $\theta\in[0,\pi]$, $\phi\in[0,2\pi]$ and $\psi\in[0,2\pi]$.

It is convenient to introduce a basis for SO(3) via the left-invariant Maurer-Cartan 1-forms, which are related to the Euler angles through the relations
\begin{align}
&\sigma_1=-\sin\psi {\bf d}\theta +\cos\psi\sin\theta {\bf d}\phi, \nonumber\\
&\sigma_2=\cos\psi {\bf d}\theta +\sin\psi\sin\theta {\bf d}\phi, \\
&\sigma_3={\bf d}\psi +\cos\theta {\bf d}\phi, \nonumber
\end{align}
and which have the property
\begin{equation}
{\bf d}\sigma_i=\frac{1}{2}\varepsilon_{ijk}\sigma_j\wedge\sigma_k.
\end{equation}
In terms of these 1-forms, provided\footnote{The fact that the Atiyah-Hitchin metric has $\psi\in[0,2\pi]$ is due to the discrete symmetry (\ref{eq:antipodal}), which swaps the positions of the identical monopoles.} $\psi\in[0,4\pi]$ the metric on a unit radius $S^3$ is given by $d\Omega_3^2=\frac14({\sigma_1}^2+{\sigma_2}^2+{\sigma_3}^2)$ and the metric on a unit radius $S^2$ is given by $d\Omega_2^2={\sigma_1}^2+{\sigma_2}^2$. The most general Euclidean four dimensional SO(3) invariant metric is given by
\begin{equation}
ds^2=f^2(r)dr^2+a^2(r){\sigma_1}^2+b^2(r){\sigma_2}^2+c^2(r){\sigma_3}^2, \label{eq:AH}
\end{equation}
where, for notational convenience, the explicit $r$-dependence in $f,a,b,c$ will hereafter be dropped.  As shown in \cite{Gibbons:1979xn} the above metric satisfies the vacuum Einstein equations and is self-dual, which in four dimensions ensures it is hyper-K\"{a}hler, if $a$, $b$, and $c$ satisfy the differential equations
\begin{equation}
\frac{da}{dr}=\frac{f}{2bc}\left((b-c)^2-a^2\right),
\end{equation}
along with the two equations obtained by cyclically permuting $a$, $b$, and $c$.  The function $f$ defines the radial coordinate and hence can be freely chosen.  Imposing $a=b=c$ and choosing $f=-1$, the constraint equation reads $\frac{da}{dr}=\frac12$, whose solution is $ds^2=dr^2+\frac{r^2}{4}({\sigma_1}^2+{\sigma_2}^2+{\sigma_3}^2)$, which is the flat metric on $\mathbb{R}^4$.  Choosing the condition $a=b\ne c$ with $f=-\frac a r$, the resulting constraint equations read $\frac{da}{dr}=\frac{1}{2r}(2a-c)$ and $\frac{dc}{dr}=\frac{c^2}{2ar}$.  The explicit solution to these equations is $a=r\sqrt{1+2N/r}$ and $c=\frac{2N}{\sqrt{1+2N/r}}$, where $N$ is a constant of integration.  The resulting metric is
\begin{equation}
ds^2=\left(1+\frac{2N}{r}\right)\big(dr^2+r^2({\sigma_1}^2+{\sigma_2}^2)\big)+4N^2\left(1+\frac{2N}{r}\right)^{-1}{\sigma_3}^2, \label{eq:TaubNUT}
\end{equation}
which is the Euclidean Taub-NUT metric with NUT charge $N$.  The same metric results if any two of the metric functions are chosen to be equal, along with a suitable rotation of the axes.  Imposing the condition that none of the functions are equal yields the Atiyah-Hitchin metric.  

\begin{figure}[htp]
  \begin{center}
\includegraphics[scale=0.569]{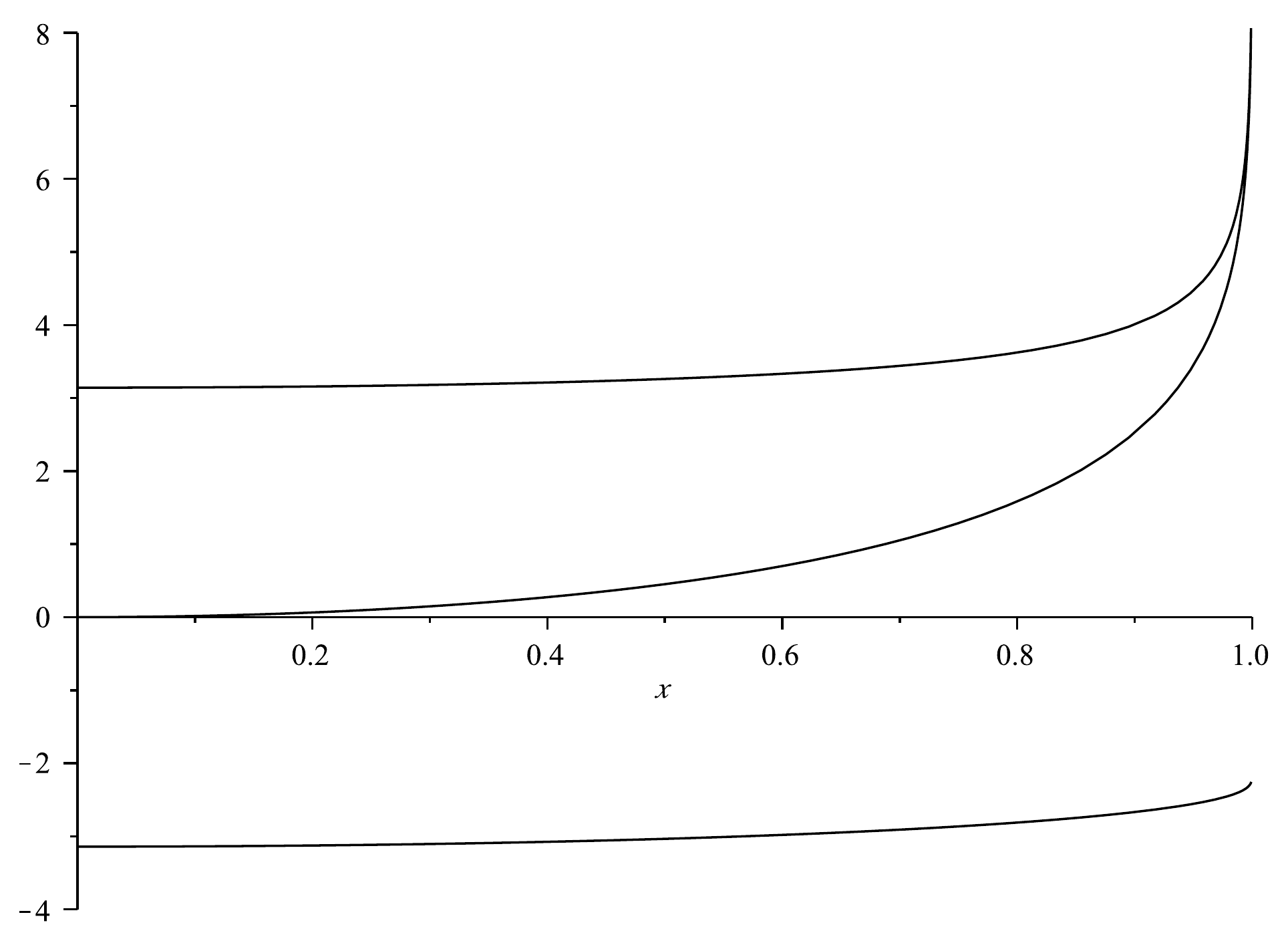}
\put(-120,103){$a(x)$}
\put(-120,148){$b(x)$}
\put(-120,33){$c(x)$}
\caption[Atiyah-Hitchin Metric Functions]{Behaviour of the Atiyah-Hitchin metric functions $a,b,c$ in terms of $x\equiv\sin\left(\frac{\gamma}{2}\right)$.}
\label{fig:abc}
\end{center}
\end{figure}

In \cite{Atiyah:1985dv}, Atiyah and Hitchin used the radial coordinate $\eta$, defined by taking $f=abc$, to obtain the original form of their solution.  If the choice $f=-b/r$ is made instead, the solution simplifies considerably.  I set
\begin{equation}
r=2nK\left(\sin\left(\frac{\gamma}{2}\right)\right), \label{eq:r}
\end{equation}
where $K(k)$, and $E(k)$ encountered shortly, are the complete elliptic integrals of the first and second kind respectively:
\begin{equation}
K(k)=\int_0^{\pi/2}{\frac{dy}{\sqrt{1-k^2\sin^2y}}},\quad\quad E(k)=\int_0^{\pi/2}{\sqrt{1-k^2\sin^2y}dy}.
\end{equation}
As $\gamma$ takes value in the range $[0,\pi]$, $r$ takes on values in the range $[n\pi,\infty)$.  If the following functions are defined
\begin{align}
w_1=bc, \quad\quad\quad\quad\>\> w_2=ca, \quad\quad\quad\quad\>\> w_3=ab, \label{eq:w} \\
\Upsilon\equiv\frac{dr}{d\gamma}=\frac{2nE\left(\sin(\frac{\gamma}{2})\right)}{\sin(\gamma)}-\frac{nK\left(\sin(\frac{\gamma}{2})\right)\cos(\frac{\gamma}{2})}{\sin(\frac{\gamma}{2})},
\end{align}
then the solutions for $w_1$, $w_2$ and $w_3$ are given by
\begin{align}
&w_1=-r\Upsilon\sin(\gamma)-r^2\cos^2\left(\frac{\gamma}{2}\right), \nonumber\\
&w_2=-r\Upsilon\sin(\gamma), \label{eq:wsols}\\
&w_3=-r\Upsilon\sin(\gamma)+r^2\sin^2\left(\frac{\gamma}{2}\right). \nonumber 
\end{align}
The metric functions $a$, $b$, and $c$, obtained by solving (\ref{eq:w}) and (\ref{eq:wsols}), are shown in figure \ref{fig:abc} and now take the explicit form
\begin{align}
&a=\sqrt{\frac{r\Upsilon\sin(\gamma)\left(r\sin^2(\frac{\gamma}{2})-\Upsilon\sin(\gamma)\right)}{r\cos^2(\frac{\gamma}{2})+\Upsilon\sin(\gamma)}}, \\
&b=\sqrt{\frac{\left(r\cos^2(\frac{\gamma}{2})+\Upsilon\sin(\gamma)\right)r\left(r\sin^2(\frac{\gamma}{2})-\Upsilon\sin(\gamma)\right)}{\Upsilon\sin(\gamma)}}, \\
&c=-\sqrt{\frac{r\Upsilon\sin(\gamma)\left(r\cos^2(\frac{\gamma}{2})+\Upsilon\sin(\gamma)\right)}{r\sin^2(\frac{\gamma}{2})-\Upsilon\sin(\gamma)}}.
\end{align}

The Taylor expansions $K(k)\approx \frac{\pi}{2}(1+\frac{1}{4}k^2)$ and $E(k)\approx \frac{\pi}{2}(1-\frac{1}{4}k^2)$ valid near $k=0$ can be used to obtain approximate forms for the metric functions near $r=n\pi$:
\begin{align}
a={}&2(r-n\pi)\left(1-\frac{1}{4n\pi}(r-n\pi)+{\cal O}\big((r-n\pi)^2\big)\right), \nonumber\\
b={}&n\pi\left(1+\frac{1}{2n\pi}(r-n\pi)+{\cal O}\big((r-n\pi)^2\big)\right), \label{eq:approx}\\
c={}&-n\pi\left(1-\frac{1}{2n\pi}(r-n\pi)+{\cal O}\big((r-n\pi)^2\big)\right).\nonumber
\end{align}
Note that $a\rightarrow 0$ as $r\rightarrow n\pi$.  This can be better understood if a rotation of the axes is performed such that
\begin{equation}
\sigma_1=d\tilde\psi +\cos\tilde\theta d\tilde\phi, \qquad \sigma_2=-\sin\tilde\psi d\tilde\theta +\cos\tilde\psi\sin\tilde\theta d\tilde\phi, \qquad \sigma_3=\cos\tilde\psi d\tilde\theta +\sin\tilde\psi\sin\tilde\theta d\tilde\phi. \label{eq:rotate}
\end{equation}
Using these axes along with the radial coordinate $\tilde r=r-n\pi$, the metric takes the form
\begin{equation}
ds^2\approx d\tilde r^2+4\tilde r^2(d\tilde\psi +\cos\tilde\theta d\tilde\phi)^2+(n\pi)^2(d\tilde\theta^2+\sin^2\tilde\theta d\tilde\phi^2).
\end{equation}
The last term is the metric on a two-sphere of radius $n\pi$, while the first two terms give the metric of a flat plane provided $\tilde\psi\in[0,\pi]$.  In terms of the original Euler angles, this corresponds to the identification of points by the discrete symmetry\cite{Gibbons:1986df}:
\begin{equation}
I: \quad\quad \theta\sim\pi-\theta,\quad\quad \phi\sim\pi+\phi, \quad\quad \psi\sim-\psi, \label{eq:antipodal}
\end{equation}
which is actually an isometry.  Such a symmetry identifies antipodal points on the two-sphere parametrised by $(\theta,\phi)$ and is responsible for restricting $\psi\in[0,2\pi]$.  With this identification, the three dimensional SO(3) orbits smoothly collapse to a two-sphere at $r=n\pi$ and so the centre of the Atiyah-Hitchin metric is a bolt.  In the BPS monopole picture, the bolt corresponds to when the monopoles coincide\cite{Gibbons:1986df}.

The opposite limit, corresponding to large monopole separation, is that of $r\rightarrow\infty$.  In this limit the metric functions $a$, $b$, and $c$ become
\begin{align}
a={}& r\sqrt{1-2n/r}+{\cal O}(e^{-r/n}), \nonumber\\
b={}& r\sqrt{1-2n/r}+{\cal O}(e^{-r/n}), \label{eq:abclarge}\\
c={}& -\frac{2n}{\sqrt{1-2n/r}}+{\cal O}(e^{-r/n}), \nonumber
\end{align}
which yields the same metric as (\ref{eq:TaubNUT}) with the substitution $N\rightarrow-n$: asymptotically the Atiyah-Hitchin metric describes a Taub-NUT space with negative NUT charge.  This is not surprising since at asymptotic infinity $a=b$ and it was previously shown that if any two of the metric functions are equal, the constraint equations determine the resulting metric to be Taub-NUT.

The last items necessary to document are the positive orientation and the volume element of the Atiyah-Hitchin metric.   The orientation of the base space must be such that the hyper-K\"ahler 2-forms are anti-self-dual.  Such a positive orientation for Taub-NUT is given by $Ndr\wedge\sigma_1\wedge\sigma_2\wedge\sigma_3$ where $N$ is the NUT charge\cite{Gauntlett:2002nw}. Since the Atiyah-Hitchin metric is asymptotically Taub-NUT with negative NUT charge, the following vierbein is established
\begin{equation}
{\bf e}^1=-f{\bf d}r,\quad\quad {\bf e}^2=a\sigma_1, \quad\quad {\bf e}^3=-c\sigma_3,\quad\quad
{\bf e}^4=b\sigma_2,
\end{equation}
where the minus signs are introduced because $f<0$ and $c<0$.  Finally, the volume 4-form is $\sigma_{\cal B}={\bf e}^1\wedge {\bf e}^2\wedge {\bf e}^3\wedge {\bf e}^4=fabc\sin\theta {\bf d}r\wedge {\bf d}\theta\wedge {\bf d}\phi\wedge {\bf d}\psi.$

\section{Solution with an Atiyah-Hitchin Base} 
\label{sec:SUSY}

Not all four dimensional hyper-K\"ahler manifolds possess a triholomorphic Killing vector field, so not all supersymmetric solutions of five dimensional minimal supergravity will be specified simply by four harmonic functions on $\mathbb R^3$.  The Atiyah-Hitchin solution is just such a space, so more work is required to obtain solutions with this base space.  Unlike other hyper-K\"ahler spaces such as ${\mathbb R}^4$, Taub-NUT, and Eguchi-Hanson, the Atiyah-Hitchin metric is rather complicated in its dependence upon the radial coordinate and some of the results presented here will have implicit radial dependence.   The first subsection is devoted to solving the differential equations for $H$ and $\omega$, after which I give their approximations in the limits $r\approx n\pi$ and $r\rightarrow\infty$.  Note that I am explicitly choosing $H=H(r)$ since it satisfies Poisson's equation on the base and is not separable when it is also a function of angle; more solutions with an Atiyah-Hitchin base space exist, corresponding to other choices of $H$, but such solutions are not studied here.

\subsection{Finding the Equations and Their Asymptotes}
\label{subsec:Asymptotes}

First, a solution of equation~(\ref{eq:Gomega}) must be found that also satisfies ${\bf G^+=\alpha d\Omega}$.  This requires a suitable ansatz for both the 1-forms $\omega$ and ${\bf \Omega}$; in this chapter I consider
\begin{eqnarray}
&&\omega=\Psi(r)\sigma_3, \label{eq:omega}\\
&&{\bf \Omega}=h(r)\sigma_3, \label{eq:Omega}
\end{eqnarray}
where $\Psi$ and $h$ are arbitrary functions.  The authors of \cite{Bena:2007ju} used a different ansatz than above to construct supergravity solutions with an Atiyah-Hitchin base.  Although the solution presented herein is distinct, I draw upon some results of \cite{Bena:2007ju} to write the present solution analytically.  Equation~(\ref{eq:Omega}) along with ${\bf G^+=\alpha d\Omega}$ gives
\begin{equation}
{\bf G^+}=\frac32n\chi\left(\frac{h'}{fc}{\bf e}^1\wedge {\bf e}^3+\frac{h}{ab}{\bf e}^2\wedge {\bf e}^4\right),
\end{equation}
where $\chi=2\alpha/3n$ is a dimensionless constant and a prime denotes differentiation with respect to $r$.  The self-duality of ${\bf G^+}$ requires $\frac{-h'}{fc}=\frac{h}{ab}$, hence $h$ is given by
\begin{equation}
h=\exp\left(-\int{\frac{fcdr}{ab}}\right),   \label{eq:h1}
\end{equation}
which is neither very illuminating nor convenient to work with.  However in \cite{Bena:2007ju}, a judicious choice of radial coordinate was used to find an analytic form for $h$; recasting back in terms of $r$ the result is
\begin{equation}
h=\frac{r^2\sqrt{\sin(\gamma/2)}}{ab},   \label{eq:h2}
\end{equation}
where I have imposed $h(r\rightarrow\infty)=1$ since any constant factors can be absorbed into $\chi_0$.  ${\bf G^+}$ can now be written explicitly as
\begin{equation}
{\bf G^+}=\frac32n\chi\left(\frac{h}{ab}\right)(-{\bf e}^1\wedge {\bf e}^3+{\bf e}^2\wedge {\bf e}^4).\label{eq:Gplus}
\end{equation}

With this solution for ${\bf G^+}$ equation~(\ref{eq:Poisson}) can be used to find the differential equation for $H$, which yields
\begin{equation}
\frac{d}{dr}\left(\frac{abc}{f}\frac{dH}{dr}\right)=2n^2{\chi}^2\left(h^2\frac{fc}{ab}\right).
\end{equation}
Equation~(\ref{eq:h1}) implies $h^2\frac{fc}{ab}=-\frac{1}{2}\frac{d}{dr}(h^2)$ so the first integration is trivial.  The second order differential equation thus reduces to a first order equation and the solution is given by
\begin{equation}
H=\delta -n^2\lambda\eta-n^2{\chi}^2\int{\frac{h^2fdr}{abc}},
\label{eq:H}
\end{equation}
where $\lambda$ and $\delta$ are dimensionless constants of integration and $\eta$, which is the original radial coordinate employed by Atiyah and Hitchin, is related to $r$ via
\begin{equation}
\eta\equiv\int{\frac{fdr}{abc}}.
\end{equation}
Following \cite{Bena:2007ju}, one can shown that 
\begin{equation}
\frac{d}{dr}\left(\frac{1}{ab}\right)=\frac{f}{abc}(1-h^2), \label{eq:hsq}
\end{equation}
so the last term in (\ref{eq:H}) can be integrated exactly, giving
\begin{equation}
H=\delta -n^2\mu\eta+\frac{n^2\chi^2}{ab}, \label{eq:HFinal}
\end{equation}
where $\mu\equiv\lambda+\chi^2$.  It is important to remember that $\eta$ cannot be written in an analytic form in terms of $r$, nevertheless it is easy to numerically integrate.

The only thing left to find is $\omega$; using the ansatz (\ref{eq:omega}) yields
\begin{align}
&{\bf d\omega}=\frac{\Psi'}{fc}{\bf e}^1\wedge {\bf e}^3+\frac{\Psi}{ab}{\bf e}^2\wedge {\bf e}^4, \\
\star_4&{\bf d}\omega=\frac{-\Psi'}{fc}{\bf e}^2\wedge {\bf e}^4-\frac{\Psi}{ab}{\bf e}^1\wedge {\bf e}^3.\nonumber
\end{align}
In accord with equation~(\ref{eq:Gomega}) it is seen that
\begin{equation}
{\bf G^+}=\frac{H^{-1}}{2}\left(\frac{-\Psi'}{fc}+\frac{\Psi}{ab}\right)(-{\bf e}^1\wedge {\bf e}^3+{\bf e}^2\wedge {\bf e}^4).
\end{equation}
Equating this with (\ref{eq:Gplus}), after rearranging and multiplying through by $h$, the ordinary differential equation for $\Psi$ is found to be
\begin{equation}
(h\Psi)'=\frac32n\chi H(h^2)'.
\end{equation}
The above can be integrated analytically, which is more apparent by changing coordinates from $r$ to $\eta$ and breaking up $H$ as follows:
\begin{equation}
\Psi=\frac{n\ell}{h}+\frac{3n\chi}{2h}\left[\int{(\delta-n^2\mu\eta)\frac{dh^2}{d\eta}d\eta}-2n^2\chi^2\int{\frac{h^2c^2}{ab}d\eta}\right], \label{eq:Psi1}
\end{equation}
where $\ell$ is a constant of integration.  The first integrand can be integrated by parts to yield:
\begin{align}
\int{\left(\delta-n^2\mu\eta\right)\frac{dh^2}{d\eta}d\eta}={}&\left(\delta-n^2\mu\eta\right)h^2+n^2\mu\int{h^2d\eta}\nonumber\\
={}&\left(\delta-n^2\mu\eta\right)h^2+n^2\mu\left(\eta-\frac{1}{ab}\right),
\end{align}
where in the second step I have used the result (\ref{eq:hsq}).  Again, following \cite{Bena:2007ju} it can be shown that 
\begin{equation}
\frac{h^2c^2}{ab}=-\frac{1}{6}\frac{d}{d\eta}\left(\frac{2c^2}{a^2b^2}-\frac{bc+ac}{a^2b^2}\right),
\end{equation}
so the second integrand of (\ref{eq:Psi1}) can be solved analytically.  The full solution for $\Psi$ then takes the form
\begin{equation}
\Psi=\frac{n\ell}{h}+\frac{3n\chi}{2}h\left(\delta-n^2\mu\eta\right)+\frac{3n^3\chi\mu}{2h}\left(\eta-\frac{1}{ab}\right)+\frac{n^3\chi^3}{2ha^2b^2}\left(2c^2-c(a+b)\right).
\end{equation}

With the solutions for $h$, $H$ and $\Psi$, their leading order behaviour near $r=n\pi$ can be obtained; using the approximations (\ref{eq:approx}) I find

\begin{align}
h\approx{}&\left(\frac{r}{n\pi}-1\right)^{-1/2}+{\cal O}\left(\sqrt{\frac{r}{n\pi}-1}\right), \label{eq:hsmall}\\
\eta\approx{}&\frac{1}{(n\pi)^2}\ln\left(\frac{r}{n\pi}-1\right)+{\cal O}\left(\frac{r}{n\pi}-1\right), \\
H\approx{}&1-\frac{5\chi^2}{4\pi^2}-\frac{\mu}{\pi^2}\ln{\left(\frac{r}{n\pi}-1\right)}+\frac{\chi^2}{2\pi^2}\left(\frac{r}{n\pi}-1\right)^{-1}+{\cal O}\left(\left(\frac{r}{n\pi}-1\right)\right), \\
\frac{\Psi}{n}\approx{}& -\frac{\chi^3}{8\pi^2}\left(\frac{r}{n\pi}-1\right)^{-3/2}+\frac{3\chi}{2}\left(1-\frac{\mu}{2\pi^2}+\frac{37\chi^2}{24\pi^2}-\frac{\mu}{\pi^2}\ln\left(\frac{r}{n\pi}-1\right)\right)\left(\frac{r}{n\pi}-1\right)^{-1/2}\nonumber\\
&+{\cal O}\left(\sqrt{\frac{r}{n\pi}-1}\right),
\end{align}
Similarly using the approximations (\ref{eq:abclarge}), I find the large-$r$ approximations to leading order to be
\begin{align}
h\approx{}&1+\frac{2n}{r}, \label{eq:hlarge}\\
\eta\approx{}&-\frac{1}{2nr}, \\
H\approx{}&1+\frac{n\mu}{2r}+\frac{n^2\chi^2}{r^2}, \\
\frac{\Psi}{n}\approx{}&j+\frac{2n\left(3\chi-2j\right)}{r},
\end{align} 
where $j=\ell+3\chi/2$ and I have fixed $\delta=1$ because it can be absorbed into the definition of $t$ and $r$.  Note that there is no gauge freedom to set $\Psi=0$ asymptotically: the constant term cannot be absorbed into the definition of $t$ by introducing $d\tau=dt+nj\sigma_3$ because such a transformation is not integrable.  This has important consequences for the global structure, which will be expanded on in section \ref{sec:Properties}.

\subsection{Generic Properties of the Solutions}
\label{sec:Properties}

In the last section, I solved the equations of motion to obtain a supersymmetric five dimensional solution with an Atiyah-Hitchin base space.  Although the solution takes a surprisingly simple form, a fully analytic solution is intractable because $\eta$ is defined as an integral of elliptic functions.  Fortunately this is unimportant since a large amount of information can be derived from the form of the solution with only qualitative knowledge about the behaviour of the metric functions.  This section is devoted to such an approach.

The full five dimensional metric takes the explicit form
\begin{equation}
ds^2=-\frac{1}{H^2}\left(dt+\Psi\sigma_3\right)^2+H\left(f^2dr^2+a^2{\sigma_1}^2+b^2{\sigma_2}^2+c^2{\sigma_3}^2\right),  \label{eq:full}
\end{equation}
however it will also prove convenient to work with the lapse-shift form, given by:
\begin{equation}
ds^2=-{\cal N}^2dt^2+{\cal G}\left(\sigma_3-\frac{\Psi}{H^2{\cal G}}dt\right)^2+H\big(f^2dr^2+a^2{\sigma_1}^2+b^2{\sigma_2}^2\big), \label{eq:lapse}
\end{equation}
with the lapse function ${\cal N}$ and the function $\cal G$ given by
\begin{align}
{\cal N}^2={}&\frac{1}{H^2}\left(1+\frac{\Psi^2}{H^2{\cal G}}\right)=\frac{c^2}{H{\cal G}}, \label{eq:Lapse} \\
{\cal G}={}&Hc^2-\frac{\Psi^2}{H^2}.
\end{align}

\subsubsection{Regions with Closed Time-like Curves}

From the metric form of (\ref{eq:lapse}) it is apparent that something special happens when ${\cal G}$ becomes negative since $g_{\psi\psi}$ also turns negative.  Although $\partial_{\psi}$ is not a Killing vector of the full metric, by examining a congruence of null geodesics I show that the region where ${\cal G}<0$ is one in which closed time-like curves (CTCs) are indeed present.   Consider a tangent vector to null geodesics
\begin{equation}
k=\dot{t}\partial_{t}+\dot{r}\partial_{r}+\dot{\theta}\partial_{\theta}+\dot{\phi}\partial_{\phi}+\dot{\psi}\partial_{\psi},  \label{eq:kGeneral}
\end{equation}
where a dot refers to differentiation with respect to some affine parameter, $\lambda$.  Along with the time-like Killing vector $V^\alpha$, there are three rotational Killing vectors from the SO(3) invariance of the base space:
\begin{align}
\xi_{\phi}={}&\partial_{\phi},\\
\xi_{1}={}&\sin\phi\partial_{\theta}+\cot\theta\cos\phi\partial_{\phi}-\frac{\cos\phi}{\sin\theta}\partial_{\psi}\nonumber,\\
\xi_{2}={}&\cos\phi\partial_{\theta}-\cot\theta\sin\phi\partial_{\phi}+\frac{\sin\phi}{\sin\theta}\partial_{\psi}\nonumber.
\end{align}
The corresponding conserved energy and angular momenta along the null geodesics with tangent vector $k$ are
\begin{eqnarray}
-E=V\cdot k,\>\>\>\>\>\>\>\>\>\>\>\>\>
L_{\phi}=\xi_{\phi}\cdot k,\>\>\>\>\>\>\>\>\>\>\>\>\>
L_1=\xi_{1}\cdot k,\>\>\>\>\>\>\>\>\>\>\>\>\>
L_2=\xi_{2}\cdot k. \label{eq:constraints}
\end{eqnarray}
One can consistently impose $\dot{\theta}=\dot{\phi}=0$, which is obtained by setting all three angular momenta proportional to one another: $L_1=-\tan\theta\cos\phi L_\phi$ and $L_2=\tan\theta\sin\phi L_\phi$.  Focus can thus be safely restricted to the effective metric on the $(t,r,\psi)$ hypersurface
\begin{equation}
d\tilde{s}^2=-\frac{1}{H^2}dt^2-2\frac{\Psi}{H^2}dtd\psi+{\cal G}d\psi^2+Hf^2dr^2. \label{eq:effective}
\end{equation}
In this effective metric, $\partial_{\psi}$ is a Killing vector and its conserved quantity corresponds to the remaining conserved angular momentum in the full metric.  $\partial_{\psi}$ is time-like when ${\cal G}<0$ and it is for this reason that the surface defined by ${\cal G}=0$ is a boundary beyond which CTCs are present.  This boundary, denoted the \emph{velocity of light surface} in the literature\cite{Cvetic:2005zi}, will hereafter be labeled $r_{ctc}$.  

To analyse the properties of the velocity of light surface, I consider locally non-rotating null geodesics by setting the angular momentum to zero.  Using the remaining constraints of (\ref{eq:constraints}), as well as the property that $k$ is null, I find the tangent vector to the family of null geodesics to be given by
\begin{equation}
{k_{\pm}}=E_{\pm}\left(\frac{H{\cal G}}{c^2}\partial_{t}\pm \frac{\sqrt{{\cal G}}}{fc}\partial_{r}+\frac{\Psi}{Hc^2}\partial_{\psi}\right)\label{eq:k}
\end{equation}
where $+$ and $-$ represent outgoing and ingoing geodesics respectively.
At the velocity of light surface, $\frac{dr}{d\lambda}=\pm \frac{\sqrt{{\cal G}}}{fc}E_{\pm} \rightarrow 0$, meaning the null congruence cannot cross this surface.  The expansion scalar, $\Theta={k^{\alpha}}_{;\alpha}$, of the congruence is given by
\begin{equation}
\Theta_{\pm}=\pm\left(\frac{{\cal G}'}{cf\sqrt{{\cal G}}}\right)E_{\pm} \label{eq:expansion}
\end{equation}
where a prime denotes differentiation with respect to $r$.  In a neighbourhood of $r_{ctc}$, ${\cal G}'>0$, $f$ and $c$ are all well behaved and hence the expansion scalar diverges: this indicates that the ingoing (outgoing) congruence is converging to (diverging from) a caustic at $r_{ctc}$.  

\begin{figure}
     \centering
     \subfigure[$\mu=3$, $\chi=0.25$, $j=0$]{\includegraphics[width=3.1 in]{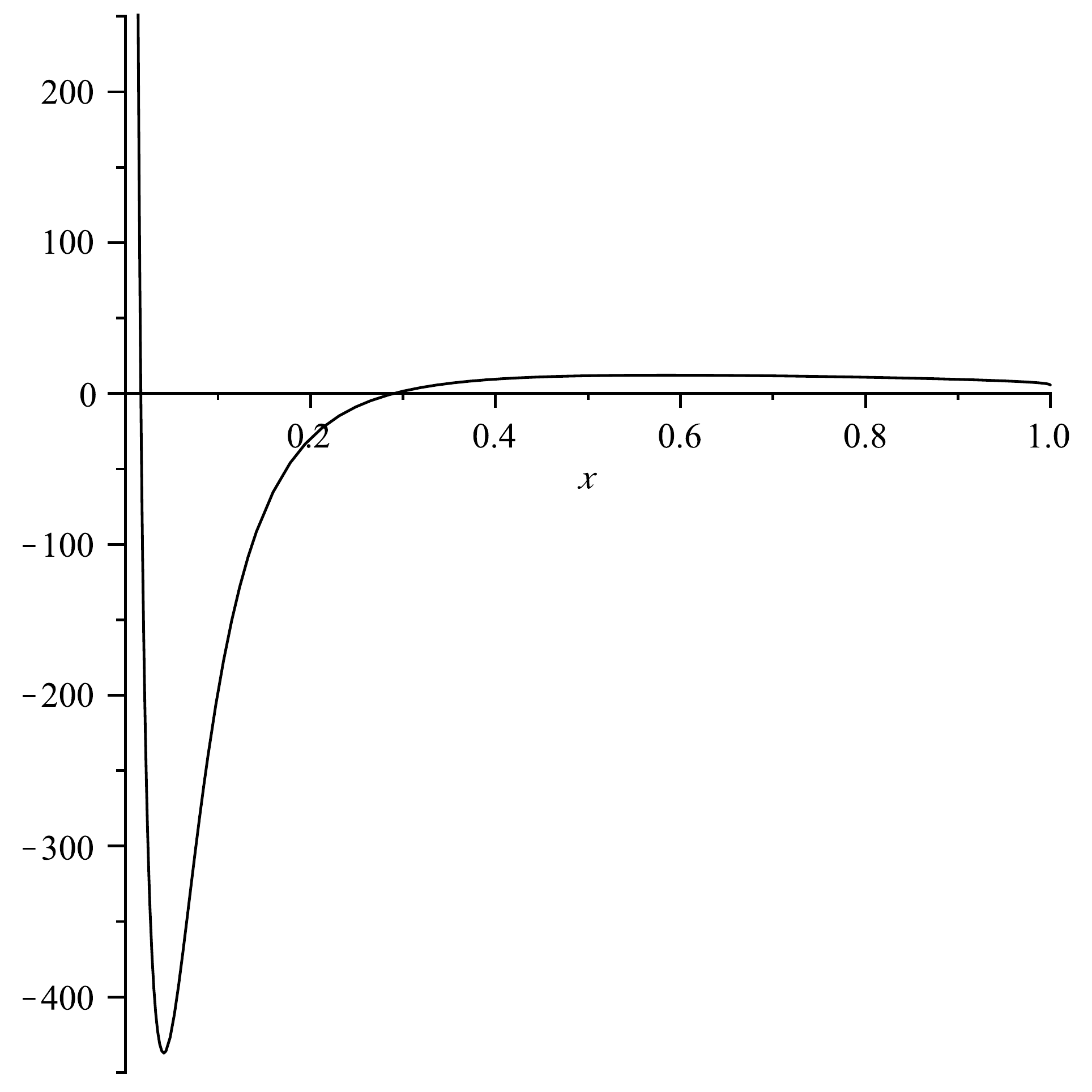}}
     \,
     \subfigure[$\mu=-3$, $\chi=0$, $j=0$]{\includegraphics[width=3.1 in]{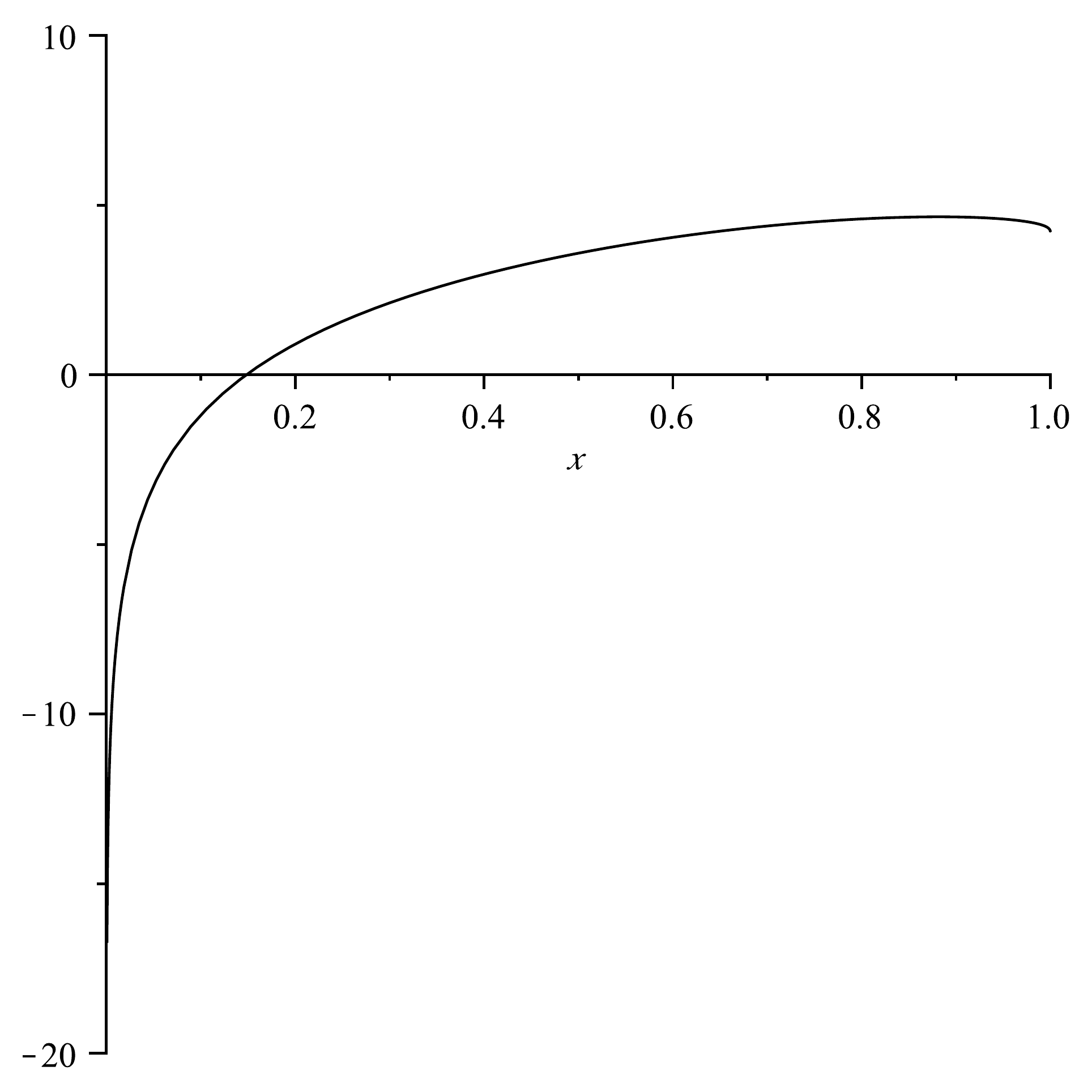}}
     \caption[Development of Closed Time-like Curves.]{Representative plots of ${\cal G}$ for the specified choices of parameters in terms of $x=\sin(\gamma/2)$: $x=0$ corresponds to $r=n\pi$ and $x=1$ corresponds to $r=\infty$.  Both plots demonstrate the existence of at least one radius, $r_{ctc}$, at which ${\cal G}$ changes sign, although there may exist multiple such surfaces. $\cal G$ diverges to positive infinity as $x\rightarrow0$ if $\chi\ne0$ or if $\chi=0,~\mu>0$, while it diverges to negative infinity if $\chi=0,~\mu<0$.  In the former case, the centre of the space-time is free of closed time-like curves while they are present at the centre in the latter case.}
     \label{fig:G}
\end{figure}

This velocity of light surface is directly analogous to what appears in G\"{o}del space-times \cite{Boyda:2002ba,Godel:1949ga,Harmark:2003ud} and solutions describing black holes embedded in G\"{o}del space-times \cite{Behrndt:2004pn,Behrndt:2005he,Gimon:2003ms,Kerner:2007jk}, hereafter called BH-G\"{o}del solutions.  There are, however, a few very important distinctions that should be pointed out.  The first is that the G\"{o}del solutions are homogeneous, meaning that the existence of CTCs outside of the velocity of light surface implies there exist CTCs through every point in the space-time.  The present solution, on the other hand, is not homogeneous because the Atiyah-Hitchin bolt imposes the notion of a centre.  This is similar to the broken homogeneity of the BH-G\"{o}del solutions in which the black hole defines the centre.  The second distinction is that the present solution describes an inverted G\"{o}del-like solution in the sense that the space-time constructed here contains no CTCs for $r>r_{ctc}$, as can clearly be seen from the plot of $\cal G$ in figure~\ref{fig:G}.  The G\"{o}del solutions are (naively) CTC-free for $r<r_{ctc}$ while the BH-G\"{o}del solutions are CTC-free for $r_H<r<r_{ctc}$ where $r_H$ denotes the horizon of the black hole.  

A null ray emanating from the origin in a G\"{o}del space-time (or the horizon in a BH-G\"{o}del space-time) travels out to the velocity of light surface where it forms a caustic and then returns to the origin (or horizon) in finite affine parameter \cite{Boyda:2002ba,Gimon:2003ms,Godel:1949ga}.  In the present solution the null ray is emitted from infinity, travels inward to the velocity of light surface where it forms a caustic and then returns to infinity.  This process is also done in finite affine parameter as can be seen by integrating $\frac{dr}{d\lambda}= \pm E_{\pm}\frac{\sqrt{{\cal G}}}{fc}$ in the vicinity of $r_{ctc}$.  It is sufficient to show that the null ray can travel from some finite $r_1>r_{ctc}$ to $r_2=r_{ctc}$ in finite affine parameter.  To this end, I Taylor expand $f$, $c$ and ${\cal G}$ around $r=r_{ctc}$:
\begin{align}
f\approx{}& f_c+f'_c(r-r_{ctc})+..., \\
c\approx{}& c_c+c'_c(r-r_{ctc})+..., \\
{\cal G}\approx{}&{\cal G}'_c(r-r_{ctc})+...,
\end{align}
and integrate $\Delta\lambda=\int_{r_1}^{r_{ctc}}\frac{-fc}{\sqrt{{\cal G}}}dr$
to find
\begin{equation}
\Delta\lambda=\frac{2f_cc_c}{\sqrt{{\cal G}'_c}}\sqrt{r_1-r_{ctc}}+...,
\end{equation}
which is clearly finite since ${\cal G}'_c>0$ by assumption.  

The Atiyah-Hitchin space-time constructed here describes a type of G\"{o}del-like solution in which the region absent of CTCs includes spatial infinity.  The velocity of light surface sections off the region of space-time containing CTCs but since it is a time-like hyper-surface, null rays and time-like observers can still penetrate into the region with CTCs.  The global structure of such space-times was considered in some detail in \cite{Cvetic:2005zi}.  In particular, it was noted that the space-time comes to an end wherever the lapse function (\ref{eq:Lapse}) vanishes; such a place defines a Killing horizon referred to as a \emph{pseudo-horizon}.  Rewriting the lapse function as
\begin{equation}
{\cal N}^2=\frac{c^2}{H\left(Hc^2-\frac{\Psi^2}{H^2}\right)}
\end{equation}
it is easy to see that a pseudo-horizon only forms when $H\rightarrow\pm\infty$ or $H\rightarrow 0$.  I have already shown that near $r=n\pi$, $H\rightarrow\infty$ for non-zero values of $\chi$ and $\mu$, so $r=n\pi$ is just such a pseudo-horizon.  In general, if the surface gravity of the pseudo-horizon is non-zero, it is necessary to identify the $t$-coordinate with an appropriate real period.  The generator of the pseudo-horizon is
\begin{equation}
\xi=\partial_t+\Omega\partial_\psi \qquad \textrm{where} \qquad \Omega\equiv\left.\frac{\Psi}{H^2G}\right|_{r=n\pi}
\end{equation}
and the surface gravity is given by
\begin{equation}
\kappa^2=\left.-\frac{1}{2}\nabla_\mu\xi_\nu\nabla^\mu\xi^\nu\right|_{r=n\pi}.
\end{equation}
Using the form of the metric (\ref{eq:smallAHmetric}) near $r=n\pi$, it is easy to see that $\Omega=0$ and I find the surface gravity of the pseudo-horizon to vanish.  The same structure has been observed in the case of the limiting BMPV solution and the corresponding object was referred to as a \emph{repulson}\cite{Gibbons:1999uv}; an examination of geodesics showed that they could not penetrate the horizon.  The solution constructed here is just such a repulson object and I will show shortly that it is also a conical singularity.

\subsubsection{Parameter Restrictions and Asymptotic Charges}

  Using the large-$r$ approximations for the various metric functions, the full metric as $r\rightarrow\infty$ takes the explicit form
\begin{align}
ds^2\approx{}&-\left(1-\frac{n\mu}{r}\right)dt^2+\left(1+\frac{n(\mu-4)}{2r}\right)(dr^2+r^2d{\Omega_2}^2)\label{eq:metricasymptote}\\
&+n^2\left(4-j^2+\frac{8n+5n\mu j^2-12n\chi j}{r}\right){\sigma_3}^2-2n\left(j+\frac{6n\chi-2nj-n\mu j}{r}\right)dt\sigma_3, \nonumber
\end{align} 
It was noted at the end of section \ref{subsec:Asymptotes} that $j$ cannot be absorbed into the definition of $t$, so if $j\ne 0$ then the space-time is rotating at infinity.  Furthermore, recall that a velocity of light surface appears whenever ${\cal G}=0$; imposing the condition that the space-time is CTC-free asymptotically implies:
\begin{equation}
-2\le j\le 2,
\end{equation}
where equality above leads to another velocity of light surface forming at spatial infinity.  From this inequality $\chi$ is interpreted as being a parameter contributing to a twisting of the space-time; the sign of $\chi$ dictates the direction of the extra twisting, whereas $H$ is independent of the sign of $\chi$.  Furthermore, the Riemann tensor vanishes as $r\rightarrow\infty$ and the solution is asymptotically described by a twisted U(1) fibration over $M_4/{\mathbb Z}_k$, \emph{i.e.} four dimensional Minkowski space-time with antipodal points identified on the two-sphere. 

To calculate the energy and angular momenta, I foliate the space-time by a family of space-like hypersurfaces, $\Sigma_t$, perpendicular to the orbits of $V$.  In this way, it is observers who are locally non-rotating at infinity -- so-called ``zero angular momentum" observers -- who measure these quantities.  The time-like normal to these hypersurfaces is $n^\alpha=HV^\alpha$ and the radial normal to the boundary, $S_t\equiv\partial\Sigma_t$, of these hypersurfaces is $r_\alpha dx^\alpha=\sqrt{H}fdr$.  Asymptotically $\xi_{\psi}^\alpha\equiv\partial_\psi$ is a Killing symmetry so the energy, $E$, and the angular momenta $J_\phi,~J_\psi$ are given by the Komar formulae:
\begin{align}
&E=-\frac{1}{8\pi G_5}\lim_{r\rightarrow\infty}\left[\oint_{S_t}{-2V^{\beta;\alpha}n_{[\alpha}r_{\beta]}\sqrt{\sigma}d\theta d\phi d\psi}\right],\\
&J_{\phi/\psi}=\frac{1}{16\pi G_5}\lim_{r\rightarrow\infty}\left[\oint_{S_t}{-2\xi_{\phi/\psi}^{\beta;\alpha}n_{[\alpha}r_{\beta]}\sqrt{\sigma}d\theta d\phi d\psi}\right],
\end{align}
where $\sqrt{\sigma}=-H^{3/2}abc\sin\theta$ is the determinant of the metric induced on $S_t$.   Similarly, the Bianchi identity ${\bf dF}=0$ and the equation of motion (\ref{eq:FEOM}) imply the following electric and magnetic charges
\begin{align}
Q_E={}&\frac{1}{16\pi G_5}\lim_{r\rightarrow\infty}\left[\oint_{S_t}{\star{\bf F}-\frac{2}{\sqrt{3}}{\bf F}\wedge{\bf A}}\right],\\
Q_M={}&\frac{1}{4\pi}\lim_{r\rightarrow\infty}\left[\oint_{S^2}{\bf F}\right].
\end{align}
One might expect the magnetic charge to vanish since $(\theta,\phi,\psi)$ parametrise a squashed $S^3$ and there does not seem to be a natural choice for the $S^2$, over which to integrate $\bf F$.  Recall, however, that the centre of the Atiyah-Hitchin metric is a bolt, at which the SO(3) orbits collapse to a non-minimal two-sphere.  Therefore, as long as the $S^2$ is chosen to encompass the bolt, the definition for the magnetic charge makes sense.  This is accomplished by taking the two-sphere parametrised by $(\theta,\phi)$ in the limit $r\rightarrow\infty$.  
The ADM energy, angular momenta, electric charge, and magnetic charge are then found to be
\begin{align}
&E=\frac{\pi n^2\mu}{G_5}, \label{eq:M}\\
&J_{\psi}= \frac{\pi n^3}{2G_5}\left(6\chi-2j-j\mu\right),\label{eq:J}\\
&J_{\phi}= 0,\\
&Q_E=\frac{\sqrt{3}}{8G_5}\pi n^2\big(\chi(\chi- j)-\mu\big),\\
&Q_M=\frac{\sqrt{3}}{4}n\big(\chi-j\big).
\end{align}
The total energy can be positive, negative or zero depending on the value of $\mu$.  Recall that $\mu=\lambda+\chi^2$ so $\lambda$, which loosely speaking plays the role of the mass of some gravitating object, can contribute positively or negatively to the total energy while $\chi$, which is related to the gauge field strength, always contributes positively to the total energy as it should.  If $j=0$, the space-time is not rotating asymptotically and it can be seen from (\ref{eq:J}) that all the angular momentum is due to the twisting supplied by $\chi$.  The fact that $J_{\phi}=0$ is related to the fact that locally non-rotating null geodesics have $\dot{\phi}=0$, \emph{i.e.} there is no frame dragging around the $\phi$-axis.

\subsubsection{Naked Singularities}

The small-$r$ approximations along with the rotated axes of (\ref{eq:rotate}) can be used to write down the asymptotic form of the metric near $r=n\pi$.  Assuming $\chi\ne0$ and defining a new radial coordinate via
\begin{equation}
\frac{r}{n\pi}-1=\frac{\tilde r^2}{2n^2\chi^2},
\end{equation}
the metric to lowest order then takes the form
\begin{align}
ds^2\approx{}& -\frac{\pi^4\tilde r^4}{n^4\chi^8}\left(dt-\frac{n^4\chi^6}{2\sqrt{2}\pi^2\tilde r^3}\tilde\sigma_2\right)^2+d\tilde r^2+\tilde r^2\tilde\sigma_3^2+\frac{n^4\chi^4}{\tilde r^2}\big(\tilde\sigma_1^2+\tilde\sigma_2^2\big). \label{eq:smallAHmetric}
\end{align}
Recall that the identification by the discrete symmetry (\ref{eq:antipodal}) is to ensure that the centre of the Atiyah-Hitchin metric is smooth.  Under the same considerations, the $(\tilde r,\tilde\psi)$ sector of the above limit is the metric of a flat plane provided $\tilde\psi\in[0,2\pi]$.  However, the family of five dimensional solutions should be viewed as excitations above the trivial solution
\begin{equation}
ds^2=-dt^2+f^2dr^2+a^2\sigma_1^2+b^2\sigma_2^2+c^2\sigma_3^2,  \label{eq:trivialAH}
\end{equation}
which necessarily has $\tilde\psi\in[0,\pi]$.  The pseudo-horizon at $r=n\pi$ is therefore an orbifold fixed-point since the $\tilde\psi$-cycle has a ${\mathbb Z}_2$ identification.  Note that in this limit, the radius of the $S^2$ is diverging.  

When $\chi=0$ there is no convenient way to write the metric in terms of proper distance like above, but upon examining the Ricci and Kretschmann scalars, one can see that the centre is still singular.  For instance, for $\chi\ne0$, as well as $\chi=\mu=0$ with $j\ne0$, the Ricci scalar behaves like ${\cal R}\sim \tilde r^{-2}$, while for $\chi=0$, $\mu\ne0$ it behaves like ${\cal R}\sim \left(\tilde r^2\log^3\tilde r\right)^{-1}$.  The $\tilde\sigma_3$-cycle still pinches off at $r=n\pi$ for $\chi\ne0$ but not in a smooth manner, hence there is a conical singularity at $r=n\pi$ for all choices of parameters, except the very specific choice $\chi=\mu=j=0$, which corresponds to the trivial solution (\ref{eq:trivialAH}). 

The singularity at $r=n\pi$ is not the only singularity potentially present in the space-time: if $H\rightarrow 0$ at some radius $r_s$, the metric turns singular.  Using the proper radial distance in this limit, spatial slices perpendicular to the orbits of $V$ look like
\begin{equation}
d_\Sigma^2\approx d\tilde r^2+A\tilde r^{2/3}d\tilde\Omega_3^2,
\end{equation}
where $A$ is a constant and $d\tilde\Omega_3^2$ is the metric of a squashed 3-sphere.  It is evident that such spatial slices have vanishing volume but do not pinch off smoothly in the $\tilde r\rightarrow 0$ limit.  The metric is thus singular at the point $r_s$ where it forms a pseudo-horizon.  Recall that the solution is valid as long as $H\ge0$ and is globally defined, so it is not surprising that the space-time develops a singularity where $H=0$.  Indeed, the Ricci scalar expanded in inverse powers of $H$ is
\begin{equation}
{\cal R}=\left(\frac{(\Psi^2)'}{2fabc}\right)\frac{1}{H^{4}}-\left(\frac{(H')^2}{2f^2}\right)\frac{1}{H^{3}}+{\cal O}\left(\frac{1}{H^2}\right).
\end{equation}
Near $r=r_s\ne n\pi$, $H\approx H'_s(r-r_s)$ while all the other metric functions are non-zero.  For a generic choice of parameters, $\Psi\ne 0$ and the curvature diverges like $(r-r_s)^{-4}$.  The specific choice of $\chi=j=0$ and $\mu<0$ leads to $\Psi=0$ and $H'\ne0$: in this case the curvature diverges like $(r-r_s)^{-3}$, which is the least singular behaviour possible.  For this choice of parameters there is no rotation, meaning there is no velocity of light surface and hence no CTCs.  

For more generic choices of parameters these singularities occur behind a velocity of light surface.  This can be seen by considering the following three possibilities: 
\begin{itemize}
\item[$(i)$] ${\cal G}\rightarrow-\infty$ because of the $-H^{-2}$ dependence, 
\item[$(ii)$] ${\cal G}<0$ and finite if $\Psi\propto H$,
\item[$(iii)$] ${\cal G}=0$ if $\Psi\propto H^p$ where $p>1$.
\end{itemize}
The latter is impossible to arrange because of the form of the solution for $\Psi$ along with the behaviour of the metric functions $a$, $b$ and $c$.  It is therefore concluded that ${\cal G}<0$ whenever $H=0$, hence the velocity of light surface is necessarily outside of the singular point.  In examining equation~(\ref{eq:HFinal}) it is clear that the only way for $H$ to become negative is if $\mu$ is sufficiently large and negative.  Since $E\propto \mu$, imposing a positivity on the energy removes these particular singularities from the spectrum.

\section{Discussion}
\label{sec:Conclusion}

Using the Atiyah-Hitchin base space, I have constructed a three-parameter family of solutions to five dimensional minimal supergravity, which possess mass, angular momentum, electric charge and magnetic charge.  The Atiyah-Hitchin metric does not admit a triholomorphic Killing vector field and hence there is no Gibbons-Hawking form of the metric to be exploited.  Instead I followed the analysis in \cite{Bena:2007ju} to find a semi-analytic solution; while a fully explicit analytic solution is intractable, it is simple enough to numerically integrate easily. 

By considering the general form of the space-time metric and using arguments based on the structures of the metric functions I have shown that these solutions describe space-times in which conically singular pseudo-horizons are present.  Such pseudo-horizons are either at the bolt of the Atiyah-Hitchin base space or at a radius where $H\rightarrow0$.  The solutions were also shown to typically include a region of CTCs where the effective Killing vector $\partial_{\psi}$ turns time-like.  There is only a small subset of parameter space in which CTCs are not present.

In these solutions I have taken the scalar function $H$ to be a function of radial coordinate only.  This is because such a choice is the easiest case to consider; there surely are further solutions corresponding to different choices of $H$.  These are typically much more difficult to construct because of the Poisson equation that $H$ must satisfy, namely $\Delta H(x^a)=\frac{4}{9}({\bf G^+})^2$.  In general $H$ cannot be made a separable function of $x^a$, hence a completely general solution would be extremely difficult to find.  However $H$ can be made separable asymptotically where the Atiyah-Hitchin metric approaches the Euclidean Taub-NUT metric, so although the full solution might be intractable, some of its properties can nevertheless be discerned.

The solution constructed in this chapter bears some similarity to the G\"{o}del  and BH-G\"{o}del solutions previously constructed in \cite{Behrndt:2004pn,Behrndt:2005he,Boyda:2002ba,Gauntlett:2002nw,Gimon:2003ms,Godel:1949ga,Harmark:2003ud,Kerner:2007jk}.   One key difference is that the G\"{o}del space-time is homogeneous and hence has CTCs through every point, whereas the present solution does not have this property because of the existence of the singularity or the bolt at $r=n\pi$.  One rather striking and interesting feature of the solution constructed herein is that unlike the (BH-)G\"{o}del solutions, the region containing CTCs is contained entirely within the bulk of the space-time and spatial infinity is seemingly free of pathologies.  It is conjectured in \cite{Behrndt:2004pn} that whenever CTCs develop in the bulk, the dual CFT is pathological and not well-defined. The idea is that the CFT metric itself develops CTCs and one would thus not be able to make sense of a quantum field theory on a space-time with CTCs.  This argument, however, considers the near horizon geometry of black holes that develop CTCs outside the horizon and presumes that the near horizon geometry possesses CTCs asymptotically.  It is not surprising that the dual CFT should suffer pathologies in such a scenario.  The space-time constructed in this chapter, however, does not suffer from CTCs at infinity so it would be interesting to see if the same argument holds here. If there is a holographic interpretation of this solution, it would be an interesting counter-example to study if the dual CFT were free of pathologies.


\chapter{Hidden Symmetries: Constructing the General Black String}


I have just demonstrated in the previous chapter how generating supersymmetric solutions to five dimensional minimal supergravity is a well-formulated problem.  However, the construction of general non-supersymmetric solutions is considerably more complicated.  In particular, the construction of a black ring solution that describes thermal excitations over the supersymmetric black ring is a significantly involved problem.  Unlike the case of topologically spherical black holes\cite{Cvetic:1996xz}, one cannot add a sufficient number of charges on a neutral black ring by applying familiar solution generating techniques.  For example,  one cannot add three independent M2-brane charges to an otherwise neutral ring by applying boosts and string dualities\cite{Elvang:2003mj}. Adding the third M2-brane charge typically requires applying a boost along the direction of a Kaluza-Klein monopole in an intermediate step and such a boost is incompatible with the identifications imposed by the Kaluza-Klein monopole fibration. In effect, one ends up generating pathological space-times.

A variant of this problem also arises in the straight string limit when one tries to add three M2-brane charges on top of a black string carrying three M5-brane charges \cite{Compere:2009zh,Elvang:2004xi}. This and related problems have resisted various attempts to construct a variety of non-extremal black string and black ring solutions \cite{Bouchareb:2007ax,Compere:2009zh,Elvang:2003mj, Elvang:2004xi,Gal'tsov:2009da}.  Using a hidden symmetry in five dimensional minimal supergravity, the authors of \cite{Compere:2009zh} shed some light on this problem from a group theoretic perspective but a systematic solution was not offered.  When five dimensional minimal supergravity is dimensionally reduced to three dimensions, the resulting theory is three dimensional Einstein gravity coupled to eight scalar fields.  The eight scalars parametrise the coset $G_2/(SL(2,{\mathbb R})\times SL(2,{\mathbb R}))$, where $G_2$ is the smallest of the exceptional Lie groups.  The generators of $G_2$ can be used to transform the scalars of a known seed solution, which then produces new solutions upon uplifting back to five dimensions.  In this chapter, which is based on the work published in \cite{Compere:2010fm}, I extend the investigations of \cite{Compere:2009zh} and present a solution of the previously mentioned problem.  That is I explicitly construct the most general black string solution of five dimensional ungauged minimal supergravity, which has five independent parameters, namely magnetic one-brane charge,  smeared electric zero-brane charge, boost along the string direction, energy above the BPS bound, and rotation in the space transverse to the string.  

Upon setting the magnetic or electric charge to zero, the general string solution reduces to the spinning non-extremal string solutions of \cite{Compere:2009zh, Tanabe:2008vz}.  The general string admits a four-parameter family of extremal black strings corresponding to maximal rotation in the transverse space, however other known extremal black strings can further be recovered by taking additional limits. In one limit the three-parameter supersymmetric string of \cite{Kim:2010bf} is recovered, while in another the three-parameter non-supersymmetric string of \cite{Kim:2010bf} is recovered. When the tensionless condition is imposed, the general black string describes the infinite radius limit of the most general black ring of minimal supergravity.

On the microscopic side, a statistical mechanical discussion of the entropy for the yet-to-be found black ring that describes thermal excitations over the supersymmetric ring of \cite{Elvang:2004rt} is already proposed in the literature \cite{Larsen:2005qr}.  The validity of the entropy formula in \cite{Larsen:2005qr}, however, depends on the absence of finite ring radius corrections, which is not guaranteed \emph{\`a priori}.  In the straight string limit this subtlety goes away, so it is expected that the entropy formula in \cite{Larsen:2005qr} reproduces the Bekenstein-Hawking entropy in the near extremal limit of the general black string constructed in this chapter. This indeed turns out to be the case once one has made precise the connection between quantities defined near the horizon and at infinity.

An important open issue in the black ring literature has been why various black ring solutions must obey a number of bounds on various parameters in order to be smooth. For example, for the supersymmetric ring of minimal supergravity, the electric charge is bounded from below \cite{Elvang:2004rt}: the lower bound is in terms of the magnetic dipole charge and the radius of the ring. Curiously enough, the bound persists in the infinite radius limit. Although, a complete understanding of these issues is not offered here, I will show that the general black string solution does shed light on some of these puzzles. The extremal tensionless limits of the non-rotating string provides a working phase diagram for a class of extremal black rings. This phase diagram suggests that as the above mentioned bound on the electric charge is violated, a non-supersymmetric branch of extremal black rings emerges; this non-supersymmetric branch has not yet been found. On this branch one can continuously take the electric charge to zero, meaning this non-supersymmetric branch is connected to a magnetic dipole ring; the dipole ring is such that in the infinite radius limit the boost along the string completely breaks supersymmetry. Additionally, the phase diagram suggests the existence of another branch of \emph{supersymmetric} black rings. This branch is also connected to a magnetic dipole ring that \emph{preserves} supersymmetry in the infinite radius limit.  Supersymmetric black rings on this branch are also not known.

\section{Hidden $G_2$ Symmetry in 5D Supergravity}
\label{sec:setup}

In this section I review the hidden $G_2$ symmetry of five dimensional minimal supergravity and outline how it can be used to generate new solutions to the equations of motion.  Since interest is in constructing a black string solution, I first compactify along the string direction to construct the corresponding four dimensional theory.  I then compactify the remaining four dimensional theory along the time-like direction to obtain a three dimensional Euclidean theory whose matter content carries the $G_2$ symmetry.  This method of approaching the problem turns out to be very systematic as there are a number of theorems \cite{Breitenlohner:1987dg} to bank on concerning black holes in the intermediate four dimensional theory, which is a single modulus  ${\cal N}=2,~D=4$ supergravity.  I begin with a brief review of the key features of the $G_2$ group.  

\subsection{The $G_2$ Group}

\begin{figure}[htp]
  \begin{center}
\includegraphics[scale=0.75]{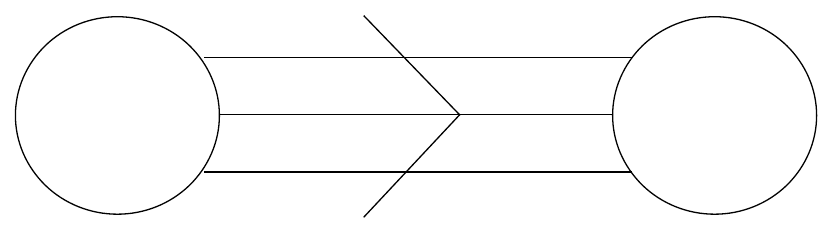}
\caption[$G_2$ Dynkin Diagram]{The Dynkin diagram of $G_2$.}
\label{fig:dynkin}
\end{center}
\end{figure}

$G_2$ is the smallest of the exceptional Lie groups, with rank 2 and dimension 14.  Its Dynkin diagram, shown in figure \ref{fig:dynkin}, consists of two nodes joined by three directed edges, where each node corresponds to a triple of Chevalley generators $\{H_a,E_a,F_a\}$ for $a=1,2$ and the three directed edges indicate three large and three small positive roots.  
The Cartan subalgebra, $\mathfrak h$, of ${\mathfrak g}_{2}$ is spanned by the $H_a$, meaning $[H_a,H_b]=0$ for all $a,b$.  The decomposition ${\mathfrak g}_2={\mathfrak h}\oplus{\mathfrak n}$ for some subalgebra $\mathfrak n$ implies there exists the split real form ${\mathfrak g}_{2(2)}$, which is used throughout the rest of the chapter.  The Cartan matrix, defined via $[H_a,E_b]=C_{ab}E_b$, is given by
\begin{equation}
C_{ab}=
\left[ {\begin{array}{cc}
 \>\>\>2& -3 \\
 -1& \>\>\>2
\end{array} } \right],  \label{eq:CartanMatrix}
\end{equation}
implying that the two positive simple roots, $\vec\alpha_a$, which live in the dual ${\mathfrak h}^{\star}$ of $\mathfrak h$, are associated to the positive generators $E_a$; in this basis, the root vectors are simply the columns of the Cartan matrix.  The negative generators, $F_a$, similarly satisfy $[H_a,F_b]=-C_{ab}F_b$ for the same Cartan matrix (\ref{eq:CartanMatrix}), and have associated with them the negative root $-\vec\alpha_a$.  Furthermore, the positive and negative generators satisfy the commutation relations
\begin{equation}
[E_a,F_b]=\delta_{ab}H_a.  \label{eq:EFHComm}
\end{equation}

The dimension of $G_{2(2)}$ is 14 but the above discussion only covers six of the generators.  The remaining four positive generators, $E_i$, and four negative generators, $F_i$, can be obtained by taking commutators of the $E_a$ and $F_a$:
\begin{align}
E_3={}&[E_1,E_2], \quad\quad\quad E_4=[E_3,E_2], \quad\quad\quad E_5=[E_4,E_2], \quad\quad\quad E_6=[E_1,E_5]  \label{eq:ECommutator}\\
F_3={}&-[F_1,F_2], \quad\quad F_4=-[F_3,F_2], \quad\quad\>\> F_5=-[F_4,F_2], \quad\quad\>\> F_6=-[F_1,F_5],  \nonumber
\end{align}
such that the Serre relations are satisfied: $[E_1,E_j]=0$ and $[E_2,E_k]=0$, where $j=3,4,6$ and $k=5,6$, along with similar relations for $F$.  Thus, the full set of positive generators, $\{E_a,E_i\}$, span a nilpotent subalgebra ${\mathfrak n}_+$ of dimension six.  Furthermore, to each $E_i$ there is a corresponding positive root, $\vec\alpha_i$, which is a linear combination of the simple roots $\vec\alpha_a$.  Similarly, the full set of negative generators, $\{F_a,F_i\}$, span another nilpotent subalgebra ${\mathfrak n}_-$ of dimension six and to each of the $F_i$ there is a corresponding negative root, $-\vec\alpha_i$.  Identifying ${\mathfrak n}={\mathfrak n}_+\oplus {\mathfrak n}_-$, the previous decomposition is seen to be ${\mathfrak g}_{2(2)}={\mathfrak h}\oplus{\mathfrak n}_+\oplus {\mathfrak n}_-$ and all of the generators have been exhausted.

The Chevalley basis, \emph{i.e.} the basis above in which the entries in the Cartan matrix (\ref{eq:CartanMatrix}) are integers, is inconvenient: a more natural basis is provided by the dimensional reduction itself of five dimensional minimal supergravity to three dimensions.  It is given by the following, whose $7\times7$ representation is given explicitly in Appendix \ref{AppendixG2}
\begin{align}
h_1={}&\frac{1}{\sqrt{3}}H_2, \quad\quad\quad\> h_2=H_2+2H_1,\nonumber\\
e_1={}&E_1,\quad\quad\quad\quad\quad e_2=\frac{1}{\sqrt{3}}E_2, \quad\quad\quad\> e_3=\frac{1}{\sqrt{3}}E_3, \nonumber\\
e_4={}&\frac{1}{\sqrt{12}}E_4,\quad\quad\quad e_5=\frac{1}{6}E_5, \quad\quad\quad\quad e_6=\frac{1}{6}E_6, \label{eq:hefDefs}
\end{align}
\begin{align}
f_1={}&F_1,\quad\quad\quad\quad\quad f_2=\frac{1}{\sqrt{3}}F_2, \quad\quad\quad\> f_3=\frac{1}{\sqrt{3}}F_3, \nonumber\\
f_4={}&\frac{1}{\sqrt{12}}F_4,\quad\quad\quad f_5=\frac{1}{6}F_5, \quad\quad\quad\quad f_6=\frac{1}{6}F_6. \nonumber
\end{align}
It can be explicitly verified that this change of basis transforms (\ref{eq:CartanMatrix}) to
\begin{equation}
\tilde C_{ab}=
\left[ {\begin{array}{cc}
 -\sqrt{3}& \frac{2}{\sqrt{3}} \\
 \>\>\>\>\>\>\>1& 0
\end{array} } \right],
\end{equation}
where now $[h_a,e_b]=\tilde C_{ab}e_b$.  The corresponding positive simple roots are then seen to be
\begin{equation}
\vec\alpha_1=\left(-\sqrt{3},1\right), \quad\quad\quad\quad \vec\alpha_2=\left(\frac{2}{\sqrt{3}},0\right). \label{eq:SimpleRoots}
\end{equation}
\begin{figure}[htp]
  \begin{center}
\includegraphics[scale=0.6]{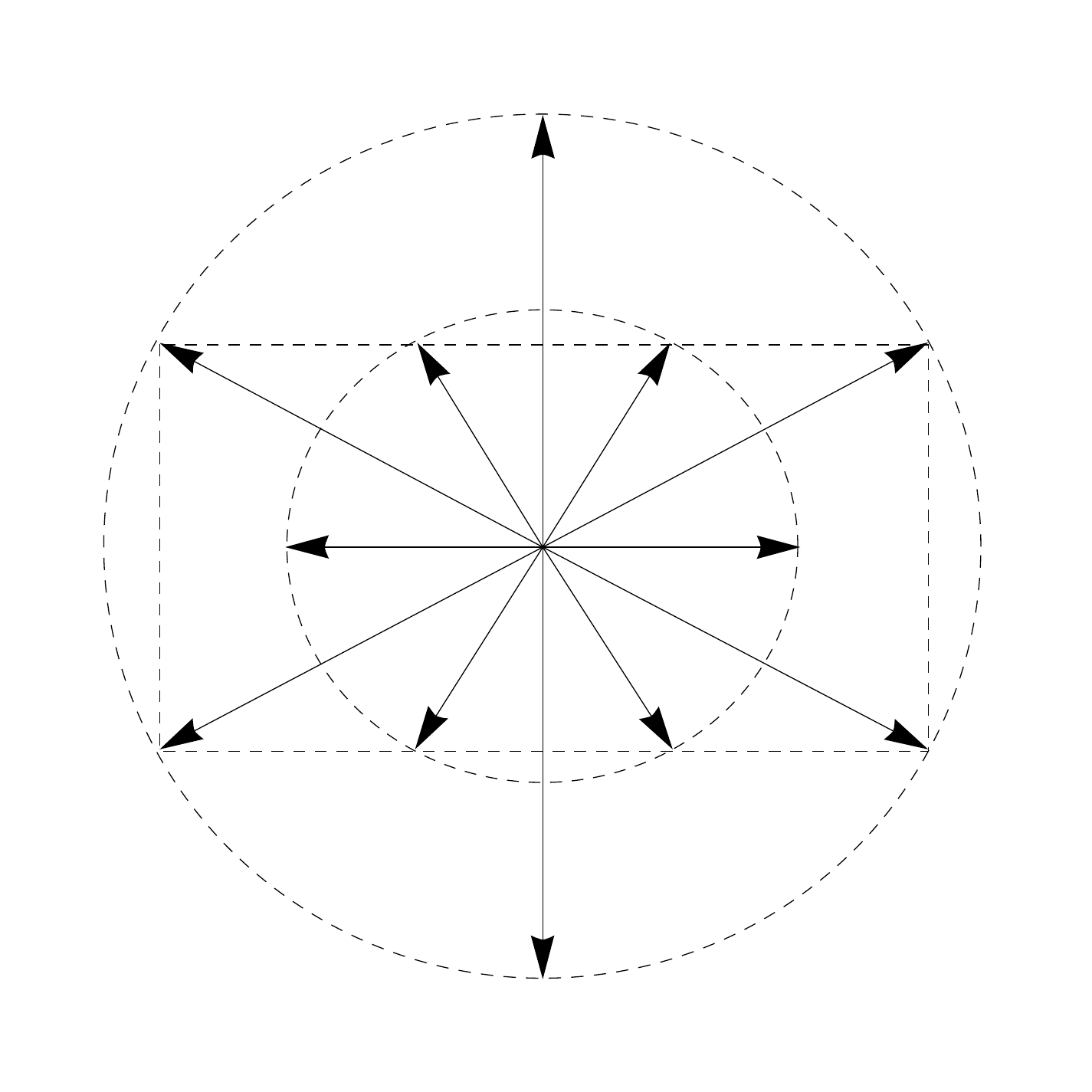}
\put(-225,172){$\vec\alpha_1$}
\put(-62,122){$\vec\alpha_2$}
\put(-163,174){$\vec\alpha_3$}
\put(-100,174){$\vec\alpha_4$}
\put(-36,171){$\vec\alpha_5$}
\put(-130,226){$\vec\alpha_6$}
\caption[$G_2$ Root Structure]{The positive and negative roots of $G_{2(2)}$.}
\label{fig:root}
\end{center}
\end{figure}
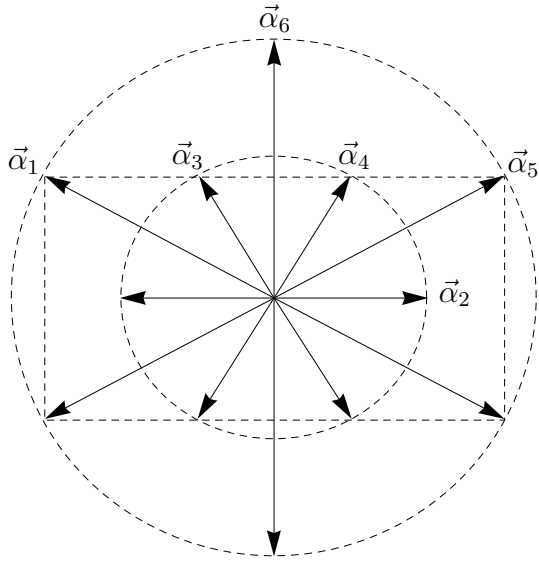

\noindent Using the definitions (\ref{eq:ECommutator}), the remaining positive roots $[h_a,e_i]=\vec\alpha_i e_i$ are obtained by the following linear combinations of $\vec\alpha_1$ and $\vec\alpha_2$:
\begin{equation}
\vec\alpha_3=\vec\alpha_1+\vec\alpha_2, \quad\quad\quad \vec\alpha_4=\vec\alpha_1+2\vec\alpha_2, \quad\quad\quad \vec\alpha_5=\vec\alpha_1+3\vec\alpha_2, \quad\quad\quad \vec\alpha_6=2\vec\alpha_1+3\vec\alpha_2. \label{eq:OtherRoots}
\end{equation}
Similarly, the negative roots corresponding to the set of negative generators, $\{F_a,F_i\}$, are simply $\{-\vec\alpha_a,-\vec\alpha_i\}$.  This root structure is depicted in figure \ref{fig:root} which clearly shows the three large and three small positive roots geometrically.

From the decomposition ${\mathfrak g}_{2(2)}={\mathfrak h}\oplus{\mathfrak n}_+\oplus {\mathfrak n}_-$, and equations (\ref{eq:CartanMatrix}), (\ref{eq:EFHComm}) and (\ref{eq:ECommutator}), clearly there is a symmetry between the positive and negative nilpotent subalgebras.  This can be expressed by an involutive isomorphism, $\tau$, that maps elements of ${\mathfrak n}_+$ to elements of ${\mathfrak n}_-$ and vice versa.  In the case at hand, the $G_{2(2)}$ symmetry comes from the dimensional reduction of five dimensional minimal supergravity to three dimensions and the isomorphism $\tau$ depends on the signature of the reduced dimensions.  If $\epsilon_j$ denotes the signature of the $j$th reduced direction, the appropriate isomorphism is given by\cite{Compere:2009zh}
\begin{align}
\tau(h_1)={}&-h_1, \quad\quad\quad\quad\>\> \tau(h_2)=-h_2,\nonumber\\
\tau(e_1)={}&-\epsilon_1\epsilon_2f_1, \quad\quad\quad \tau(e_2)=-\epsilon_1f_2, \quad\quad\quad \tau(e_3)=-\epsilon_2f_3, \label{eq:tau} \\
\tau(e_4)={}&-\epsilon_1\epsilon_2f_4, \quad\quad\quad \tau(e_5)=-\epsilon_2f_5, \quad\quad\quad \tau(e_6)=-\epsilon_1f_6, \nonumber
\end{align}
with the action of $\tau$ on the generators $f_i$ given by the requirement that $\tau^2$ have eigenvalue +1.  This involutive isomorphism admits the decomposition ${\mathfrak g}_{2(2)}=\tilde{\mathfrak k}\oplus\tilde{\mathfrak p}$ where $\tilde{\mathfrak k}=\left\{x\in{\mathfrak g}_{2(2)}|\tau(x)=x\right\}$ is the Lie algebra of the subgroup $\tilde K$ invariant under the involution, and its orthogonal complement $\tilde{\mathfrak p}=\left\{x\in{\mathfrak g}_{2(2)}|\tau(x)=-x\right\}$ is the Lie algebra of the subgroup $\tilde P$ anti-invariant under the involution.  Clearly $\tilde{\mathfrak k}$ is spanned by the set $\{k_i\}=\{e_i+\tau(e_i)\}$ while $\tilde{\mathfrak p}$ is spanned by the set $\{p_i\}=\{e_i-\tau(e_i)\}$, the latter of which contains the Cartan subalgebra $\mathfrak h$.  Since the dimensional reductions are performed such that $\epsilon_1=+1$, $\epsilon_2=-1$, explicitly here and in what follows the basis for $\tilde{\mathfrak k}$ is
\begin{align}
k_1=e_1+f_1, \quad\quad\quad k_2=e_2-f_2, \quad\quad\quad k_3=e_3+f_3,\label{eq:kElements}\\
k_4=e_4+f_4, \quad\quad\quad k_5=e_5+f_5, \quad\quad\quad k_6=e_6-f_6,\nonumber
\end{align}
while the basis for $\tilde{\mathfrak p}$ is $h_1,h_2$ along with
\begin{align}
p_1=e_1-f_1, \quad\quad\quad p_2=e_2+f_2, \quad\quad\quad p_3=e_3-f_3,\label{eq:pElements}\\
p_4=e_4-f_4, \quad\quad\quad p_5=e_5-f_5, \quad\quad\quad p_6=e_6+f_6.\nonumber
\end{align}

In the case that $\epsilon_1=\epsilon_2=+1$, $\tilde{\mathfrak k}$ is the maximal compact subalgebra and is isomorphic to ${\mathfrak so}(4)$.  In the present case, $\epsilon_1=+1$, $\epsilon_2=-1$ and $\tilde{\mathfrak k}$ is called a `pseudo-compact' subalgebra and is isomorphic to ${\mathfrak sl}(2,{\mathbb R})\oplus{\mathfrak sl}(2,{\mathbb R})$.\footnote{In \cite{Compere:2009zh} the dimensional reduction was performed such that $\epsilon_1=-1$, $\epsilon_2=+1$, which requires a different $\tau$ according to (\ref{eq:tau}).  In terms of the dimensionally reduced theory, this corresponds to a nontrivial redefinition of the fields but the pseudo-compact subalgebra $\tilde{\mathfrak k}$ is still isomorphic to ${\mathfrak sl}(2,{\mathbb R})\oplus{\mathfrak sl}(2,{\mathbb R})$.}  This is more easily seen by taking the basis

\begin{align}
k_h={}&\frac{3}{4}k_1+\frac{\sqrt{3}}{4}k_4,\quad\quad\quad\quad\quad \bar k_h=\frac{1}{4}k_1-\frac{\sqrt{3}}{4}k_4,\nonumber\\
k_e={}&-\frac{\sqrt{3}}{4}k_2-\frac{3}{4}k_6,\quad\quad\quad\>\>\> \bar k_e=-\frac{\sqrt{3}}{4}k_2+\frac{1}{4}k_6,\\
k_f={}&\frac{\sqrt{3}}{4}k_3+\frac{3}{4}k_5,\quad\quad\quad\quad\quad \bar k_f=\frac{\sqrt{3}}{4}k_3-\frac{1}{4}k_5,\nonumber
\end{align}
in which case the generators satisfy the ${\mathfrak sl}(2,{\mathbb R})$ commutation relations $[k_f,k_e]=k_h$, $[k_e,k_h]=k_f$, and $[k_h,k_f]=-k_e$.  The same commutators apply for the barred generators while all commutators of the form $[k,\bar k]$ vanish.  The dimensional reduction of five dimensional minimal supergravity describes three dimensional Euclidean gravity coupled to a non-linear sigma model whose target space is the coset $G_{2(2)}/\tilde K$.  The generators (\ref{eq:kElements}) thus play a vital role in the reduced theory and have been interpreted as charging transformations in the five dimensional theory\cite{Compere:2009zh}.

\subsection{Reduction Along the String: ${\cal N}=2,~D=4$ Supergravity}
\label{RedAlongSt}

The action for five dimensional minimal ungauged supergravity is given by (\ref{eq:UngaugedSUGRA}), however in the next subsections I will be dimensionally reducing the theory so it is more convenient to work directly in terms of the Lagrangian density
\begin{equation}
\mathcal{L}_5 = R_5 \star_5  \mathbf{1} - \frac{1}{2}\star_5  {\bf F} \wedge {\bf F} +\frac{1}{3\sqrt 3} {\bf F} \wedge {\bf F} \wedge {\bf A}. \label{5dsugra}
\end{equation}
I first reduce along the string direction, denoted $z$, by assuming $z$ to be a compact Killing direction in the five dimensional space-time. Using the standard Kaluza-Klein ansatz to yield a four dimensional Lagrangian in Einstein frame, the five dimensional metric and gauge potential is written
\begin{align}
ds^2_5 &= e^{\frac{1}{\sqrt 3}\phi_1} ds^2_4 + e^{-\frac{2}{\sqrt 3}\phi_1}( dz+ A_1)^2, \label{eqn:metric5to4d} \\
{\bf A} &= {\bf A}_2 + \chi_2  {\bf d}z, \label{pot5to4d}
\end{align}
From this ansatz one can show that the resulting four dimensional Lagrangian takes the form\cite{Cremmer:1999du}
\begin{equation}
\begin{split}
\mathcal{L}_4 &= R_4 \star_4 \mathbf{1} - \frac{1}{2} \star_4 {\bf d} \phi_1 \wedge  {\bf d}\phi_1 - \frac{1}{2} e^{\frac{2}{\sqrt{3}} \phi_1} \star_4  {\bf d}\chi_2 \wedge  {\bf d}\chi_2  - \frac{1}{2} e^{-\sqrt{3}\phi_1}\star_4 {\bf F}_1 \wedge {\bf F}_1 \\
&\quad\, \: - \: \frac{1}{2} e^{-\frac{1}{\sqrt{3}}\phi_1} \star_4 {\bf F}_2 \wedge {\bf F}_2 + \frac{1}{\sqrt{3}} \, \chi_2 \,  {\bf dA}_2 \wedge  {\bf dA}_2,
\label{lagrangian}
\end{split}
\end{equation}
where
\begin{align}
\label{eqn:4dfieldstrengths}
{\bf F}_1 &= {\bf dA}_1, &
{\bf F}_2 &= {\bf dA}_2 - {\bf d}\chi_2 \wedge {\bf A}_1.
\end{align}
The resulting theory is ${\cal N}=2,~D=4$ supergravity, where the scalars $\phi_1$ and $\chi_2$ parametrise an $\mathrm{SL}(2, \mathbb{R})/\mathrm{U}(1)$ coset.  The five dimensional solutions are classified according to the corresponding black holes in this four dimensional theory so it will prove very useful to interpret solutions from this four dimensional perspective.  To this end, the various charges associated to these four dimensional black holes are of direct importance.  The equations of motion for the potentials ${
\bf A}_1$ and ${\bf A}_2$ are
\begin{align}
&{\bf d}\big( e^{-\sqrt{3} \phi_1} \star_4 {\bf F}_1)+e^{-\frac{1}{\sqrt{3}} \phi_1} \star_4 {\bf F}_2\wedge {\bf d}\chi_2 =0, \label{eqn:eom4A1}\\
&{\bf d}\beta_2 \equiv {\bf d}\big( e^{-\frac{1}{\sqrt{3}} \phi_1} \star_4 {\bf F}_2 - \frac{2}{\sqrt{3}} \chi_2 {\bf d A}_2 \big) = 0, \label{eqn:eom4A2}
\end{align}
however using \eqref{eqn:eom4A2} one can rewrite \eqref{eqn:eom4A1} as the closure of a form $\beta_1$:
\begin{equation}\label{eqn:eom4A1-1}
{\bf d}\beta_1 \equiv {\bf d}\big( e^{-\sqrt{3} \phi_1} \star_4 {\bf F}_1+e^{-\frac{1}{\sqrt{3}} \phi_1} \star_4 {\bf F}_2 \chi_2 - \frac{1}{\sqrt{3}} {\bf d A}_2 \chi_2^2\big)=0.
\end{equation}
The closed 2-forms $\beta_1$ and $\beta_2$ imply conserved electric charges in asymptotically flat four dimensional space-times as integrals over a two-sphere at spatial infinity $S^2_\infty$
\begin{align}
Q_{1} &= \frac{1}{4 \pi}  \int_{S^2_{\infty}}  \beta_1,&
Q_{2}  &= \frac{1}{4 \pi \sqrt{3}}  \int_{S^2_{\infty}} \beta_2.  \label{eq:ElectricCharges}
\end{align}
In a similar fashion, the Bianchi identities for ${\bf A}_1$ and ${\bf A}_2$
\begin{equation}
{\bf d F}_1 = 0, \qquad\qquad
{\bf d}({\bf F}_2 + {\bf d} \chi_2 \wedge {\bf A}_1 ) = 0,  \label{eq:MagneticCharges}
\end{equation}
imply the conserved magnetic charges
\begin{align}
P_1 &=- \frac{1}{4 \pi}  \int_{S^2_{\infty}} {\bf F}_1, &
\label{P1P2}
P_2 &=  \frac{1}{4 \pi \sqrt{3}}  \int_{S^2_{\infty}} {\bf F}_2 + {\bf d}\chi_2 \wedge {\bf A}_1.
\end{align}
Note the minus sign in the definition of $P_1$: the sign convention used is the one in which the static extremal black hole carrying positive $P_1$ and $Q_2$ charges
is supersymmetric.  From the M-theory point of view the electromagnetic charges correspond to the following brane charges:
\vskip 0.2cm
\begin{center}
\begin{tabular}{|c|c|}\hline
$Q_1$ & Kaluza-Klein momentum (P) along the M-theory circle  \\ \hline
$P_1$ & Kaluza-Klein-monopole (KKM) along the M-theory circle \\ \hline
$Q_2$ & M2$^3$ (with three equal M2-brane charges)  \\ \hline
$P_2$ &  M5$^3$ (with three equal M5-brane charges) \\ \hline
\end{tabular}
\end{center}
\vskip 0.2 cm

\noindent Thus, the $(Q_1, P_2)$ system with $Q_1,\,P_2 > 0$ corresponds to M5$^3$ -- P, which is BPS at extremality.  The $(Q_2, P_1)$ system with $Q_2,\,P_1 > 0$ corresponds to M2$^3$ -- KKM, which is also BPS at extremality.

Two scalar charges can also be defined for $\phi_1$ and $\chi_2$ as the radial derivatives of these fields at spatial infinity.  The condition that both the scalars vanish at spatial infinity is imposed, which is a consequence of the five dimensional Kaluza-Klein boundary conditions accompanied by a natural gauge choice.  With this condition imposed, the scalar charges can simply be defined as
\begin{align}
\Sigma_s &= \lim_{r\to\infty}\frac{r\,\phi_{1}(r)}{\sqrt3},&
\Xi&= \lim_{r\to\infty}\frac{r\,\chi_{2}(r)}{\sqrt3}.
\end{align}
According to (\ref{eqn:metric5to4d}) and (\ref{pot5to4d}), from the five dimensional perspective the scalar charge $\Sigma_s$ is related to the asymptotic value of the modulus, while the scalar charge $\Xi$ plays the role of a magnetic flux around the string direction.

\subsection{Reduction on Time}
\label{sec:reductionontime}

Since the desired solution is stationary, it is assumed that there exists a time-like Killing vector, $\partial_t$, which commutes with $\partial_z$. This allows the reduction of the four dimensional theory to three dimensions over this time-like Killing vector. The standard Kaluza-Klein ansatz for this reduction is
\begin{align}
\label{eqn:metric4dto3d}
ds^2_4  &= e^{\phi_2}ds^2_3  - e^{-\phi_2}(dt + \omega)^2, \\
\label{eqn:metricMaxwellto3d}
{\bf A}_1 & =  {\bf B}_1 + \chi_1 {\bf d}t, \\
\label{eqn:Maxwellto3d}
{\bf A}_2 & =  {\bf B}_2+ \chi_3 {\bf d}t.
\end{align}
As an inventory check at this point, note that the three dimensional theory contains five scalars $\phi_1,\phi_2,\chi_1,\chi_2,\chi_3$ and three 1-forms $\omega,B_1,B_2$.  Following the notation of \cite{Compere:2009zh}, associated to the axions $\chi_i$ and the 1-forms, the following field strengths are defined:
\begin{equation}
{\cal F}_{(1)}={\bf d}\chi_1, \qquad\qquad\qquad {\bf F}_{(1)}^1={\bf d}\chi_2, \qquad\qquad\qquad  {\bf F}^2_{(1)}={\bf d}\chi_3-\chi_1{\bf d}\chi_2
\end{equation}
\begin{equation}
\begin{split}
&{\cal F}_{(2)}^1={\bf d B}_1+\omega\wedge{\bf d}\chi_1, \quad\quad\quad\quad {\cal F}^2_{(2)}={\bf d}\omega, \\
 &{\bf F}_{(2)}={\bf d B}_2-{\bf d}\chi_2\wedge({\bf B}_1-\chi_1\omega)-{\bf d}\chi_3\wedge\omega, \label{eq:FDualisable}
 \end{split}
\end{equation}
where a subscript $(n)$ denotes the field strength is an $n$-form.  In three dimensions, the field strengths (\ref{eq:FDualisable}) can be dualised into 1-form field strengths to define three more axions $\chi_4,\chi_5,\chi_6$.  This is achieved by\cite{Compere:2009zh}
\begin{align}
e^{-\frac{1}{\sqrt{3}}\phi_1-\phi_2}\star_3{\bf F}_{(2)}\equiv G_{(1)}^4={}&{\bf d}\chi_4+\frac{1}{\sqrt{3}}(\chi_2{\bf d}\chi_3-\chi_3{\bf d}\chi_2),\\
e^{-\sqrt{3}\phi_1-\phi_2}\star_3{\cal F}_{(2)}^1\equiv G_{(1)}^5={}&{\bf d}\chi_5-\chi_2{\bf d}\chi_4+\frac{1}{3\sqrt{3}}\chi_2(\chi_3{\bf d}\chi_2-\chi_2{\bf d}\chi_3),\\
e^{-2\phi_2}\star_3{\cal F}_{(2)}^2\equiv G_{(1)}^6={}&{\bf d}\chi_6-\chi_1{\bf d}\chi_5+(\chi_1\chi_2-\chi_3){\bf d}\chi_4\\
&+\frac{1}{3\sqrt{3}}(\chi_1\chi_2-\chi_3)(\chi_2{\bf d}\chi_3-\chi_3{\bf d}\chi_2).\nonumber
\end{align}
Defining $\vec\phi=(\phi_1,\phi_2)$, in terms of the three dimensional fields the Lagrangian now takes the simple form
\begin{align}
{\cal L}={}&R_3\star_3 1-\frac12\star_3{\bf d}\vec\phi\wedge{\bf d}\vec\phi+\frac12e^{\vec\alpha_1\cdot\vec\phi}\star_3{\cal F}_{(1)}\wedge{\cal F}_{(1)}-\frac12e^{\vec\alpha_2\cdot\vec\phi}\star_3{\bf F}_{(1)}^1\wedge{\bf F}_{(1)}^1 \nonumber\\
&+\frac12e^{\vec\alpha_3\cdot\vec\phi}\star_3{\bf F}_{(1)}^2\wedge{\bf F}_{(1)}^2+\frac12e^{\vec\alpha_4\cdot\vec\phi}\star_3G_{(1)}^4\wedge G_{(1)}^4+\frac12e^{\vec\alpha_5\cdot\vec\phi}\star_3G_{(1)}^5\wedge G_{(1)}^5  \label{eq:3dLagrangian}\\
&-\frac12e^{\vec\alpha_6\cdot\vec\phi}\star_3G_{(1)}^6\wedge G_{(1)}^6,\nonumber
\end{align}
where the dilaton couplings, $\vec\alpha_i$, are precisely the root vectors (\ref{eq:SimpleRoots}) and (\ref{eq:OtherRoots}) of $G_{2(2)}$.  This Lagrangian describes three dimensional Euclidean gravity coupled to eight scalars; the scalars parametrise the pseudo-Riemannian coset $\mathrm{G}_{2(2)}/\tilde{K}$, with
\begin{align}
\tilde{K} = \mathrm{SL}(2, {\mathbb{R}}) \times \mathrm{SL}(2, {\mathbb{R}}).
\end{align}
To see this, first let $\mathcal{V}$ be the coset representative for the coset $\mathrm{G}_{2(2)}/\tilde{K}$.  This is constructed by exponentiating the Cartan and positive root generators with the scalars as coefficients by\cite{Compere:2009zh}
\begin{equation}
{\cal V}=e^{\frac12\phi_1h_1+\frac12\phi_2h_2}e^{\chi_1e_1}e^{-\chi_2e_2+\chi_3e_3}e^{\chi_6e_6}e^{\chi_4e_4-\chi_5e_5}. \label{eq:CosetRep}
\end{equation}
${\cal V}$ transforms under a global $G_{2(2)}$ transformation, $g$, and a local $\tilde K$ transformation, $k$, as ${\cal V}\rightarrow k{\cal V}g$.  A Lie algebra element, $v$, is constructed from the coset representative by $v={\bf d}{\cal VV}^{-1}$, which can be decomposed as $v={\cal K}+{\cal P}$, where ${\cal K}\in \tilde{\mathfrak k}$ and ${\cal P}\in\tilde{\mathfrak p}$.  Recall that the pseudo-compact subalgebra $\tilde{\mathfrak k}$ is invariant under the involution $\tau$ while $\tilde{\mathfrak p}$ is anti-invariant, thus
\begin{equation}
{\cal K}\equiv\frac12\big(v+\tau(v)\big), \qquad\qquad\qquad {\cal P}\equiv\frac12\big(v-\tau(v)\big).
\end{equation}
One can then write the Lagrangian (\ref{eq:3dLagrangian}) as\cite{Henneaux:2007ej}
\begin{equation}
{\cal L}=R_3\star_3 1-\frac12{\mathrm {Tr}}\big(\star_3{\cal P}\wedge{\cal P}\big),
\end{equation}
which is manifestly invariant under transformations $g$ and $k$ since ${\cal P}$ is invariant under global $G_{2(2)}$ transformations and transforms as ${\cal P}\rightarrow k{\cal P}k^{-1}$ under local $\tilde K$ transformations.

The group action on the coset representative, ${\cal V}\rightarrow k{\cal V}g$, is rather complicated to deal with practically, so instead a matrix $\mathcal{M}$ is defined via
\begin{eqnarray}
\mathcal{M} \equiv (\mathcal{V}^{\sharp})  \mathcal{V}, \label{M}
\end{eqnarray}
where $\sharp$ stands for the generalised transposition, which is
defined on the generators of $\mathfrak{g}_{2(2)}$ by
\begin{eqnarray}
x^\sharp \equiv -  \tau (x),\, \qquad  \forall \, x \in \mathfrak{g}_{2(2)}
\label{generalizedtransposition}
\end{eqnarray}
for the involution $\tau$ defined in (\ref{eq:tau}).  Since $k$ is an exponentiation of generators invariant under the involution, $k^{\sharp}=k^{-1}$ and the matrix $\mathcal{M}$ transforms simply as
\begin{eqnarray}
\mathcal{M} \rightarrow g^\sharp \, \mathcal{M} \, g \hspace{1cm} \mbox{when} \hspace{1cm}\mathcal{V} \rightarrow  k\mathcal{V} \, g, \hspace{1cm}  k\in \tilde K, \>\> g \in G_{2(2)}.
\end{eqnarray}
Acting with $G_{2(2)}$ transformations thus transforms the matrix $\cal M$, which in turn transforms the scalars.  Upon uplifting the transformed scalars back to five dimensions, new solutions are obtained.

There are two more charges that characterise the four dimensional solutions: the mass and the NUT charge.  Using the reduction ansatz (\ref{eqn:metric4dto3d}) these charges are calculated explicitly in terms of the three dimensional fields. Following \cite{Bossard:2008sw}, the Komar mass and NUT charge are defined as
\begin{align}
M &= \frac{1}{8 \pi G_4} \int_{S^2_{\infty}} \star_4 {\bf K}, & N &=  \frac{1}{8 \pi G_4} \int_{S^2_{\infty}}  {\bf K},
\end{align}
where ${\bf K}={\bf dg}$, ${\bf g}=g_{\mu\nu} \kappa^\mu {\bf d}x^\nu$, and  $\kappa^\mu = \partial_t$. From the metric ansatz (\ref{eqn:metric4dto3d}) I find
\begin{equation}
{\bf K} = \partial_{\nu} g_{t \mu} {\bf d}x^{\nu} \wedge {\bf d}x^{\mu},
\end{equation}
which yields
\begin{equation}
M = -\frac{1}{8 \pi G_4} \int_{S^2_{\infty}} \partial_r e^{-\phi_2} \,\star_4 ({\bf d}r \wedge {\bf d}t),
\end{equation}
and $\star_4 ( {\bf d} r \wedge {\bf d} t) = - e^{\phi_2}r^2 \sin(\theta)  {\bf d} \theta \wedge {\bf d} \phi$.
Imposing the condition of asymptotic flatness implies $\phi_2 (r)= 2 G_4 M/r+ \mathcal{O}\left(r^{-2}\right)$, thus
\begin{align}
G_4 M = \lim_{r\to\infty}\frac{r\phi_{2}(r)}{2}.
\end{align}
Similarly, the NUT charge can be expressed as 
\begin{equation}
N = -\lim_{r\to\infty}\frac{r\chi_{6}(r)}{2G_4}.
\end{equation}

To summarise, the ansatz
\begin{align}
ds^{2}_{5} &=
e^{\frac{1}{\sqrt3} \phi_{1}+\phi_{2}}ds^{2}_{3}
-e^{\frac{1}{\sqrt3} \phi_{1}-\phi_{2}}  (dt+\omega)^2
+e^{-\frac{2}{\sqrt3} \phi_{1}} ( dz + \chi_1 dt+B_1 )^{2},\label{eqn:canonicalform}
\\
\label{eqn:canonicalMaxwell}
{\bf A}&={\bf B}_2+\chi_{3}{\bf d}t + \chi_{2}{\bf d}z,
\end{align}
describes stationary solutions of single modulus ${\cal N}=2,~D=4$ supergravity uplifted to five dimensions. It follows from four dimensional asymptotic flatness that the electric and magnetic charges (\ref{eq:ElectricCharges}) and (\ref{eq:MagneticCharges}) can  also be expressed in terms of asymptotic values of the scalars as
\begin{align}
Q_1&=   \lim_{r \to \infty} \,r \chi_1(r),& 
Q_2&=    \lim_{r \to \infty} \, \frac{r \chi_3(r)}{\sqrt{3}},\nonumber\\
P_1&= \lim_{r \to \infty} \,r \,\chi_5(r),& 
P_2&= -\lim_{r \to \infty} \, \frac{r\, \chi_4(r)}{\sqrt{3}}.
\end{align}

\subsection{Five Dimensional Charges}

For black strings the spatial part of the metric at infinity is $\mathbb{R}^{3} \times S^1$. For asymptotically Kaluza-Klein space-times,
the five dimensional ADM mass $M_5$, linear momentum $P_z$ along the string direction, and ADM tension $\mathcal T$ are defined as \cite{Harmark:2004ch,Myers:1999psa}
\begin{eqnarray}
M_5 &=& 2 \pi R \: T_{tt}  = \frac{\pi R}{2G_5}(2c_{tt}-c_{zz}),\\
\mathcal T &=& -T_{zz} = \frac{1}{4G_5}( c_{tt}-2c_{zz}),\\
P_z &=& 2\pi R \: T_{tz} = \frac{\pi R}{2G_5} c_{tz},
\end{eqnarray}
where $c_{tt}, c_{tz},$ and $c_{zz}$ are the leading corrections of the metric at infinity
\begin{eqnarray}
g_{tt} \simeq - 1 + \frac{c_{tt}}{r}, \qquad g_{zz} \simeq  1 + \frac{c_{zz}}{r}, \qquad g_{tz} \simeq \frac{c_{tz}}{r}.
\end{eqnarray}
One can re-express these charges in terms of the four dimensional charges as
\begin{eqnarray}
M_5 = M,\qquad
\mathcal T  = \frac{1}{8 \pi R G_4}( 3\Sigma_s +2 G_4 M ),\qquad
P_z = \frac{Q_1}{4G_4},
\end{eqnarray}
where I have used $G_5 = (2\pi R) G_4$.
The five dimensional electric charge in geometrised units for asymptotically Kaluza-Klein spaces is
\begin{eqnarray}
Q_\mathrm{E} \equiv \frac{1}{16\pi G_5}\int_{S^2 \times \mathbb R}\star_{5}{\bf F} =\frac{\sqrt{3}}{4 G_4}Q_2,
\end{eqnarray}
where the Chern-Simons term has been set to zero using the boundary conditions at infinity. The magnetic one-brane charge is
\begin{equation}
Q_\mathrm{M} \equiv \frac{1}{4\pi} \int_{S^2} {\bf F} = \sqrt{3} P_2.
\end{equation}

\section{Constructing the General String}
\label{sec:construction}

In this section I explicitly construct the five-parameter family of black strings and analyse their physical properties. To construct the solution, I will proceed in two steps: in the first step I construct an appropriate `charge matrix', and in the second step I use the group element $k \in \tilde K$ used to generate the charge matrix to construct the general string.  I begin by briefly reviewing the concept of a charge matrix.

\subsection{The Charge Matrix}
\label{charge_matrix}

Due to the work of Breitenlohner, Maison, and Gibbons \cite{Breitenlohner:1987dg} it is known that single centre spherically symmetric black holes for a wide class of four dimensional gravity theories correspond to geodesic segments on coset manifolds $G/{\tilde K}$.  A geodesic on the coset manifold is completely specified by its starting point $p \in G/{\tilde K}$ and its velocity at $p$, which is the conserved Noether charge $\mathcal{Q} \in \mathfrak{g}$ taking values in the Lie algebra $\mathfrak{g}$ of $G$. From the four dimensional space-time viewpoint, the starting position $p$ of the geodesic is associated to the values of the moduli at spatial infinity, \emph{i.e.} it determines the asymptotic structure, while the velocity $\mathcal{Q}$ at the point $p$ is associated to the four dimensional conserved charges.

A general action of $G$ on a given solution acts both on the position and the velocity of the corresponding geodesic.  From a space-time viewpoint, however, the asymptotic structure should be fixed so the subgroup of $G$ that is of particular importance for generating solutions with additional charges is the one that keeps the starting point $p$ fixed and acts only on the velocity $\cal Q$.  This is exactly the subgroup $\tilde K$: the full set of transformations of the conserved charge $\mathcal{Q}$ is contained in a single $\tilde K$-orbit.  Indeed it was shown in \cite{Breitenlohner:1987dg} that one can generate the full class of single-centre non-extremal spherically symmetric black holes by acting on the Schwarzschild solution with $\tilde K$.  It was also shown in \cite{Breitenlohner:1987dg} that acting with $\tilde K$ on the Kerr solution, one can generate the full class of single-centre non-extremal rotating black holes. In other words, \emph{all} single-centre non-extremal black holes with requisite symmetry lie in a single $\tilde K$-orbit containing the Kerr black hole.

Assuming asymptotic flatness and that the geodesic corresponding to a given black hole starts at the identity coset, then the matrix $\mathcal{M}$ admits an expansion in powers of the radial coordinate \begin{equation}
\mathcal{M} = 1 + \frac{ \mathcal{Q}}{r} + \ldots,
\end{equation}
where $\mathcal{Q} \in \tilde{\mathfrak{p}}$.
 The matrix $\mathcal{Q}$ is also called the ``charge matrix" because it is related to the four dimensional conserved charges corresponding to the scalars $\phi_i,\chi_i$ defined previously.  Note that information about the rotation in the four dimensional space-time is not contained in the charge matrix since rotation would enter at ${\cal O}(r^{-2})$ in the expansion of $\mathcal{M}$.  
 
Following \cite{Breitenlohner:1987dg}, it was shown in \cite{Bossard:2009at} that for any four dimensional asymptotically flat non-extremal axisymmetric solution, the charge matrix satisfies
\begin{equation}
\mathcal{Q}^3 - \frac{\mathrm{tr}(\mathcal{Q}^2)}{4} \mathcal{Q} = 0. \label{characterstic}
\end{equation}
In terms of the four dimensional electromagnetic and scalar charges,  the most general charge matrix for ${\cal N}=2~D=4$ supergravity is written as\cite{Kim:2010bf}
\begin{equation}
\mathcal{Q}=-4 G_4 Mh_1-\big(\Sigma_s+2G_4 M\big)h_2-Q_1 p_1+\Xi p_2-Q_2 p_3+\tfrac12 P_2 p_4+\tfrac16 P_1p_5+\tfrac13 G_4 N p_6. \label{generalchargematrix}
\end{equation}
where $\{h_i,p_i\}$ span the subalgebra $\tilde{\mathfrak p}$, \emph{i.e.} they are the generators anti-invariant under the involution $\tau$.
From this, it is easy to see the Schwarzschild solution simply corresponds to
\begin{equation}
\mathcal{Q}^\mathrm{Schw}=-2  m (2 h_1+h_2),
\end{equation}
where $m= G_4 M$, and $M$ is the ADM mass of the Schwarzschild solution. To construct the general black string solution, I  now proceed in two steps.  First I act on $\mathcal{Q}^\mathrm{Schw}$ with a carefully chosen element $k \in \tilde K$
  \begin{equation}
   -2 m   k^{-1} \cdot (2 h_1 + h_2) \cdot k
  \end{equation}
to yield a charge matrix with all of the desired asymptotic charges.  Acting with the same $k$ on $\mathcal{M}^\mathrm{Schw}$ then allows one to uplift to five dimensions to obtain the non-rotating string with all of the desired charges.   In the second step I act with the \emph{same} $k$ on $\mathcal{M}^\mathrm{Kerr}$ and uplift to five dimensions to obtain the most general black string.  This two step decomposition is somewhat artificial but it helps to disentangle space-time physics from group theory.  The first problem can be solved by systematically studying the action of various generators of $\tilde K$ on the general charge matrix \eqref{generalchargematrix}. On the technical side one has to overcome three main obstacles: 
\begin{itemize}
\item[$(i)$] When acting with $\tilde K$ the resulting charge matrix is in general parametrised in an unilluminating manner, so one has to choose parameters judiciously.
\item[$(ii)$] In order to obtain
regular black holes, one has to force the NUT charge to vanish, $N=0$.
\item[$(iii)$] In order for the solution to correspond to a black string of five dimensional minimal supergravity one also has to force the Kaluza-Klein monopole charge to vanish $P_1=0$.
\end{itemize}
Although the actions of the generators $k_i$ were identified in \cite{Compere:2009zh} as adding individual particular charges to the Kerr solution, when acting with successive generators the charges generated become mixed, and criteria $(ii)$ and $(iii)$ become nontrivial to enforce in practice.  Nevertheless, all of the criteria are satisfied by the following element of the group $\tilde K$:
\begin{eqnarray}
k = e^{-\frac{1}{2}\log\alpha \, k_3}e^{{\frac12} \beta k_4}e^{- \beta k_1}e^{-\frac{1}{2}\log(\tilde \delta/\alpha) k_3}e^{-\pi k_2}e^{-\gamma k_1}, \label{generatork}
\end{eqnarray}
with
\begin{equation}
\alpha^3 = \frac{\tilde \delta^2 (3 + \tilde \delta)}{(1 + 3 \tilde \delta)}.
\end{equation}
This group element was found by a process involving a considerable amount of trial and error, followed by an explicit verification of the physical conditions mentioned above. The final charge matrix indeed satisfies \eqref{characterstic} and the non-extremality parameter is simply
\begin{equation}
c^2 \equiv \frac{1}{4}\mathrm{tr}(\mathcal{Q}^2) = 4  m^2.
\end{equation}
There are only three independent parameters in the group element $k$, namely, $\beta, \tilde \delta$ and $\gamma$. Choosing $\tilde \delta = \exp(2 \delta)$ I find that the charge matrix is parametrised as:
\begin{align}
G_4 M ={}& m \left( -\frac{2 c_\beta c_\gamma s_\beta s_\gamma}{\sqrt{1 + 3 c_\delta^2}}+ {\frac12}(c_\beta^2 + s_\beta^2) (c_\delta^2 (1 +  c_\gamma^2) + s_\delta^2 (s_\gamma^2 + c_\gamma^2))\right), \label{charge1}\\
Q_1 ={}& 2 m\left( \frac{2 c_\beta s_\beta (c_\gamma^2 + s_\gamma^2)}{\sqrt{1 + 3 c_\delta^2}}- c_\gamma s_\gamma (c_\beta^2 + s_\beta^2)(1 + 3 s_\delta^2)\right),\\
Q_2 ={}& 2m c_\delta s_\delta \left(  \frac{2 s_\gamma s_\beta c_\beta}{\sqrt{1 + 3 c_\delta^2}} + c_\gamma (c_\beta^2 + s_\beta^2)\right),\\
P_2 ={}&
\frac{4m c_\beta   s_\beta c_\delta^2 }{\sqrt{1 + 3 c_\delta^2}},\\
\Sigma_s ={}&\frac{4m c_\beta c_\gamma s_\beta s_\gamma}{\sqrt{1 + 3 c_\delta^2}} - m(c_\beta^2 + s_\beta^2) \left(s_\gamma^2 c_\delta^2 + s_\gamma^2 s_\delta^2 + c_\gamma^2 s_\delta^2\right), \\
\Xi ={}& - 2m c_\delta  s_\delta \left( \frac{2  c_\beta c_\gamma  s_\beta }{\sqrt{1 + 3 c_\delta^2}} + (c_\beta^2 + s_\beta^2)s_\gamma\right), \\
N = {}& 0, \qquad P_1 = 0,
\label{charge8}
\end{align}
where I have defined $c_\alpha \equiv \cosh \alpha$ and $s_\alpha \equiv \sinh \alpha$ for $\alpha=\beta,~\delta,~\gamma$ in order to avoid needless notational clutter.

\subsection{The General Black String Solution}
\label{sec:monster}

To explicitly construct the most general black string, I now act with the element $k$ on the matrix $\mathcal M^{\mathrm{Kerr}}$ to extract the transformed scalars, which are then used to uplift the solution to five dimensions.  The seed solution is the direct product of the Kerr geometry with a line along the $z$-direction. This metric can be written in the convenient form
\begin{equation}
ds^2 = -\frac{\Delta_2}{\Sigma}(dt+\omega^\mathrm{seed}_\phi d\phi)^2 +\Sigma \, ds_\mathrm{base}^2 + dz^2, \hspace{1cm} \epsilon_{r x \phi tz} = + \Sigma,\label{Kerr}
\end{equation}
where
\begin{eqnarray}
ds^2_\mathrm{base} &=& \left( \frac{dr^2}{\Delta}+\frac{dx^2}{1-x^2} \right) + \frac{\Delta (1-x^2)}{\Delta_2}d\phi^2,
\end{eqnarray}
\begin{equation}
\omega^\mathrm{seed}_\phi = 2amr\frac{1-x^2}{\Delta_2},
\end{equation}
and
\begin{equation}
\Delta = r^2 - 2m r+a^2,\qquad \Delta_2 = r^2 - 2 m r +a^2 x^2 ,\qquad \Sigma = r^2 + a^2 x^2.  \label{eq:basefcns}
\end{equation}
For convenience the directed cosine $x = \cos \theta$ is used instead of the polar coordinate $\theta$.  Upon reducing this seed solution to three dimensions, constructing the matrix $\mathcal M^\mathrm{Kerr}$, acting with the group element $k$ and lifting back to five dimensions, I find the most general asymptotically flat black string solution of five dimensional minimal supergravity. The solution is endowed with five independent parameters: magnetic one-brane charge corresponding to the parameter $\beta$, smeared electric zero-brane charge corresponding to the parameter $\delta$, boost along the string direction\footnote{This isn't quite true.  See the pertinent comments at the end of section \ref{thermo}} corresponding to the parameter $\gamma$, energy above the BPS bound corresponding to the parameter $m$, and rotation in the transverse space corresponding to the parameter $a$.  The steps for proceeding have been explicitly laid out in section~\ref{sec:reductionontime} and the intermediate details of the solution construction process are rather cumbersome.  To avoid cluttering the presentation, I simply quote the final result here and push the details of the construction to Appendix~\ref{app:ScalarTransform}.  The solution takes a simpler form when $\gamma=0$: the line element and gauge potential in this case can be written
\begin{equation}
ds^2 = (F_1+\Delta_2) \left[ -  \frac{\Delta_2}{\xi} (dt + \omega_{\phi} d \phi)^2 + ds^2_\mathrm{base}\right] +\frac{\xi}{(F_1+\Delta_2)^2} \left( dz + \hat A_{t} dt + \hat A_{\phi} d \phi \right)^2, \label{eq:monster}
\end{equation}
\begin{align}
A_t ={}& -\sqrt{3} \frac{F_3}{(F_1+\Delta_2)},\qquad\qquad  A_z = -\sqrt{3}\frac{F_2}{(F_1+\Delta_2)},\\
 A_\phi ={}& \frac{2\sqrt{3}m c_{\delta}^2}{\Delta_2}\left( \frac{2s_{\beta}c_{\beta}\Delta x}{\sqrt{1+3c_{\delta}^2}}-a(1-x^2)s_{\delta}\left(2ms_{\beta}^2\left(\frac{6c_{\beta}^2 c_{\delta}^2} {1+3c_{\delta}^2}-1\right)+r(c_{\beta}^2+s_{\beta}^2)\right)\right)\nonumber\\ &-\sqrt{3}\bar A_\phi \frac{F_2}{(F_1+\Delta_2)}-\sqrt{3}\omega_\phi \frac{F_3}{(F_1+\Delta_2)},
\label{monster_metric}
\end{align}
where I have defined the following functions
\begin{align}
\hat A_\phi ={}&  \bar A_\phi + \frac{k}{\xi}\omega_\phi, \qquad \qquad
\hat A_t  =\frac{k}{\xi},\\
\omega_\phi ={}&  2am \frac{1-x^2}{\Delta_2}c_{\delta}^3\left(2ms_{\beta}^2\left(c_{\beta}^2+s_{\beta}^2-\frac{4c_{\beta}^2}{1+3c_{\delta}^2}\right)+r(c_{\beta}^2+s_{\beta}^2)\right),\\
\bar A_\phi ={}& -\frac{4am(1-x^2)s_{\beta}c_{\beta}c_{\delta}^3(r+2s_{\beta}^2m)}{\Delta_2\sqrt{1+3c_{\delta}^2}},\\
\xi ={}& (F_4+\Delta_2) (F_1+\Delta_2) -F_2^2,\qquad \qquad k = F_5(F_1+\Delta_2) -F_2 F_3.
\end{align}
The remaining five functions $F_1,F_2,F_3,F_4,F_5$ are linear in $r$ and $x$ and are given by
\begin{align}
F_1 ={}& 2mc_{\delta}^2\left(2ms_{\beta}^2\left(s_{\beta}^2+\frac{s_{\delta}^2c_{\beta}^2}{1+3c_{\delta}^2}\right)+r(c_{\beta}^2+s_{\beta}^2)+\frac{2axs_{\delta}s_{\beta}c_{\beta}}{\sqrt{1+3c_{\delta}^2}}\right),\nonumber \\
F_2 ={}& 2mc_{\delta}s_{\delta}\left(\frac{2ms_{\beta}c_{\beta}}{\sqrt{1+3c_{\delta}^2}}\left(-1+(c_{\beta}^2+s_{\beta}^2)c_{\delta}^2\right)+\frac{2rs_{\beta}c_{\beta}}{\sqrt{1+3c_{\delta}^2}}+axs_{\delta}(c_{\beta}^2+s_{\beta}^2)\right),\nonumber \\
F_3 ={}& 2mc_{\delta}\left(2ms_{\delta}s_{\beta}^2\left(\frac{c_{\beta}^2(1+c_{\delta}^2)}{1+3c_{\delta}^2}-s_{\beta}^2\right)-r(c_{\beta}^2+s_{\beta}^2)s_{\delta}+\frac{2axs_{\beta}c_{\beta}}{\sqrt{1+3c_{\delta}^2}}\right), \label{eq:monsterFfields}\\
F_4 ={}& 2m\left(2m\left((c_{\beta}^2s_{\delta}^2+s_{\beta}^2c_{\delta}^2)^2+\frac{s_{\delta}^2s_{\beta}^2c_{\beta}^2}{1+3c_{\delta}^2}\right)+r(c_{\beta}^2+s_{\beta}^2)(c_{\delta}^2+s_{\delta}^2)-\frac{2axs_{\delta}s_{\beta}c_{\beta}}{\sqrt{1+3c_{\delta}^2}}\right),\nonumber \\
F_5 ={}& 2m\left(\frac{2ms_{\beta}c_{\beta}}{\sqrt{1+3c_{\delta}^2}}\left(-1+(c_{\beta}^2+s_{\beta}^2)c_{\delta}^4\right)+\frac{2rs_{\beta}c_{\beta}}{\sqrt{1+3c_{\delta}^2}}-axs_{\delta}^3(c_{\beta}^2+s_{\beta}^2)\right).  \nonumber
\end{align}
The parameter $\gamma$ can be added by simply applying a boost along the $z$-direction to this solution
\begin{equation}
t \rightarrow t \cosh \gamma - z \sinh \gamma, \qquad z \rightarrow - t \sinh \gamma + z \cosh \gamma.
\end{equation}
The range of the parameters $\beta$ and $\delta$ can be restricted to non-negative values without loss of generality. The solution with parameters $(m,a,\beta,\delta,\gamma)$ are related to the solution with parameters $(m,-a,-\beta,\delta,-\gamma)$ with the change of coordinates $\phi \rightarrow -\phi$, $z \rightarrow -z$ and to the solutions of parameters $(m,a,\beta,-\delta,\gamma)$ with the change of coordinates $\phi \rightarrow -\phi$, $z \rightarrow -z$, $t \rightarrow -t$, $x \rightarrow -x$.

\section{Thermodynamics}

In this section, I discuss the thermodynamics of the general black string.  First the asymptotic and near-horizon quantities are calculated and the first law of thermodynamics is shown to hold.  Then the Bekenstein-Hawking entropy in the decoupling limit is explicitly shown to be reproduced by microstate counting in the Maldacena-Strominger-Witten CFT.

\subsection{Thermodynamics}
\label{thermo}

With the explicit solution above, I now turn to the thermodynamics of the general black string solution in the boosted frame, \emph{i.e.} $\gamma \neq 0$. Expressions for the asymptotic charges and horizon quantities simplify substantially with the introduction of the three following functions
\begin{equation}
G[X,Y] = \frac{-2s_{\beta}c_{\beta}XY}{\sqrt{1+3c_{\delta}^2}}+\frac{1}{4}(c_{\beta}^2+s_{\beta}^2)\big(X^2-2Y^2+3X^2(c_{\delta}^2+s_{\delta}^2)\big),
\end{equation}
\begin{equation}
H[X,Y] = \frac{2s_{\beta}c_{\beta}Y}{\sqrt{1+3c_{\delta}^2}}+(c_{\beta}^2+s_{\beta}^2)X,
\end{equation}
\begin{equation}
F[X,Y]=2mc_{\delta}^3\left(\left(c_{\beta}^4r_++s_{\beta}^4r_-+\frac{6ms_{\beta}^2c_{\beta}^2s_{\delta}^2}{1+3c_{\delta}^2} \right)X - \frac{2s_{\beta}c_{\beta}(c_{\beta}^2r_++s_{\beta}^2r_-)}{\sqrt{1+3c_{\delta}^2}}Y\right),
\end{equation}
where
\begin{equation}
r_\pm = m \pm \sqrt{m^2-a^2}
\end{equation}
are the inner and outer horizons.
In terms of these functions, the ADM mass, ADM tension, linear momentum along the string direction, angular momentum in the transverse space, horizon area, and angular and linear velocities at the outer horizon are found to be
\begin{align}
M_5 ={}& \frac{2\pi Rm}{G_5}G[c_\gamma,s_\gamma],\qquad \mathcal T =- \frac{m}{G_5}G[s_\gamma,c_\gamma],\label{eq:T2}\\
P_z ={}& \frac{m\pi R}{G_5} \left(-\frac{2s_{\beta}c_{\beta}(s_{\gamma}^2+c_{\gamma}^2)}{\sqrt{1+3c_{\delta}^2}}+c_{\gamma}s_{\gamma}(c_{\beta}^2+s_{\beta}^2)(1+3s_{\delta}^2)\right),\\
J_\phi ={}&  \frac{2\pi R m a c_{\delta}^3}{G_5}H[c_\gamma,-s_\gamma], \qquad A_{\mathrm{H}}=8\pi^2R \, F[c_\gamma,s_\gamma],\label{eq:A}\\
\Omega_{\phi} ={}& \frac{a}{F[c_\gamma,s_\gamma]},\qquad v_z=\frac{F[s_\gamma,c_\gamma]}{F[c_\gamma,s_\gamma]}.
\end{align}
The temperature can be calculated from the surface gravity, which results in
\begin{equation}
T=\frac{r_+ - r_-}{4 \pi F[c_\gamma,s_\gamma]}.\label{eq:T}
\end{equation}
This, of course, vanishes for extremal solutions where $a=\pm m$.  When $a=0$ and the other parameters are finite, the function $F[X,Y]\propto m^2$ and so the temperature scales in the usual way for a Schwarzschild string as $T\sim m^{-1}$.

The electric and magnetic charges are
\begin{equation}
Q_\mathrm{E} = \frac{\sqrt{3}m\pi Rc_{\delta}s_{\delta}}{G_5}H[c_\gamma,s_\gamma],  \qquad\qquad\qquad  Q_\mathrm{M} = \frac{4\sqrt{3}mc_{\delta}^2c_{\beta}s_{\beta}}{\sqrt{1+3c_{\delta}^2}}.  \label{eq:EMCharges}
\end{equation}
To calculate the chemical potentials associated with these two charges I follow \cite{Copsey:2005se}. To this end, the gauge field must be carefully defined: when the magnetic charge is non-zero, the gauge field asymptotically behaves as
\begin{equation}
A_\phi\sim Q_\mathrm{M} x +{\mathcal O}(r^{-1}), \quad\quad A_z\sim{\mathcal O}(r^{-1}), \quad\quad A_t\sim{\mathcal O}(r^{-1}),
\end{equation}
which contains Dirac strings since the angular component is not well-defined simultaneously on both the north and the south pole of the $S^2$.  To overcome this, regular gauge potentials are introduced on two patches that smoothly cover the poles.  The boundary between the two patches is denoted by $E$, which is taken to be the surface of constant $t$ and $\theta=\pi/2$.  The angular parts of the gauge potentials on the two patches then satisfy the boundary conditions
\begin{align}
A_\phi^\mathrm{North}\sim{}& Q_\mathrm{M}(x-1)+{\mathcal O}(r^{-1}),\\
A_\phi^\mathrm{South}\sim{}& Q_\mathrm{M}(x+1)+{\mathcal O}(r^{-1}).
\end{align}
Ultimately a quantity of the form $\xi^\mu A_\mu$ is required to be continuous across $E$, where $\xi^\mu=\partial_t+\Omega_\phi \partial_\phi+v_z \partial_z $ is the generator of the horizon.  To this end, some constant $\Lambda$ is introduced on each patch such that
\begin{equation}
\Lambda^\mathrm{North}=\Omega_\phi Q_\mathrm{M},\quad\quad\qquad \Lambda^\mathrm{South}=-\Omega_\phi Q_\mathrm{M}
\end{equation}
and it is clear that $\xi^\mu A_\mu+\Lambda$ is indeed continuous across the boundary $E$.  The electric and magnetic potentials can now be computed properly. The electric potential that appears in the first law is given by
\begin{equation}
\Phi_{E}=(\xi^\mu A_\mu+\Lambda)\Big{|}_{r_+},
\end{equation}
which yields the result
\begin{equation}
\Phi_\mathrm{E}=\frac{2\sqrt{3}mc_{\delta}^2s_{\delta}}{F[c_{\gamma},s_{\gamma}]}\left(r_+c_{\beta}^2\left(\frac{6c_{\delta}^2s_{\beta}^2}{1+3c_{\delta}^2}+1\right)+r_-s_{\beta}^2\left(\frac{6c_{\delta}^2c_{\beta}^2}{1+3c_{\delta}^2}-1\right)\right).
\end{equation}

In \cite{Copsey:2005se} a general expression for the magnetic potential was given in the Hamiltonian form, which can be expressed in Lagrangian form as\cite{Compere:2009zh}
\begin{align}
\Phi_\mathrm{M}={}&\frac1{8\pi}\int_E(d^3x)_{\mu\nu}(2\xi^\mu F^{\alpha\nu}\partial_\alpha\phi-\Omega_\phi F^{\mu\nu})+\frac1{8\sqrt{3}\pi}\int_E F_{\alpha\beta}\partial_\gamma\phi(\Phi){\bf d}x^\alpha\wedge {\bf d}x^\beta\wedge {\bf d}x^\gamma,
\end{align}
where $(\Phi)=A_\rho\xi^\rho+\Lambda$ and $(d^3x)_{\mu\nu}=\frac1{2!3!}\epsilon_{\mu\nu\alpha\beta\gamma}{\bf d}x^\alpha\wedge {\bf d}x^\beta\wedge {\bf d}x^\gamma$.  Although one could in principle use this to calculate the magnetic potential for the general black string, in practice such a computation is extremely involved and not very illuminating. Instead I use the Smarr relation to guess the magnetic potential in terms of the other physical quantities: the Smarr relation takes the form
\begin{equation}
M=\frac32\left(\frac1{4G_5}T A +\Omega_\phi J_\phi+v_z P_z\right)+\frac12{\mathcal T}(2\pi R)+\Phi_\mathrm{E} Q_\mathrm{E}+\frac12\Phi_\mathrm{M} Q_\mathrm{M}.
\end{equation}
Upon solving for $\Phi_\mathrm{M}$, it is a straightforward exercise, but a nontrivial check, to verify that the first law
\begin{equation}
dM=\frac1{4G_5}TdA+\Omega_\phi dJ_\phi+ \frac{v_z}{R} d\left(P_z R\right)+2\pi {\mathcal T}dR+\Phi_\mathrm{E} d Q_\mathrm{E}+\Phi_\mathrm{M} dQ_\mathrm{M}
\end{equation}
holds. Stated differently, one can assume that the first law holds and derive the magnetic potential $\Phi_\mathrm{M}$ from the first law. The fact that it is possible and nontrivial to do so is because the first law involves six differential equations, one for the variation of each parameter $(m,a,\beta,\delta,\gamma,R)$. All of these equations give the same answer for $\Phi_\mathrm{M}$. Moreover, one then checks that the Smarr relation holds. This procedure of using the first law to derive a physical quantity has been used previously in the literature, for example in \cite{Chong:2005hr}.

Now that all of the conserved quantities have been presented, I discuss the role of the parameter $\gamma$ in the general solution. This parameter was introduced as a boost performed after generating the electric and magnetic charges, however the final solution has a non-zero linear momentum $P_z$ when $\gamma = 0$. This means that the solution has  already been boosted along the $z$-direction as a result of the intricate action of the $\mathfrak g_{2(2)}$ generators.  It is, however, straightforward to find the boost $\gamma = \gamma_c(\beta,\delta)$ one has to apply to reach the non-boosted frame. Solving for $P_z = 0$, I find
\begin{equation}
\gamma_c(\beta,\delta) = \frac{1}{2} \tanh^{-1}\left[ \frac{2 \tanh 2 \beta}{(1+ 3 s_\delta^2)\sqrt{1 + 3 c_\delta^2}}\right] .\label{amitabhsigma}
\end{equation}
In order to keep the boost as a free parameter, I then simply define the new boost parameter $\sigma$ as
\begin{equation}
\sigma \equiv \gamma_c(\beta,\delta) - \gamma,
\end{equation}
in which case the linear momentum  $P_z$ then takes the form
\begin{equation}
P_z = m c_{\sigma }s_{\sigma } \left(\frac{  \left(3 s_{\delta }^2+1\right)
   \left(\left(s_{\beta }^2+1\right)
   \sqrt{\frac{3 s_{\delta }^2}{2}+4} \left(3 s_{\delta
   }^2+2\right)-\frac{32 c_{\beta }^2 s_{\beta
   }^2}{\sqrt{3 c_{\delta }^2+1} \left(2 s_{\beta
   }^2+1\right) \left(3 s_{\delta
   }^2+1\right)}\right)}
{4 G_4 \sqrt{27 c_{\delta }^4
   s_{\delta }^2+\frac{4}{\left(s_{\beta
   }^2+1\right){}^2}}} \right),
\end{equation}
\emph{i.e.} $P_z$ is proportional to $c_\sigma s_\sigma$.  Similarly, examining (\ref{eq:EMCharges}) it can be seen that $Q_E$ is proportional to $c_\delta s_\delta$, while $Q_M$ is proportional to $c_\beta s_\beta$.  Therefore $\delta,~\beta,~\sigma$ are the parameters controlling the electric charge, magnetic charge and linear momentum respectively.  It is more convenient to work with $\gamma$ than $\sigma$, however.

\subsection{Decoupling Limit}
\label{decoupling}

From an M-theory perspective, the general black string corresponds to intersecting M5-branes and M2-branes, where the M5-branes wrap various four-cycles of the $T^6$ and all have a world-volume direction along the string direction.  In the low energy limit, corresponding to small distances between branes, the dynamics on the branes decouples from the bulk and the system behaves like a free gas of M5-branes and M2-branes.  This eliminates coupling terms to bulk modes and thus drastically simplifies calculations.  This limit is known as the decoupling limit.  Following \cite{Cvetic:1999ja,Maldacena:1997re}, the decoupling limit of the black string  \eqref{eq:monster}--\eqref{monster_metric} is obtained by taking the five dimensional Planck length to zero, $\ell_p \rightarrow 0$, while keeping the following quantities fixed:
\begin{itemize}
\item[$(i)$] The rescaled coordinate, $\tilde r$, and the variables, $\tilde m$ and $\tilde a$, defined via
\begin{eqnarray}
r \rightarrow \tilde r {\frac{\ell_p^3}{R^2}}, \hspace{1cm}m \rightarrow \tilde{m} {\frac{\ell_p^3}{R^2}}, \hspace{1cm} a \rightarrow \tilde{a}  {\frac{\ell_p^3}{R^2}}. \label{eq:decouple1}
\end{eqnarray}
The radius of the string direction, $R$, is introduced in these rescalings to keep the quantities $\tilde r$, $\tilde m$, $\tilde a$ dimensionless.  Likewise, the coordinates $t$ and $z$ are rescaled according to
\begin{equation}
\tau  = \frac{n \ell_p t }{R}, \qquad\qquad  \sigma = \frac{z}{R}.
\end{equation}
\item[$(ii)$]  The number, $n$, of M5-branes, and the number, $N$, of M2-branes.   Following \cite{Emparan:2006mm}, assuming that the supergravity solution can be modeled by free M2- and M5-branes, the number of branes is then given by the following quantisation
\begin{equation}
n= \frac{2}{\sqrt{3}} \left( \frac{\pi}{4 G_5} \right)^{1/3} Q_\mathrm{M} = \frac{2Q_\mathrm{M}}{\sqrt{3} \ell_p}, \qquad N = \frac{1}{\sqrt{3}} \left( \frac{4G_5}{\pi} \right)^{1/3} Q_\mathrm{E} = \frac{\ell_p Q_\mathrm{E}}{\sqrt{3}}.  
\end{equation}
\item[$(iii)$] The quantisation of the linear momentum, $P_z$, in integer units, $n_p$, of $\frac{1}{R}$,
\begin{equation}
n_p = R P_z = \frac{2 \pi R^2 Q_1}{4 G_5} = \frac{2 R^2 Q_1}{\ell_p^3}.
\end{equation}
\end{itemize}

In terms of the parameters $\beta, \delta, \gamma$ of the general string, this limit is formally obtained via
\begin{equation}
\ell_p^2 e^{2\beta} \rightarrow \frac{n R^2}{\tilde m}, \qquad \ell_p^{-1}\delta e^{\gamma} \rightarrow \frac{2 N}{3 n R}, \qquad \ell_p^{-2}e^{-2\gamma} \rightarrow  \frac{B}{2 n R^2},\label{eq:decouple2}
\end{equation}
where
\begin{equation}
B=  n_p + \frac{3 N^2}{4n} + \sqrt{4 \, \tilde{m}^2 + \left( n_p + \frac{3 N^2}{4n}\right)^2}  \label{bigB}
\end{equation}
is non-negative. This decoupling limit is a near-extremal limit since the temperature \eqref{eq:T} vanishes as $\ell_p\rightarrow0$. Moreover, it also corresponds to a near-horizon limit since $r\rightarrow 0$ and $r_{\pm} \rightarrow 0$ as $\ell_p \rightarrow 0$. In this near-extremal, near-horizon limit, the general string \eqref{eq:monster}--\eqref{monster_metric} reduces to a Ba\~nados-Teitelboim-Zanelli (BTZ) black hole\cite{Banados:1992wn}, with AdS radius $R_{AdS}$, times a two-sphere of radius $R_{S^2}$
\begin{equation}
ds^2_{\mbox{\tiny decoupling}} =  ds^2_{\mathrm{BTZ}} + R_{S^2}^2d\tilde\Omega_2^2,
\end{equation}
while the gauge field admits a magnetic flux on the $S^2$
\begin{equation}
{\bf A} = \sqrt{3}R_{\mathrm{S}^2}\cos\theta {\bf d} \tilde\phi.
\end{equation}
This factorised form is obtained by introducing a new radial coordinate, $\rho$, and shifting to a co-moving frame with angular coordinate, $\tilde \phi$, according to
\begin{equation}
\tilde r = \tilde m - \frac{n_p}{2}-\frac{3 N^2}{8n}+\frac{\tilde a^2 - 2 \tilde m^2}{B} + \frac{n}{2 l_p^2}\rho^2,
\end{equation}
\begin{equation}
\tilde \phi = \phi -  \frac{2 \sqrt{2} \tilde a}{R \sqrt{n^3 B}} \left( t - z \right).
\end{equation}
With these definitions, the BTZ metric takes the standard form
\begin{align}
&ds^2_{\mathrm{BTZ}} = - {\mathfrak N}^2 d \tau^2 + {\mathfrak N}^{-2} d\rho^2 + \rho^2 (d \sigma + N_\sigma d \tau)^2, \\
&{\mathfrak N}^2 = \frac{\rho^2}{R^2_\mathrm{AdS}} - 8 G_3 M_3 +  \frac{16 G_3^2 J_3^2}{\rho^2}, \quad\qquad N_\sigma =  \frac{4G_3 J_3}{\rho^2},
\end{align}
while the metric on the $S^2$ is simply
\begin{equation}
d\tilde\Omega_2^2 = d\theta^2 + \sin^2\theta d\tilde \phi^2.
\end{equation}
In the decoupled geometry, the radii of the three dimensional AdS space and the two-sphere are $ R_{\mathrm{AdS}} = 2 R_{\mathrm{S}^2} = n \ell_p,$ while the mass and angular momentum parameters of the BTZ black hole are,
\begin{eqnarray}
M_3 &=& \frac{1}{4 n^2 \ell_p}\left( 4 n n_p + 3 N^2 + \frac{ 8 n(2 \tilde{m}^2  - \tilde{a}^2)}{B} \right),  \\
J_3 &=&   n_p + {\frac{3 N^2}{4 n}} + \frac{ 2 \tilde{a}^2}{B}.
\end{eqnarray}
In arriving at these expressions the effective three dimensional Newton's constant has been related to the five dimensional one via
\begin{equation}
G_3  = \frac{G_5}{A_2}  = \frac{\ell_p}{4 n^2}, \label{eq:G5}
\end{equation}
where $A_2$ is the area of the two-sphere.

In terms of the unrestricted parameters of the general black string, the Bekenstein-Hawking entropy is given by
\begin{equation}
S_{BH}=\frac{4\pi^2Rmc_\delta^3}{G_5}\left(\left(c_\beta^4r_++s_\beta^4r_-+\frac{6ms_\beta^2c_\beta^2s_\delta^2}{1+3c_\delta^2}\right)c_\gamma-\frac{2s_\beta c_\beta(c_\beta^2r_++s_\beta^2r_-)}{\sqrt{1+3c_\delta^2}}s_\gamma\right).
\end{equation}
Upon carefully taking the decoupling limit (\ref{eq:decouple1}) and (\ref{eq:decouple2}), using (\ref{eq:G5}) to write $G_5=\pi\ell_p^3/4$ and finally sending $\ell_p\rightarrow0$, the entropy yields a finite result, which can be written
\begin{equation}
S_{BH}=2\pi\left(\sqrt{\frac{n^3B}{2}}+\sqrt{\frac{2n^3(\tilde m^2-\tilde a^2)}{B}}\right). \label{eq:BHEntropy}
\end{equation}

In section \ref{matching} I will show that this entropy exactly matches the entropy via a statistical-mechanical counting based on the Maldacena-Strominger-Witten CFT.

\subsection{Matching with the CFT prediction}
\label{matching}

The eleven dimensional M-theory picture of the black string constructed in this chapter corresponds to M-branes wrapping various cycles of the $T^6$, on which the theory is compactified to five dimensions.  Specifically, three equal number sets of M2-branes wrap the canonical (12), (34) and (56) two-cycles, while three equal number sets of M5-branes wrap the four-cycles orthogonal to the (12), (34) and (56) two-cycles.  Furthermore, all of the M5-branes wrap the Kaluza-Klein direction and hence intersect along the string, along which linear momentum is also carried.  This brane intersection can be succinctly represented by the following table:
\begin{eqnarray}
\begin{array}{c c c c c c c c c c c c}
     & t   &   z_1  &    z_2&   z_3  &   z_4  &   z_5   & z_6    &     z &    r   &   \theta  & \phi  \\
  \mathrm{M2} & -   &    -   &      -& \sim  &  \sim & \sim   & \sim  &  \sim &  \cdot &   \cdot   & \cdot \\
  \mathrm{M2}  & -   & \sim  & \sim &   -    &   -    & \sim   & \sim  &  \sim &  \cdot &   \cdot   & \cdot \\
  \mathrm{M2}  & -   & \sim  & \sim & \sim  &  \sim & -       & -      &  \sim &  \cdot &   \cdot   & \cdot \\
  \mathrm{M5}   & -   & \sim  & \sim & -      &  -     &   -     &   -    &  -    &  \cdot &   \cdot   & \cdot \\
  \mathrm{M5}  & -   &    -   &      -& \sim  &  \sim &  -      &    -   &  -    &  \cdot &   \cdot   & \cdot \\
  \mathrm{M5}  & -   &    -   &      -&  -     &   -    & \sim   & \sim  &   -   &  \cdot &   \cdot   & \cdot \\
  \mathrm{P} & -   & \sim  & \sim & \sim  &  \sim & \sim   & \sim &- &  \cdot &   \cdot   & \cdot \\
\end{array}
\label{M2M5}
\end{eqnarray}
where, following the conventions of \cite{Peet:2000hn}, a $-$ indicates a world-volume direction of the brane, a $\sim$ indicates that the brane is smeared in that direction, while a dot $\cdot$ indicates that the brane is point-like in that direction.

In the decoupling limit, in which the theory on the branes decouples from the bulk, the intersection of the M5-branes along $z$ leads to an \emph{effective string} in five dimensions whose world-sheet theory is the two dimensional $(4,0)$ Maldacena-Strominger-Witten (MSW) CFT\cite{Maldacena:1997de} as proposed in \cite{Bena:2004tk,Cyrier:2004hj,Emparan:2004wy}.  In \cite{Larsen:2005qr}, working under the hypothesis that near-extremal black \emph{rings} can be identified with black strings in the straight string limit, an entropy formula of the near-extremal black ring was proposed.  This identification is motivated by the fact that the $S^1\times S^2$ topology of the horizon of the ring is the same as that of a black string, which is well described by the MSW CFT.   The analysis of \cite{Larsen:2005qr} extends beyond the simplifying scenario of minimal supergravity, in which all three sets of M5-branes carry the same charge and similarly for the M2-branes, so it is expected that the CFT result should reproduce the Bekenstein-Hawking entropy (\ref{eq:BHEntropy}) of the previous subsection.  Here I show that this is indeed the case.  A clear and in depth derivation of the results is provided in \cite{Larsen:2005qr} so in what follows I will simply outline the salient features and quote the important results.

The CFT expression for the entropy is given by Cardy's formula
\begin{equation}
S = 2 \pi \left( \sqrt{ \frac{c \, h_{\mathrm{L}}^{\mathrm{irr}}}{6}}+\sqrt{ \frac{c \, h_{\mathrm{R}}^{\mathrm{irr}}}{6}}\right), \label{entropylarsen}
\end{equation}
where the central charge, $c$, is
\begin{equation}
c= 6 n^3, \label{central}
\end{equation}
$n$ being the (equal) number of M5-branes along the four-cycles.  In \eqref{entropylarsen}, $h_{\mathrm L}^{\mathrm{irr}}$ and $h_{\mathrm R}^{\mathrm{irr}}$ are respectively the left and right moving oscillator levels of the CFT. In the MSW CFT, M2-branes are realised as charged excitations.  As a result, the oscillator levels are shifted by an amount proportional to the M2-brane charges:
\begin{eqnarray}
h_{\mathrm L}^{\mbox{\tiny irr}} &=&\frac{\epsilon + n_p}{2},\\
h_R^{\mbox{\tiny irr}} &=& \frac{\epsilon - n_p}{2}- \frac{3 N^2}{4 n} -  \frac{J^2}{4 n^3},
\end{eqnarray}
where $N$ is the (equal) number of M2-branes wrapping the two-cycles, $\epsilon$ is the excitation energy, $n_p$ is the momentum quantum number along the $z$-direction, and $J$ is the quantum number under a U(1) subgroup of the SU(2) R-symmetry group. The SU(2) R-symmetry is interpreted as the rotation group in the four dimensional space-time transverse to the string; $J/2$ is simply the angular momentum on the $S^2$.  With respect to reference \cite{Larsen:2005qr} I have set $q^1=q^2=q^3=n$ and $Q_1=Q_2=Q_3=N$ since the general black string under consideration is a solution of minimal supergravity.

In order to facilitate a match to the Bekenstein-Hawking entropy, the CFT quantities need to be expressed in terms of the supergravity quantities.  From the quantisation condition of the magnetic charge:
\begin{eqnarray} 
n= \frac{2}{\sqrt{3}\ell_p} Q_M,
\end{eqnarray}
which implies that a single M5-brane carries the charge
$Q_M^\mathrm{unit}=\sqrt{3}\ell_p/2$.  The variable $\epsilon$ is interpreted as the energy above the BPS ground state, quantised in units of $\frac{1}{R}$. 
The mass of a single M5-brane is $M^\mathrm{unit}_{\mathrm{M5}} = \frac{R}{\ell_p^2}$, so $\epsilon$ is found to be
\begin{eqnarray}
\epsilon = R (M_{5} - 3 M_{\mathrm{ M5}}),
\hspace{1cm}
\mbox{where}
\hspace{1cm}
M_{\mathrm{M5}}= n  M^\mathrm{unit}_{\mathrm{M5}}  = \frac{n R}{\ell_p^2}.
\end{eqnarray}
In the strict decoupling limit, I find $\epsilon$ to explicitly take the following form in terms of the supergravity variables of the previous subsection:
\begin{eqnarray}
\epsilon =  \frac{3 N^2}{4n} + \sqrt{4 \tilde m^2  + \left(n_p + \frac{3 N^2}{4n}\right)^2} = B - n_p,
\end{eqnarray}
where $B$ is given by \eqref{bigB}. 
The half-integer quantised angular momentum, $J_\phi$, is half the integer quantised charge $J$:
\begin{equation}
\frac{J^2}{4 n^3} = \frac{J_{\phi}^2}{n^3} = \frac{2 \tilde a^2}{B}.
\end{equation}
Putting this all together, I find the left and right moving oscillator levels to be
\begin{align}
h_\mathrm{L}^{\mathrm{irr}} ={}&  \frac{B}{2}, \nonumber \\
h_\mathrm{R}^{\mathrm{irr}} ={}& \frac{2 (\tilde m^2-\tilde a^2)}{B}. \label{conformalweights2}
\end{align}
When these are put into the statistical entropy formula (\ref{entropylarsen}), the expression (\ref{eq:BHEntropy}) is exactly recovered.  Thus, the Bekenstein-Hawking entropy of the general string in the decoupling limit is reproduced by the MSW CFT, as expected.

\section{Limits of the General Black String}
\label{sec:limits}

The general black string is both electrically and magnetically charged and has finite temperature.  For a general choice of parameters $m,~a,~\delta,~\beta,~\gamma$, the solution is quite complicated; in this section I discuss simpler solutions that can be obtained either in the zero-temperature limits or in the limits when either the electric or the magnetic charge vanishes.

\subsection{Extremal Limits}
\label{extremal_limits}

Any regular extremal limit with finite entropy per unit length  has $F[c_\gamma,s_\gamma]$ finite in \eqref{eq:A}. Since the temperature has the form \eqref{eq:T}, extremality is therefore equivalent to $r_+ = r_-$, which is achieved either by the maximal rotation in the transverse space, equivalent to $a = \pm m$, or when $a$ and $m$ both go to zero at different rates while still maintaining the condition $r_+ = r_-$. After careful analysis it can be shown there are three distinct extremal limits:
\begin{itemize}
\item[$(i)$] $a = \pm m ,\quad m > 0$,
\item[$(ii)$] $ a = 0, \quad m\rightarrow 0, \quad \beta \rightarrow \infty, \quad$ such that
\begin{equation}
m e^{2 \beta} = 2 P,
\end{equation}
with $P$ finite and  keeping $\gamma$ and $\delta$ arbitrary free parameters.
\item[$(iii)$] $ a = 0, \quad m\rightarrow 0, \quad  \beta \to \infty ,\quad \delta \to 0, \quad  \gamma \to \infty,  \quad$  such that
\begin{equation}
m e^{2 \beta} = 2q, \qquad  m e^{\gamma} = \frac{4 \sqrt{-\Diamond}}{q}, \qquad \frac{\delta}{m} = \frac{Q}{3\sqrt{-\Diamond}},
\end{equation}
with $q$, $Q$ and $\Diamond$ finite, and $\Diamond < 0$.
\end{itemize}
In case $(i)$, the string displays maximal rotation in the space transverse to the string direction; its explicit metric, gauge field and thermodynamics are trivially obtained from the previous sections. Case $(ii)$ is the three-parameter supersymmetric $\mathrm{M2^3-M5^3-P}$ string found in \cite{Kim:2010bf}\footnote{The fact that the parameters $\gamma$ and $\delta$ of the general string should be left free in the supersymmetric limit can be expected from the analysis of \cite{Elvang:2004xi, Emparan:2008qn}, where in the context of five dimensional U(1)$^3$ supergravity a similar thing happens.} that is described in more detail below. Finally, case $(iii)$ corresponds to the non-supersymmetric extremal string of \cite{Kim:2010bf} with independent $\mathrm{M2^3-M5^3-P}$ charges.

\subsubsection{Supersymmetric Limit}
\label{app:extremal1}

In four dimensions, the maximal ${\cal N}=8$ supergravity theory is known to possess a global $E_{7(7)}$ symmetry\cite{Cremmer:1979up}.  This symmetry acts only on the scalars and vectors of the theory, leaving the space-time geometry invariant.  As such, it was shown in \cite{Kallosh:1996uy} that the $E_{7(7)}$ invariant quartic polynomial, denoted $\Diamond$, is proportional to the square of the entropy of the black hole.  Furthermore, $\Diamond$ is related to the electromagnetic charges of the black hole and if $\Diamond>0$ the black hole is supersymmetric, otherwise it is not.  Recall that the black strings of five dimensional minimal supergravity can be viewed as black hole solutions to the four dimensional ${\cal N}=2$ theory -- obtained by reducing along the string direction -- which is a consistent truncation of the ${\cal N}=8$ theory.  Hence the $E_{7(7)}$ quartic invariant decends to a function of the electromagnetic charges in the ${\cal N}=2$ theory.  In the notation of section \ref{RedAlongSt} the result is\cite{Kim:2010bf}
\begin{equation}
\Diamond\big(Q_1,Q_2,P_1,P_2\big)=3\big(Q_2P_2\big)^2+6Q_1Q_2P_1P_2-\big(Q_1P_1\big)^2+4Q_2^3P_1+4Q_1P_2^3.
\end{equation}
Whether the black string is supersymmetric thus simply reduces to a question of whether the quartic invariant of the corresponding four dimensional black hole is positive.  For the black holes to correspond to solutions of five dimensional minimal supergravity, $P_1=0$ so this condition reduces to $\Diamond=3\big(Q_2P_2\big)^2+4Q_1P_2^3>0$.

The string in limit $(ii)$ is the $\mathrm{M2^3-M5^3-P}$ supersymmetric string of \cite{Kim:2010bf} and is connected to the supersymmetric $\mathrm{M5}^3-\mathrm{P}$ string. In ${\cal N}=2,~D=4$ language the corresponding black hole solution was previously known in the literature \cite{Shmakova:1996nz}.  Since these limits are themselves supersymmetric solutions to five dimensional minimal supergravity, the discussion in section \ref{sec:solutions} applies.  That is, the five dimensional metric and field strength take the form
\begin{align}
\label{eqn:metric5dto4dEuclidean}
&ds^2 = -\frac1{H^2} (dt + \omega)^2+H ds_{\cal B}^2,\\
&{\bf F} = -\sqrt{3} {\bf d}\left(\frac{1}{H}({\bf d}t+\omega)\right)+\frac{2}{\sqrt{3}} {\bf G}^+,\label{eqn:FieldStrength}
\end{align}
where ${\bf G}^+=\frac{1}{2H}\big({\bf d}\omega+\star_4{\bf d}\omega\big)$ and the base space $\cal B$ contains a tri-holomorphic Killing vector, $\partial_z$, implying it can be put into Gibbons-Hawking form
\begin{equation}
\label{eqn:GibbonsHawkingBaseSpace}
ds_{\cal B}^2= f^{-1}\left( dz + \vec a\cdot d\vec x \right)^2 + f d\vec x\cdot d\vec x,
\end{equation}
where $\nabla\times\vec a=\nabla f$, meaning $f$ is a harmonic function on ${\mathbb R}^3$.  Furthermore, the connection 1-form $\omega$ is decomposed as
\begin{equation}
\label{eqn:generalomega}
\omega = \omega_z ({\bf d} z+ \vec a\cdot{\bf d}\vec x) + \vec\omega\cdot {\bf d}\vec x ,
\end{equation}
where $\omega_z$ is a function on $\mathbb{R}^3$, whereas $\vec\omega$ is identically zero for the supersymmetric limit $(ii)$ of the general black string.  Since the solution describes a five dimensional black string, the harmonic function $f$ is a constant, hence one can fix $\vec a = 0$. The functions $\omega_z$ and $H$ are then written in terms of $f$ and three additional functions, $K,~L,~M,$ harmonic on ${\mathbb R}^3$, as
\begin{equation}
\label{eqn:fharmonic}
H= K^2 f^{-1}+L, \qquad \omega_z = K^3f^{-2}+\frac{3}{2} K L f^{-1}+ M.
\end{equation}
For the supersymmetric limit, the harmonic functions are given by
\begin{align}
f ={}&\frac{Q}{S},\qquad\qquad\>\> K =  \frac{\Delta}{S}+ \frac{q}{2r},\label{harm1} \\
L  ={}&\frac{Q}{S}+ \frac{Q}{r}, \qquad M =  -\frac{1}{2}\left(\frac{\Delta}{S}+ \frac{3}{2r}(2\Delta -q)\right),\label{harm2}
\end{align}
where $S = \sqrt{Q^2+\Delta ^2}$ and the three independent parameters are given by
\begin{align}
q ={}& \frac{4 P c_\delta^2 }{\sqrt{1 + 3 c_\delta^2}},\\
Q={}& 2 P c_\delta s_\delta \left( c_\gamma + \frac{s_\gamma}{\sqrt{1 + 3 c_\delta^2}}\right),\\
\Delta ={}& \frac{2P}{3} \left( \frac{c_\gamma^2 + s_\gamma^2 + 3 c_\delta^2}{\sqrt{1+ 3 c_\delta^2}} - c_\gamma s_\gamma (1+ 3 s_\delta^2)\right).
\end{align}
In terms of these parameters, the non-zero four dimensional charges are
\begin{align}
Q_2 &= Q, & P_2 &= \frac{q}{2}, & Q_1 &= 3 \Delta -\frac{3 q}{2}, \label{eq:SUSYCharges}\\  
G_4 M &= \frac{3}{4} S,&\Sigma &= -\frac{S^2 + 2 \Delta (\Delta -q)}{2 S}, & \Xi &= \frac{Q (\Delta -q)}{S}.
\end{align}
Note that $Q_2$, $P_2$ and $Q_1$ can be identified with the numerators of $1/r$ terms (up to numerical factors) in the harmonic functions \eqref{harm1}-\eqref{harm2}.  In this parametrisation, the quartic invariant is given by
\begin{equation}
\Diamond = \frac{3}{4} q^2 \left(Q^2 -q^2+2 \Delta  q\right),
\end{equation}
where $q, \Delta, Q$ are restricted to ensure the positivity of the quartic invariant. The ADM tension for the supersymmetric string is
\begin{equation}
{\cal T} = \frac{1}{8 \pi R G_4 }\left(3 \Sigma_s  + 2G_4 M \right) = \frac{3 }{4 G_5}\frac{(q-\Delta ) \Delta}{ \sqrt{Q^2+\Delta ^2}}.
\label{tension}
\end{equation}
Note that the supersymmetric string becomes tensionless either when $\Delta =0$ or when $\Delta = q$.  This has important consequences for supersymmetric black \emph{rings}, which will be elaborated on in section~\ref{sec:discussion}.

The above metric is presented in Gibbons-Hawking form in order to facilitate comparison with the formalism presented in \cite{Gauntlett:2002nw}, however as pointed out in \cite{Kim:2010bf}, as a result of this gauge choice the matrix $\cal M$ does not go to the identity at spatial infinity.  To return to the coordinates of the general string (\ref{eq:monster}), in which case $\cal M$ does asymptote to the identity, the following shift of the coordinates is required in \eqref{eqn:metric5dto4dEuclidean}--\eqref{eqn:generalomega}
\begin{equation}
z \rightarrow  z + \frac{\Delta }{\sqrt{Q^2+\Delta ^2}} \, t .
\end{equation}
Finally note that the solution expressed in the coordinates $t^\prime =\lambda t$, $z^\prime = z/\lambda$ still has the form \eqref{eqn:metric5dto4dEuclidean}--\eqref{eqn:generalomega} with the harmonic functions \eqref{harm1}--\eqref{harm2} transformed as
\begin{equation}
f^\prime = f/\lambda,\qquad L^\prime = \lambda L,\qquad K^\prime = K,\qquad M^\prime =\lambda^2 M .
\end{equation}
One can then more easily take the $Q \rightarrow 0$ limit after having performed the above change of coordinates with $\lambda = Q/S$.

\subsubsection{Non-Supersymmetric Limit}
\label{app:extremal2}

The limit $(iii)$ yields the three-parameter extremal \emph{non-supersymmetric} black string of minimal supergravity with independent $\mathrm{M2^3-M5^3-P}$ charges, which is most conveniently presented as\cite{Kim:2010bf}
\begin{align}
ds^2 = {}&V^2 ds^2_3 + 2 V^{-1} dt dz + V^{-4} g dz^2, \label{nonsusymetric3} \\
{\bf A} ={}& - \frac{\sqrt{3}}{2} q \cos \theta {\bf d} \phi - \frac{ \sqrt{3} Q}{r}\left(1+ \frac{q}{4r} \right) V^{-2} {\bf d}z, \label{nonsusymaxwell3}
\end{align}
where
\begin{eqnarray}
V &=& 1 + \frac{q}{2r}, \\
g &=& 1+\frac{3 \Delta }{r}-\frac{3\left(q^2-3 \Delta  q+Q^2\right) }{2 r^2} + \frac{9 q^2 \Delta -4 q \left(q^2+Q^2\right) }{4r^3} -\frac{\Diamond}{4r^4}.
\end{eqnarray}
The non-zero charges in this parametrisation are
\begin{equation}
G_4 M = \frac{3 \Delta }{4}, \quad Q_1= \frac{3}{2} (q-2 \Delta ), \quad Q_2 = Q, \quad P_2  = \frac{q}{2}, \quad \Sigma = q-\frac{3 \Delta }{2}, \quad \Xi = -Q,
\end{equation}
and the quartic invariant for the corresponding four dimensional black hole is found to be
\begin{equation}
\Diamond = - \frac{3}{4} q^2 \left(2 \Delta  q - q^2 - Q^2\right), \label{quarticnonsusy}
\end{equation}
where the parameters $q, \Delta, Q$ are restricted to ensure that the quartic invariant is strictly \emph{negative}.
The ADM tension for the solution is
\begin{equation}
\mathcal{T} = \frac{3}{8\pi R G_4} (q - \Delta),
\end{equation}
where now there is only the one condition, $\Delta = q$, for which the string becomes tensionless.  Again, the metric and gauge field as presented in \eqref{nonsusymetric3}--\eqref{nonsusymaxwell3} are such that the matrix $\mathcal{M}$ does not asymptote to the identity at spatial infinity.  If the $z$ coordinate is shifted according to $z \rightarrow z - t$, the resulting solution is identical to limit $(iii)$ of the general string \eqref{eq:monster}, in which case the matrix $\cal M$ does go to the identity at spatial infinity.

\subsection{Non-extremal Limits}
\label{electric}
\label{magnetic}

\subsubsection*{Spinning $\mathrm{M2}^3$ String}

The boosted spinning $\mathrm{M2}^3$ string is obtained by setting the magnetic charge to zero, \emph{i.e.} it is obtained by the limit $\beta \rightarrow 0$. The space-time fields are presented in \cite{Compere:2009zh}, where  a detailed discussion of physical properties of this solution is also included.  The details are omitted here, however it can be verified that the general string \eqref{eq:monster} indeed reproduces the results of \cite{Compere:2009zh} in the $\beta\rightarrow 0$ limit.  In this limit, $\gamma$ can directly be regarded as the boost parameter along the string. When $\gamma$ is non-zero the extremal limit of the string is never supersymmetric, however when $\gamma=0$ the solution admits a further limit to the supersymmetric $\mathrm{M2}^3$ string. It is obtained by taking $a=0$, $m \rightarrow 0,~ \delta \rightarrow \infty$ such that $m e^{2 \delta} = 2 Q. $ In this limit, the conserved charges are
\begin{equation}
G_4 M = \frac{3Q}{4}, \qquad Q_2 = Q, \qquad \Sigma = -\frac{Q}{2},
\end{equation}
which corresponds to the $q\rightarrow0,~\Delta\rightarrow0$ limit of the supersymmetric limit $(ii)$ above.  The quartic invariant of the corresponding four dimensional black hole gives $\Diamond=0$, meaning the entropy vanishes; this is indeed observed for the supersymmetric M2$^3$ string.  The charge matrix in this case becomes nilpotent and belongs to the $\mathcal{O}_2$ nilpotent $\tilde K$-orbit in the nomenclature of \cite{Kim:2010bf}.

\subsubsection*{Spinning $\mathrm{M5}^3$ String}

The boosted spinning $\mathrm{M5}^3$ string is obtained by setting the electric charge to zero, \emph{i.e.} it is obtained by the limit $\delta \rightarrow 0$.  The space-time fields and a detailed discussion of the physical properties are again presented in \cite{Compere:2009zh}, so these details are omitted here, but it can be verified that the results of \cite{Compere:2009zh} are reproduced in the limit $\delta\rightarrow0$.  A stability analysis of the limit $a\rightarrow0,~\gamma\rightarrow0$ of this magnetic black string will be studied in detail in chapter 4.
In the present limit, the boost parameter along the string is,
\begin{equation}
\sigma=\beta - \gamma.
\end{equation}
The string admits a supersymmetric limit even when the boost parameter $\sigma$ is non-zero.  It is obtained by taking the limit $a = 0$, $ m \rightarrow 0$, $\beta \rightarrow \infty,$ $\sigma \rightarrow \infty$ such that $m e^{2 \beta} = 2 P,$ and  $m e^{2 \sigma} = 2 Q.$ The conserved charges in this limit are
\begin{equation}
G_4 M = \frac{|Q| + 3|P|}{4}, \qquad Q_1 = Q, \qquad P_2 = P, \qquad \Sigma = \frac{|P|-|Q|}{2},
\end{equation}
and the corresponding charge matrix is nilpotent. The quartic invariant of the corresponding four dimensional black hole is
\begin{equation}
\Diamond = 4 QP^3.
\end{equation}
For $PQ> 0$ the quartic invariant is strictly positive and the corresponding string is supersymmetric; the charge matrix in this case belongs to the $\mathcal{O}_{3K}$ orbit in the nomenclature of \cite{Kim:2010bf}.  This string solution for which $Q = 3 P$ corresponds to the infinite radius limit of the extremal dipole black ring of \cite{Emparan:2004wy}.  The solution for which $Q = - 3 P$ also corresponds to the infinite radius limit of the extremal dipole black ring of \cite{Emparan:2004wy}, but with opposite sense of rotation in the plane of the ring.  For $PQ< 0$ the quartic invariant is strictly negative and the corresponding string completely breaks supersymmetry; the charge matrix in this case belongs to the $\mathcal{O}_{4K}'$ orbit in the nomenclature of \cite{Kim:2010bf}. When $Q= 0,$ $P \neq 0$ the charge matrix belongs to the $\mathcal{O}_2$ orbit, and finally, when $Q \neq 0,$ $P = 0$ the charge matrix belongs to the $\mathcal{O}_1$ orbit \cite{Kim:2010bf}.

\section{Discussion and implications for black rings}
\label{sec:discussion}

 In this chapter I have presented the most general black string solution to ungauged minimal supergravity in five dimensions, which has five independent parameters: energy above the BPS bound, angular momentum in the four dimensional transverse space, linear momentum along the string, smeared electric zero-brane charge, and magnetic one-brane charge.  The solution admits three separate extremal limits, one of which is supersymmetric and the other two of which are non-supersymmetric.  The physical properties and thermodynamics of the general black string were analysed and it was shown that the first law indeed holds.  In the near-extremal limit, the entropy of the general black string was shown to exactly match the CFT prediction of \cite{Larsen:2005qr}.

This general black string solution has a number of implications for yet-to-be-found black rings in five dimensional ungauged minimal supergravity; in this section I briefly discuss some of these important points.

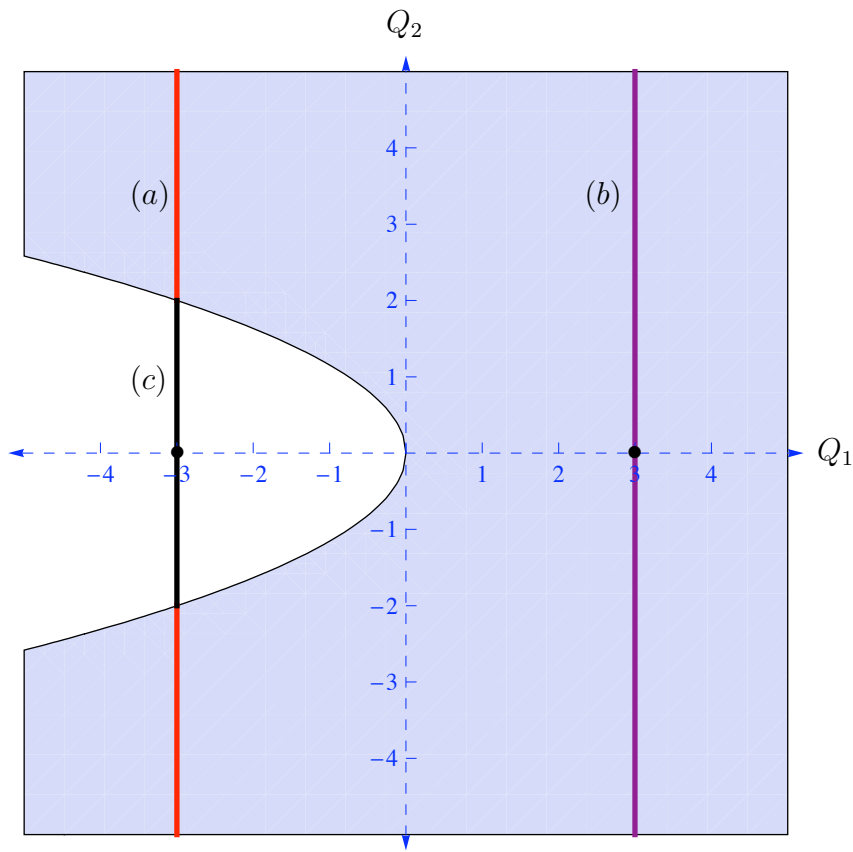
\begin{figure}[htp]
  \begin{center}
\includegraphics[scale=1.0]{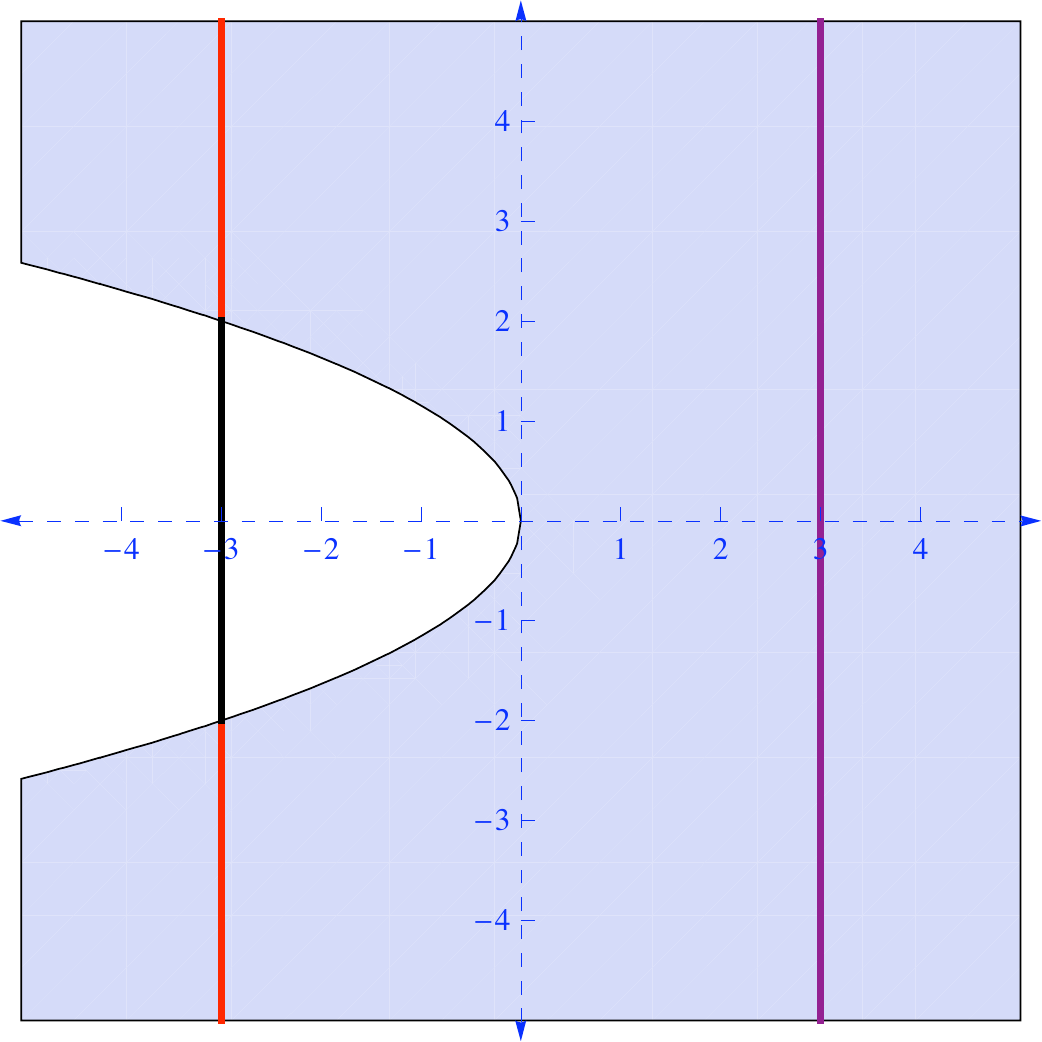}
\put(-255,245){$(a)$}
\put(-255,175){$(c)$}
\put(-240,148){$\bullet$}
\put(-67,148){$\bullet$}
\put(-83,245){$(b)$}
\put(5,148){$Q_1$}
\put(-158,310){$Q_2$}
\caption[Phase Diagram for Extremal Black Rings]{A working phase diagram for a class of extremal black rings of five dimensional ungauged minimal supergravity. It is obtained from imposing a tensionless condition on the extremal black strings of section \ref{app:extremal1} according to the assumptions of the blackfold approach. Rotation in the transverse space vanishes in this diagram, so the corresponding black ring will also have vanishing angular momentum along the $S^2$. The vertical axis is the M2$^3$ charge $(Q_2)$ and the horizontal axis is the boost charge $(Q_1)$; the M5$^3$ charge is held fixed ($P_2=1$). The shaded region corresponds to the parameter space in which the quartic invariant is positive $\Diamond = 3 Q_2^2 + 4 Q_1 >0$ and hence the black strings are supersymmetric.  Similarly, the unshaded region corresponds to the parameter space with negative quartic invariant and the black strings are non-supersymmetric.  The tensionless condition is satisfied only on the two vertical lines, which thus show the allowed parameter space for extremal black rings for a fixed value of the dipole charge.  Branch $(a)$ corresponds to the supersymmetric black ring of \cite{Elvang:2004rt}. Black rings corresponding to branches $(b)$ and $(c)$ are yet to be discovered.}
\label{fig:phase}
\end{center}
\end{figure}

\subsubsection*{New extremal branches of black rings}

In section \ref{app:extremal1} it was noted that there are two tensionless limits of the three-parameter supersymmetric black string, \emph{i.e.} extremal case $(ii)$, and that there is one tensionless limit of the three-parameter extremal non-supersymmetric black string, \emph{i.e.} extremal case $(iii)$. The blackfold approach \cite{Emparan:2009cs, Emparan:2009at, Emparan:2009vd} suggests that all tensionless black strings describe the infinite radius limit of some black ring; since the black strings are all extremal, the corresponding black rings will also be extremal. Furthermore, since considerations of \cite{Kim:2010bf} ensure that all possible extremal black strings of minimal ungauged supergravity with no rotation in the transverse space have been exhausted, the tensionless extremal black strings of section \ref{app:extremal1} provide a working phase diagram for a class of extremal black rings of this theory. To draw this phase diagram I collect important properties of the extremal tensionless black strings from section \ref{app:extremal1}. This is summarised in the table below. For concreteness I only consider $q > 0$; $q < 0$ can be dealt with similarly.
\vskip 0.3cm
\begin{center}
\begin{tabular}{|c|c|c|}
\hline
$(a)$:  $\Delta = 0$ susy & $(b)$:  $\Delta = q$ susy & $(c)$:  $\Delta = q$ non-susy \\ \hline
$Q_2 = Q$ & $Q_2 = Q$ & $Q_2 = Q$ \\ 
\hline
$P_2 = q/2$ & $P_2 =q/2$ & $P_2 = q/2$ \\ 
\hline
$Q_1 = -3q/2$ & $Q_1 =  +3q/2$ & $Q_1 = -3q/2$ \\ 
\hline
$|Q_2| > q $ & no condition & $|Q_2|< q$ \\ 
\hline
$M_5 = \sqrt{3} |Q_E|$ & $M_5 = \sqrt{3} |Q_E| \sqrt{1 + q^2/Q^2}$ & $M_5 = 3 \pi R q/2G_5$  \\ \hline
\end{tabular}
\end{center}
\vskip 0.2cm
From this table one can immediately draw a number of conclusions.
The BPS string in case $(a)$ is the black string of \cite{Bena:2004wv}, which in turn corresponds to the infinite radius limit of the supersymmetric black ring of \cite{Elvang:2004rt}. For this string, the electric charge $Q$ is bounded from below $|Q| > q$; as this bound is violated, solution $(a)$ ceases to be smooth. Precisely when solution $(a)$ becomes singular, the tensionless non-supersymmetric solution, $(c)$, takes over. Branch $(c)$ is completely smooth: one can continuously take the electric charge to zero. Thus the corresponding black ring is connected to an extremal dipole ring, represented as the black dot on branch $(c)$ in figure \ref{fig:phase}. This dipole ring is such that in the infinite radius limit, the boost along the string completely breaks supersymmetry. Black ring solutions on this branch are not currently known. Case $(b)$, on the other hand, indicates that there is another branch of extremal black rings of  minimal supergravity and that this branch is also connected to a dipole ring, represented as the black dot on branch $(b)$ in figure \ref{fig:phase}.  For this dipole ring, boost along the string in the infinite radius limit \emph{preserves} supersymmetry.  Black ring solutions on this branch are also not currently known.  Within the assumptions of the blackfold approach, the phase diagram in figure \ref{fig:phase} thus predicts two new classes of extremal black rings, one of which is supersymmetric, the other of which is non-supersymmetric but is continuously connected to the known class that is supersymmetric.

\subsubsection*{Most general black ring:}

\begin{figure}[htp]
  \begin{center}
    \subfigure{\label{fig:non-susy}\includegraphics[scale=0.3]{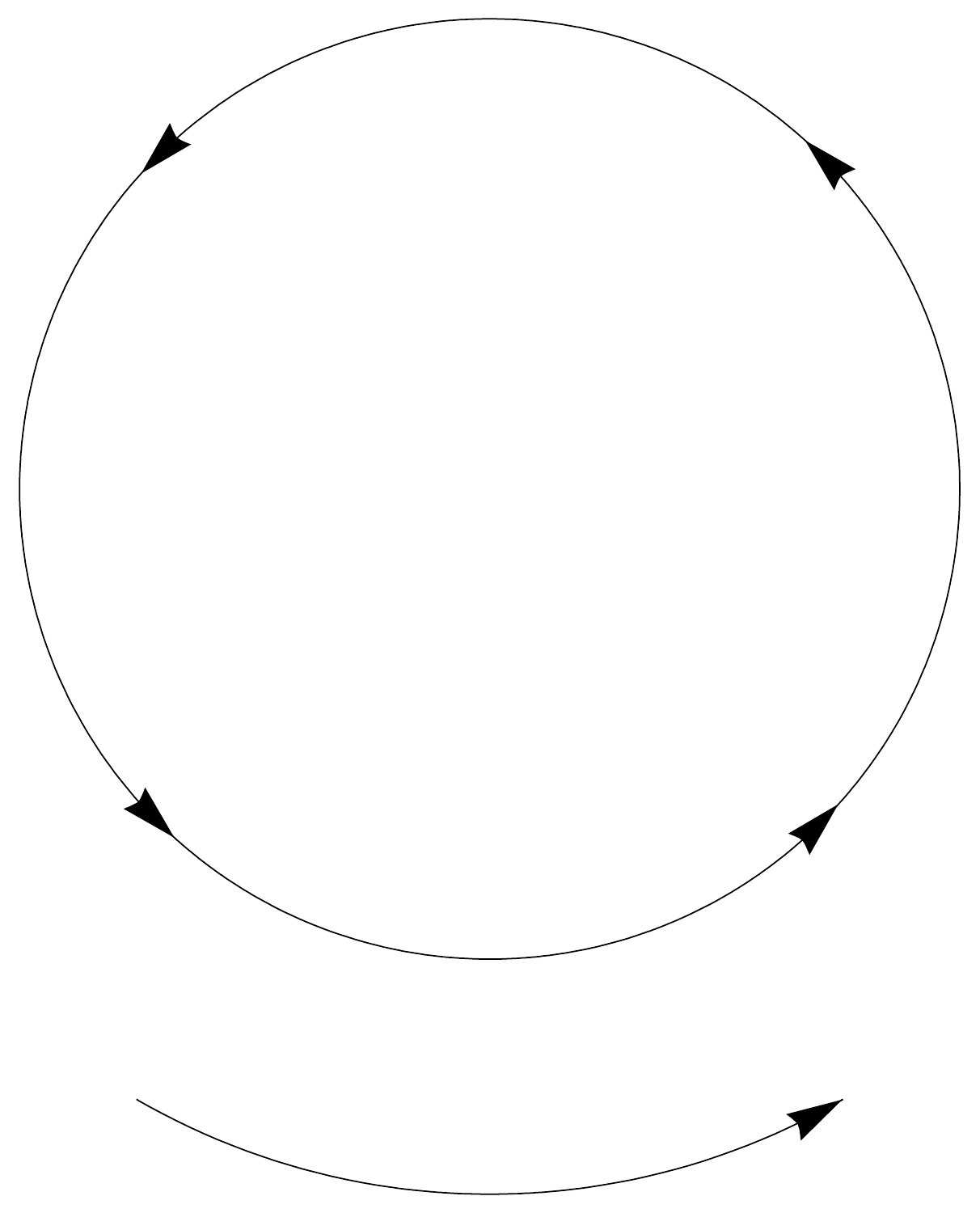}}
    \hskip 2.5cm
   \subfigure{\label{fig:susy}\includegraphics[scale=0.3]{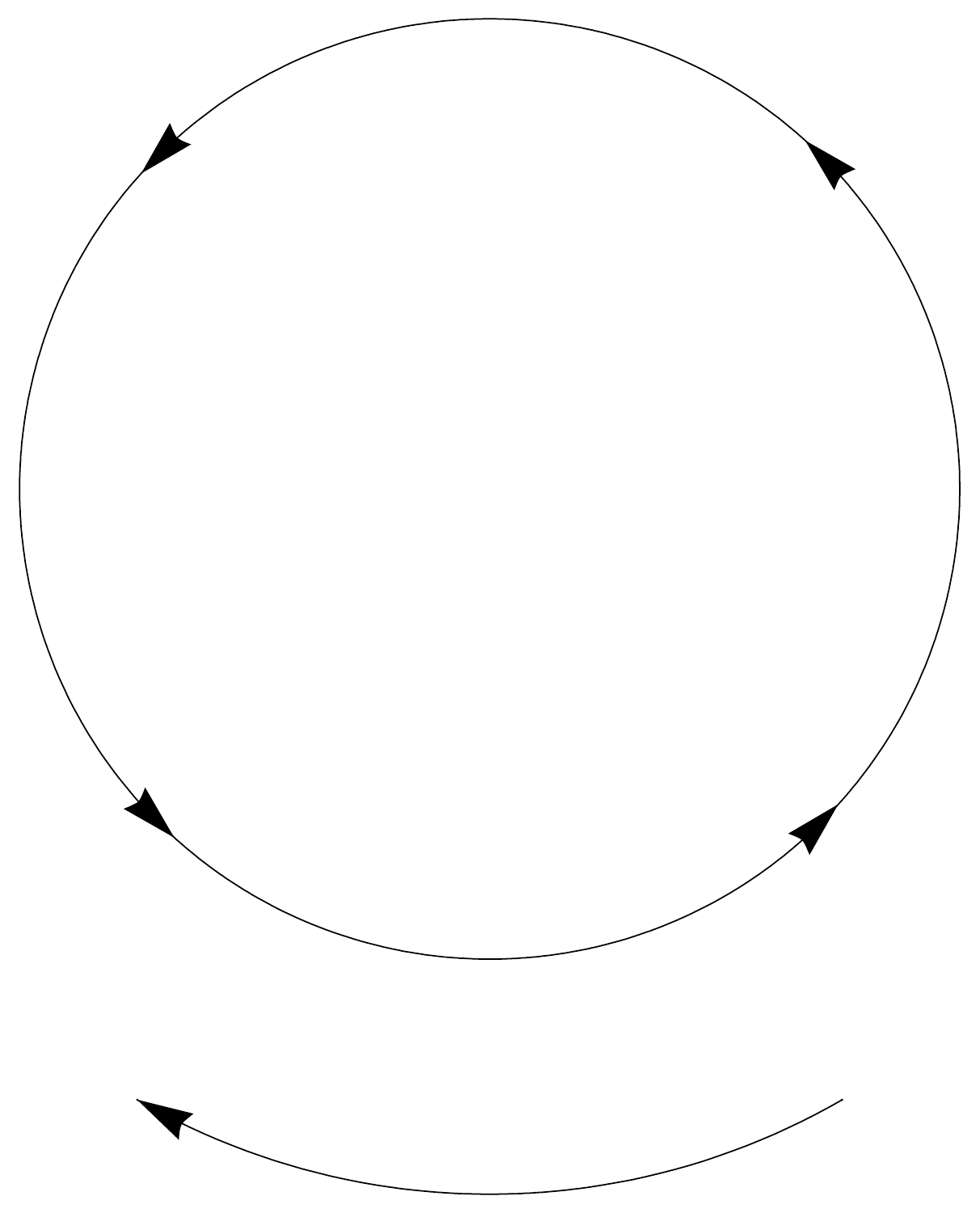}} \\
  \end{center}
  \caption[Aligned and Anti-Aligned Black Rings]{The two branches of tensionless strings bent into rings. The aligned string (left) has the same orientation for the momentum along the string and the magnetic charge, while the anti-aligned string (right) has the opposite relative orientation.}
  \label{fig:rings}
\end{figure}

A five-parameter family of general black ring solutions is conjectured to exist in minimal supergravity \cite{Elvang:2004xi}. For ranges of parameters for which this family would admit the infinite radius limit, the corresponding string solution would be contained in the general string presented in this chapter.  One can show that for the general black string, the tension \eqref{eq:T2} can be set to zero for two distinct values of the boost parameter when the remaining parameters are left arbitrary and finite. The two distinct branches of tensionless strings are distinguished by the relative sign of $Q_1$ and $P_2$, or in five dimensional language, by the property that the orientation of the momentum along the string is aligned or anti-aligned with the orientation of the magnetic charge $Q_\mathrm{M}$. I refer to these two branches as the \emph{aligned} tensionless string and the \emph{anti-aligned} tensionless string. The former contains branch $(b)$ of the tensionless extremal strings while the latter contains branches $(a)$ and $(c)$ of figure \ref{fig:phase}.  Aligned and anti-aligned rings are depicted in figure \ref{fig:rings}, in which the direction along the circle corresponds to the orientation of the magnetic charge while the arrow segments beneath indicate the boost direction.

Without loss of generality, the magnetic charge is taken to be positive, so the aligned and anti-aligned tensionless strings have $P_z > 0$ and $P_z <0$ respectively.  The boost parameter $\gamma$ in these two tensionless limits is rather complicated for general $\beta,\delta$ so I refrain from writing it out explicitly.  Nevertheless, these values can be easily implemented in programs such as \emph{Mathematica}.  I find that the linear momenta satisfy $P_z^{\mathrm{anti-aligned}}=-P_z^{\mathrm{aligned}}$ for arbitrary $\beta,\delta$ and that the magnitude of the linear momentum is bounded from below
\begin{equation}
\frac{\sqrt{2}}{3} M^{\mathrm{neut}}_5 \leq |P_z|,
\end{equation}
where $M_5^\mathrm{neut} = \frac{3\pi m R}{G_5}$ is the mass of the neutral tensionless black string (\emph{i.e.} $\delta = \beta = 0$) and equality only holds in the neutral limit. The fact that the magnitude of the linear momentum is bounded from below indicates that the two branches are always separated. Furthermore for a fixed value of the mass and magnetic charge, the two branches of tensionless strings have different electric charges, which always obey
\begin{equation}
Q_\mathrm{E}^\mathrm{aligned} \geq Q_\mathrm{E}^\mathrm{anti-aligned}.  \label{eq:QIneq}
\end{equation}
The equality holds only when there is no magnetic charge or when the electric charge simply vanishes.  This can be understood by the coupling of an object with angular momentum, $\vec L$, to a magnetic field, $\vec B$: the coupling energy is $E\sim -\vec L\cdot \vec B$.  In the present case, making the connection $\vec L \rightarrow P_z$ and $\vec B\rightarrow Q_M$, the aligned string has a lower energy than the anti-aligned string at fixed electric charge.  The electric charge contributes positively to the overall energy of the string, so at fixed \emph{mass} and magnetic charge, the aligned string must have a higher electric charge to compensate for its lower coupling energy.  When $Q_M=0$ there is no coupling and the inequality (\ref{eq:QIneq}) turns to equality.

The presence of two distinct tensionless limits offers two possibilities for the most general black ring solution. The first is that there are two distinct classes of general black rings, each admitting an infinite radius limit which rotate in different directions relative to the orientation of the dipole charge. The second is that in the regime where the blackfold approach cannot be trusted, the two branches merge. If the latter case is true it would imply the existence of a black ring with no angular momentum along the $S^1$; for this to be stable against collapse, the electric charge would necessarily need to be non-zero.


\chapter{Stabilising Kaluza-Klein Vacua with Magnetic Flux}


In the last two chapters, I constructed solutions to five dimensional minimal supergravity but I did not consider possible instabilities in these solutions.  The Atiyah-Hitchin solution of chapter 2 is expected to be stable because it is BPS, however the five parameter black string of chapter 3 is not supersymmetric in general.  In this chapter, I analyse the stability properties of the magnetic black string, which is the limit $a=\delta=\gamma=0$ of the general black string.

Since the seminal papers of Hawking and Page \cite{Hawking:1982dh} and of Gregory and Laflamme \cite{Gregory:1993vy}, the stability of black objects, particularly in various higher dimensional gravity theories, has become a particularly active vein of research.  In the modern language of the AdS/CFT correspondence, the Hawking-Page phase transition between AdS black holes and thermal AdS, which is thermodynamic in nature, is dual to a confinement-deconfinement phase transition in the large $N$ limit of the boundary gauge theory \cite{Witten:1998zw}.  Meanwhile the qualitatively unrelated Gregory-Laflamme instability, which manifests as unstable modes in linearised metric perturbations at the horizon, has been shown to be analogous to the Rayleigh-Plateau instability of classical membranes \cite{Cardoso:2006ks} through the fluid/gravity correspondence.  Thus there are two seemingly unrelated but equally important forms of stability pertaining to black holes\footnote{There are other types of black hole instabilities, one of which will be explored in detail in chapter 6.}: thermodynamic and perturbative.  The Gubser-Mitra conjecture \cite{Gubser:2000mm}, also known as the correlated stability conjecture, asserts that there exists a correlation between the thermodynamic and perturbative stabilities of black branes, the idea being that Gregory-Laflamme instabilities arise when it is entropically favourable for the brane to bifurcate into a series of black holes.  However, this conjecture has since been shown to be violated in some cases, such as dilatonic magnetically charged black p-branes, which have unstable modes in the dilaton even when the specific heat is positive \cite{Friess:2005zp}.

The perturbative (Gregory-Laflamme) stability of non-dilatonic black strings with magnetic charge and mass parameters $P$ and $m$, consistent with the notation throughout this chapter, has been demonstrated to hold in the parameter range $P\le m\le\frac3{2\sqrt{2}}P$, which corresponds to the range in which the heat capacity is positive~\cite{Miyamoto:2007mh}.   While this was regarded by the author of \cite{Miyamoto:2007mh} as evidence that the Gubser-Mitra conjecture is satisfied for these objects, a proper thermodynamic stability analysis was not carried out because it was unclear what the magnetic black string could phase transition to.  For neutral and electrically charged black strings, there exist spherical black holes that are (un)charged under the same gauge field as the string, so under a Gregory-Laflamme instability the black string is able to bifurcate into a series of black holes while conserving charge.  This is not true for strings carrying a topological magnetic charge because such a charge must have flux on a sphere of the same dimension as the rank of the gauge field.   This means that spherical magnetically charged black holes are charged under a larger rank gauge field than the black string, so charge conservation must be violated if the string bifurcates.  The question is then begged: do there exist other objects with the same topological charge as the black string?  In this chapter, which is based on the work published in \cite{Stotyn:2011tv}, I demonstrate that if one considers antiperiodic boundary conditions for fermions, that is to say supersymmetry-breaking boundary conditions, there indeed exists a two-parameter family of topological solitons that carries the same magnetic charge as the black string. This two-parameter family is subdivided into ``small" and ``large" solitons, which I will show contribute dramatically to the thermodynamic stability of magnetic black strings.  

It is, however, well known that asymptotically Kaluza-Klein space-times with supersymmetry breaking boundary conditions have rather fatal instabilities themselves.  For instance, in \cite{Witten:1981gj} Witten showed that in empty space with a Kaluza-Klein direction, ``bubbles of nothing" can nucleate and expand exponentially fast, eating up the entire space-time.  Brill and Horowitz \cite{Brill:1991qe} then expanded on the work of Witten by considering an asymptotically Kaluza-Klein space-time with initial data on a Cauchy surface of time symmetry.  They demonstrated that such space-times with antiperiodic fermionic boundary conditions are unstable to the nucleation of dynamic bubbles of arbitrary negative energy, which again expand and consume the space-time.  These results hold for vacuum Einstein and Einstein-Maxwell with electric charge as well as magnetic charge with flux around the periodic dimension.  Further investigations have revealed that topological solitons carrying the appropriate Maxwell charges in these theories generically correspond to perturbatively unstable static bubbles \cite{Dine:2006we,Sarbach:2003dz,Sarbach:2004rm} and are therefore a subset of the dynamic bubbles considered in \cite{Brill:1991qe}. 

However, bubbles carrying a topological magnetic charge are not covered by the above analysis.  In this chapter I will demonstrate that the addition of a topological magnetic charge locally stabilises the Kaluza-Klein vaccum\footnote{It should be understood that this is a local vacuum.  There are bubbles with arbitrary negative energy and so a true vacuum cannot be defined.}, which is a small static magnetically charged topological soliton.

\section{Magnetically Charged Bubbles}

In this section, I introduce the magnetic black string and construct static magnetically charged bubble solutions, as well as the more general dynamic bubbles.

\subsection{Large and Small Topological Solitons}

The metric and gauge potential of the static, magnetically charged black string of five dimensional ungauged minimal supergravity \cite{Miyamoto:2007mh}, which can be obtained by taking the limit $\delta=\gamma=a=0$ of the general black string (\ref{eq:monster})-(\ref{eq:monsterFfields}), is given by:
\begin{align}
&ds^2=-f(r)dt^2+\frac{dr^2}{f(r)h(r)}+h(r)dz^2+r^2d\Omega_2^2, \label{eq:Metric}\\
&{\bf A}=-\sqrt{3}P\cos\theta {\bf d}\phi,\\
&f(r)=1-\frac{r_+}r, \quad\quad\quad\quad h(r)=1-\frac{r_-}r,
\end{align}
where $d\Omega_2^2$ is the metric on a unit two-sphere, and the outer and inner horizons are located at $r_\pm=m\pm\sqrt{m^2-P^2}$.  The magnetic charge and the ADM mass of the black string are related to the parameters $m$ and $P$ by
\begin{equation}
Q_M=\sqrt{3}P,\qquad 
M_{BS}=\frac{\pi R_s}{2G_5}\left(3m+\sqrt{m^2-P^2}\right), \label{eq:BSMass}
\end{equation}
where $R_s$ corresponds to the radius of the compact $z$ direction.  The extremal black string is obtained by setting $m^2=P^2$ and takes the explicit form
\begin{equation}
ds^2=\left(1-\frac{P}{r}\right)\big(-dt^2+dz^2\big)+\frac{dr^2}{\left(1-\frac{P}{r}\right)^2}+r^2d\Omega_2^2, \label{eq:MetricExtremal}
\end{equation}
which is identified as the extremal M$5^3$ string of \cite{Kim:2010bf}, but is not supersymmetric if antiperiodic boundary conditions for the fermions are imposed.

Observe that $\partial_t$ and $\partial_z$ are Killing directions and that the gauge field has the form ${\bf A}=A_\phi {\bf d}\phi$.  This permits the following Wick rotation of (\ref{eq:Metric}): $t\rightarrow iz$ and $z\rightarrow it$.  This operation yields the metric and gauge potential:
\begin{align}
&ds^2=-\tilde f(r)dt^2+\frac{dr^2}{\tilde f(r)\tilde h(r)}+\tilde h(r)dz^2+r^2d\Omega_2^2, \label{eq:MetricSol}\\
&{\bf A}=-\sqrt{3}P \cos\theta{\bf d}\phi, \\
&\tilde f(r)=1-\frac{r_c}r, \quad\quad\quad\quad \tilde h(r)=1-\frac{r_s}r,
\end{align}
where the critical radii are now $r_s=\mu+\sqrt{\mu^2-P^2}$ and $r_c=\mu-\sqrt{\mu^2-P^2}$, such that $r_s>r_c$.  A new mass parameter, $\mu$, has been introduced to stress that it is distinct from $m$.  The magnetic charge\footnote{Despite the space being simply connected, the magnetic charge is supported for these solitons because the gauge field carries flux on a non-minimal two-sphere at $r=r_s$} and the ADM mass of this solution are 
\begin{equation}
Q_M=\sqrt{3}P,\qquad
M_{Sol}=\frac{\pi R_s}{2G_5}\left(3\mu-\sqrt{\mu^2-P^2}\right). \label{eq:SolitonMass}
\end{equation}

The signature of the metric in the region $r_c<r<r_s$ is $(---++)$, so $r_s$ represents an inner boundary to the space-time.  These solutions represent topological solitons if the condition that there is no conical singularity at $r=r_s$ is imposed.  That is, the Kaluza-Klein direction can be written as $z=R_s\sigma$ such that $\sigma$ has period $2\pi$ and
\begin{equation}
R_s=2 \sqrt{\frac{r_s^3}{r_s-r_c}}\label{eq:Rs}
\end{equation}
is the asymptotic radius of the Kaluza-Klein dimension.  Rearranging (\ref{eq:Rs}) for $r_s$, I find $r_s^2=\left(1\pm\sqrt{1-16P^2/R_s^2}\right)R_s^2/8$, meaning that a given compactification radius admits a large soliton ($+$ sign) and a small soliton ($-$ sign).  Furthermore these two solitons degenerate at a minimum compactification radius of $R_s=4P$.  These solitons also admit an ``extremal" limit obtained by taking $\mu^2=P^2$.  Such an extremal limit again yields the metric of equation~(\ref{eq:MetricExtremal}), as could be anticipated since this metric is invariant under Wick rotation of $t$ and $z$.

\subsection{Dynamic Bubbles}

Having constructed the static bubble solutions, I now construct general bubble solutions that are dynamic.  Well-defined instantons are required in this construction because ultimately the phase transitions between the black string and these bubbles will be mediated semiclassically.  This means the time derivatives of the metric must be small close to the moment of nucleation, which is guaranteed by choosing a Cauchy surface, $\Sigma$, at a moment of time symmetry so that the extrinsic curvature vanishes:
\begin{equation}
ds_{\Sigma}^2=\frac{dr^2}{f(r)h(r)}+r^2d\Omega^2+h(r)R_s^2d\sigma^2 ,\label{eq:InitialData}
\end{equation}
where, to ensure the static bubbles are contained in the initial data, I choose $f(r)=1-\frac{r_c}r$.  $h(r)$ is then determined by the Hamiltonian constraint
\begin{equation}
R_\Sigma=\frac14F_{ab}F^{ab},
\end{equation}
where $R_\Sigma$ is the Ricci scalar of the Cauchy surface and $F_{ab}F^{ab}=\frac{6P^2}{r^4}$ is the square of the gauge field projected onto the Cauchy surface.  With this choice of $f(r)$, the Ricci scalar is
\begin{equation}
R_\Sigma=-\frac{2r(r-r_c)h''+(8r-7r_c)h'+4(h-1)}{2r^2},
\end{equation}
where a prime denotes a derivative with respect to $r$.  The function $h(r)$ satisfying the Hamiltonian constraint is
\begin{equation}
h(r)=1-\frac{r_s}{r}+\frac{\sqrt{f(r)}}{r^2}C_1+\frac{2(r-3r_c)+3\sqrt{f(r)}r_c\ln\left(r-\frac{r_c}{2}+r\sqrt{f(r)}\right)}{r^2}C_2, \label{eq:h}
\end{equation}
where $C_1$ and $C_2$ are constants of integration  and I have used $r_sr_c=P^2$ in the second term.  

Next, $h(r)$ is required to have a simple zero at $r_s$, meaning $C_1$ is determined in terms of the other variables as
\begin{equation}
C_1=\frac{-2(r_s-3r_c)-3\sqrt{f(r_s)}r_c\ln\left(r_s-\frac{r_c}{2}+r_s\sqrt{f(r_s)}\right)}{\sqrt{f(r_s)}}C_2. \label{eq:C1}
\end{equation}
Combining (\ref{eq:h}) and (\ref{eq:C1}), near $r=r_s > r_c$ the function $h(r)$ behaves like
\begin{equation}
h(r)=\frac{r_s-r_c+2C_2}{r_s(r_s-r_c)}(r-r_s)+.... \label{eq:hPrime}
\end{equation}
Examining (\ref{eq:InitialData}), in order for the boundary $r=r_s$ to be free of conical singularities, $R_s$ must be constrained via
\begin{equation}
h'(r_s)\sqrt{f(r_s)}=\frac{2}{R_s},
\end{equation}
which can be solved for $C_2$:
\begin{equation}
C_2=\frac12\sqrt{r_s^2-P^2}\left(\frac{2r_s}{R_s}-\frac{\sqrt{r_s^2-P^2}}{r_s}\right).
\end{equation}
From (\ref{eq:h}) and (\ref{eq:C1}) it is clear that when $C_2=0$ the static solution (\ref{eq:MetricSol}) is recovered.  This occurs when $r_s^2=P^2$, which is the extremal string, or when $R_s=2\sqrt{\frac{r_s^3}{r_s-r_c}}$ which is the relation (\ref{eq:Rs}) defining the static solitons.  When $R_s<4P$, $C_2$ has no zeroes, meaning there are no static bubbles, which confirms the previous finding.

\section{Perturbative Stability}

By construction, when the bubble nucleates the time derivative of the metric vanishes, meaning that $\partial_t$ is approximately Killing and the five dimensional metric momentarily takes the form of (\ref{eq:MetricSol}) with $\tilde f(r)=1-\frac{r_c}{r}$ and $\tilde h(r)$ given by equation~(\ref{eq:h}).  This allows one to construct an ADM energy sufficiently close to the moment of time symmetry. The physical process is that a static black string with a well-defined ADM energy tunnels to a bubble configuration by releasing energy in the form of radiation; the remaining energy will be the ADM energy of the bubble.  Away from the moment of nucleation, the bubble will generally expand or contract so the ADM energy will act as a potential for the dynamic bubble.  The asymptotic forms of the metric functions are $f(r)=1-\frac{r_c}{r}$ and $h(r)\approx1-\frac{r_s-2C_2}{r}$, which I find to yield the ADM energy
\begin{equation}
M_5=\frac{\pi R_s}{2G_5}\left(\frac{P^2}{r_s}+2r_s-\frac{2r_s}{R_s}\sqrt{r_s^2-P^2}\right). \label{eq:BubbleMass}
\end{equation}
Note that if $R_s=2\sqrt{\frac{r_s^3}{r_s-r_c}}$, corresponding to the static soliton where $C_2=0$, the above mass agrees with (\ref{eq:SolitonMass}).  Furthermore, if $r_s=P$, corresponding to the extremal string (\ref{eq:MetricExtremal}), the above mass (\ref{eq:BubbleMass}) agrees with the extremal limit of (\ref{eq:BSMass}).

\begin{figure}
\centering
\includegraphics[width=4.5 in]{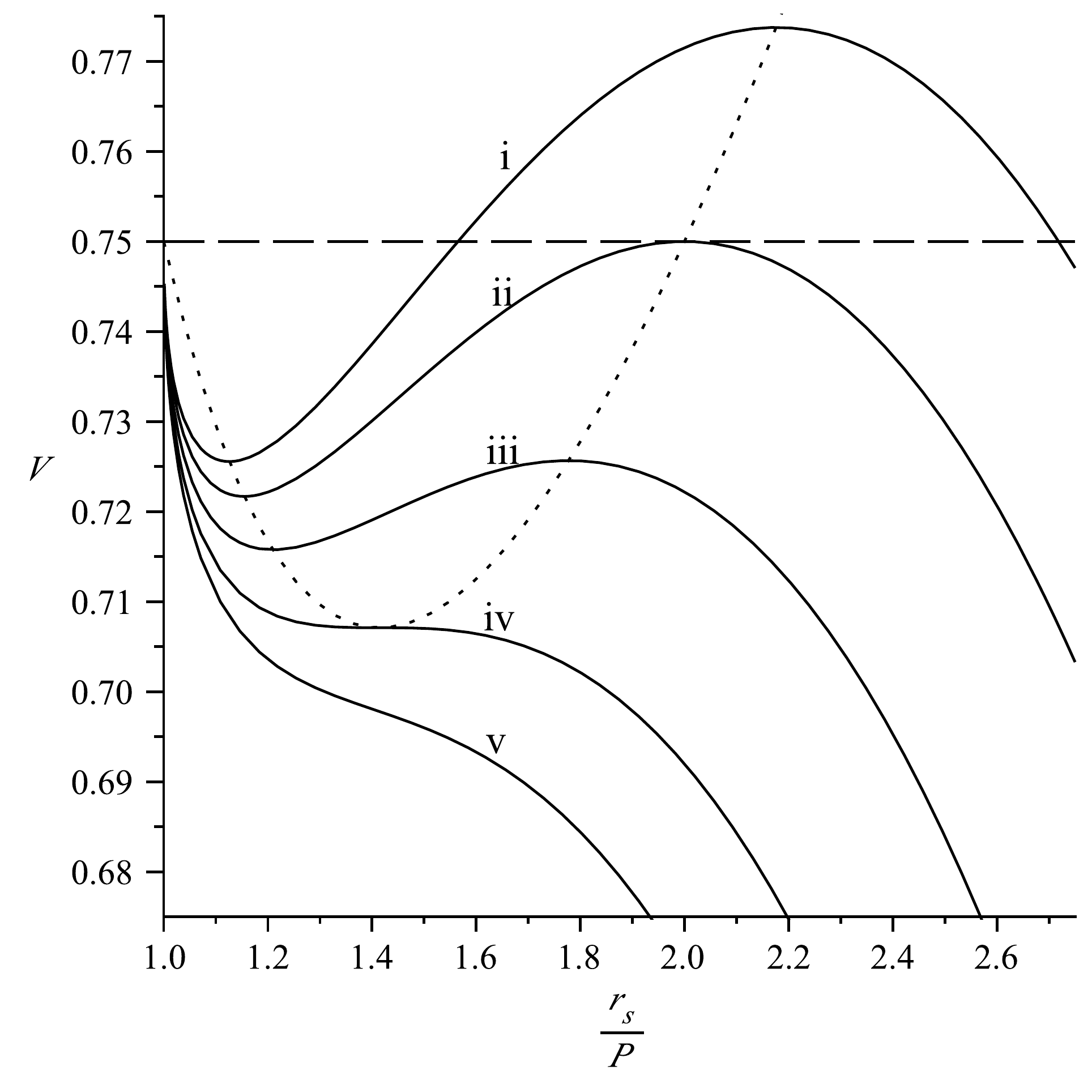}
\caption[Potential Energy Curve for Kaluza-Klein Bubbles]{The solid lines are the potential $V(r_s)$ for (i) $R_s=4.9P$, (ii) $R_s=\frac{8P}{\sqrt{3}}$, (iii) $R_s=4.3P$, (iv) $R_s=4P$ and (v) $R_s=3.8P$.  The dotted line shows the energy of the static solitons: the intersection of the dotted line with a potential curve always occurs at an extremum of the potential.  The dashed line shows the energy of the extremal black string.  \label{fig:Potential}}
\end{figure}

In figure~\ref{fig:Potential}, I plot the potential $V(r_s)\equiv\frac{M_5G_5}{2\pi R_s P}$ for different values of $R_s/P$ (solid lines) as well as plot the position of $r_s$ for large and small static solitons (dotted line) and the energy of the extremal black string (dashed line).  Wherever the dotted line crosses a solid curve, the potential has an extremum corresponding to a static soliton.  The minima represent the small solitons, the maxima represent the large solitons and the inflection point on line (iv) corresponds to where the small and large solitons degenerate at $R_s=4P$.  Line (v) has $R_s<4P$ and hence has no static solutions.  

From this potential diagram, it is apparent that the large solitons are perturbatively unstable while the small solitons are perturbatively stable.  In \cite{Miyamoto:2007mh}, it was found that the magnetic black string is stable against metric perturbations only if the mass parameter lies in the range $P\le m\le\frac3{2\sqrt{2}}P$.  Wick rotating $t\rightarrow iz$ and $z\rightarrow it$ to get the corresponding soliton, this range corresponds to $P\le \mu\le\frac3{2\sqrt{2}}P$ which is the parameter range of $\mu$ defining small solitons.  $\mu=\frac3{2\sqrt{2}}P$ corresponds to the minimum compactification radius, $R_{min}=4P$, where large and small solitons degenerate.  The Gregory-Laflamme stability of the magnetic black string is thus directly related to the perturbative stability of the topological solitons.  Furthermore, although the exact evolution of the dynamic bubbles is not known, the small static bubble will remain perturbatively stable and the large one unstable.  This is because although the potential may evolve in time, the positions of the minima and maxima in figure~\ref{fig:Potential} are time independent.

Recall that in setting up the initial data on the Cauchy slice, I chose a specific form for $f(r)$ and determined $h(r)$ via the Hamiltonian constraint.  Such initial data was found to contain the static soliton solutions, one of which is locally stable within this phase space slice.  To verify that the small soliton is truly stable, I choose an orthogonal slice through phase space by fixing $h(r)=1-\frac{r_s}{r}$ and now determine $f(r)$ via the Hamiltonian constraint.  It is conceivable that in this orthogonal slice, the small soliton might be unstable.  For this choice of initial data with $f(r)$ vanishing at some radius $r_c<r_s$ such that $r_cr_s=P^2$, there is a unique solution given by $f(r)=1-\frac{r_c}{r}$.  This initial data yields only the static solution so there is no orthogonal direction in phase space to which the small soliton could be unstable.  It is safe to conclude, then, that the small soliton is indeed locally stable.

\section{Thermodynamic Stability}

To determine the thermodynamic phase structure, I use the semiclassical Euclidean path integral approach to quantum gravity in the canonical ensemble, in which the partition function is given by
\begin{equation}
Z=\int{[dg]e^{-{\cal I}[g]}},
\end{equation}
where ${\cal I}[g]$ is the on-shell classical Euclidean action of a solution with metric $g$ and the path integral is taken over all metrics in the same topological class.  In the case at hand, there are two such metrics, namely the magnetic black string and the magnetic bubble (soliton).  The partition function for this two level system can be decomposed as $Z=Z_b+Z_s$, where 
\begin{equation}
Z_b=e^{-{\cal I}_b}, \quad\quad\quad\quad Z_s=e^{-{\cal I}_s},
\end{equation}
and subscripts $b$ and $s$ denote the black string and soliton respectively.  The probability of the system being in each phase is then given by
\begin{equation}
P_b=\frac{Z_b}{Z_b+Z_s},\quad\quad\quad\quad P_s=\frac{Z_s}{Z_b+Z_s}.
\end{equation}
Classically only one phase can exist at any particular time, which will be the phase with the higher probability.  Therefore, if phase $i$ is to dominate over phase $j$,
\begin{equation}
Z_i>Z_j \quad\quad\quad {\mathrm {or~equivalently}} \quad\quad\quad {\cal I}_i<{\cal I}_j. \label{eq:FreeEnergyInequality}
\end{equation}
In the canonical ensemble, the free energy of phase $i$ is defined as $F_i\equiv-\beta^{-1}\log Z_i=\beta^{-1}{\cal I}_i$ so the inequality (\ref{eq:FreeEnergyInequality}) is really a statement about the free energies of the two phases.  Thus, to determine the thermodynamic phase structure, the actions of the black string and the soliton need to be computed.  

The action is Euclidean Einstein-Maxwell with a Gibbons-Hawking counter-term and a Mann-Marolf counter-term \cite{Mann:2005yr,Mann:2006bd}, 
\begin{equation}
{\cal I}=-\frac{1}{16\pi G_5}\int_{\cal M}{d^5x\sqrt{g}\bigg({\cal R}-\frac{1}{4}F_{\mu\nu}F^{\mu\nu}\bigg)}-\frac{1}{8\pi G_5}\int_{\partial \cal M}{d^4x\sqrt{\gamma}\left(K-\hat K\right)},
\end{equation}
where $g$ is the determinant of the metric on the manifold $\cal M$, ${\cal R}$ is the Ricci scalar, $\gamma$ is the determinant of the metric on $\partial {\cal M}$, $K$ is the trace of the extrinsic curvature of $\partial {\cal M}$ in the metric $g$, and $\hat K\equiv \hat K^{ij}\gamma_{ij}$ is the Mann-Marolf counter-term which satisfies the implicit equation ${\cal R}_{\partial{\cal M}}={\hat K}^2-\hat K_{ij}\hat K^{ij}$ with ${\cal R}_{\partial{\cal M}}$ the Ricci scalar of the boundary geometry.
The effect of the Mann-Marolf counter-term, $\hat K$, is to renormalise the action without relying on ambiguous background subtractions.  Both instantons have a metric $ds^2=f(r)d\tau^2+ds_{\Sigma}^2$ where $ds_{\Sigma}^2$ is the initial data (\ref{eq:InitialData}): the black string is given by $f_b(r)=1-\frac{r_+}r$, $h_b(r)=1-\frac{r_-}r$ while the solitons are given by $f_s(r)=1-\frac{r_c}{r}$ and $h_s(r)$ as in equation (\ref{eq:h}).  For the black string instanton to be smooth, $\tau$ must be chosen to have periodicity $\beta=\frac{4\pi r_+^2} {\sqrt{r_+^2-P^2}}$ while the compactification radius is arbitrary.  Similarly, the soliton instanton is automatically smooth because the periodicity of $\tau$ is arbitrary and the compactification radius has already been chosen to remove conical singularities at $r_s$.  In order for phase transitions between the two instantons to be possible, the instantons must match asymptotically, which means the periodicity of $\tau$ and the compactification radius must be the same for both instantons.  Explicitly calculating the actions, I find
\begin{align}
&{\cal I}_b=\frac{\beta\pi R_s}{2G_5}\left(\frac{2P^2}{r_+}+r_+\right), \\
&{\cal I}_s=\frac{\beta\pi R_s}{2G_5}\left(\frac{2P^2}{r_s}+r_s-2C_2\right).
\end{align}   

The phase transition should be described by a local transition at the horizon,\footnote{Indeed, the modern string theoretic viewpoint of localised tachyon condensation causing a topology change will be discussed in section \ref{Tachyon}.} so a black string with horizon radius $r_+$ and a bubble with radius $r_s=r_+$ are considered.  The relative free energy per unit length, ${\cal F}\equiv \frac{F_b-F_s}{2\pi R_s}$ determines the phase structure: if ${\cal F}>0$ then the bubble is preferred, if ${\cal F}<0$ then the black string is preferred, with the phase transition boundary corresponding to ${\cal F}=0$.  A simple calculation yields
\begin{equation}
{\cal F}=\frac{C_2}{2G_5},
\end{equation}
meaning that the black string is thermodynamically favoured over a bubble of the same size as long as $C_2<0$.  I find generically that the condition $C_2<0$ occurs whenever the potential has positive slope, which is the range of parameters lying between the small and large static solitons.  Alternatively this corresponds to the region in which the nucleated bubble would be initially contracting (see figure~\ref{fig:Potential}).  Therefore a black string of horizon radius $r_+$ will tunnel to a bubble of equal size as long as the nucleated bubble is initially expanding; otherwise it will be thermodynamically stable against such a transition.

The extremal black string, corresponding to $r_s=P$ in figure~\ref{fig:Potential}, is always unstable to decay to a bubble and has the lowest energy in the family of black strings, so a natural question to ask is whether such an extremal string is stable against forming a bubble that will expand out to consume the entire space-time.  From figure~\ref{fig:Potential} it is clear that lines (iii), (iv) and (v) will allow this since the bubble will form arbitrarily close to $r_s=P$, giving it enough energy to continue down the potential indefinitely.  However, line (ii) has a maximum at the same energy as the extremal string so as long as $R_s>\frac8{\sqrt{3}}P$, as in line (i), the extremal string is stable, at least classically.  In this case the nucleated bubble will oscillate between the two turning points, settling down to the static small soliton by emitting gravitational radiation.  Strictly speaking quantum tunneling through the potential barrier is possible and should be considered.  However imposing arbitrarily weak supersymmetry-breaking boundary conditions means $R_s$ is taken to be large compared to the magnetic charge.   In this case, the height and thickness of the potential barrier become arbitrarily large so that tunnelling through the barrier can be ignored.  The small soliton then effectively becomes the local vacuum, provided the black string's mass is sufficiently small compared to $R_s$.  

I now combine the thermodynamic and perturbative stabilities of the magnetic black string into a unified picture.  In \cite{Miyamoto:2007mh} it was found that magnetic black strings are perturbatively stable so long as $P\le r_+\le\sqrt{2}P$.  Here I have found that the magnetic black strings are thermodynamically stable in the range $r_1<r_+<r_2$, where $r_{2/1}=\left(R_s/2\sqrt{2}\right)\big(1\pm\sqrt{1-16P^2/R_s^2}\big)^{1/2}$ are the radii of the small and large solitons, given the compactification radius.  As long as $R_s>4P$ there is an overlap region $r_1<r_+<\sqrt{2}P$ in which the black string is both perturbatively and thermodynamically stable.  However, there are also regions of parameter space in which the black string is perturbatively stable but thermodynamically unstable, $P<r_+<r_1$, and vice versa, $\sqrt{2}P<r_+<r_2$, meaning the Gubser-Mitra conjecture is \emph{not} satisfied in general for non-dilatonic magnetically charged black strings, contrary to the claim in \cite{Miyamoto:2007mh}.  It is still unclear what the black string evolves or transitions to when the Gregory-Laflamme instability sets in because the magnetic black string cannot bifurcate without violating charge conservation.  Therefore, there must exist other objects with the same topological magnetic charge yet to be discovered.  Furthermore, such objects ought not to care about boundary conditions on fermions since the Gregory-Laflamme instability is insensitive to this.

\section{Closed String Tachyon Condensation as a Phase Transition Mechanism}
\label{Tachyon}

Apart from antiperiodic boundary conditions for the fermions, no stringy considerations have been required in the analysis up to this point. Such considerations are, however, quite relevant to the physics of 
 Kaluza-Klein bubbles.  For instance, when the size of the periodic dimension shrinks below the string scale, strings wrapped around this cycle become tachyonic\cite{Rohm:1983aq}.  Furthermore,  when this tachyon condensation is localised, it induces a topology changing transition \cite{Adams:2005rb}.  

For the black strings considered in this chapter, the size of the periodic dimension at arbitrary radius is controlled by the value of $h(r)$, from which it is clear that the circle shrinks to zero at the inner horizon (recall $h(r_-)=0$).  One must also in principle be careful about $\alpha'$ corrections, however the curvature at the outer horizon goes like $1/r_+^2$ which will remain much smaller than the string scale, even at extremality, provided $P^2\gg \ell_s^2$.  Restricting attention to sufficiently large magnetic charges, what is required in order to nucleate a bubble is for tachyon condensation to take place just outside the horizon.  Using the relation $r_-r_+=P^2$, this condition becomes
\begin{equation}
\frac{\ell_s^2}{R_s^2}=\left(1-\frac{P^2}{r_+^2}\right). \label{eq:tachyoncond}
\end{equation}
In order to avoid tachyon condensation at large $r$, one must also require $R_s^2\gg\ell_s^2$, which from (\ref{eq:tachyoncond}) implies that the black string is near extremality, $r_+^2\approx P^2$.

These stringy considerations provide a nice verification of the above findings that the near-extremal black string will always nucleate a bubble.  In the extremal limit, the circle shrinks to zero at the degenerate horizon, so tachyon condensation takes over before extremality is reached and a bubble is nucleated.  Sufficiently far from extremality it becomes increasingly difficult to nucleate a bubble if the size of the extra dimension is appropriately large.  For the static bubble of minimum compactification radius ($R_s=4P$, $r_+=\sqrt{2}P$) for which all nucleated bubbles will expand out to infinity, I find that $\ell_s^2=8P^2$ which is inconsistent with the requirement that $\alpha'$ corrections can be ignored; one must choose $R_s^2\gg P^2$ which means any ``catastrophic" bubbles can be ignored.  Stated concisely, stringy analysis suggests that magnetic black strings will Hawking evaporate until they approach extremality sufficiently closely, at which point tachyon condensation will cause the circle to pinch off.  This destroys the horizon, leaving a bubble which settles down to the small static configuration considered above.  Similar results were found in \cite{Horowitz:2005vp} where black strings with F1 charge and black strings with F1 and NS5 charges also settle down to stable static bubbles.  It warrants special mention, however, that the current analysis makes use only of the  Einstein-Maxwell equations and does not require the properties of other string-inspired fields, such as the dilaton, other gauge fields or exotic brane charges.

The ability to construct magnetically charged solitons from magnetic branes seems to be generic and directly carries over into ten dimensional Einstein-Maxwell-Dilaton theory.  The dilaton has a nontrivial effect on the space-time structure, for instance making the singularity for the charged black hole space-like and as well as eliminating the problematic inner Cauchy horizon\cite{Garfinkle:1990qj}.  It was shown in \cite{Gregory:1994tw} that supersymmetric black p-branes are perturbatively stable, i.e. do not suffer the Gregory-Laflamme instability.  This result is insensitive to fermionic boundary conditions and holds for extremal non-supersymmetric black p-branes as well.  Further arguments were given in favour of their thermodynamic stability \cite{Gregory:1994tw}; however
these were based on comparing the entropies of black p-branes to spherical black holes but these objects are charged under different rank gauge fields.    In much the same way as black p-branes with RR charge were considered in \cite{Horowitz:2005vp}, it would be interesting to revisit the question of thermodynamic stability of black p-branes with magnetic charge to see if they suffer a similar fate.

\chapter{Phase Transitions in AdS$/{\mathbb Z}_k$ space-times}

For the remainder of this thesis I change focus to the study of black hole stability in five and seven dimensional gauged supergravity theories, \emph{i.e.} theories which include a negative cosmological constant.  Asymptotically anti-de Sitter (AdS) space-times do not suffer the same instability toward nucleation of bubbles of arbitrary negative energy seen in the last chapter because the positive energy theorem ensures there exists a lowest energy ground state\cite{Abbott:1981ff,Woolgar:1994ar}.  Asymptotically globally AdS space-times have a ground state that is simply the AdS vacuum, however this is not true for asymptotically locally AdS space-times.  For example, if one of the dimensions is periodically identified and antiperiodic boundary conditions for fermions are chosen, there exists a local vacuum state known as the AdS soliton\cite{Horowitz:1998ha}.  Similarly, the orbifolds AdS$/{\mathbb Z}_k$ in odd space-time dimensions, where the quotient by the subgroup ${\mathbb Z}_k$ means one of the cycles on the sphere is modded out by an integer $k$, have a local vacuum state known as the Eguchi-Hanson soliton\cite{Clarkson:2005qx,Clarkson:2006zk,Copsey:2007hw}.  Since antiperiodic boundary conditions for fermions have been chosen, these space-times do not admit a supercovariantly constant Killing spinor and hence cannot be supersymmetric.  Because of this the AdS vacuum is no longer supersymmetric, which allows for solutions of lower energy; indeed both the AdS soliton and the Eguchi-Hanson soliton represent local vacua for metrics of their respective asymptotic classes and have energies lower than AdS.

The thermodynamic stability of black holes in AdS geometries was first considered by Hawking and Page in \cite{Hawking:1982dh}.  It was found that at fixed temperature there exist both small and large black holes, the former having negative specific heat, hence being unstable to Hawking evaporation, and the latter having positive specific heat, hence being in thermal equilibrium with the surrounding heat bath.  Furthermore, by considering thermal AdS as the ground state it was found that there is a range of large black holes which are thermodynamically unstable, despite having positive specific heat.  Using the AdS/CFT correspondence, this phase transition between thermal AdS and the AdS black hole was found to be dual to a confinement-deconfinement phase transition in the large $N$ limit of ${\cal N}=4$ super Yang-Mills theory \cite{Witten:1998qj}.  It was then shown that there is no analogous phase transition between the thermal AdS geometry and AdS black holes with Ricci flat horizons \cite{Birmingham:1998nr,Mann:1999bt,Vanzo:1997gw}.  However, inspired by the AdS soliton being of lowest energy in its asymptotic class, it was shown in \cite{Surya:2001vj} that there is indeed an analogous phase transition between Ricci flat black holes and the AdS soliton.  This transition was also shown to be dual to a confinement-deconfinement phase transition in the dual gauge theory.  

Since the Eguchi-Hanson soliton (EHS) has been shown to be the ground state in its asymptotic class, it must contribute to the phase behaviour of asymptotically AdS$/{\mathbb Z}_k$ space-times.  Following this reasoning, the authors of \cite{Hikida:2006qb,Hikida:2007pr} performed a detailed study in five dimensions of the phase behaviour of the large $N$ limit of ${\cal N}=4$ super Yang-Mills theory on ${\mathbb R}\times S^3/{\mathbb Z}_k$ with U$(N)$ gauge symmetry, which is the holographic dual of asymptotically AdS$/{\mathbb Z}_k$ space-times. It was found that the EHS and the empty orbifold AdS$/{\mathbb Z}_k$ (OAdS) correspond to local vacua of the gauge theory and furthermore that the vacuum corresponding to the EHS is at a lower energy.  In this chapter, which is based on the work published in \cite{Stotyn:2009ff}, I analyse the gravitational side of the AdS/CFT correspondence by studying the full phase structure of the EHS, the OAdS geometry, and the Schwarzschild-AdS$/{\mathbb Z}_k$ black hole in odd space-time dimensions $D\ge5$.  The three dimensional case is not treated because the EHS only exists for $D\ge 5$.

\section{Solitons and Black Holes}

I begin by presenting the three known static uncharged asymptotically AdS/$\mathbb{Z}_k$ space-times in $D=2n+1$ dimensions and review their basic features.  All three metrics have the general form 
\begin{equation}
ds^2=-f_i(r)dt^2+\frac{dr^2}{f_i(r)h_i(r)}+\frac{r^2}{2n}\sum_{j=1}^{n-1}d\Omega_j^2+\left(\frac{r}{n}\right)^2h_i(r)\left(d\psi+\sum_{j=1}^{n-1}\cos\theta_jd\phi_j\right)^2,  \label{eq:OrbMetric}
\end{equation}
where $d\Omega_j^2=d\theta_j^2+\sin^2\theta_jd\phi_j^2$ and the subscript $i=s,a,b$ denotes the EHS, OAdS geometry and black hole respectively.  The angular coordinates take value in the range $\theta_j\in[0,\pi]$, $\phi_j\in[0,2\pi]$ and $\psi$ has period $\eta_i$.  The metric functions $f_i(r)$ and $h_i(r)$ take the following form for the three metrics:
\begin{align}
f_s(r)={}&1+\frac{r^2}{\ell^2}, \quad\quad\quad\quad\quad\quad\quad\>\>\> h_s(r)=1-\left(\frac{a}{r}\right)^{2n},\\
f_a(r)={}&1+\frac{r^2}{\ell^2}, \quad\quad\quad\quad\quad\quad\quad\>\>\> h_a(r)=1,\\
f_b(r)={}&1-\frac{\mu}{r^{2n-2}}+\frac{r^2}{\ell^2}, \quad\quad\quad\quad h_b(r)=1,
\end{align}
where $\ell$ is related to the cosmological constant via $\Lambda=-n(2n-1)/\ell^2$.  For the EHS to be free of conical singularities at the soliton radius, $r=a$, the condition $\eta_s=2\pi/\sqrt{f_s(a)}$ must be imposed.  Furthermore, to remove Misner strings from the poles of the sphere, the condition $k\eta_s=4\pi$ must be satisfied for some integer $k$ \cite{Clarkson:2005qx,Clarkson:2006zk}.   These two restrictions are summarised by a relation between the soliton radius and the AdS length:
\begin{equation}
a^2=\ell^2\left(\frac{k^2}{4}-1\right),  \label{eq:aRelation}
\end{equation}
from which it is clear $k\ge3$ to yield a real value for the soliton radius.  The only restriction on $\eta_a$ and $\eta_b$ is the removal of Misner strings, which similarly requires $\eta_a=\eta_b=4\pi/k$ and again $k\ge3$ is chosen to allow transitions between all three solutions.  There is a further subtlety here regarding the orbifold parameter $k$.  The spatial boundary of AdS$_D/{\mathbb Z}_k$ is $S^{D-2}/{\mathbb Z}_k$, which has no cycle the fermions can wrap around since any closed loop is deformable to a point.  This means the overall phase shift of fermions must be an integer multiple of $2\pi$ upon traversing $k$ times around the $\psi$-cycle.  Furthermore, antiperiodic boundary conditions for the fermions are required, meaning traversing the $\psi$-cycle once yields a phase shift of $\pi$.  The only way to have both conditions met is by restricting $k$ to be even.

The black hole has an event horizon at $r=r_+$ defined by $f_b(r_+)=0$.  The mass parameter, $\mu$, is inconvenient so the horizon radius is used instead, in terms of which $\mu=r_+^{2n-2}\left(1+r_+^2/\ell^2\right)$.  Smooth instantons can be constructed for all three solutions by introducing Euclidean time $\tau=it$ and identifying $\tau$ with period $\beta_i$.  For the EHS and the OAdS geometry $\beta$ can take any value, whereas for the black hole the elimination of conical singularities requires
\begin{equation}
\beta_b=\frac{2\pi r_+}{n-1+n\frac{r_+^2}{\ell^2}}.
\end{equation}

\section{Computing the Euclidean Actions}

It was seen in the last chapter that in the canonical ensemble with $N$ discrete levels, the partition function for semiclassical gravity can be written as $Z=\sum_{i=1}^{N}{Z_i}$ where $Z_i=e^{-{\cal I}_i}$ and ${\cal I}_i$ is the on-shell Euclidean action.  The phase that dominates is then the one with the highest probability, which is equivalently the phase with the lowest free energy, $F_i=\beta^{-1}{\cal I}_i$.  Thus, to determine the phase structure, the Euclidean actions for the three solutions must be evaluated.  However, if one attempts to compute the Euclidean action for a non-compact manifold one obtains an infinite answer due to the infinite volume, which is clearly undesirable.  

Recall that in chapter 4, a Mann-Marolf boundary counter-term was added to the action, which renormalises the action to produce a finite value.  Here I will use a similar holographic renormalisation scheme by inserting appropriate boundary counter-terms.  Even though this must be done for each space-time dimension separately and it is difficult to find expressions for the renormalised action valid for arbitrary dimension, I find that when \emph{differences} in the actions are considered, the solutions can indeed be written for general space-time dimension.

The action under consideration is the Einstein-Hilbert action with a Gibbons-Hawking counter-term and a renormalisation counter-term:
\begin{equation}
{\cal I}=-\frac1{16\pi G_D}\int_{\cal M}{d^Dx\sqrt{g}(R_D-2\Lambda)}-\frac1{8\pi G}\int_{\partial{\cal M}}{d^{D-1}x\sqrt{\gamma}\left(K-\frac{\Theta}{\ell}\right)} \label{eq:action1}
\end{equation}
where $g$ is the determinant of the Euclidean metric on $\cal M$, $R_D$ is the Ricci scalar, $\gamma$ is the determinant of the Euclidean metric on $\partial {\cal M}$, $K$ denotes the trace of the extrinsic curvature of $\partial {\cal M}$ in the metric $g$, and $\Theta$ is a dimensionless function of the boundary geometry.  Specifically $\Theta$ can only be a function of the boundary Riemann tensor, its contractions and their derivatives in order to leave the equations of motion invariant \cite{Balasubramanian:1999re,Emparan:1999pm,Henningson:1998gx,Mann:1999pc}.  $\partial {\cal M}$ is taken to be surfaces of constant $r$ with unit normal $n^\alpha=\sqrt{fh}\partial_r$.  With this choice, it is easy to see from (\ref{eq:OrbMetric}) that $\sqrt{\gamma}=\sqrt{f_ih_i}\sqrt{g}$ where $\sqrt{g}\propto r^{2n-1}$.  This allows the combination of (\ref{eq:action1}) into a single integral if the $r$ integration in the first term is performed out to the boundary radius $R\equiv r\big|_{\partial {\cal M}}$.  From the equations of motion, the Ricci scalar is given simply by $R_D=-\frac{2n(2n+1)}{\ell^2}$, so the first term in (\ref{eq:action1}) yields
\begin{equation}
\frac{1}{8\pi G_D}\int_{\partial {\cal M}}{\sqrt{\sigma}d^{D-1}x\left(\frac{R^{2n}}{\ell^2}-\frac{r_i^{2n}}{\ell^{2}}\right)},
\end{equation}
where $\sqrt{g}\equiv r^{2n-1}\sqrt{\sigma}$ so that $\sigma$ is independent of $r$, and $r_i$ is the inner boundary of each instanton.  A calculation of the extrinsic curvature gives
\begin{align}
\sqrt{fh}R^{2n-1}K={}&\frac{r^{2n-1}}{2}\left(f_i'h_i+f_ih_i'+\frac{2(2n-1)}{r}f_ih_i\right)\bigg|_{r=R}\nonumber\\
={}&\left\{
{\begin{array}{ll}
2n\frac{R^{2n}}{\ell^2}+(2n-1)R^{2n-2}, & \qquad {\mathrm {for~OAdS,}}\\
2n\frac{R^{2n}}{\ell^2}+(2n-1)R^{2n-2}-n\mu, & \qquad {\mathrm {for~black~hole,}}\\
2n\frac{R^{2n}}{\ell^2}+(2n-1)R^{2n-2}-n\frac{a^{2n}}{\ell^2}, & \qquad {\mathrm {for~EHS.}}
\end{array}} \right.
\end{align}

An algorithm for calculating $\Theta$ for general space-time dimension in terms of a series expansion in the AdS radius was put forth in \cite{Kraus:1999di} and the explicit form in three and five dimensions has been constructed in \cite{Balasubramanian:1999re}.  Furthermore, in \cite{Henningson:1998gx} it was shown that such a $\Theta$ indeed exists in seven dimensions.  A more thorough discussion of how to write down $\Theta$ explicitly in any dimension was given in \cite{Clarkson:2002uj}.  The series expansion in the AdS radius effectively amounts to a series expansion in inverse powers of the boundary radius so in what follows I choose the convenient basis
\begin{equation}
\Theta=A+B\frac{l^2}{R^2}+C\frac{l^4}{R^4}+....
\end{equation}
Bringing all of the results together thus far, the constants $A,B,C,...$ are fixed by requiring the action to be finite in the limit $R\rightarrow\infty$.  Furthermore, only the minimum number of terms in the expansion for $\Theta$ are kept such that a finite action is obtained; higher order terms contribute at ${\cal O}(1)$ and are thus otherwise unconstrained.   This ``quick and dirty" method of finding $\Theta$ agrees exactly with what one gets from the algorithm of \cite{Kraus:1999di} with the added benefit that it easily generalises to any number of dimensions.  

It was shown in \cite{Clarkson:2005qx} that in five dimensions the EHS is perturbatively the solution of lowest energy within its asymptotic class, although no such analysis has been performed for the higher dimensions.  Nevertheless, it is expected that the five dimensional result remains valid in higher dimensions, hence I take the EHS to be the thermal background.  Of ultimate interest, then, is the relative action $I_i= {\cal I}_i-{\cal I}_s$.  In taking this difference, care must be taken to ensure that the periodicities of $\tau$ and $\psi$ match between solutions: the condition $\beta=\beta_b$ must be imposed for all solutions whereas the periodicities of $\psi$ already match since they are all $4\pi/k$ by construction.  I find the relative actions to be
\begin{align}
I_s={}&0, \\
I_a={}&\frac{2^{n-3}\pi^{n-1}}{\ell^2n^nkG_D}\beta_b a^{2n}, \label{eq:Ia} \\
I_b={}&\frac{2^{n-3}\pi^{n-1}}{\ell^2n^nkG_D}\beta_b\left[a^{2n}+r_+^{2n}\left(\frac{\ell^2}{r_+^2}-1\right)\right]. \label{eq:Ib}
\end{align}

\section{Thermodynamic Phase Structure}

In this section, I work out the phase structure of the EHS, OAdS and black hole solutions by considering the EHS as the thermal background.  Note that there is only one thermodynamic parameter, $r_+$.  Strictly speaking, for the EHS and OAdS solutions it makes no sense to speak of a horizon radius, however for these solutions one can regard $\beta$ as the thermodynamic parameter and consider $r_+(\beta)$.  Note that $a=a(\ell,k)$ is fixed since $\ell$, $n$ and $k$ are all held fixed as they represent the AdS radius, the space-time dimension and the topology respectively and cannot change.\footnote{In \cite{Hikida:2007pr} it was shown that in three dimensions ($n=1$) there is a topological transition from spaces with orbifold parameter $k'$ to spaces with orbifold parameter $k<k'$ eventually ending in the stable vacuum of AdS$_3$.  This was later shown in \cite{Stotyn:2012ap} to be due to the fact that AdS$_3/{\mathbb Z}_k$ is identical to a conical singularity in the AdS$_3$ mass gap with mass parameter $M=-1/k^2$.  This subtlety is specific to three dimensions and does not affect the analysis here.}

The action for the OAdS geometry (\ref{eq:Ia}), and hence its free energy relative to the EHS, ${\cal F}_a=\beta_b^{-1} I_a$, is always positive, meaning the OAdS geometry is always an unstable phase.  That is to say there is no order parameter that can be tuned to bring about a continuous phase transition between the OAdS and EHS geometries.  This supports the holographic analysis of \cite{Hikida:2007pr}, which concludes that both the EHS and OAdS solutions are holographically dual to local vacuum states in the gauge theory with the EHS corresponding to the lowest energy ground state.  In fact, the thermodynamic energy, $\langle E\rangle=\frac{\partial I}{\partial \beta_b}$, of the OAdS geometry relative to the EHS is given by
\begin{equation}
\langle E_a\rangle=\frac{2^{n-3}\pi^{n-1}a^{2n}}{\ell^2n^nkG},
\end{equation}
which is strictly positive.  The OAdS geometry is unstable and, from a semiclassical viewpoint, it lowers the free energy by tunneling to the EHS geometry at the rate $\Gamma\sim e^{-\beta_b \langle E\rangle_a}=e^{-I_a}$.  From a complementary string theoretic viewpoint, the mechanism for the phase transition is closed string tachyon condensation:  the centre of the OAdS geometry is an orbifold fixed point and closed string tachyons become localised around it, inducing a topology changing transition to the EHS geometry\cite{Hikida:2007pr}.  Finally the entropy, $S=\beta_b\langle E\rangle-I$, and the heat capacity, $C=-\beta^2\frac{\partial^2 I}{\partial \beta^2}$, both vanish for the OAdS geometry as they ought to for a horizonless space-time.

To analyse the thermodynamic behaviour of the black hole it is convenient to introduce the Hawking-Page action,
\begin{equation}
I_{HP}\equiv {\cal I}_b-{\cal I}_a=I_b-I_a,  \label{eq:HP}
\end{equation}
so that the black hole action can be written $I_b=I_a+I_{HP}$.  I find the thermodynamic energy of the black hole to be
\begin{equation}
\langle E_b\rangle=\langle E_a\rangle+\frac{2^{n-3}\pi^{n-1}}{n^nkG_D}(2n-1)\mu,\label{eq:Eb}
\end{equation}
where $\mu=\mu(r_+)$ is the black hole mass parameter.  The utility of introducing (\ref{eq:HP}) is immediately clear: the first term in (\ref{eq:Eb}) is due to $I_a$ and the second term is due to $I_{HP}$.  It has already been shown that the entropy and heat capacity vanish for $I_a$ so these quantities for the black hole are governed by the Hawking-Page action alone.  As expected for the entropy, I find
\begin{equation}
S_b=\frac{2^{n-1}\pi^n}{n^nkG}r_+^{2n-1}=\frac{A_b}{4G_D},
\end{equation}
where $A_b$ is the area of the black hole horizon.  Similarly, I find the heat capacity for the black hole to be
\begin{equation}
C_b=\frac{(2n-1)2^n\pi^{n+1}r_+^{2n}}{\beta_bn^nkG_D\left(n\frac{r_+^2}{\ell^2}-(n-1)\right)},
\end{equation}
which diverges at $r_+=r_0$, where $r_0^2=\frac{n-1}n \ell^2$, signifying a first-order phase transition.  In direct analogy with the original analysis of Hawking and Page in \cite{Hawking:1982dh}, $r_0$ is precisely the radius at which $\frac{\partial\beta_b}{\partial r_+}=0$ and hence represents the boundary between small and large black holes; $\beta_b(r_0)$ gives the lowest possible temperature for the black hole.

I now discuss the thermodynamic stability of the black hole in detail.  The OAdS geometry has been shown to always be unstable to decay to the EHS since its free energy is strictly positive.  Likewise, to determine the thermodynamic stability of the black hole, the sign of the black hole action (\ref{eq:Ib}) is important; if (\ref{eq:Ib}) turns negative the black hole will be the thermodynamically favoured phase.  This will happen at some $r_+=r_1$ such that $I_b(r_1)=0$, signifying the cross over from positive to negative free energy.  Defining $\chi_1\equiv r_1^2/\ell^2$ and using the relation (\ref{eq:aRelation}), the condition for a phase boundary becomes
\begin{equation}
\left(\frac{k^2}4-1\right)^n-\chi_1^{n-1}(\chi_1-1)=0. \label{eq:fchi}
\end{equation}
This equation is a simple polynomial of degree $n$, and imposing $\chi_1>1$ guarantees at least one real solution due to the mean value theorem: the action evaluated at $r_+=r_0$ is positive,
\begin{equation}
\left(\frac{k^2}4-1\right)^n-\left(\frac{r_0}\ell\right)^{2n-2}\left(\frac{r_0^2}{\ell^2}-1\right)=\left(\frac{k^2}4-1\right)^n+\frac{(n-1)^{n-1}}{n^n}>0, \label{eq:chir0}
\end{equation}
whereas the action for sufficiently large $r_+=R$ is negative,
\begin{equation}
\left(\frac{k^2}4-1\right)^n-\left(\frac{R}\ell\right)^{2n-2}\left(\frac{R^2}{\ell^2}-1\right)\rightarrow-\left(\frac{R}\ell\right)^{2n}<0.
\end{equation}
Therefore at least one phase boundary $r_1$ exists, however it is also necessary to establish whether it is the only real solution to (\ref{eq:fchi}).  To this end, it can be easily verified that a function defined as the left hand side of (\ref{eq:fchi}) has a single maximum located at $r_0$, where $I_b(r_0)>0$ from (\ref{eq:chir0}).  The only remaining option for another phase boundary is if $r_+<r_0$, however the action evaluated at $r_+=0$, \emph{i.e.} the action of the OAdS geometry, is also positive.  Thus $r_1>r_0$ exists and is the only phase boundary between the black hole and the EHS.

Since the free energy of the OAdS geometry is positive while that of the black hole can be positive or negative, there exists another critical point, $r_+=r_{HP}$, above which the free energy of the black hole is lower than the OAdS geometry and below which it is higher:  this happens at $r_{HP}=\ell$ and is where the Hawking-Page action (\ref{eq:HP}) vanishes.  It is easily verified that $r_0<\ell<r_1$.

There are thus three critical horizon radii of an AdS$/{\mathbb Z}_k$ black hole, namely $r_0$, which separates small and large black holes, $\ell$, which separates black holes (un)stable with respect to the OAdS geometry, and $r_1$, which separates black holes (un)stable with respect to the EHS.  Small black holes, corresponding to $r_+<r_0$, have negative specific heat and hence cannot be in thermal equilibrium with their surroundings; they will evaporate to the OAdS geometry and eventually transition to the EHS.  Large black holes in the range $r_0<r_+<\ell$ have positive specific heat, meaning they are at least locally thermodynamically stable, however their free energy is greater than the OAdS geometry, so they will tend to evaporate and eventually transition to the EHS.  Large black holes in the range $\ell<r_+<r_1$ have positive specific heat and are thermodynamically favoured over the OAdS geometry so they are not expected to evaporate but rather remain in thermal equilibrium with the surrounding heat bath.  Such black holes still have positive free energy, however, and are unstable to decay to the EHS.  The process of evaporation would raise the free energy so the only transition available to these black holes is to tunnel to the EHS.  Large black holes in the range $r_+>r_1$ have positive specific heat and negative free energy and are thus truly stable.  Similarly, an EHS submersed in a heat bath with inverse temperature $\beta<\beta(r_1)$ has a higher free energy than a large black hole at the same temperature and will thus form a stable black hole.  Finally, a thermal OAdS space-time with inverse temperature $\beta<\beta(\ell)$ has a higher free energy than a large black hole at the same temperature so one might expect it to form a black hole.  It is much more likely, however, for closed string tachyon condensation around the orbifold fixed point to first induce a transition to the EHS, which would then either be stable if $\beta>\beta(r_1)$ or unstable and form a black hole if $\beta<\beta(r_1)$.

\section{Discussion and the Confinement-Deconfinement Transition}

In this chapter the thermodynamic behaviour of Eguchi-Hanson solitons and orbifold AdS and Schwarzschild-AdS spaces has been examined in detail, expanding on the dual gauge theory work in \cite{Hikida:2007pr} and generalising to arbitrary odd dimension.  I have given a thorough analysis on the gravity side of the gauge/gravity correspondence and have left discussions related to the holographic dual description to \cite{Hikida:2007pr} and the appropriate references therein as the details are rigorously worked out.  The gravity description presented here very strongly supports the idea that the OAdS geometry and the EHS are both local vacua with the EHS being the ground state.  Indeed, the OAdS geometry is the end point of black hole evaporation, signifying a local vacuum, but the OAdS geometry has a positive free energy and hence tunnels to the zero free energy EHS configuration.  This instability in the OAdS geometry is further supported by the development of closed string tachyons around the orbifold fixed point, causing a topology changing transition to the EHS.  This indicates that the EHS is at least a lower energy local vacuum than the OAdS geometry.  This is consistent with previous results in \cite{Clarkson:2006zk}, which demonstrated that the EHS in five dimensions is perturbatively the lowest energy solution in its asymptotic class. While such an analysis remains to be done for the EHS in higher dimensions it is anticipated that the same outcome holds based on the results obtained in this chapter.

I have shown that the expected results for the entropy and the heat capacity are obtained for the black hole, however a richer thermodynamic structure than in asymptotically globally AdS spaces has been uncovered.  Firstly, there are three phases, two of which are vacua, present in this analysis, whereas the cases studied in \cite{Hawking:1982dh,Surya:2001vj} each contain only two phases and consequently one vacuum.  Secondly, although all of the general elements of the original Hawking-Page analysis are present, such as the existence of large and small black holes as well as a range of large black holes that are unstable despite the positivity of their specific heat, the existence of the soliton widens the parameter range of unstable large black holes.  The phase behaviour of black holes can effectively be split up into a Hawking-Page part and a soliton part: if the black hole is unstable against the OAdS geometry, the Hawking-Page part dictates that it will evaporate toward the OAdS vacuum and eventually transition to the EHS.  If the black hole is stable against the OAdS geometry, the soliton part of the phase behaviour becomes important.  Thus, the EHS only has an effect on the phase behaviour of sufficiently large black holes, except of course in being the ground state and hence the ultimate end state of Hawking evaporation.

Finally, I show that the phase transitions considered in this chapter correspond to a confinement-deconfinement phase transition in the dual gauge theory using the same topological arguments originally proposed in \cite{Witten:1998qj}.  The expectation value of the temporal Wilson loop operator, $<W(C)>$, is an order parameter for the confinement-deconfinement phase transition, where $C$ is a closed loop wrapping the $\tau$-cycle.  In the large $N$ limit of the gauge theory, $<W(C)>\propto e^{-A({\cal D})}$ where $A({\cal D})$ is the smallest regularised world-area of the string world-sheet, ${\cal D}$, with boundary $C$.  The (Euclidean) dual gauge theory lives on the conformal boundary, $S^1\times S^{D-2}/{\mathbb Z}_k$ and the partition function receives contributions from all three metrics, so the behaviour of $<W(C)>$ for all three solutions is needed.  The topologies of the three instantons are
\begin{align}
&S^1\times B^{D-1}/{\mathbb Z}_k  \qquad\qquad  {\mathrm {for~the~EHS~and~the~OAdS~geometry,}}\nonumber \\
&B^2\times S^{D-2}/{\mathbb Z}_k  \qquad\qquad {\mathrm {for~the~black~hole.}} \nonumber
\end{align}
For the EHS and the OAdS geometry $C$ must lie along the $S^1$, which is non-contractible.  There is thus no world-sheet with this boundary $C$, so $<W(C)>=0$ for the EHS and the OAdS geometry, meaning they correspond to a confinement phase.  On the other hand, for the black hole $C$ must lie along the $B^2$, {\emph{i.e.} it is the boundary of a disk.  Therefore $<W(C)> \ne 0$ for the black hole, meaning it corresponds to a deconfinement phase.  

The thermodynamics of black holes in asymptotically locally or globally AdS space-times, such as those considered in this chapter, is an example where the phase structure is understood reasonably well on both sides of the gauge/gravity correspondence.  In chapter \ref{HairyBH} below, on the gravitational side of the correspondence I analyse a perturbative instability for rotating black holes, which currently does not have a solid interpretation on the gauge theory side.

\chapter{Hairy Black Holes with One Killing Field}
\label{HairyBH}

In general, constructing analytic solutions to Einstein's equations is not an easy task; indeed in the first two chapters I used two rather sophisticated solution generating methods that heavily exploit the underlying symmetries of the space-time and matter fields.  A less sophisticated method but common strategy is to assume a set of symmetries by inputting a suitable metric and matter field ansatz into the field equations.  This is not guaranteed \emph{\`a priori} to produce a consistent solution but there exist theorems on space-time structure which aid in choosing an appropriate ansatz.  For instance, the rigidity theorem states that in any number of dimensions a stationary space-time must admit at least two Killing vectors since it must also be axisymmetric \cite{Hawking:1971vc,Hollands:2006rj,Moncrief:2008mr}, topological censorship demands that four dimensional black holes can only have a spherical topology \cite{Friedman:1993ty}, and the no-hair theorem states that black holes in space-times with non-negative cosmological constant have no hair \cite{Bhattacharya:2007ap,Ruffini:1971xx} when coupled to numerous matter fields.

Despite the rigidity theorem, four dimensional static magnetically charged Reissner-N\"ordstrom black holes with broken rotational symmetry were constructed perturbatively in situations where the horizon radius is close to the critical value for the onset of an instability toward the development of a nonzero massive vector meson field just outside the horizon \cite{Ridgway:1995ke}.  Such black holes have vector meson hair and possess discrete rotational symmetries but are not axisymmetric.  They sidestep the requirement of rotational symmetry because that requirement is a consequence of the lack of higher mass multipole moments implied by the no-hair theorem.  These solutions were the first example of static black holes invariant under a single Killing vector field.

For space-times with a negative cosmological constant the no-hair theorem no longer applies because the reflecting boundary conditions of AdS can support a nontrivial matter field.  Despite this, all known asymptotically AdS black hole solutions had at least two Killing vectors, regardless of the presence of hair. 
In \cite{Dias:2011at} a novel solution was constructed that considered five dimensional Einstein gravity  with negative cosmological constant minimally coupled to a massless complex doublet scalar field, describing ``lumpy'' scalar hair co-rotating with a black hole.  This configuration is motivated by a superradiant scattering instability for rotating AdS black holes: scalar field modes $\Phi\sim e^{-i(\omega t-m\phi)}$ can increase their amplitude by scattering off the horizon of a rotating black hole with horizon angular velocity $\Omega_H$ so long as $\omega/m<\Omega_H$.  In asymptotically AdS space-times, this leads to an instability since the amplified mode is reflected back to the horizon by the AdS boundary conditions, leading to a second scattering and a further increase in amplitude.  However, superradiant scattering mines rotational energy from the black hole until $\omega/m=\Omega_H$ and the end result is a lump of scalar field co-rotating with the black hole.  When this condition is met, the space-time metric and scalar fields are collectively invariant under a single Killing vector, which is tangent to the generators of the horizon.  In addition to these hairy black holes, there are solitonic scalar field configurations invariant under a single Killing vector field, which describe smooth geometries whose self-gravitation is balanced by centrifugal rotation; such solutions are known as boson stars.  Both of these single Killing vector solutions bypass the rigidity theorem because they are not stationary but rather harmonic in time.

In this chapter, which is based on the work published in \cite{Stotyn:2011ns}, I extend the results of \cite{Dias:2011at} to arbitrary odd space-time dimension $D\ge5$ and construct analytic boson star and black hole solutions perturbatively in the dimensionless scalar field amplitude parameter $\epsilon\ll1$.  This is done by a cohomogeneity-1 ansatz, \emph{i.e.} metric functions of the radial coordinate only, along with a judicious choice of scalar fields whose stress tensor shares the symmetries of the metric.  This ensures that the resulting equations of motion form a set of coupled ordinary differential equations instead of a system of coupled partial differential equations.  The solutions cannot be written in a form valid for arbitrary dimension so I give explicit results in five and seven dimensions;\footnote{Explicit solutions in nine and eleven dimensions are also available in \cite{Stotyn:2011ns} but for the sake of brevity, I only give the five and seven dimensional results here.} I include five dimensions because various technical discrepancies are found with the results of \cite{Dias:2011at}; these discrepancies are explicitly discussed when they appear in section \ref{NearRegion}.  Furthermore, I present arguments for why these hairy black hole configurations are believed to be stable with respect to certain classes of scalar field perturbations but are unstable with respect to others, leading to the development of structure on smaller and smaller scales.

\section{Setup}\label{Setup}

In this section the model for constructing hairy black holes and boson stars in arbitrary odd dimension is introduced and the resulting equations of motion are given.  A vital ingredient to this construction is choosing an ansatz for the scalar fields such that the matter stress tensor shares the same symmetries as the metric.

\subsection{Metric and Scalar Field Ansatz}

The theory under consideration is $D=n+2$ dimensional Einstein gravity with negative cosmological constant minimally coupled to an $\frac{n+1}{2}$-tuplet complex scalar field
\begin{equation}
S=\frac{1}{16\pi G_D}\int{d^Dx\sqrt{-g}\left(R_D+\frac{n(n+1)}{\ell^2}-2\big|\nabla\vec{\Pi}\big|^2\right)},\label{eq:action}
\end{equation}
where $R_D$ is the Ricci scalar and I take the usual convention $\Lambda=-\frac{n(n+1)}{2\ell^2}$.  In order to obtain the desired symmetries, namely that the matter stress tensor has the same symmetries as the metric, the $n$-sphere must possess a Hopf-fibration, thus only odd dimensions with $n\ge3$ are considered.  The proposed metric and scalar field ans\"atze are
\begin{equation}
ds^2=-f(r)g(r)dt^2+\frac{dr^2}{f(r)}+r^2\left(h(r)\big(d\chi+A_idx^i-\Omega(r) dt\big)^2+g_{ij}dx^idx^j\right),  \label{eq:metric}
\end{equation}
\begin{equation}
\Pi_i=\Pi(r) e^{-i\omega t}z_i, \quad\quad\quad\quad i=1...\frac{n+1}{2}, \label{eq:ScalarField}
\end{equation}
where $z_i$ are complex coordinates such that $dz_id\bar{z}_i$ is the metric of a unit $n$-sphere.  An explicit and convenient choice for the $z_i$ is

\begin{equation}
z_i=\left\{{\begin{array}{lll}
e^{i(\chi+\phi_i)}\cos\theta_i\displaystyle\prod_{j<i}\sin\theta_j  & \qquad\quad i=1...\frac{n-1}{2}, \\
e^{i\chi}\displaystyle\prod_{j=1}^{\frac{n-1}{2}}\sin\theta_j  & \qquad\quad   i=\frac{n+1}{2},
\end{array} } \right.
\end{equation}
in which case $dz_id\bar{z}_i=(d\chi+A_idx^i)^2+g_{ij}dx^idx^j$ is the Hopf fibration of the unit $n$-sphere, where
\begin{equation}
A_idx^i=\sum_{i=1}^{\frac{n-1}2}{\cos^2\theta_i\left[\prod_{j<i}\sin^2\theta_j\right]d\phi_i}
\end{equation}
and $g_{ij}$ is the metric on a unit $\mathbb{CP}^{\frac{n-1}{2}}$.  In these coordinates the scalar fields are manifestly single-valued on the space-time since $\chi$ and $\phi_i$ have period $2\pi$ while the $\theta_i$ have period $\frac{\pi}{2}$.  To verify the ansatz at this point, if $n=3$ and the coordinate transformation $\chi=\psi-\frac{\phi}{2},$ $\theta=\frac{\vartheta}2$ is performed, the ansatz considered in \cite{Dias:2011at} is exactly recovered.

The form of the scalar fields is crucial to this construction and was first considered in \cite{Hartmann:2010pm}: it is clear from equation (\ref{eq:ScalarField}) that the scalar fields can be viewed as coordinates on ${\mathbb C}^\frac{n+1}{2}$.  Given that $\Pi(r)$ is a function of $r$ only, for each value of $r$, $\vec{\Pi}$ traces out a round $n$-sphere with a  time-varying but otherwise constant  phase.  On the other hand, constant-$r$ surfaces in the metric (\ref{eq:metric}) correspond to squashed rotating $n$-spheres.  The stress tensor for the scalar field takes the form
\begin{equation} \label{eq:Tab}
T_{\mu\nu}=\left(\partial_\mu\vec{\Pi}^*\partial_\nu\vec{\Pi}+\partial_\mu\vec{\Pi}\partial_\nu\vec{\Pi}^*\right)-g_{\mu\nu}\left(\partial_\rho\vec{\Pi}\partial^\rho\vec{\Pi}^*\right),
\end{equation}
which has the same symmetries as the metric (\ref{eq:metric}) since the first term is the pull-back of the round metric of the $n$-sphere and the second term is proportional to $g_{\mu\nu}$.

Although the matter stress tensor has the same symmetries as the metric, the scalar fields themselves do not.  Indeed, the metric (\ref{eq:metric}) is invariant under $\partial_t$, $\partial_\chi$ as well as the rotations of $\mathbb{CP}^{\frac{n-1}{2}}$, while the scalar field (\ref{eq:ScalarField}) is only invariant under the combination
\begin{equation}
K=\partial_t+\omega\partial_\chi. \label{eq:KV}
\end{equation}
Therefore, any solution with nontrivial scalar field will only be invariant under the single Killing vector field given by (\ref{eq:KV}).

\subsection{Equations of Motion}

The equations of motion resulting from the action (\ref{eq:action}) are $G_{\mu\nu}-\frac{n(n+1)}{2\ell^2}g_{\mu\nu}=T_{\mu\nu}$ and $\nabla^2\vec{\Pi}=0$ which ought to have nontrivial solutions by virtue of the matter stress tensor possessing the same symmetries as the metric.  Indeed, inserting the ansatz (\ref{eq:metric}) and (\ref{eq:ScalarField}) into the equations of motion yields a system of five coupled second order ODEs
\begin{equation}
\begin{split}
f''-\frac{6f'}{fr}\left(\frac{rf'}{6}-\frac f6+\Xi\right) + \frac{4h'}{r}+\frac{n^2-1}{r^2}+\frac{n^2-1}{\ell^2}+\frac{8\Pi'\Pi}{r}+\frac{4\Pi^2(\omega-\Omega)^2}{fg}&\\-\frac{8\Xi^2}{fr^2}
-\frac{4\Pi^2\left(1+\frac{(n-1)}2h\right)}{hr^2}-\frac{2(n-3)\Xi}{r^2}&=0,\label{eq:fEq}
\end{split}
\end{equation}
\begin{equation}
\begin{split}
g''-g'\left(\frac{4\Xi}{fr}+\frac{g'}{g}-\frac{1}{r}\right)-4g\left(\frac{\big(\Xi r^{\frac{n-1}2}\sqrt{h}\big)'}{fr^{\frac{n+1}2}\sqrt{h}}+\frac{\frac{(n-1)}2h^2-\Pi^2}{fhr^2}+\frac{3(n+1)}{2f\ell^2}\right.&\\
\left.-\frac{(n-3)\Xi}{2fr^2}\right)-\frac{8\Pi^2(\omega-\Omega)^2}{f^2}-\frac{hr^2\Omega'^2}{f}&=0,\label{eq:gEq}
\end{split}
\end{equation}
\begin{equation}
h''+\frac{h'}{r}-\frac{2h'}{fr}\left(\Xi+\frac{frh'}{2h}\right)+\frac{h^2r^2\Omega'^2}{fg}+\frac{4(1-h)}{fr^2}\left(\Pi^2+\frac{(n+1)}{2}h\right)=0, \label{eq:hEq}
\end{equation}
\begin{equation}
\Omega''+\frac{4\Pi^2}{fhr^2}(\omega-\Omega)+\Omega'\left(\frac{f'}{f}+\frac{2h'}{h}+\frac{2\Xi}{fr}+\frac{2n+1}{r}\right)=0, \label{eq:OmegaEq}
\end{equation}
\begin{equation}
\Pi''-\frac{2\Pi'}{fr}\left(\Xi-\frac{f}{2}\right)+\frac{\Pi(\omega-\Omega)^2}{f^2g}-\frac{\big(1+(n-1)h\big)\Pi}{fhr^2}=0, \label{eq:PiEq}
\end{equation}
where $\Xi=h+\Pi^2-\frac{n+1}{2}-\frac{(n+1)r^2}{2\ell^2}$ and a $'$ denotes differentiation with respect to $r$.  In addition to these second order ODEs, the Einstein equations further impose two first order ODEs in the form of constraint equations, $C_1=0$ and $C_2=0$. Explicitly, these are
\begin{equation}
C_1=\frac{(f^2ghr^{2(n-1)})'}{fr^{2n-3}}+4gh\Xi,
\end{equation}
\begin{equation}
\begin{split}
C_2={}&\frac{\Pi^2(\omega-\Omega)^2}{f^2g}+\Pi'^2-\frac{r^2h\Omega'^2}{4fg}+\frac{\Xi(r^{n+1}h)'}{fhr^{n+2}}+\frac{(hf)'h'}{4fh^2}+\frac{n(fhr^{n-1})'}{2fhr^n}\\
&+\frac{\frac{(n-1)}{2}h^2-\Pi^2}{fhr^2}+\frac{n+1}{2f\ell^2}.
\end{split}
\end{equation}

Inserting $n=3$ into the above equations of motion yields the five dimensional equations of motion in \cite{Dias:2011at}.  Additionally, it is interesting to note the presence of terms proportional to $n-3$ in equations (\ref{eq:fEq}) and (\ref{eq:gEq}), which are therefore absent in five dimensions; one might expect such terms to change the physics of the higher dimensional solutions, though at the perturbative level this may not be apparent.  Before moving on, I emphasise that the above equations of motion are \emph{exact} for arbitrary odd dimension $D\ge5$, however in what follows the equations must be solved for each dimension separately, which I do explicitly here in five and seven dimensions.

\section{Rotating Boson Stars: a Warm-up}\label{Boson}

In this section, I present the boundary conditions that define a boson star and use these to construct such solutions as perturbations around the global AdS background.  Boson stars are smooth horizonless solutions describing rotating solitonic clumps of the massless scalar field.  Furthermore, they can be viewed as a warm-up problem to constructing the hairy black holes.  The perturbative expansion is carried out in orders of the scalar field condensate parameter, $\epsilon$, and results are given up to order $\epsilon^6$.  As a perturbative construction, these results will only be valid for small energies and angular momenta.

\subsection{Boson Star Boundary Conditions} \label{BosonConditions}

\subsubsection*{Boson Star Origin}

Boson stars are smooth, horizonless geometries, which means that all metric functions must be regular at the origin.  Furthermore, due to the slow physical rotation of points as $r\rightarrow0$, surfaces of constant $t$ in the vicinity of the origin ought to be described by round $n$-spheres with $r$ being the proper radial distance.  To find the boundary condition on $\Pi$, (\ref{eq:PiEq}) is multiplied by $r^2$ from which it can be seen that $\Pi$ must vanish at the origin in order to yield consistent equations of motion.  Thus, the boundary conditions at the boson star origin take the form
\begin{eqnarray}
\left.f\right|_{r \rightarrow 0} = 1 + \mathcal{O}(r^2), \quad 
\left.g\right|_{r \rightarrow 0} = g(0)+\mathcal{O}(r), \quad 
\left.h\right|_{r \rightarrow 0} = 1 + \mathcal{O}(r^2), \label{eq:OriginBC}\\
\left.\Omega\right|_{r \rightarrow 0} = \Omega(0)+\mathcal{O}(r), \quad 
\left.\Pi\right|_{r \rightarrow 0} = \mathcal{O}(r).\quad\quad\quad\quad\quad \nonumber
\end{eqnarray}

\subsubsection*{Boson Star Asymptotics}

In order to simplify the asymptotic boundary conditions, a residual gauge freedom is first exploited.  It is straightforward to show that the transformation
\begin{eqnarray} \label{PsiGaugeFreedom}
\chi \rightarrow \chi + \lambda t, \quad\quad \Omega \rightarrow \Omega + \lambda, \quad\quad \omega \rightarrow \omega - \lambda
\end{eqnarray}
for some arbitrary constant $\lambda$, leaves both the metric (\ref{eq:metric}) and scalar field (\ref{eq:ScalarField}) unchanged.  This gauge invariance is used to pick a frame which is not rotating at infinity, \emph{i.e.} it is used to set $\Omega \rightarrow 0$ asymptotically.

In the $r \rightarrow \infty$ limit, the boundary conditions for the boson star will asymptote to the AdS solution with corrections for mass and angular momentum.  To ensure the solution asymptotically describes a mass monopole, an $r^{-(n-1)}$ correction is imposed for $f$.  Next, requiring the solutions to have finite masses and angular momenta means $g,h$ and $\Omega$ must each have corrections with an $r^{-(n+1)}$ fall-off.  These considerations determine the boundary conditons for $f,g,h,$ and $\Omega$ up to some constants $C_f, C_h,$ and $C_\Omega$.  The remaining boundary condition is set by requiring $\Pi$ to be normalizable, which means it must decay like $r^{-(n+1)}$.  Explicitly, the asymptotic boundary conditions are given by
\begin{align} \label{eq:AsymBC}
\left.f\right|_{r \rightarrow \infty}={}& \frac{r^2}{\ell^2} + 1 + \frac{C_f \ell^{n-1}}{r^{n-1}} + \mathcal{O}(r^{-n}), \quad\>\> 
\left.g\right|_{r \rightarrow \infty} = 1 - \frac{C_h \ell^{n+1}}{r^{n+1}} + \mathcal{O}(r^{-(n+2)}), \nonumber \\ 
\left.h\right|_{r \rightarrow \infty}={}& 1 + \frac{C_h \ell^{n+1}}{r^{n+1}}+ \mathcal{O}(r^{-(n+2)}), \quad\quad 
\left.\Omega\right|_{r \rightarrow \infty} = \frac{C_{\Omega} \ell^{n}}{r^{n+1}} + \mathcal{O}(r^{-(n+2)}),\\ 
\left.\Pi\right|_{r \rightarrow \infty}={}& \frac{\epsilon \ell^{n+1}}{r^{n+1}} + \mathcal{O}(r^{-(n+2)}).\nonumber
\end{align}
Here and in what follows, $\epsilon$ provides a dimensionless measure of the amplitude of the scalar field.

\subsection{Perturbative Boson Star}\label{PertBS}

With the appropriate boundary conditions and the equations of motion at hand, I now explicitly construct perturbative boson stars, starting with expanding the fields in terms of the scalar field amplitude, $\epsilon$, as follows:
\begin{equation}
F(r,\epsilon)=\sum_{i=0}^m{\tilde{F}_{2i}(r)\epsilon^{2i}}, \quad\quad\quad \Pi(r,\epsilon)=\sum_{i=0}^m{\tilde{\Pi}_{2i+1}(r)\epsilon^{2i+1}}, \quad\quad\quad \omega(\epsilon)=\sum_{i=0}^m{\tilde{\omega}_{2i}\epsilon^{2i}}, \label{eq:FieldExpansion}
\end{equation}
where $F=\{f,g,h,\Omega\}$ is shorthand for the metric functions in (\ref{eq:metric}).  The metric functions are expanded in even powers of $\epsilon$ while the scalar fields are expanded in odd powers.  This allows a perturbative expansion as follows: start with the global AdS background at order $i=0$, then introduce nontrivial scalar fields without back-reacting on the metric by solving (\ref{eq:PiEq}).  The full set of equations of motion will then be satisfied up to order $\epsilon$.  At the next order, $i=1$, the scalar fields $\tilde{\Pi}_1(r)$ then source corrections to the gravitational fields $\tilde{F}_2(r)$, which in turn back-react on the scalar fields producing the corrections $\tilde{\Pi}_3(r)$.  The equations of motion will then be satisfied up to order $\epsilon^3$.  The perturbative solution can, in principle, be obtained by this bootstrapping procedure up to arbitrary order, $i$.  Note that the frequency must also be expanded in even powers of $\epsilon$.  This is because at the linear order, the frequency is determined by the scalar field alone but at the next order, the scalar field back reacts on the metric, inducing nontrivial frame-dragging effects, which in turn have a nontrivial effect on the rotation of the scalar field.  In practice, these corrections to $\omega$ are found by imposing the boundary conditions.

The global AdS solution is given by (\ref{eq:metric}) with
\begin{equation}
f_0=1+\frac{r^2}{\ell^2},\quad\quad\quad g_0=1,\quad\quad\quad h_0=1,\quad\quad\quad \Omega_0=0.
\end{equation}
In this background, I find the most general massless scalar field solution to (\ref{eq:PiEq}), consistent with the asymptotic boundary conditions (\ref{eq:AsymBC}), to be:
\begin{equation}
\Pi_1(r)=\frac{r\ell^{n+1} }{(r^2+\ell^2)^{\frac{n+2}{2}}}{_2F_1}\left[\frac{n+2-\omega \ell}{2},\frac{n+2+\omega\ell}{2};\frac{n+3}{2};\frac{\ell^2}{r^2+\ell^2}\right],   \label{eq:Pi1}
\end{equation}
where $_2F_1$ is the hypergeometric function.  Now in order to satisfy the boundary conditions at the origin (\ref{eq:OriginBC}), $\omega$ must further be restricted to
\begin{equation}
\omega\ell=n+2+2k,\quad\quad\quad\quad k=0,1,2,...,  \label{eq:kRadialModes}
\end{equation}
where the non-negative integer, $k$, describes the various possible radial modes of the scalar field.  Although in principle, any radial profile can be built up out of a linear combination of the radial modes, this introduces multiple frequency parameters, $\omega_k$, which is inconsistent with the existence of the Killing vector field (\ref{eq:KV}), as well as the scalar field ansatz (\ref{eq:ScalarField}).  Furthermore, the mode $k=0$ yields the ground state while higher modes represent excited states.  Therefore, in what follows $k=0$ is taken to be the only mode present, in which case (\ref{eq:Pi1}) simplifies to
\begin{equation}
\Pi_1(r)=\frac{r\ell^{n+1} }{(r^2+\ell^2)^{\frac{n+2}{2}}}. \label{eq:Piepsilon}
\end{equation}

Proceeding up the perturbative ladder, (\ref{eq:Piepsilon}) and the field expansions (\ref{eq:FieldExpansion}) are inserted into the equations of motion, which are then expanded in $\epsilon$ and solved at order $\epsilon^2$.  In general, the solutions contain two constants of integration, which are uniquely fixed by the boundary conditions.  These fields, $\tilde{F}_2(r)$ are then inserted into the equation of motion for $\Pi(r)$ to find the $\epsilon^3$ correction to the scalar fields.  This process can, in principle, be taken to arbitrary order in $\epsilon$. In practice, however, the expressions become rather unwieldy at higher orders making this increasingly difficult to accomplish.  Up to order $\epsilon^6$, I find the general solutions to take the form
\begin{equation}
f(r)=1+\frac{r^2}{\ell^2}-\frac{r^2\ell^{n-1}f_{n;2,0}}{(r^2+\ell^2)^{n+1}}\epsilon^2-\frac{r^2\ell^{n-1}f_{n;4,0}}{(r^2+\ell^2)^{2n+3}}\epsilon^4+{\cal O}(\epsilon^6),\label{eq:fExpansion}
\end{equation}
\begin{equation}
g(r)=1-\frac{2\ell^{2n+2}\big((n+1)r^2+\ell^2\big)}{n(r^2+\ell^2)^{n+2}}\epsilon^2-\frac{\ell^{n+1}g_{n;4,0}}{(r^2+\ell^2)^{2n+4}}\epsilon^4+{\cal O}(\epsilon^6),
\end{equation}
\begin{equation}
h(r)=1+\frac{r^2\ell^{n+1}h_{n;4,0}}{(r^2+\ell^2)^{2n+3}}\epsilon^4+{\cal O}(\epsilon^6),
\end{equation}
\begin{equation}
\Omega(r)=\frac{\ell^{n}\Omega_{n;2,0}}{(r^2+\ell^2)^{n+1}}\epsilon^2+\frac{\ell^{n}\Omega_{n;4,0}}{(r^2+\ell^2)^{2n+3}}\epsilon^4+{\cal O}(\epsilon^6),
\end{equation}
\begin{equation}
\Pi(r)=\frac{r\ell^{n+1} }{(r^2+\ell^2)^{\frac{n+2}{2}}}\epsilon+\frac{r\ell^{n+3}\Pi_{n;3,0}}{(r^2+\ell^2)^{\frac{3n+4}{2}}}\epsilon^3+\frac{r\ell^{n+3}\Pi_{n;5,0}}{(r^2+\ell^2)^{\frac{5(n+2)}{2}}}\epsilon^5+{\cal O}(\epsilon^7), \label{eq:PiExpansion}
\end{equation}
where the fields $\{f_{n;s,0},g_{n;s,0},h_{n;s,0},\Omega_{n;s,0},\Pi_{n;s,0}\}$ are simple polynomials in $r$; in this notation $s$ labels the order in $\epsilon$.\footnote{The purpose of the ``$0$'' index on these coefficients will become clear in the next section where the perturbative black hole fields involve a second pertubation parameter.}  These fields are catalogued in Appendix~\ref{app:fields} for $n=3,5$ up to order $\epsilon^6$.  Similarly, the explicit corrections to $\omega_n$ for the boson star can be found by taking the $r_+\rightarrow0$ limit of the general expressions for $\omega_n$ in Appendix~\ref{app:thermo}.

\section{Perturbative Black Holes}\label{BlackHole}

I now turn to the more involved problem of constructing hairy black hole solutions.  The presence of a horizon at $r=r_+$ provides an additional length scale, or rather another perturbative expansion parameter, $r_+ / \ell$, on top of the scalar field amplitude, $\epsilon$, making this a two-parameter family of solutions.  One might expect no stable solutions with scalar hair since it was shown in \cite{Astefanesei:2003qy,Pena:1997cy} that one cannot have black holes inside boson stars.  However this applies to static black holes and it was pointed out in \cite{Dias:2011at} that this prohibition is removed if the black hole and the boson star are co-rotating, since the attraction to the black hole is balanced by centrifugal repulsion.  Similar to the boson star case then, to find these black hole solutions the fields must be expanded in powers of both the amplitude of the scalar field condensate, $\epsilon$, and the dimensionless horizon radius, $r_+/\ell$.  

As pointed out in \cite{Dias:2011at}, inserting this double expansion into the equations of motion results in a set of differential equations that cannot be simultaneously solved everywhere in the space-time.  This is essentially because one cannot simultaneously have full control over both perturbation parameters asymptotically and near the horizon; $\epsilon$ is intrinsically an asymptotic expansion parameter while $r_+/\ell$ is intrinsically a near horizon expansion parameter. This problem is circumvented by applying a matched asymptotic expansion, which involves splitting the space-time into two regions -- a far-region, $r \gg r_+$, and a near-region, $r_+ \le r \ll \ell$ -- and solving the equations of motion in each separately; this process will be made explicit in section \ref{MatchingConditions}.  At each order there will be four arbitrary constants of integration, two from each region.  The boundary conditions can be used to fix one of these in each region, while the remaining two constants can be fixed by matching the solutions where the two regions overlap: $r_+ \ll r \ll \ell$.   This admits a unique solution valid in the entire space-time that satisfies both boundary conditions.   This analysis is carried out in the following section and results are given up to $\mathcal{O}(\epsilon^m (r_{+}/\ell)^n)$ where $m+n \le 6$.  As in the boson star case only the ground state hairy black holes will be considered and these results will again only be valid for small energies and angular momenta.

The double expansion of the fields can be interpreted as placing a black hole inside a rotating boson star, or alternatively as placing a nontrivial scalar field around a small rotating black hole.  Consequently, the limit $r_+\rightarrow0$ should recover the boson star of the previous section and similarly, the limit $\epsilon\rightarrow0$ should recover an asymptotically AdS rotating black hole, \emph{i.e.} a Myers-Perry-AdS black hole \cite{Gibbons:2004js,Gibbons:2004uw,Hawking:1998kw}.  The two-parameter family of hairy black hole solutions is then seen to be connected to a two-parameter family of Myers-Perry-AdS solutions.  For the hairy black hole, the frequency is uniquely determined in terms of $r_+$ and $\epsilon$, so in the $\epsilon\rightarrow0$ limit, the hairy black hole joins with a one-parameter subset of Myers-Perry-AdS black holes, whose horizon angular velocity is identified with the $\epsilon\rightarrow0$ limit of $\omega$. To phrase this in terms of the space of solutions, consider the $(\epsilon,\omega,r_+)$ octant: the one-parameter family of boson stars corresponds to a line in the $(\epsilon,\omega)$ plane, the Myers-Perry-AdS black holes correspond to the $(\omega,r_+)$ plane, while the hairy black holes correspond to a sheet through the $(\epsilon,\omega,r_+)$ octant.  This sheet meets the $(\epsilon,\omega)$ plane on the line defining boson stars and it meets the $(\omega,r_+)$ plane where it joins with the Myers-Perry-AdS family of solutions.  The double expansion of the fields can also be represented schematically in the following table:
\begin{center}
\renewcommand{\arraystretch}{1.5}
\begin{tabular}{ | c |  c | c | c | c| c | }
\hline
Order   &  $\epsilon^0$   &   \quad \quad $\epsilon^2$ \quad \quad & \quad \quad $\epsilon^4$ \quad \quad & $\cdots$ &   $\sum_{i=0}^{\infty} \left(\epsilon\right)^{2i}$  \\ \hline
$\left(\frac{r_+}{\ell}\right)^0$  &      Global AdS  		&    \multicolumn{3}{c|}{$\rightarrow$ Perturbative Boson Star $\rightarrow$} &    Exact Boson Star    \\ \hline
$\left(\frac{r_+}{\ell}\right)^2$  &      $\downarrow$  &  \multicolumn{3}{c|}{\quad $\searrow$ \quad \bf{Perturbative} \quad $\searrow$ \quad} &       \\ \cline{1-1}
$\left(\frac{r_+}{\ell}\right)^4$  &      Pert. MP-AdS BH  &   \multicolumn{3}{c|}{\bf{Hairy}} & $\vdots$      \\ \cline{1-1}
$\vdots$                           &      $\downarrow$    &   \multicolumn{3}{c|}{\quad $\searrow$ \quad \bf{Black Hole} \quad $\searrow$ \quad} &       \\ \hline
$\sum_{i=0}^{\infty} \left(\frac{r_+}{\ell}\right)^{2i}$  &     Exact MP-AdS BH       &   \multicolumn{3}{c|}{$\cdots$}  &   Exact Hairy BH    \\ \hline
\end{tabular}
\end{center}

Before continuing, I explicitly present the Myers-Perry-AdS solution with equal angular momenta in the two planes of rotation.  In terms of the metric ansatz (\ref{eq:metric}), the metric functions take the form \cite{Dias:2011at}
\begin{equation}
\begin{split}
f=\frac{r^2}{\ell^2}+1-\frac{r_M^{n-1}}{r^{n-1}}&\left(1-\frac{a^2}{\ell^2}\right)+\frac{r_M^{n-1}a^2}{r^{n+1}}, \qquad\quad h=1+\frac{r_M^{n-1}a^2}{r^{n+1}},\\ 
&g=\frac1{h(r)},\qquad\quad \Omega=\frac{r_M^{n-1}a}{r^{n+1}h}, \label{eq:MPAdS}
\end{split}
\end{equation}
where $r_M$ and $a$ are related to the outer horizon, $r_+$, and the angular velocity of the horizon, $\Omega_H=\omega$, by
\begin{equation}
r_M^{n-1}=\frac{r_+^{n+1}(r_+^2+\ell^2)}{r_+^2\ell^2-a^2(r_+^2+\ell^2)}, \quad\quad\quad a=\frac{r_+^2\ell^2\omega}{r_+^2+\ell^2}.
\end{equation}
Small black holes are then obtained by Taylor expanding the metric functions (\ref{eq:MPAdS}) in powers of $r_+/\ell$.

\subsection{Black Hole Boundary Conditions}

\subsubsection*{Black Hole Horizon}

The hairy black holes constructed here are assumed to be non-extremal, so the location of the non-degenerate outermost horizon is defined to be $r=r_{+}$; consequently, $f$ must have a simple zero at $r_+$.  The scalar field and all other metric functions must remain finite in the vicinity of the horizon.  Furthermore, multiplying equation (\ref{eq:PiEq}) by $f^2$ shows that the equations of motion remain consistent across the horizon provided $\Omega(r_+)=\omega$, which is the statement that the black hole and scalar field are co-rotating.  The boundary conditions at the black hole horizon are thus
\begin{equation}
f\big|_{r \rightarrow r_{+}} = \mathcal{O}(r-r_{+}), \quad \quad
g\big|_{r \rightarrow r_{+}} = g(r_+)+\mathcal{O}(r-r_+), \quad \quad
h\big|_{r \rightarrow r_{+}} = h(r_+)+\mathcal{O}(r-r_+), \nonumber
\end{equation}
\begin{equation}\label{eq:BHBC}
\Omega\big|_{r \rightarrow r_{+}} = \omega+\mathcal{O}(r-r_+), \quad \quad
\Pi\big|_{r \rightarrow r_{+}} =\Pi(r_+)+\mathcal{O}(r-r_+).
\end{equation}

\subsubsection*{Black Hole Asymptotics}

The asymptotic boundary conditions for the black hole will be identical to those of the boson star since both are globally AdS with next-to-leading order terms accounting for mass and angular momentum.  Thus the same asymptotic boundary conditions (\ref{eq:AsymBC}) are applied to the black hole as well.

\subsection{Perturbative Hairy Black Hole}

As discussed previously, for the hairy black hole the fields must be perturbatively expanded both in terms of $\epsilon$ and $r_+ / \ell$.  Furthermore, this must be done in the asymptotic (far) region, as well as the near horizon (near) region.

\subsubsection{Far Region}

In the far region, the fields are expanded as follows:
\begin{equation}
\begin{split}
&F^{out}(r,\epsilon,r_+) = \sum_{i=0}^m \sum_{j=0}^n \tilde{F}^{out}_{2j, 2i}(r) \epsilon^{2j} \left(\frac{r_+}{\ell}\right)^{2i},\\
&\Pi^{out}(r,\epsilon,r_+) = \sum_{i=0}^m \sum_{j=0}^n \tilde{\Pi}^{out}_{2j+1, 2i}(r) \epsilon^{2j+1} \left(\frac{r_+}{\ell}\right)^{2i}, \label{eq:FarFieldExpansion}
\end{split}
\end{equation}
where $F^{out}=\{f^{out},g^{out},h^{out},\Omega^{out}\}$ is shorthand for the metric functions in the far-region.  As discussed in section \ref{PertBS}, the scalar frequency $\omega$ has a similar expansion
\begin{equation}
\omega(\epsilon,r_+) = \sum_{i=0}^m \sum_{j=0}^n \tilde{\omega}_{2j, 2i} \epsilon^{2j} \left(\frac{r_+}{\ell}\right)^{2i}, \label{eq:BHomegaExpansion}
\end{equation}
where this expansion holds in both the outer and inner regions since $\omega$ is a globally defined constant. 

The perturbative expansion then proceeds similar to the boson star case except that instead of the background being the global AdS solution it is taken it to be the Myers-Perry-AdS black hole (\ref{eq:MPAdS}) expanded in powers of $r_+ / \ell\ll1$.  The field coefficients $\tilde{F}^{out}_{0, 2i}(r)$ can then be immediately written down up to arbitrary order while trivially satisfying the field equations and asymptotic boundary conditions.  With the complete $r_+ / \ell$ expansion in hand, the goal is then to introduce a nontrivial scalar field and  solve the field equations  in powers of $\epsilon$.  As before, this is done by first inserting the fields $\tilde{F}^{out}_{0, 2i}(r)$ into equation (\ref{eq:PiEq}) and solving the order $\epsilon$ equation to determine $\tilde{\Pi}^{out}_{1,2i}$ order by order in $r_+/\ell$.  The presence of the scalar field then sources the gravitational fields at order $\epsilon^2$, which are then determined by solving equations (\ref{eq:fEq})--(\ref{eq:OmegaEq}) for $\tilde{F}^{out}_{2, 2i}(r)$ order by order in $r_+/\ell$.  At the next order in $\epsilon$, there is a back-reaction on the scalar field, which is then similarly solved for from  equation (\ref{eq:PiEq}) order by order in $r_+/\ell$.  This iteration process continues with gravitational field corrections at every even order in $\epsilon$ and scalar field corrections at every odd order in $\epsilon$.  Due to the powers of $r_+/\ell$ that appear in the Myers-Perry-AdS solution, this procedure is carried out up to $\mathcal{O}(\epsilon^0 (r_{+}/\ell)^{n+3})$, $\mathcal{O}(\epsilon^2 (r_{+}/\ell)^{n+1})$ and $\mathcal{O}(\epsilon^4 (r_{+}/\ell)^{n-1})$.  Applying the asymptotic boundary conditions, I find the fields to have the structure:

\begin{equation}
\begin{split}  \label{eq:fFarRegion}
f^{out}(r)={}&\left[\frac{r^2}{\ell^2}+1-\frac{\ell^{n-1}}{r^{n-1}}\frac{r_+^{n-1}}{\ell^{n-1}}-\frac{\big((n+2)^2+1\big)\ell^{n-1}}{r^{n-1}}\frac{r_+^{n+1}}{\ell^{n+1}}+{\cal O}\left(\frac{r_+^{n+3}}{\ell^{n+3}}\right)\right]\\
&+\epsilon^2\left[-\frac{r^2\ell^{n-1}f_{n;2,0}}{(r^2+\ell^2)^{n+1}}+\frac{\ell^{n-1}f_{n;2,2}^{out}}{r^{n-3}(r^2+\ell^2)^{n+2}}\frac{r_+^{n-1}}{\ell^{n-1}}+{\cal O}\left(\frac{r_+^{n+1}}{\ell^{n+1}}\right)\right]\\
&+\epsilon^4\left[-\frac{r^2\ell^{n-1}f_{n;4,0}}{(r^2+\ell^2)^{2n+3}}+{\cal O}\left(\frac{r_+^{n-1}}{\ell^{n-1}}\right)\right],
\end{split}
\end{equation}
\begin{align}  \label{eq:gFarRegion}
g^{out}(r)={}&\left[1+{\cal O}\left(\frac{r_+^{n+3}}{\ell^{n+3}}\right)\right]+\epsilon^2\left[-\frac{2\ell^{2n+2}\big((n+1)r^2+\ell^2\big)}{n(r^2+\ell^2)^{n+2}}+\frac{g_{n;2,2}^{out}}{r^{n-3}(r^2+\ell^2)^{n+3}}\frac{r_+^{n-1}}{\ell^{n-1}}\right.\nonumber\\
&\left.+{\cal O}\left(\frac{r_+^{n+1}}{\ell^{n+1}}\right)\right]+\epsilon^4\left[-\frac{\ell^{n+1}g_{n;4,0}}{(r^2+\ell^2)^{2n+4}}+{\cal O}\left(\frac{r_+^{n-1}}{\ell^{n-1}}\right)\right],
\end{align}
\begin{equation}
h^{out}(r)=\left[1+{\cal O}\left(\frac{r_+^{n+3}}{\ell^{n+3}}\right)\right]+\epsilon^2\left[{\cal O}\left(\frac{r_+^{n+1}}{\ell^{n+1}}\right)\right]+\epsilon^4\left[\frac{r^2\ell^{n+1}h_{n;4,0}}{(r^2+\ell^2)^{2n+3}}+{\cal O}\left(\frac{r_+^{n-1}}{\ell^{n-1}}\right)\right], \label{eq:hFarRegion}
\end{equation}
\begin{align}  \label{eq:OmegaFarRegion}
\ell\Omega^{out}(r)={}&\left[\frac{(n+2)\ell^{n+1}}{ r^{n+1}}\frac{r_+^{n+1}}{\ell^{n+1}}+{\cal O}\left(\frac{r_+^{n+3}}{\ell^{n+3}}\right)\right]+\epsilon^2\left[\frac{\ell^{n+1}\Omega_{n;2,0}}{(r^2+\ell^2)^{n+1}}+\frac{\Omega_{n;2,2}^{out}}{r^{n-3}(r^2+\ell^2)^{n+2}}\frac{r_+^{n-1}}{\ell^{n-1}}\right.\nonumber\\
&\left.+{\cal O}\left(\frac{r_+^{n+1}}{\ell^{n+1}}\right)\right]+\epsilon^4\left[\frac{\ell^{n+1}\Omega_{n;4,0}}{(r^2+\ell^2)^{2n+3}}+{\cal O}\left(\frac{r_+^{n-1}}{\ell^{n-1}}\right)\right],
\end{align}
\begin{align}
\Pi^{out}(r)={}&\epsilon\left[\frac{r\ell^{n+1} }{(r^2+\ell^2)^{\frac{n+2}{2}}}+\frac{\ell^{n+1}\Pi^{out}_{n;1,n-1}}{r^{n-2}(r^2+\ell^2)^{\frac{n+4}{2}}}\frac{r_+^{n-1}}{\ell^{n-1}}+\frac{\ell^{n+1}\Pi_{n;1,n+1}^{out}}{r^n(r^2+\ell^2)^{\frac{n+4}{2}}}\frac{r_+^{n+1}}{\ell^{n+1}}+{\cal O}\left(\frac{r_+^{n+3}}{\ell^{n+3}}\right)\right]\nonumber\\
&+\epsilon^3\left[\frac{r\ell^{n+3}\Pi_{n;3,0}}{(r^2+\ell^2)^{\frac{3n+4}{2}}}+\frac{\Pi_{n;3,n-1}^{out}}{r^{n-2}(r^2+\ell^2)^{\frac{3(n+2)}{2}}}\frac{r_+^{n-1}}{\ell^{n-1}}+{\cal O}\left(\frac{r_+^{n+1}}{\ell^{n+1}}\right)\right]\\
&+\epsilon^5\left[\frac{r\ell^{n+3}\Pi_{n;5,0}}{(r^2+\ell^2)^{\frac{5(n+1)}{2}}}+{\cal O}\left(\frac{r_+^{n-1}}{\ell^{n-1}}\right)\right],\nonumber
\end{align}
where the fields $F_{n;s,0}$ are the boson star fields of section \ref{PertBS} and $F^{out}_{n;s,t}$ are new fields which enter at order in $(r_+/\ell)^t$.  In solving for these fields, one of the arbitrary constants from each second order ODE was fixed by the asymptotic boundary conditions while the other constant was fixed by the matching condition.  The double expansion of $\omega$, which is also obtained by matching the inner and outer solutions, is catalogued in Appendix~\ref{app:thermo} and the fields with the matching condition already imposed are catalogued in Appendix~\ref{app:fields}.  
An explicit discussion of the matching procedure is postponed to section \ref{MatchingConditions}. Note that taking the $r_+\rightarrow0$ limit of the above yields the boson star fields of section \ref{PertBS} as is required.

\subsubsection{Near Region}\label{NearRegion}

Since these solutions are constructed perturbatively, it is assumed that the energy and angular momenta are low, so by construction $r_+\ll\ell$.  In the far region expansion both the perturbation parameters are small compared to any other scale since $r/\ell\gg1$ by assumption, whereas $\epsilon\ll1$ and $r_+/\ell\ll1$.  In the near region, however, the scale of $r$ can become comparable to $r_+$ so it is necessary to introduce a new radial coordinate, $z\equiv \ell r /r_+ $, such that the horizon is located at $z=\ell$.  This ensures that $z/\ell\ge1$ is always large with respect to $r_+/\ell\ll1$ so that the fields can be safely expanded in powers of $r_+/\ell$.  Thus, the double field expansion in the near region is done using $z$ as the radial coordinate.

The gravitational and scalar field expansions are given by
\begin{equation}
\begin{split}
&F^{in}(z,\epsilon,r_+) = \sum_{i=0}^m \sum_{j=0}^n \tilde{F}^{in}_{2j, 2i}(z) \epsilon^{2j} \left(\frac{r_+}{\ell}\right)^{2i}, \\
&\Pi^{in}(z,\epsilon,r_+) = \sum_{i=0}^m \sum_{j=0}^n \tilde{\Pi}^{in}_{2j+1, 2i+1}(z) \epsilon^{2j+1} \left(\frac{r_+}{\ell}\right)^{2i+1}, \label{eq:NearFieldExpansion}
\end{split}
\end{equation}
where $F^{in}=\{f^{in},g^{in},h^{in},\Omega^{in}\}$ are the metric functions in the near-region.  Recall, however, that the frequency $\omega$ is still given by equation (\ref{eq:BHomegaExpansion}) since it is defined globally.

Again, the perturbative solutions start at order $\epsilon^0$ with the Myers-Perry-AdS solution, with gravitational fields $\tilde{F}^{in}_{0,2i}$.  These are inserted into the equations of motion, appropriately transformed to equations of $z$, nontrivial scalar fields are added by expanding to order $\epsilon$ and solving the equations order by order in $r_+/\ell$.  Once all orders in $r_+/\ell$ have been calculated up to the desired cutoff, the matching conditions must be imposed before continuing up the perturbative ladder.  Next, the scalar fields at order $\epsilon$ source the gravitational fields at order $\epsilon^2$.  These have to be solved order by order in $r_+/\ell$ and matched to the far region expansions before the process is continued.  The result of this procedure up to ${\cal O}(\epsilon^a(r_+/\ell)^b)$ such that $a+b\le6$, with the matching conditions already imposed are found to be

\begin{align}
f^{in}(z)&=\left[1-\frac{\ell^{n-1}}{z^{n-1}}+\left\{\frac{z^2}{\ell^2}-\big((n+2)^2+1\big)\frac{\ell^{n-1}}{z^{n-1}}+(n+2)^2\frac{\ell^{n+1}}{z^{n+1}}\right\}\frac{r_+^2}{\ell^2}\right.\label{eq:fNearRegion}\\
&\left.+\bigg(2(n+2)\omega_{n;0,2}\ell+(n+2)^2\big((n+2)^2-1\big)\bigg)\left(\frac{\ell^{n+1}-\ell^{n-1}z^2}{z^{n+1}}\right)\frac{r_+^4}{\ell^4}+{\cal O}\left(\frac{r_+^6}{\ell^6}\right)\right]\nonumber\\
&+\epsilon^2\left[\frac{r_+^2}{\ell^2}\frac{\ell^{n-1}}{z^{n-1}}\left(\int_{1}^{z/\ell}{\left(\log\left[1-\frac{1}{x^2}\right]f^{in}_{n;2,2}(x)dx\right)}-\log\left[1-\frac{\ell^2}{z^2}\right]\int_{1}^{z/\ell}{f^{in}_{n;2,2}(x)dx}\right)\right.\nonumber\\
&\left.+{\cal O}\left(\frac{r_+^4}{\ell^4}\right)\right]+\epsilon^4{\cal O}\left(\frac{r_+^2}{\ell^2}\right),\nonumber
\end{align}
\begin{align}
g^{in}(z)={}&\bigg[1-(n+2)^2\frac{\ell^{n+1}}{z^{n+1}}\frac{r_+^2}{\ell^2}+\bigg\{-\bigg(2(n+2)\omega_{n;0,2}\ell+(n+2)^2\big((n+2)^2-1\big)\bigg)\frac{\ell^{n+1}}{z^{n+1}}\nonumber\\
&+(n+2)^4\frac{\ell^{2n+2}}{z^{2n+2}}\bigg\}\frac{r_+^4}{\ell^4}+{\cal O}\left(\frac{r_+^6}{\ell^6}\right)\bigg]+\epsilon^2\bigg[-\frac{2}{n}+\frac{r_+^2}{\ell^2}\bigg\{K^{g}_{n;2,2}+g^{in}_{n;2,2} \nonumber\\
&+\frac{4\Gamma^4\big[\frac{n}{n-1}\big]}{n\Gamma^2\big[\frac{n+1}{n-1}\big]}\int_1^{z/\ell}{x\left({P^{\>'}_{\frac1{n-1}}}[2x^{n-1}-1]\right)^2dx}\bigg\}+{\cal O}\left(\frac{r_+^4}{\ell^4}\right)\bigg]  \label{eq:gNearRegion} \\
&+\epsilon^4\left[{-\frac{g_{n;4,0}(0)}{\ell^{3n+7}}+\cal O}\left(\frac{r_+^2}{\ell^2}\right)\right],\nonumber
\end{align}
\begin{align}
\ell\Omega^{in}(z)&=\bigg[(n+2)\frac{\ell^{n+1}}{z^{n+1}}+\left\{\ell\omega_{n;0,2}\frac{\ell^{n+1}}{z^{n+1}}+(n+2)^3\big(z^{n+1}-\ell^{n+1}\big)\frac{\ell^{n+1}}{z^{2n+2}}\right\}\frac{r_+^2}{\ell^2}\nonumber\\
&+\bigg\{\ell\omega_{n;0,4}\frac{\ell^{n+1}}{z^{n+1}}+\big(z^{n+1}-\ell^{n+1}\big)(n+2)^2\bigg(\big(3\ell\omega_{n;0,2}+(n+2)\big((n+2)^2-1\big)\big)\frac{\ell^{n+1}}{z^{2n+2}}\nonumber\\
&+(n+2)^3\frac{\ell^{2n+2}}{z^{3n+3}}\bigg)\bigg\}\frac{r_+^4}{\ell^4}+{\cal O}\left(\frac{r_+^6}{\ell^6}\right)\bigg]+\epsilon^2\bigg[\big(\ell\Omega_{n;2,0}(0)-\ell\omega_{n;2,0}\big)\left(1-\frac{\ell^{n+1}}{z^{n+1}}\right)\nonumber\\
&+\ell\omega_{n;2,0}+\frac{r_+^2}{\ell^2}\bigg\{\omega_{n;2,2}+\int_{1}^{z/\ell}\bigg[\frac{K^{\Omega}_{n;2,2}}{x^{n+2}}+\frac{1}{x^{n+2}}\int_{1}^{x}\left[\frac{4(n+2)}{y^{n+2}(y^{n-1}-1)}\bigg(\Omega^{in}_{n;2,2}\right.\nonumber\\
&+y^{2n}\bigg(\frac{n+3}{2}-y^{n+1}\bigg)\frac{\Gamma^4\big[\frac{n}{n-1}\big]}{\Gamma^2\big[\frac{n+1}{n-1}\big]}\big(P_{\frac1{n-1}}[2y^{n-1}-1]\big)^2 \label{eq:OmegaNearRegion} \\
&\left.\left.-\frac{(n^2-1)y^{n+1}}{4(y^{n-1}-1)}\int_1^y{f^{in}_{n;2,2}(w)dw}\right)\right]dy\bigg]dx\bigg\}+{\cal O}\left(\frac{r_+^4}{\ell^4}\right)\bigg]+\epsilon^4\bigg[\ell\omega_{n;4,0}\nonumber\\
&+\big(\ell\Omega_{n;4,0}(0)-\ell\omega_{n;4,0}\big)\left(1-\frac{\ell^{n+1}}{z^{n+1}}\right)+{\cal O}\left(\frac{r_+^2}{\ell^2}\right)\bigg],\nonumber
\end{align}
\begin{align}
h^{in}(z)={}&\bigg[1+(n+2)^2\frac{\ell^{n+1}}{z^{n+1}}\frac{r_+^2}{\ell^2}+\bigg(2(n+2)\omega_{n;0,2}\ell+(n+2)^2\big((n+2)^2-1\big)\bigg)\frac{\ell^{n+1}}{z^{n+1}}\frac{r_+^4}{\ell^4}\nonumber\\
&+{\cal O}\left(\frac{r_+^6}{\ell^6}\right)\bigg]+\epsilon^2\left[h_{n;2,2}^{in}\frac{r_+^2}{\ell^2}+{\cal O}\left(\frac{r_+^4}{\ell^4}\right)\right]+\epsilon^4{\cal O}\left(\frac{r_+^2}{\ell^2}\right),\label{eq:hNearRegion}
\end{align}
\begin{align}
\Pi^{in}(z)&=\epsilon\left[\frac{\Gamma^2\big[\frac{n}{n-1}\big]}{\Gamma\big[\frac{n+1}{n-1}\big]}P_{\frac{1}{n-1}}\bigg[2\frac{z^{n-1}}{\ell^{n-1}}-1\bigg]\frac{r_+}{\ell}+\frac{r_+^3}{\ell^3}\left\{K^{\Pi}_{n;1,3}P_{\frac{1}{n-1}}\left[\frac{2z^2}{\ell^2}-1\right]+\frac{2\Gamma^2\big[\frac{n}{n-1}\big]}{(n-1)\Gamma\big[\frac{n+1}{n-1}\big]}\times\right.\right.\nonumber\\
&\times\bigg[Q_{\frac{1}{n-1}}\bigg[\frac{2z^{n-1}}{\ell^{n-1}}-1\bigg]\int_1^{z/\ell}\frac{P_{\frac1{n-1}}\big[2x^{n-1}-1\big]}{x^3\left(\sum_{j=0}^{\frac{n-3}2}{x^{2j}}\right)^2}\bigg(P_{\frac1{n-1}}\big[2x^{n-1}-1\big]\tilde{s}_n(x)\label{eq:PiNearRegion}\\
&+nP_{\frac{n}{n-1}}\big[2x^{n-1}-1\big]\hat{s}_n(x)\bigg)dx-P_{\frac1{n-1}}\bigg[\frac{2z^{n-1}}{\ell^{n-1}}-1\bigg]\int_1^{z/\ell}\frac{Q_{\frac{1}{n-1}}\big[2x^{n-1}-1\big]}{x^3\left( \sum_{j=0}^{\frac{n-3}2}{x^{2j}}\right)^2}\times\nonumber\\
&\left.\left.\left.\times\bigg(P_{\frac1{n-1}}\big[2x^{n-1}-1\big]\tilde{s}_n(x)+nP_{\frac{n}{n-1}}\big[2x^{n-1}-1\big]\hat{s}_n(x)\bigg)dx\right]\right\}+{\cal O}\left(\frac{r_+^5}{\ell^5}\right)\right]\nonumber\\
&+\epsilon^3\bigg[K^{\Pi}_{n;3,1}P_{\frac{1}{n-1}}\left[2\frac{z^{n-1}}{\ell^{n-1}}-1\right]\frac{r_+}{\ell}+{\cal O}\left(\frac{r_+^3}{\ell^3}\right)\bigg]+\epsilon^5\left[{\cal O}\left(\frac{r_+}{\ell}\right)\right].\nonumber
\end{align}
where the constants $K^{g}_{n;2,2}$, $K^{\Omega}_{n;2,2}$ and $K^{\Pi}_{n;1,3}$ will be calculated in the next subsection and the constants $K^{\Pi}_{n;3,1}$ in equation (\ref{eq:PiNearRegion}) are given by 
\begin{equation}
K^{\Pi}_{3;3,1}=\frac{55\pi}{288},\quad\quad\quad K^{\Pi}_{5;3,1}=\frac{1067\Gamma^2\big[\frac{5}{4}\big]}{900\sqrt{\pi}},\quad\quad\quad K^{\Pi}_{7;3,1}=\frac{6403\Gamma^2\big[\frac{7}{6}\big]}{14112\Gamma\big[\frac{4}{3}\big]}.  \label{eq:Kn31}
\end{equation}
The functions $P_\nu[y]$ and $Q_\nu[y]$ are the Legendre functions of the first and second kind, respectively, the functions $\tilde{s}_n(x)$ are given in Appendix~\ref{app:fields} and
\begin{equation}
\hat{s}_n(x)=x^{n+1}\left(\displaystyle \sum_{j=0}^{\frac{n-3}{2}}{(j+1)x^{n-3-2j}}\right)+(n+2)^2\left(\displaystyle \sum_{j=0}^{\frac{n-3}{2}}{(j+1)x^{2j}}\right).
\end{equation}

There are a few discrepancies with \cite{Dias:2011at} -- the $n=3$ case -- that warrant mentioning.  In the order $\epsilon(r_+/\ell)^3$ term of equation (\ref{eq:PiNearRegion}), the matching conditions were used to determine that the constants $K^{\Pi}_{n;1,3}$ are non-zero, contrary to what is assumed in \cite{Dias:2011at}.  Next, although $Q_{\frac{1}{n-1}}[y]$ is complex for $y>1$, it can be explicitly verified that close to the horizon and in the large-$z$ limit, the explicit procedure for which is discussed shortly, the imaginary part of (\ref{eq:PiNearRegion}) vanishes, whereas it does not vanish for the solution quoted in \cite{Dias:2011at}.  Furthermore, the authors found no order $\epsilon^3(r_+/\ell)$ term in equation (\ref{eq:PiNearRegion}), whereas I find this term to be non-zero with the constants $K^{\Pi}_{n;3,1}$ given by (\ref{eq:Kn31}).  Despite these discrepancies  in the near-horizon scalar field, the stress energy tensor is unaffected at the perturbative order probed.  Similarly, although these discrepancies affect the temperature, the first law of thermodynamics is nevertheless satisfied to the appropriate perturbative order for the present $n=3$ solution as well as the solution of \cite{Dias:2011at}.  A discussion of the thermodynamics of these solutions is presented in section \ref{Thermo}. Finally, although the solutions for $f^{in}$ and $g^{in}$ at order $\epsilon^2(r_+/\ell)^2$ above look different than the corresponding fields in \cite{Dias:2011at}, it can be verified that the above solution for $n=3$ agrees with \cite{Dias:2011at} term by term in the near horizon expansion.

\subsubsection{Matching Region}\label{MatchingConditions}

A crucial and often difficult step in this analysis involves matching the near region fields to the far region fields in order to ensure a valid solution everywhere.  The heuristic procedure is as follows: take a small-$r$ expansion of the outer region fields, \emph{i.e.} expand $F^{out}$ around $r=0$, and match these at each order in $\epsilon$ and $r_+$ to the large-$z$ expansion of the inner fields, \emph{i.e.} an expansion of $F^{in}$ around $z=\infty$.  Such a matching takes place in an area between the two regions, where $r_+\ll r\ll\infty$.  This then fixes the remaining arbitrary constants of integration in the fields as well as fixes the expansion parameters for $\omega$.  When taking the large-$z$ limit of the inner region fields, one must first pick an order in $\epsilon$, take the large-$z$ limit at that order, transform the limit back to the original radial coordinate, $r=r_+z/\ell$, and finally expand the result in powers of $r_+/\ell$.  In principle, this is a straightforward procedure; in practice it can be subtle and difficult.  This is best illustrated with an explicit example.  

Consider the large-$z$ expansion of the order $\epsilon^2$ term of $g^{in}(z)$ for $n=3$.  The difficulty arises because the large-$z$ limit of a function that is defined via an integral of the form
\begin{equation}
\int_{1}^{z/\ell}{f(x)dx} \label{eq:ConvergentIntegral}
\end{equation}
must be taken, for some function $f(x)$.  Now, if (\ref{eq:ConvergentIntegral}) converges in the limit $z\rightarrow\infty$ then this integral is easy enough to perturbatively compute.  Indeed
\begin{equation}
\int_{1}^{z/\ell}{f(x)dx}=\int_{1}^{\infty}{f(x)dx}-\int_{z/\ell}^{\infty}{f(x)dx},
\end{equation}
where now $\int_{1}^{\infty}{f(x)dx}$ is a constant that is easily computed, albeit numerically, and the second term can be evaluated by taking a Taylor series expansion of $f(x)$ asymptotically.  A problem arises, however, if the integral (\ref{eq:ConvergentIntegral}) does not converge.  This is true of the integral appearing in the above example and is, in fact, generically true for the \emph{majority} of the inner region fields above defined through integrals.  Care must be taken in order to find a way to correctly take the large-$z$ limit of these integrals.

Motivated by the above discussion for convergent integrals, if $\int_{1}^{\infty}{f(x)dx}$ diverges, I then define $Div(x)$ as the sum of the terms which cause the integral to diverge.  That is, I Taylor expand $f(x)$ at infinity and collect the terms whose integral diverges and call that $Div(x)$.  Then, by definition, $\int_{1}^{\infty}{(f(x)-Div(x))dx}$ converges and the above consideration is valid.  This allows the original integral to be rewritten as
\begin{equation}
\int_{1}^{z/\ell}{f(x)dx}=\int_{1}^{\infty}{\big(f(x)-Div(x)\big)dx}-\int_{z/\ell}^{\infty}{\big(f(x)-Div(x)\big)dx}+\int_{1}^{z/\ell}{Div(x)dx}. \label{eq:largezIntegral}
\end{equation}
Now $\int_{1}^{\infty}{\big(f(x)-Div(x)\big)dx}$ is a numerical constant, the second integral can be evaluated by Taylor expanding $(f(x)-Div(x))$ near infinity and since $Div(x)$ is a simple power series in $x$, the last integral has an exact analytic expression.  Note that extra convergent terms can be included in the definition of $Div(x)$ since such terms will consequently disappear from the first two integrals of (\ref{eq:largezIntegral}), not changing the end result.

Applying this technique to the order $\epsilon^2$ term of $g^{in}(z)$, according to equation (\ref{eq:gNearRegion}) the function is $f(x)=\frac{\pi^2}{12}x\left(P^{\>'}_{\frac12}[2x^2-1]\right)^2$.  Expanding $f(x)$ around infinity, I find 
$$
Div(x)=\frac{4x}{3}+\frac{2}{3x}-\frac{\log[x]}{2x^3}, \qquad 
\int_{1}^{\infty}{\big(f(x)-Div(x)\big)dx}=-10.3749804991685... \equiv C_1
$$
up to fifteen digits of precision.  The large-$z$ limit is now readily available since the other integrals are straightforward to compute.  I will now detail how the matching condition is employed.

The large-$z$ limit of the entire order at $\epsilon^2$ must be taken.  That is
\begin{equation}
-\frac{2}{3}+\frac{r_+^2}{\ell^2}\left\{\int^{z/\ell}_{1}{f(x)dx}+K^{g}_{3;2,2}+g^{in}_{3;2,2}\right\}\label{eq:gIn32Match}
\end{equation}
must be expanded at large $z$.  Next, this must be transformed back to the coordinate $r=zr_+/\ell$ and series expanded in powers of $r_+/\ell$.  I find the result for (\ref{eq:gIn32Match}) up to order $(r_+/\ell)^2$ to be
\begin{equation}
g^{in}_{\epsilon^2}\rightarrow-\frac{2}{3}+\frac{2r^2}{3\ell^2}+\frac{r_+^2}{\ell^2}\left\{-\frac{19}{24}+C_1+K^{g}_{3;2,2}-\frac{2}{3}\log\left[\frac{r_+}{\ell}\right]+\frac{2}{3}\log\left[\frac{r}{\ell}\right]\right\}.\label{eq:gInMatch}
\end{equation}
This is now to be matched with the small-$r$ limit of the order $\epsilon^2$ term of $g^{out}(r)$.  Direct calculation using the fields in Appendix~\ref{app:fields} (with the constant of integration, $C_2$, not yet fixed) yields
\begin{equation}
g^{out}_{\epsilon^2}\rightarrow-\frac{2}{3}+\frac{2r^2}{3\ell^2}+\frac{r_+^2}{\ell^2}\left\{-\frac{C_2\ell^4}{4r^4}+\frac{C_2\ell^2}{r^2}+\frac{C_2}{2}+3C_2\log\left[\frac{r}{\ell}\right]-\frac{83}{9}+\frac{2}{3}\log\left[\frac{r}{\ell}\right]\right\}.\label{eq:gOutMatch}
\end{equation}
The order $(r_+/\ell)^0$ terms cancel so I continue to the order $(r_+/\ell)^2$ terms: equation (\ref{eq:gOutMatch}) has a term of order $r^{-4}$ while equation (\ref{eq:gInMatch}) does not, meaning $C_2=0$.  The $\log[r/\ell]$ terms cancel and what is left is a condition on the constant $K^{g}_{3;2,2}$.  Upon explicit calculation, I find
\begin{equation}
K^{g}_{3;2,2}=-\frac{607}{72}-C_1+\frac{2}{3}\log\left[\frac{r_+}{\ell}\right],
\end{equation}
which has a numerical value of $1.94442...+\frac{2}{3}\log\left[\frac{r_+}{\ell}\right]$.  This result differs from the value of $\frac{635}{504}+\frac{7}{8}\log[2]+\frac{2}{3}\log\left[\frac{r_+}{\ell}\right]\approx1.86642...+\frac{2}{3}\log\left[\frac{r_+}{\ell}\right]$ in \cite{Dias:2011at} well beyond the margins of numerical error.  This discrepancy arises because in \cite{Dias:2011at} an approximation scheme\footnote{This was confirmed through private communication with the authors of \cite{Dias:2011at}.} was used to evaluate the large-$z$ limit of these integrals, whereas the results presented here are exact.

Following the procedure outlined above for $n=5$, I find the corresponding constant to be
\begin{equation}
K^g_{5;2,2}=9.020251036253986.
\end{equation}
Similar considerations show that $K^{\Pi}_{n;1,3}$ are given by
\begin{equation}
\begin{split}
K^{\Pi}_{3;1,3}=-14.36212300918522-\frac{21\pi}{8}\log\left[\frac{r_+}{\ell}\right], \qquad K^{\Pi}_{5;1,3}=-&5.48986673419516.
\end{split}
\end{equation}

Evaluation of the constants $K^{\Omega}_{n;2,2}$ in (\ref{eq:OmegaNearRegion}) requires further subtle considerations since the corresponding field is defined via a series of nested integrals.  The procedure outlined above fails under such conditions.  This is remedied by treating the inner integrals first by Taylor expanding the integrand at infinity and near the horizon, while keeping a sufficient number of terms in each series such that the two match-up at some intermediate radius.  This integral can then be evaluated directly, with the answer then becoming the integrand of the next integral.  This integrand can then be expanded in the same way and its answer used as the integrand of the next integral in the nest.  Finally, the last integral in the nest can be treated as discussed above to obtain the large-$z$ limit of the field.  From this rather tedious procedure, I find
\begin{equation}
\begin{split}
K^{\Omega}_{3;2,2}=64.6979...-10\log\left[\frac{r_+}{\ell}\right], \quad\quad K^{\Omega}_{5;2,2}=306.637...,
\end{split}
\end{equation}
where only six digits of precision have been kept due to the numerical error introduced by the repeated matching of Taylor series.  Again there is a discrepancy well beyond the bounds of numerical error with the results of \cite{Dias:2011at}, which found $K^{\Omega}_{3;2,2}=87.5209...-10\log\left[\frac{r_+}{\ell}\right]$, again because of the approximation procedure used there.

\section{Thermodynamics and Physical Properties}\label{Thermo}

Since the boson star and hairy black hole solutions are invariant under the Killing vector field (\ref{eq:KV}), they must satisfy the first law of thermodynamics, which follows from a Hamiltonian derivation of the first law \cite{Wald:1993ki}.  In this section I work out the thermodynamic quantities of the boson star and hairy black hole solutions and verify that the first law holds in each dimension up to the appropriate perturbative order.  

\subsection{Asymptotic Charges}

In an asymptotically anti-de Sitter space-time $(M,g_{\mu\nu})$ there is an ambiguity in defining ``the" asymptotic metric because certain metric functions diverge and thus there does not exist a smooth limit to the boundary.  To get around this subtlety, Penrose proposed a conformal completion $(\hat{M},\hat{g}_{\mu\nu})$, where $\hat{g}_{\mu\nu}=\tilde\Omega^2g_{\mu\nu}$ for some conformal factor $\tilde\Omega$, such that the boundary of $\hat{M}$ is reached by the smooth limit $\tilde\Omega\rightarrow0$.  In particular, by virtue of $\tilde\Omega$ vanishing (smoothly) on the boundary, $\nabla\tilde\Omega$ can be used as a radial direction near infinity and the subtlety of taking infinite distance limits in the physical metric $g_{\mu\nu}$ reduces to local differential geometry of fields in the conformal completion $\hat{g}_{\mu\nu}$ over finite distances.  Using the conformal completion with reflective boundary conditions,  Ashtekar and Das elucidated how to properly define conserved charges in asymptotically anti-de Sitter space-times \cite{Ashtekar:1999jx}, a procedure
that was extended to the rotating case in \cite{Das:2000cu}.
Furthermore, it has been pointed out in \cite{Gibbons:2004ai} that while there exist results in the literature in disagreement with the Ashtekar-Das formalism, such definitions of mass and angular momenta fail to satisfy the first law.  I therefore restrict attention to the Ashtekar-Das formalism for computing the mass and angular momentum, briefly detailing the procedure.

The conformal metric $\hat{g}_{\mu\nu}=\tilde\Omega^2g_{\mu\nu}$, where the physical metric $g_{\mu\nu}$ given in (\ref{eq:metric}), looks like
\begin{equation}
d\hat{s}^2=-\tilde{f}gdt^2+\frac{d\tilde\Omega^2}{\tilde{f}}+h\big(d\chi+A_idx^i-\Omega dt\big)^2+g_{ij}dx^idx^j, \label{eq:ConformalMetric}
\end{equation}
where the conformal factor is $\tilde\Omega=1/r$, the functions $g,h$, and $\Omega$ are the previous (physical) metric functions written in terms of, $\tilde\Omega$, and $\tilde{f}=\tilde\Omega^2f$ now admits a smooth, finite limit to the conformal boundary, $\cal I$, which is defined by $\tilde\Omega=0$.  The Weyl tensor, $\hat{C}_{\mu\nu\rho\sigma}$, of the conformal metric (\ref{eq:ConformalMetric}) vanishes on $\cal I$ but ${\bf K}_{\mu\nu\rho\sigma}\equiv\displaystyle \lim_{\rightarrow{\cal I}}\tilde\Omega^{3-D}\hat{C}_{\mu\nu\rho\sigma}$ admits a smooth limit, the electric part of which, ${\cal E}_{\mu\nu}$, is used to define conserved quantities.  Using the transformation properties of the Weyl  and Ricci tensors under conformal rescalings and imposing the Einstein equations, it can be shown that on the conformal boundary, ${\cal E}_{ab}$ satisfies the continuity equation \cite{Ashtekar:1999jx,Das:2000cu}
\begin{equation}
D^d{\cal E}_{md}=-8\pi(D-3){\bf T}_{\mu \nu}n^\mu{h^b}_m e^\nu_b,\label{eq:ContinuityEquation}
\end{equation}
where $n_\mu=\partial_\mu\tilde\Omega$ is the normal vector to $\cal I$, $D_a$ is the derivative operator compatible with the induced metric $h_{ab}$ on $\cal I$ with frame $e^\mu_a$, and ${\bf T}_{\mu\nu}=\displaystyle\lim_{\rightarrow{\cal I}}{\tilde\Omega^{2-D}T_{\mu\nu}}$ admits a smooth limit, where $T_{\mu\nu}$ is the matter stress tensor in equation  (\ref{eq:Tab}).  If there is no net flux of the matter stress tensor on $\cal I$, then the corresponding charge is conserved.  

To set this up explicitly, the electric part of the Weyl tensor is ${\cal E}_{ab}\hat{=}\ell^2{\bf K}_{\mu\rho\nu\sigma}n^\rho n^\sigma e^\mu_a e^\nu_b$ where the symbol $\hat=$ is used to denote equality on $\cal I$.  The normal $n_\mu$ to the boundary is space-like, which means $\cal I$ is time-like.  Consider now a $t$=constant slice of $\cal I$: this is a $D-2$ dimensional space-like hypersurface, $\Sigma$, with induced metric $\gamma$ and normal vector $t_a=\partial_a t$.  For any conformal Killing vector, $\xi^a$, the continuity equation (\ref{eq:ContinuityEquation}) leads to the conserved charge
\begin{equation}
Q_{\xi}=\pm\frac{1}{8\pi(n-1)}\int_{\Sigma}{{\cal E}_{ab}\xi^{a}t^{b}\mathrm{d}\Sigma},
\end{equation}
where the $\pm$ sign corresponds to a time-like/space-like conformal Killing vector respectively and $n=D-2$ is not to be confused with $n_\mu$.

Although the boson star and hairy black hole solutions are invariant under only a single Killing field, both $\partial_t$ and $\partial_\chi$ are asymptotic Killing vectors since the scalar field vanishes asymptotically.  Note that the vanishing of the scalar field at infinity ensures there will be no net flux of matter fields on the conformal boundary, so the charges defined above are strictly conserved.   $\partial_t$ and $\partial_\chi$ are then the desired conformal Killing vectors and they lead to a conserved energy and angular momentum respectively.  Upon direct calculation of the conserved charges, I find
\begin{equation}
E=\frac{(n+1)\pi^{\frac{n-1}{2}}\ell^{n-1}}{16\left(\frac{n+1}{2}\right)!}\left((n+1)C_h-nC_f\right),
\end{equation}
\begin{equation}
J=\frac{(n+1)^2\pi^{\frac{n-1}{2}}\ell^{n}C_\Omega}{16\left(\frac{n+1}{2}\right)!},
\end{equation}
where $C_f, C_h$ and $C_\Omega$ are the leading order corrections to the asymptotic fields $f, h$ and $\Omega$ appearing in the boundary conditions (\ref{eq:AsymBC}).  Using the far region field expansions (\ref{eq:fFarRegion}), (\ref{eq:hFarRegion}) and (\ref{eq:OmegaFarRegion}), along with the catalogued fields in Appendix~\ref{app:fields}, these coefficients are easily obtained.  The resulting perturbative expansions of the energy and angular momentum charges are catalogued in Appendix~\ref{app:thermo}.

\subsection{Near Horizon Quantities}

Since boson stars are horizonless geometries their thermodynamics are completely governed by the above subsection.  For the case of black holes, the presence of the horizon introduces a temperature and an entropy which also enter the first law.

The norm of the Killing vector (\ref{eq:KV}) is $\left|K\right|^2=-fg+r^2(\omega-\Omega)^2$, which is null at the horizon by virtue of the condition $\Omega_H=\omega$ and the fact that $f$ vanishes there. The Killing vector, under which the black hole is invariant, is therefore tangent to the generators of the horizon.  The event horizon is thus also a Killing horizon and has a temperature $T_H=\frac{\kappa}{2\pi}$.  For a metric of the form (\ref{eq:metric}), a straightforward calculation yields
\begin{equation}
T_H=\frac{1}{4\pi}{f'(r_+)\sqrt{g(r_+)}}. \label{eq:Temperature}
\end{equation}
Furthermore, any solution to Einstein's equations with or without cosmological constant obeys the Bekenstein-Hawking area-entropy law.  Spatial sections of the horizon have an induced metric of a squashed $n$-sphere
\begin{equation}
ds_H^2=r^2\big(h\big(d\chi+A_idx^i)^2+g_{ij}dx^idx^j\big)\bigg|_{r_+},
\end{equation}
so the entropy takes the form
\begin{equation}
S=\frac{A_n}{4}r_+^n\sqrt{h(r_+)},  \label{eq:Entropy}
\end{equation}
where $A_n$ is the area of a round unit $n$-sphere.  Now, the near region field expansions (\ref{eq:fNearRegion}), (\ref{eq:hNearRegion}) and (\ref{eq:OmegaNearRegion}), along with the fields catalogued in Appendix~\ref{app:fields} yield the perturbative expansions of (\ref{eq:Temperature}) and (\ref{eq:Entropy}); these expansions for the temperature and entropy are catalogued in Appendix~\ref{app:thermo}.

With the thermodynamic charges and potentials in Appendix~\ref{app:thermo}, it is straightforward to verify that the first law holds up to the appropriate perturbative order in each dimension.  For boson stars, obtained in the $r_+\rightarrow0$ limit, the first law takes the form
\begin{equation}
\mathrm{d}E=\omega \mathrm{d}J,
\end{equation}
while for the hairy black holes, the first law takes the form
\begin{equation}
\mathrm{d}E=T_H\mathrm{d}S+\omega \mathrm{d}J.
\end{equation}

Upon examining the thermodynamic quantities in Appendix~\ref{app:thermo}, it can be seen that the angular momentum is primarily carried by the scalar field.  Furthermore, the scalar field carries more and more of the angular momentum as the space-time dimension increases: the leading terms in the perturbative expansions are due to the scalar field at orders $\epsilon^2,...,\epsilon^{n-1}$ and the next terms are a mixture of scalar field and black hole at orders $r_+^p\epsilon^q$ where $p+q=n+1$.  Similarly, for $n=3$ the energy has contributions at the same perturbative order from the black hole and the scalar field since the leading terms in $E_3$ are of order $\epsilon^2$ and $r_+^2$.  However, for $n=5$ the majority of the energy is carried again by the scalar field: the leading term in $E_5$ is at order $\epsilon^2$ and the next terms are at orders $r_+^{4}$ and $\epsilon^{4}$.  This is a rather striking feature of the perturbative regime of these hairy black holes: in $n=3$ the thermodynamics are governed both by the scalar hair and the black hole while in $n>3$ the thermodynamics are dominated by the scalar hair.\footnote{This continues to be true for the $D=9,11$ solutions in \cite{Stotyn:2011ns}.}

\section{Discussion about Stability}\label{Discussion}

I have constructed perturbative solutions describing asymptotically AdS rotating hairy black holes and boson stars in dimensions five and seven dimensions.  Apart from several technical discrepancies, these results are in general agreement with previous results \cite{Dias:2011at} for the five dimensional case.  Such solutions describe lumpy massless scalar hair co-rotating with a black hole and are invariant under a single Killing vector field.  This is made possible by a particular choice of scalar field ansatz, whose stress tensor shares the same symmetries as the metric.  These are the first known examples of asymptotically AdS black holes that are invariant under one Killing vector; all previous AdS black holes were stationary and hence had at least two Killing vector fields.  Furthermore, these hairy black holes are the end states of a superradiant instability for rotating AdS black holes, and it is important to establish whether these end state solutions are themselves stable against perturbations.  In this section I provide arguments for their stability against certain types of perturbations and instability against others.

The hairy black hole solutions constructed herein describe scalar field modes $\Pi\sim e^{-i(\omega t-m\chi)}$ co-rotating with the black hole, where $m=1$.  Within the $m=1$ class of perturbations, it is expected that the hairy black holes are stable.  If the scalar hair is perturbed, that perturbation can be decomposed as a weighted sum of the different radial modes parametrised by the integer $k$ in (\ref{eq:kRadialModes}).  The corresponding frequency parameter for each radial mode is $\omega_k\ell=\omega_0\ell+2k$ where $\omega_0=\Omega_H$ is the ground state frequency.  In this dynamic intermediate period, each higher frequency radial mode will begin to get eaten by the horizon because they cannot undergo superradiant scattering by virtue of $\omega_k>\Omega_H$.  Perturbing the scalar field in this way will simply cause the black hole to settle back down to the ground state configuration considered in this chapter.  

However considering classes of perturbations in the full theory, that is with $m>1$, suggests that the hairy black holes are unstable toward the development of structures on smaller and smaller scales.  Recall that superradiant scattering occurs for modes as long as $\omega<m\Omega_H$.  This means that for a fixed $\omega$, the $m=1$ hairy black hole is unstable to superradiant scattering from higher $m$ modes.  The following picture then arises: the Myers-Perry-AdS black hole becomes unstable and evolves toward the $m=1$ hairy configuration constructed in this chapter.  The $m=1$ configuration then becomes unstable and evolves toward the $m=2$ configuration, which then becomes unstable and evolves toward the $m=3$ configuration, and so on.  Each higher order $m$ configuration has structure on smaller and smaller scales.  It is unclear whether this process terminates at a finite order $m$ in a truly stable configuration or whether the black hole never settles down.  It would be interesting to have an understanding of these hairy black holes from a holographic perspective to see if there are any insights to be gained.  The interpretation of these solutions from a boundary gauge theory perspective, however, is currently an open issue and certainly warrants future investigation, particularly in light of this issue of stability.

This work is a step in the direction of further understanding the instability of rotating AdS black holes, but it is far from complete.  For instance, only the perturbative regime was considered here, in which energy and angular momenta are small by construction.  In \cite{Dias:2011at}, non-perturbative results were also investigated in five dimensions and a rather rich and interesting thermodynamic structure was unveiled.  One must construct solutions numerically because of the highly non-linear and coupled equations of motion, however the physics is more interesting; for details the reader is referred to \cite{Dias:2011at}.  However as noted in section \ref{Setup}, there are terms in the equations of motion that are absent in five dimensions and it is unclear from the perturbative analysis whether these terms have any consequences for higher dimensional hairy black holes.  Furthermore, these hairy black holes are the $m=1$ superradiant end states of Myers-Perry-AdS black holes with equal angular momenta in all planes of rotation.  To fully determine the stability of arbitrarily rotating AdS black holes, it would be necessary to construct hairy black holes with independent angular momenta in the different planes of rotation.  Such solutions ought to exist, although they will have reduced symmetry and are therefore likely harder to construct analytically.

\chapter{Conclusion}

In this thesis, I have analytically constructed exact and perturbative solutions to gauged and ungauged supergravity theories in space-time dimensions $D\ge5$.  In the first three chapters, focus was placed on solutions to five dimensional ungauged minimal supergravity, which is a consistent truncation of eleven dimensional supergravity on a six-torus.  As such, general solutions to this theory are also valid solutions of eleven dimensional supergravity, and furthermore extremal and near-extremal solutions have an interpretation in terms of the dynamics of free M-branes wrapping various cycles of the six-torus.  

In chapter 2, I used a solution generating technique that exploits the existence of a Killing spinor, the bilinears of which form a scalar, a null or time-like Killing vector, and three 2-forms, to construct supersymmetric solutions to the equations of motion.  For the case of a time-like Killing vector, the full solution is a time-like fibration over a four dimensional hyper-K\"ahler base space, which I took to be the Atiyah-Hitchin metric.  The three-parameter family of solutions so constructed have mass, angular momentum, electric charge, and magnetic charge.  They describe conically singular pseudo-horizons inside a region containing closed time-like curves.  Such space-times are outwardly geodesically complete and have an asymptotic region free of pathologies described by a twisted U(1) fibration over four dimensional Minkowski space $M_4/{\mathbb Z}_2$, where the ${\mathbb Z}_2$ identifies antipodal points on the two-sphere.  This family of solutions could potentially provide an example of a well defined holographic dual theory despite the pathologies present in the bulk if such a holographic dual description indeed exists. 

In chapter 3, I constructed the most general black string solution to five dimensional ungauged minimal supergravity by exploiting a hidden $G_2$ symmetry that becomes manifest when the Lagrangian is expressed in terms of three dimensional fields.  Upon dimensional reduction, the Lagrangian becomes a combination of a three dimensional Einstein-Hilbert term and a scalar Lagrangian, where the eight scalars parametrise the coset $G_2/\tilde K$.  The pseudo-compact subgroup, $\tilde K=SL(2,{\mathbb R})\times SL(2,{\mathbb R})$, is invariant under the involutive isomorphism that defines the coset and, in five dimensional language, keeps the asymptotic structure invariant.  Acting with elements $k\in\tilde K$ then generically transform the scalars, leading to new solutions in five dimensions.  I used this technique to add independent electric and magnetic charges to the Kerr black string, yielding a black string with five independent parameters, namely mass, angular momentum, linear momentum, magnetic charge and electric charge.  I explicitly verified that the first law of thermodynamics holds for this general black string and that the Bekenstein-Hawking entropy is reproduced in the decoupling limit by a statistical microstate counting using the Maldacena-Strominger-Witten CFT.  The general five-parameter family of black strings was shown to reduce to various previously known extremal and supersymmetric black string solutions in appropriate limits.  Most interestingly, since tensionless black strings have been conjectured to correspond to infinite radius black rings, a phase diagram for extremal tensionless black strings was produced to illuminate the corresponding phase space of extremal black rings.  The previously known class of supersymmetric black rings displays a puzzling lower bound on the electric charge.  However this phase diagram suggests that as this bound is approached the supersymmetric ring smoothly transitions to an extremal non-supersymmetric ring, the full solution for which is currently not known.  Furthermore, the phase diagram suggests the existence of a previously unknown class of supersymmetric black rings that is completely disconnected from the known supersymmetric class.  This carries over to the non-extremal regime in which the general black string admits two disjoint tensionless limits corresponding to the relative orientations of the momentum along the string and the magnetic charge.  This analysis suggests that there may be two disjoint branches of general black ring solutions, in which the angular momentum along the $S^1$ is either aligned or anti-aligned with the magnetic dipole charge.

In chapter 4, I analysed the thermodynamic stability of a simple limit of the general black string of chapter 3: the magnetically charged black string.  I showed that if one considers boundary conditions that break supersymmetry, magnetically charged topological solitons are part of the thermodynamic phase space of the black string and play a key role in their thermodynamic stability.  For a given compactification radius of the Kaluza-Klein dimension there exist a small and a large static soliton and the two degenerate at a minimum compactification radius.  By setting up initial data on a Cauchy surface of time symmetry, these static topological solitons were then shown to belong to a family of dynamic ``bubbles of nothing."  Through analysis of a potential energy diagram for these dynamic bubbles, it was shown that the small solitons are perturbatively stable, while the large solitons are unstable.  Furthermore, if the radius of the compact dimension is taken to be sufficiently large compared to the magnetic charge, the small solitons effectively become a stable vacuum, which is the end state of Hawking evaporation.  The phase transition between the black string and the bubble configuration was then analysed from a string theoretic point of view: the topology changing transition is due to localised closed string tachyon condensation wrapping the Kaluza-Klein direction outside the horizon in the near extremal limit.  This string theoretic analysis is in direct agreement with the semiclassical analysis showing that there exists a parameter range in which the magnetic black strings are perturbatively stable, yet thermodynamically unstable, in direct contradiction to the Guber-Mitra conjecture.  The magnetic black string is thus an example of an object whose end-state from a thermodynamic instability is distinct from its end-state from a perturbative instability, the latter end-state still being an open question.

In the last two chapters, focus was placed on solutions to gauged supergravity theories in which the gauge fields vanish.  These theories correspond to various supergravity theories reduced along appropriate spheres and tori, an example being five dimensional gauged supergravity being the consistent truncation of type IIB supergravity on a five-sphere.  The solutions of this latter theory are particularly important since they are also AdS$_5\times S^5$ solutions to type IIB supergravity.  Such solutions arise as the near-horizon geometries of D3-brane configurations in the decoupling limit, which led Maldacena to propose the AdS/CFT conjecture in \cite{Maldacena:1997re}.

In chapter 5, I analysed the thermodynamic stability of black holes and topological solitons in gauged supergravity theories in dimensions $D\ge5$.  Asymptotically AdS space-times exhibit a natural stability for topological solitons because energies are bounded below by the positive energy theorem.  One such topological soliton is the Eguchi-Hanson soliton, which asymptotes to an orbifold AdS space-time and hence plays a nontrivial role in the thermodynamic phase structure of other orbifold AdS space-times.  In this vein, via semiclassical gravity considerations I considered the phase behaviour of the orbifold Schwarzschild-AdS black hole, the empty orbifold AdS geometry and the Eguchi-Hanson soliton.  The usual phase behaviour associated with the Hawking-Page transition was observed, in addition to new phase behaviour associated to the presence of the soliton in the phase space.  In particular, the orbifold AdS geometry is always unstable to decay to the Eguchi-Hanson soliton, the mediation of which is via closed string tachyon condensation around the orbifold fixed point from a string theoretic perspective.  This is in agreement with previous considerations from the AdS/CFT viewpoint, in which both the orbifold AdS geometry and the Eguchi-Hanson soliton are dual to distinct local vacua of the gauge theory with the Eguchi-Hanson soliton corresponding to the lower energy ground state.  Similarly, there is additional phase behaviour for the black hole above the Hawking-Page analysis: there is a range of large black holes which are stable toward the orbifold AdS geometry but which are unstable to decay to the Eguchi-Hanson soliton.  Since they are stable to the orbifold AdS geometry, such black holes will not Hawking evaporate but will rather undergo quantum tunneling to the soliton configuration.  I further used geometric arguments to show that this thermodynamic phase transition is holographically dual to a deconfinement-confinement phase transition in the boundary gauge theory, with the black hole corresponding to the deconfinement phase, and the orbifold AdS geometry and the Eguchi-Hanson soliton corresponding to confinement phases.

Finally in chapter 6, I considered another instability specific to rotating black holes in anti-de Sitter space-times and constructed meta-stable end states for this instability in the low energy and angular momentum limit.  The instability is due to superradiance, where scalar field modes incident on the horizon are reflected with a larger amplitude.  These increased amplitude modes are then reflected back to the horizon by the reflecting AdS boundary conditions and the modes undergo further superradiant scattering.  The modes increase their amplitude at the expense of the rotational energy of the black hole, suggesting an end state described by a clump of scalar hair co-rotating with the black hole.  The combination of space-time and matter fields are invariant under only a single Killing vector field.  To construct these solutions, I made use of a judicious choice for the metric and scalar fields such that the matter stress tensor shares the symmetries of the metric.  This choice yields equations of motion that are a system of coupled second order ordinary differential equations and therefore guarantee the existence of a solution.  The boundary conditions impose the physical condition that the scalar hair and the black hole are co-rotating and that there is no flux of scalar field across the horizon.  Boson star solutions were first constructed perturbatively in the asymptotic amplitude of the scalar field, $\epsilon$; these solitonic boson star solutions describe rotating non-singular clumps of scalar field whose self-gravitation is balanced by centrifugal repulsion.  Small black holes were then added to the centres of these boson star solutions, yielding the desired hairy black hole configurations, which were constructed perturbatively in $\epsilon$ as well as the dimensionless measure of the horizon radius, $r_+/\ell$.  This two-parameter family of hairy black holes is connected to a two-parameter family of Myers-Perry-AdS black holes, which are precisely the black holes unstable toward superradiant scattering.  In the full non-perturbative theory, the hairy black holes constructed are unstable to higher order perturbative modes that undergo further superradiant scattering.  However within the class of perturbative modes fixed by the considered scalar field ansatz, the perturbative hairy black holes are stable and describe the end-state of the superradiant instability of a two-parameter family of Myers-Perry-AdS black holes.

It was my goal in this thesis to convince the reader that higher dimensional supergravity theories harbour a rich spectrum of solutions with interesting stability properties, particularly the black hole and black brane solutions, whose (near-)extremal limits can be mapped to non-abelian gauge theories via the AdS/CFT conjecture.  If the solution is supersymmetric, it is locally at a minimum of the energy, so BPS solutions are necessarily stable.  If the condition of supersymmetry is relaxed, say by imposing antiperiodic boundary conditions for the fermions, then bubble solutions enter the spectrum.  In Kaluza-Klein space-times this usually leads to disastrous consequences of dynamic bubbles falling down a runaway potential as they consume the space-time.  However these bubbles can be locally stabilised by placing a magnetic flux on them.  In AdS space-times, on the other hand, the energies are bounded from below and such bubble configurations are automatically guaranteed to be at least locally stable.  To this end, a complete classification of solutions in supergravity theories would prove extremely useful.  Such classifications are currently not available, but there nevertheless is a lot that is known about these theories, particularly in five dimensions.  More importantly, however, there is even more left to find out.

\appendix
\chapter*{APPENDICES}
\addcontentsline{toc}{chapter}{APPENDICES}

\chapter{Representation of ${\mathfrak g}_{2(2)}$}
\label{AppendixG2}

In this Appendix I give the representation of $\mathfrak{g}_{2(2)}$ that is used in chapter 3; this representation is the same as that used in \cite{Compere:2009zh} and is explicitly given by:
\begin{equation} \small{
h_1 = \left(
\begin{array}{lllllll}
 \frac{1}{\sqrt{3}} & \>\>0 & \>\>0 & 0 & \>\>0 & \>\>0 & \>\>0 \\
 \>\>0 & \frac{-1}{\sqrt{3}} & \>\>0 & 0 & \>\>0 & \>\>0 & \>\>0 \\
 \>\>0 & \>\>0 & \frac{2}{\sqrt{3}} & 0 & \>\>0 & \>\>0 & \>\>0 \\
 \>\>0 & \>\>0 & \>\>0 & 0 & \>\>0 & \>\>0 & \>\>0 \\
 \>\>0 &\> \>0 & \>\>0 & 0 & \frac{-2}{\sqrt{3}} & \>\>0 & \>\>0 \\
 \>\>0 & \>\>0 & \>\>0 & 0 & \>\>0 & \frac{1}{\sqrt{3}} & \>\>0 \\
 \>\>0 & \>\>0 & \>\>0 & 0 & \>\>0 & \>\>0 & \frac{-1}{\sqrt{3}}
\end{array}
\right)
 \, , \quad\>\>\>\>
h_2 = \left(
\begin{array}{lllllll}
 1 & 0 & 0 & 0 & 0 & \>\>0 & \>\>0 \\
 0 & 1 & 0 & 0 & 0 & \>\>0 & \>\>0 \\
 0 & 0 & 0 & 0 & 0 & \>\>0 & \>\>0 \\
 0 & 0 & 0 & 0 & 0 & \>\>0 & \>\>0 \\
 0 & 0 & 0 & 0 & 0 & \>\>0 & \>\>0 \\
 0 & 0 & 0 & 0 & 0 & -1 & \>\>0 \\
 0 & 0 & 0 & 0 & 0 & \>\>0 & -1
\end{array}
\right) \, , } \nonumber \end{equation}
\begin{equation}
\small{
e_1 = \left(
\begin{array}{lllllll}
 0 & 0 & 0 & 0 & 0 & 0 & 0 \\
 0 & 0 & 1 & 0 & 0 & 0 & 0 \\
 0 & 0 & 0 & 0 & 0 & 0 & 0 \\
 0 & 0 & 0 & 0 & 0 & 0 & 0 \\
 0 & 0 & 0 & 0 & 0 & 1 & 0 \\
 0 & 0 & 0 & 0 & 0 & 0 & 0 \\
 0 & 0 & 0 & 0 & 0 & 0 & 0
\end{array}
\right) \, ,\hspace{1cm}\qquad\>\>
e_2 = \left(
\begin{array}{lllllll}
 0 & \frac{1}{\sqrt{3}} & 0 & \>\>0 & \>\>0 & 0 & \>\>0 \\
 0 & \>\>0 & 0 & \>\>0 & \>\>0 & 0 & \>\>0 \\
 0 & \>\>0 & 0 & \frac{2}{\sqrt{3}} & \>\>0 & 0 & \>\>0 \\
 0 & \>\>0 & 0 & \>\>0 & \frac{2}{\sqrt{3}} & 0 & \>\>0 \\
 0 & \>\>0 & 0 & \>\>0 & \>\>0 & 0 & \>\>0 \\
 0 & \>\>0 & 0 & \>\>0 & \>\>0 & 0 & \frac{1}{\sqrt{3}} \\
 0 & \>\>0 & 0 & \>\>0 & \>\>0 & 0 &\>\> 0
\end{array}
\right) \, ,} \nonumber \end{equation}
\begin{equation}
\small{
e_3 = \left(
\begin{array}{lllllll}
 0 & 0 & \frac{-1}{\sqrt{3}} & \>\>0 & 0 & \>\>0 & \>\>0 \\
 0 & 0 & \>\>0 & \frac{2}{\sqrt{3}} & 0 & \>\>0 & \>\>0 \\
 0 & 0 & \>\>0 & \>\>0 & 0 & \>\>0 & \>\>0 \\
 0 & 0 & \>\>0 & \>\>0 & 0 & \frac{-2}{\sqrt{3}} & \>\>0 \\
 0 & 0 & \>\>0 & \>\>0 & 0 & \>\>0 & \frac{1}{\sqrt{3}}  \\
 0 & 0 & \>\>0 & \>\>0 & 0 & \>\>0 & \>\>0 \\
 0 & 0 & \>\>0 & \>\>0 & 0 & \>\>0 & \>\> 0
\end{array}
\right)\, ,\qquad
e_4 = \left(
\begin{array}{lllllll}
 0 & 0 & 0 & \frac{-2}{\sqrt{3}} & \>\>0 & \>\>0 & \>\>0 \\
 0 & 0 & 0 & \>\>0 & \frac{2}{\sqrt{3}} & \>\>0 & \>\>0 \\
 0 & 0 & 0 & \>\>0 & \>\>0 & \frac{2}{\sqrt{3}} & \>\>0 \\
 0 & 0 & 0 & \>\>0 & \>\>0 & \>\>0 & \frac{-2}{\sqrt{3}} \\
 0 & 0 & 0 & \>\>0 & \>\>0 & \>\>0 & \>\>0  \\
 0 & 0 & 0 & \>\>0 & \>\>0 & \>\>0 & \>\>0 \\
 0 & 0 & 0 & \>\>0 & \>\>0 & \>\>0 & \>\> 0
\end{array}
\right)\, ,} \nonumber \end{equation}
\begin{equation}
\small{
e_5 = \left(
\begin{array}{lllllll}
 0 & 0 & 0 & 0 & -2 & 0 & 0 \\
 0 & 0 & 0 & 0 & \>\>0 & 0 & 0 \\
 0 & 0 & 0 & 0 & \>\>0 & 0 & 2 \\
 0 & 0 & 0 & 0 & \>\>0 & 0 & 0 \\
 0 & 0 & 0 & 0 & \>\>0 & 0 & 0 \\
 0 & 0 & 0 & 0 & \>\>0 & 0 & 0 \\
 0 & 0 & 0 & 0 & \>\>0 & 0 & 0
\end{array}
\right) ,\hspace{1.75cm}\qquad\quad
e_6 = \left(
\begin{array}{lllllll}
 0 & 0 & 0 & 0 & 0 & 2 & 0 \\
 0 & 0 & 0 & 0 & 0 & 0 & 2 \\
 0 & 0 & 0 & 0 & 0 & 0 & 0 \\
 0 & 0 & 0 & 0 & 0 & 0 & 0 \\
 0 & 0 & 0 & 0 & 0 & 0 & 0 \\
 0 & 0 & 0 & 0 & 0 & 0 & 0 \\
 0 & 0 & 0 & 0 & 0 & 0 & 0
 \end{array}
\right) ,} \nonumber \end{equation}
\begin{equation}
\small{
f_1 = \left(
\begin{array}{lllllll}
 0 & 0 & 0 & 0 & 0 & 0 & 0 \\
 0 & 0 & 0 & 0 & 0 & 0 & 0 \\
 0 & 1 & 0 & 0 & 0 & 0 & 0 \\
 0 & 0 & 0 & 0 & 0 & 0 & 0 \\
 0 & 0 & 0 & 0 & 0 & 0 & 0 \\
 0 & 0 & 0 & 0 & 1 & 0 & 0 \\
 0 & 0 & 0 & 0 & 0 & 0 & 0
\end{array}
\right) ,\hspace{1cm}\qquad\>\>
f_2 = \left(
\begin{array}{lllllll}
 \>\>0 & 0 & \>\>0 & \>\>0 & 0 & \>\>0 & 0 \\
 \frac{1}{\sqrt{3}} & 0 & \>\>0 & \>\>0 & 0 & \>\>0 & 0 \\
 \>\>0 & 0 & \>\>0 & \>\>0 & 0 & \>\>0 & 0 \\
 \>\>0 & 0 & \frac{1}{\sqrt{3}} & \>\>0 & 0 & \>\>0 & 0 \\
 \>\>0 & 0 & \>\>0 & \frac{1}{\sqrt{3}} & 0 & \>\>0 & 0 \\
 \>\>0 & 0 & \>\>0 & \>\>0 & 0 & \>\>0 & 0 \\
 \>\>0 & 0 & \>\>0 & \>\>0 & 0 & \frac{1}{\sqrt{3}} & 0
\end{array}
\right) ,} \nonumber \end{equation}
\begin{equation}
\small{
f_3 = \left(
\begin{array}{lllllll}
 \>\>0 & \>\>0 & 0 & \>\>0 & \>\>0 & 0 & 0 \\
 \>\>0 & \>\>0 & 0 & \>\>0 & \>\>0 & 0 & 0 \\
 \frac{-1}{\sqrt{3}} & \>\>0 & 0 & \>\>0 & \>\>0 & 0 & 0 \\
 \>\>0 & \frac{1}{\sqrt{3}} & 0 & \>\>0 & \>\>0 & 0 & 0 \\
 \>\>0 & \>\>0 & 0 & \>\>0 & \>\>0 & 0 & 0 \\
 \>\>0 & \>\>0 & 0 & \frac{-1}{\sqrt{3}} & \>\>0 & 0 & 0 \\
 \>\>0 & \>\>0 & 0 & \>\>0 & \frac{1}{\sqrt{3}} & 0 & 0
\end{array}
\right),\quad\>
f_4 = \left(
\begin{array}{lllllll}
 \>\>0 & \>\>\>0 & \>\>\>0 & \>\>0 & 0 & 0 & 0 \\
 \>\>0 & \>\>\>0 & \>\>\>0 & \>\>0 & 0 & 0 & 0 \\
 \>\>0 & \>\>\>0 & \>\>\>0 & \>\>0 & 0 & 0 & 0 \\
 \frac{-1}{\sqrt{3}} & \>\>\>0 & \>\>\>0 & \>\>0 & 0 & 0 & 0 \\
 \>\>0 & \frac{1}{2\sqrt{3}} & \>\>\>0 & \>\>0 & 0 & 0 & 0 \\
 \>\>0 & \>\>\>0 & \frac{1}{2\sqrt{3}} & \>\>0 & 0 & 0 & 0 \\
 \>\>0 & \>\>\>0 & \>\>\>0 & \frac{-1}{\sqrt{3}} & 0 & 0 & 0
\end{array}
\right) ,} \nonumber \end{equation}
\begin{equation}
\small{
f_5 = \left(
\begin{array}{lllllll}
 \>\>0 & 0 & 0 & 0 & 0 & 0 & 0 \\
 \>\>0 & 0 & 0 & 0 & 0 & 0 & 0 \\
 \>\>0 & 0 & 0 & 0 & 0 & 0 & 0 \\
 \>\>0 & 0 & 0 & 0 & 0 & 0 & 0 \\
 -\frac{1}{2} & 0 & 0 & 0 & 0 & 0 & 0 \\
 \>\>0 & 0 & 0 & 0 & 0 & 0 & 0 \\
 \>\>0 & 0 & \frac{1}{2} & 0 & 0 & 0 & 0
\end{array}
\right) ,\hspace{1cm}\qquad\qquad\>
f_6 = \left(
\begin{array}{lllllll}
 0 & 0 & 0 & 0 & 0 & 0 & 0 \\
 0 & 0 & 0 & 0 & 0 & 0 & 0 \\
 0 & 0 & 0 & 0 & 0 & 0 & 0 \\
 0 & 0 & 0 & 0 & 0 & 0 & 0 \\
 0 & 0 & 0 & 0 & 0 & 0 & 0 \\
 \frac{1}{2} & 0 & 0 & 0 & 0 & 0 & 0 \\
 0 & \frac{1}{2} & 0 & 0 & 0 & 0 & 0
\end{array}
\right) .} \nonumber \end{equation}

\chapter{Transforming the Scalars of the Kerr Black String}
\label{app:ScalarTransform}

In this Appendix, I show how the eight scalar fields that parametrise the $G_{2(2)}/(SL(2,{\mathbb R})\times SL(2,{\mathbb R}))$ coset are transformed when the Kerr string seed solution is acted on by the element (\ref{generatork}).  To extract the scalars, I write the Kerr string metric (\ref{Kerr})-(\ref{eq:basefcns}) in the form of the Kaluza-Klein ansatz (\ref{eqn:canonicalform}):
\begin{align}
&ds^2=-\frac{\Delta_2}{\Sigma}\big(dt+\omega_\phi^{\mathrm {seed}}d\phi\big)^2+dz^2+\frac{\Sigma}{\Delta_2}ds_3^2,\\
&\omega_\phi^{\mathrm {seed}}=\frac{2amr(1-x^2)}{\Delta_2},\\
&ds_3^2=\frac{\Delta_2}{\Delta}dr^2+\frac{\Delta_2}{1-x^2}dx^2+\Delta(1-x^2)d\phi^2.
\end{align}
Comparing this with (\ref{eqn:canonicalform}), it is immediately evident that
\begin{equation}
\phi_1=0,\quad \phi_2=-\log\left[\frac{\Delta_2}{\Sigma}\right], \quad \chi_1=0, \quad \chi_2=0, \quad \chi_3=0, \quad \chi_4=0, \quad \chi_5=0, \label{eq:KerrScalars15}
\end{equation}
where the latter two are because ${\bf B}_1=0,~{\bf B}_2=0$ and recall that ${\bf B}_1$ and ${\bf B}_2$ are dualised to $\chi_4$ and $\chi_5$ respectively.  The only scalar left is $\chi_6$ which is obtained by integrating
\begin{equation}
{\bf d}\chi_6=e^{-2\phi_2}\star_3 {\bf d}\omega_\phi^{\mathrm {seed}}.  \label{eq:chi6seed}
\end{equation}
With a positive orientation on $ds_3^2$ given by ${\bf d}r\wedge {\bf d}x\wedge {\bf d}\phi$, (\ref{eq:chi6seed}) yields
\begin{equation}
\chi_6=-\frac{2amx}{\Sigma}.  \label{eq:KerrScalars6}
\end{equation}
These scalar fields now combine into the matrix $\cal M$ according to (\ref{eq:CosetRep}) and (\ref{M}); the result is
\begin{equation}
{\cal M}^{\mathrm{seed}}=\left(
\begin{array}{ccccccc}
 \frac{\Sigma}{\Delta_2} & 0 & 0 & 0 & 0 & -\frac{4amx}{\Delta_2} & 0 \\
 0 & \frac{\Sigma}{\Delta_2} & 0 & 0 & 0 & 0 & -\frac{4amx}{\Delta_2} \\
 0 & 0 & 1 & 0 & 0 & 0 & 0 \\
 0 & 0 & 0 & 1 & 0 & 0 & 0 \\
 0 & 0 & 0 & 0 & 1 & 0 & 0 \\
 -\frac{amx}{\Delta_2} & 0 & 0 & 0 & 0 & \frac{4a^2m^2x^2}{\Delta_2\Sigma}+\frac{\Delta_2}{\Sigma} & 0 \\
 0 & -\frac{amx}{\Delta_2} & 0 & 0 & 0 & 0 & \frac{4a^2m^2x^2}{\Delta_2\Sigma}+\frac{\Delta_2}{\Sigma}
\end{array}
\right).  \label{eq:MSeed}
\end{equation}

I now act on ${\cal M}^{\mathrm{seed}}$ with the element $k$ given by (\ref{generatork}) according to $k^{-1} {\cal M}^{\mathrm{seed}}k$.  The scalars can then be extracted as particular entries of the matrix ${\cal M}^{\mathrm{new}}$ to be explained below.  The transformed matrix ${\cal M}^{\mathrm{new}}$ is far too complicated to write down here but the element $k$ is simple enough so I give its explicit form; defining $A\equiv\frac{s_{2\beta}}{\sqrt{1+3c_\delta^2}}$, $B_{\pm}\equiv1\pm c_{2\beta}c_{2\delta}$ and $C\equiv3+c_{2\delta}$, I find $k$ and $k^{-1}$ to be given by
\begin{equation}
k=\left(
\begin{array}{ccccccc}
 c_\beta^2c_\delta & \frac 12As_\delta & c_\beta^2s_\delta & -2Ac_\delta & -2s_\beta^2s_\delta & -As_\delta & 2s_\beta^2c_\delta \\
 \frac 14As_{2\delta} & \frac12B_+ & -\frac14AC & -c_{2\beta}s_{2\delta} & \frac 12AC & B_- & \frac 12As_{2\delta} \\
 c_\beta^2s_\delta & -Ac_\delta & c_\beta^2c_\delta & As_\delta & -2s_\beta^2c_\delta & 2Ac_\delta & 2s_\beta^2s_\delta \\
 -\frac14AC & -\frac12c_{2\beta}s_{2\delta} & \frac 14As_{2\delta} & c_{2\beta}c_{2\delta} & -\frac 12As_{2\delta} & c_{2\beta}s_{2\delta} & -\frac 12AC \\
 -\frac12s_\beta^2s_\delta & \frac 12Ac_\delta & -\frac12s_\beta^2c_\delta & -\frac 12As_\delta & c_\beta^2c_\delta & -Ac_\delta & -c_\beta^2s_\delta \\
 -\frac18As_{2\delta} & \frac14B_- & \frac 18AC & \frac12c_{2\beta}s_{2\delta} & -\frac14AC & \frac12B_+ & -\frac14As_{2\delta} \\
 \frac12s_\beta^2c_\delta & \frac 14As_\delta & \frac12s_\beta^2s_\delta & -Ac_\delta & -c_\beta^2s_\delta & -\frac12As_\delta & c_\beta^2c_\delta
\end{array}
\right),
\end{equation}
\begin{equation}
k^{-1}=\left(
\begin{array}{ccccccc}
 c_\beta^2c_\delta & \frac 14As_{2\delta} & -c_\beta^2s_\delta & \frac12AC & 2s_\beta^2s_\delta & -\frac12As_{2\delta} & 2s_\beta^2c_\delta \\
 \frac 12As_{\delta} & \frac12B_+ & Ac_\delta & c_{2\beta}s_{2\delta} & -2Ac_\delta & B_- & As_{\delta} \\
 -c_\beta^2s_\delta & \frac14AC & c_\beta^2c_\delta & \frac12As_{2\delta} & -2s_\beta^2c_\delta & -\frac12AC & -2s_\beta^2s_\delta \\
 Ac_\delta & \frac12c_{2\beta}s_{2\delta} & \frac 12As_{\delta} & c_{2\beta}c_{2\delta} & -As_{\delta} & -c_{2\beta}s_{2\delta} & 2Ac_\delta \\
 \frac12s_\beta^2s_\delta & -\frac 18AC & -\frac12s_\beta^2c_\delta & -\frac 14As_{2\delta} & c_\beta^2c_\delta & \frac14AC & c_\beta^2s_\delta \\
 -\frac14As_{\delta} & \frac14B_- & -\frac 12Ac_\delta & -\frac12c_{2\beta}s_{2\delta} & Ac_\delta & \frac12B_+ & -\frac12As_{\delta} \\
 \frac12s_\beta^2c_\delta & \frac 18As_{2\delta} & -\frac12s_\beta^2s_\delta & \frac14AC & c_\beta^2s_\delta & -\frac14As_{2\delta} & c_\beta^2c_\delta
\end{array}
\right).
\end{equation}
In terms of the matrix elements ${\cal M}_{ab}$, for $a,b=1,2,...,7$, the scalars are given by
\begin{equation}
\phi_1=-\frac{\sqrt3}{2}\log\left[\frac{{\cal M}_{11}{\cal M}_{22}-{\cal M}_{12}^2}{{\cal M}_{11}^2}\right], \qquad\qquad \phi_2=\frac12\log\left[{\cal M}_{11}{\cal M}_{22}-{\cal M}_{12}^2\right], \nonumber
\end{equation}
\begin{equation}
\chi_1=-\left(\frac{{\cal M}_{11}{\cal M}_{23}-{\cal M}_{12}{\cal M}_{13}}{{\cal M}_{11}{\cal M}_{22}-{\cal M}_{12}^2}\right), \qquad\qquad \chi_2=-\sqrt{3}\frac{{\cal M}_{12}}{{\cal M}_{11}}, \qquad\qquad \chi_3=-\sqrt{3}\frac{{\cal M}_{13}}{{\cal M}_{11}}, \nonumber
\end{equation}
\begin{equation}
\chi_4=-\frac{\sqrt{3}}{2}\frac{{\cal M}_{14}}{{\cal M}_{11}}, \qquad \chi_5=\frac12\frac{{\cal M}_{11}{\cal M}_{15}-{\cal M}_{12}{\cal M}_{14}}{{\cal M}_{11}^2}, \qquad \chi_6=\frac12\frac{{\cal M}_{11}{\cal M}_{16}-{\cal M}_{13}{\cal M}_{14}}{{\cal M}_{11}^2}. \nonumber
\end{equation}
It can easily be verified that (\ref{eq:KerrScalars15}) and (\ref{eq:KerrScalars6}) satisfy these conditions using the elements of (\ref{eq:MSeed}).  Upon acting with the element $k$, I find that the transformed matrix elements are given by
\begin{equation}
{\cal M}_{11}=F_1+\Delta_2, \qquad {\cal M}_{12}=F_2, \qquad {\cal M}_{13}=F_3, \qquad {\cal M}_{22}=F_4+\Delta_2, \qquad {\cal M}_{23}=F_5,
\end{equation}
where the fields $F_i$ are given in (\ref{eq:monsterFfields}).  The only missing components are ${\cal M}_{14}$, ${\cal M}_{15}$ and ${\cal M}_{16}$ so this does not yet give the solution for $\chi_4$, $\chi_5$ and $\chi_6$.  Although in principle these scalars can be found by the elements of ${\cal M}$, they would need to be dualised back into 1-forms, which requires integration that can be quite complicated in general.  Following \cite{Compere:2009zh} I introduce a new matrix-valued 1-form, $\cal N$, defined by
\begin{equation}
{\bf d}{\cal N}=\star_3{\cal M}^{-1}{\bf d}{\cal M},
\end{equation}
which transforms under global $G_{2(2)}$ transformations as ${\cal N}\rightarrow g^{-1}{\cal N}g$.  The matrix elements ${\cal N}_{ab}$ conveniently encode the 1-form fields $\omega$, ${\bf B}_1$, and ${\bf B}_2$ via
\begin{equation}
\omega=-2{\cal N}_{61}, \qquad {\bf B}_1=-2{\cal N}_{51}-2\chi_1{\cal N}_{61}, \qquad {\bf B}_2=\sqrt{3}{\cal N}_{41}-2\chi_2{\cal N}_{51}-2\chi_3{\cal N}_{61}.
\end{equation}
Upon calculating ${\cal N}^{\mathrm{seed}}$ and acting with the element $k$ to obtain ${\cal N}^{\mathrm{new}}=k^{-1}{\cal N}^{\mathrm{seed}}k$, I find the transformed elements to be\footnote{I have used the gauge freedom ${\cal N}\rightarrow {\cal N}+\Lambda$ for constant $\Lambda$ to eliminate constant terms asymptotically.}
\begin{align}
{\cal N}_{41}={}&\frac{2m c_{\delta}^2}{\Delta_2}\left( \frac{2s_{\beta}c_{\beta}\Delta x}{\sqrt{1+3c_{\delta}^2}}-a(1-x^2)s_{\delta}\left(2ms_{\beta}^2\left(\frac{6c_{\beta}^2 c_{\delta}^2} {1+3c_{\delta}^2}-1\right)+r(c_{\beta}^2+s_{\beta}^2)\right)\right), \\
{\cal N}_{51}={}&\frac{2am(1-x^2)s_{\beta}c_{\beta}c_{\delta}^3(r+2s_{\beta}^2m)}{\Delta_2\sqrt{1+3c_{\delta}^2}},\\
{\cal N}_{61}={}&-am \frac{1-x^2}{\Delta_2}c_{\delta}^3\left(2ms_{\beta}^2\left(c_{\beta}^2+s_{\beta}^2-\frac{4c_{\beta}^2}{1+3c_{\delta}^2}\right)+r(c_{\beta}^2+s_{\beta}^2)\right).
\end{align}
The transformed scalar fields $\phi_1,~\phi_2,~\chi_1,~\chi_2,~\chi_3$ and the transformed 1-forms $\omega,~{\bf B}_1,~{\bf B}_2$ now exactly yield the general black string metric (\ref{eq:monster})-(\ref{eq:monsterFfields}).

\chapter{Hairy Black Hole Fields and Thermodynamics}
\label{AppendixHBH}

\section{Conserved Charges and Thermodynamic Quantities}
\label{app:thermo}

 In this section I catalogue the thermodynamic charges and potentials entering the first law in space-time dimension $D=n+2$, for $n=3,5$.  The boson star quantities are obtained by the limit $r_+\rightarrow0$, except for the temperature, which exhibits the usual Schwarzschild-like divergence as $r_+\rightarrow0$; the boson star temperature is zero.  The constants $K^g_{n;2,2}$ appearing in the expressions for the temperatures are calculated in section \ref{MatchingConditions}.

\subsection{n=3}

\begin{align}
E_3={}&\frac{\pi\ell^2}{4}\left(\left[\frac{3}{2}\frac{r_+^2}{\ell^2}+39\frac{r_+^4}{\ell^4}+{\cal O}\left(\frac{r_+^6}{\ell^6}\right)\right]+\epsilon^2\left[\frac56+\frac{191}{48}\frac{r_+^2}{\ell^2}+{\cal O}\left(\frac{r_+^4}{\ell^4}\right)\right]+\epsilon^4\left[\frac{77951}{127008}\right.\right.\nonumber\\
&\left.+{\cal O}\left(\frac{r_+^2}{\ell^2}\right)\right]+{\cal O}(\epsilon^6)\bigg)
\end{align}
\begin{align}
J_3={}&\frac{\pi\ell^3}{2}\left(\left[5\frac{r_+^4}{\ell^4}+{\cal O}\left(\frac{r_+^6}{\ell^6}\right)\right]+\epsilon^2\left[\frac1{12}+\frac{43}{96}\frac{r_+^2}{\ell^2}+{\cal O}\left(\frac{r_+^4}{\ell^4}\right)\right]+\epsilon^4\left[\frac{83621}{1270080}+{\cal O}\left(\frac{r_+^2}{\ell^2}\right)\right]\right.\nonumber\\
&+{\cal O}(\epsilon^6)\bigg)
\end{align}

\begin{align}
\ell\omega_3={}&\left[5-3\frac{r_+^2}{\ell^2}-\left\{\frac{959}{16}-\frac{3}{2}\log\left[\frac{r_+}{4\ell}\right]\right\}\frac{r_+^4}{\ell^4}+{\cal O}\left(\frac{r_+^6}{\ell^6}\right)\right]-\epsilon^2\left[\frac{15}{28}+\frac{4469}{840}\frac{r_+^2}{\ell^2}+{\cal O}\left(\frac{r_+^4}{\ell^4}\right)\right]\nonumber\\
&-\epsilon^4\left[\frac{22456447}{35562240}+{\cal O}\left(\frac{r_+^2}{\ell^2}\right)\right]+{\cal O}(\epsilon^6)
\end{align}
\begin{align}
T_{H;3}={}&\frac{1}{\pi r_+}\left(\left[\frac12-\frac{71}{4}\frac{r_+^2}{\ell^2}-\frac{2665}{16}\frac{r_+^4}{\ell^4}+{\cal O}\left(\frac{r_+^6}{\ell^6}\right)\right]-\epsilon^2\left[\frac{1}{6}+\left\{\frac{59}{42}-\frac{K^g_{3;2,2}}4+\frac{\pi^2}{32}\right\}\frac{r_+^2}{\ell^2}\right.\right.\nonumber\\
&+\left.\left.{\cal O}\left(\frac{r_+^4}{\ell^4}\right)\right]-\epsilon^4\left[\frac{101341}{508032}+{\cal O}\left(\frac{r_+^2}{\ell^2}\right)\right]\right)
\end{align}
\begin{align}
S_3={}&r_+^3\frac{\pi^2}{2}\left(\left[1+\frac{25}{2}\frac{r_+^2}{\ell^2}+\frac{1655}{8}\frac{r_+^4}{\ell^4}+{\cal O}\left(\frac{r_+^6}{\ell^6}\right)\right]+\epsilon^2\left[\frac{265}{84}\frac{r_+^2}{\ell^2}+{\cal O}\left(\frac{r_+^4}{\ell^4}\right)\right]+\epsilon^4{\cal O}\left(\frac{r_+^2}{\ell^2}\right)\right.\nonumber\\
&+{\cal O}(\epsilon^6)\bigg)
\end{align}

\subsection{n=5}

\begin{align}
E_5={}&\frac{\pi^2\ell^4}{8}\left(\left[\frac{5}{2}\frac{r_+^4}{\ell^4}+125\frac{r_+^6}{\ell^6}+{\cal O}\left(\frac{r_+^8}{\ell^8}\right)\right]+\epsilon^2\left[\frac7{20}+\frac{3463}{480}\frac{r_+^4}{\ell^4}+{\cal O}\left(\frac{r_+^6}{\ell^6}\right)\right]\right.\nonumber\\
&+\epsilon^4\left[\frac{314018183}{2208492000}+{\cal O}\left(\frac{r_+^4}{\ell^4}\right)\right]+{\cal O}(\epsilon^6)\bigg)
\end{align}
\begin{align}
J_5={}&\frac{\pi^2\ell^5}{8}\left(\left[21\frac{r_+^6}{\ell^6}+{\cal O}\left(\frac{r_+^8}{\ell^8}\right)\right]+\epsilon^2\left[\frac1{20}+\frac{529}{480}\frac{r_+^4}{\ell^4}+{\cal O}\left(\frac{r_+^6}{\ell^6}\right)\right]+\epsilon^4\left[\frac{327248543}{15459444000}\right.\right.\nonumber\\
&\left.+{\cal O}\left(\frac{r_+^4}{\ell^4}\right)\right]+{\cal O}(\epsilon^6)\bigg)
\end{align}

\begin{align}
\ell\omega_5={}&\left[7-10\frac{r_+^4}{\ell^4}-\left\{360+\frac{320\Gamma^2\big[\frac54\big]}{\Gamma^2\big[\frac{-1}4\big]}\right\}\frac{r_+^6}{\ell^6}+{\cal O}\left(\frac{r_+^8}{\ell^8}\right)\right]-\epsilon^2\left[\frac{514}{2145}+\frac{1438545023}{96621525}\frac{r_+^4}{\ell^4}\right.\nonumber\\
&\left.+{\cal O}\left(\frac{r_+^6}{\ell^6}\right)\right]-\epsilon^4\left[\frac{60223029794867}{315024820110000}+{\cal O}\left(\frac{r_+^4}{\ell^4}\right)\right]+{\cal O}(\epsilon^6)
\end{align}
\begin{align}
T_{H;5}={}&\frac{1}{\pi r_+}\left(\left[1-\frac{95}{2}\frac{r_+^2}{\ell^2}-\frac{5985}{8}\frac{r_+^4}{\ell^4}+{\cal O}\left(\frac{r_+^6}{\ell^6}\right)\right]-\epsilon^2\left[\frac{1}{5}+\left\{\frac{10559}{2145}-\frac{K^g_{5;2,2}}2\right.\right.\right.\nonumber\\
&\left.\left.\left.+\frac{2\Gamma^4\big[\frac{5}{4}\big]}{\pi}\right\}\frac{r_+^2}{\ell^2}+{\cal O}\left(\frac{r_+^4}{\ell^4}\right)\right]-\epsilon^4\left[\frac{997904461}{5521230000}+{\cal O}\left(\frac{r_+^2}{\ell^2}\right)\right]\right)
\end{align}
\begin{align}
S_5={}&r_+^5\frac{\pi^3}{4}\left(\left[1+\frac{49}{2}\frac{r_+^2}{\ell^2}+\frac{6447}{8}\frac{r_+^4}{\ell^4}+{\cal O}\left(\frac{r_+^6}{\ell^6}\right)\right]+\epsilon^2\left[\frac{19831}{4290}\frac{r_+^2}{\ell^2}+{\cal O}\left(\frac{r_+^4}{\ell^4}\right)\right]\right.\nonumber\\
&\left.+\epsilon^4{\cal O}\left(\frac{r_+^2}{\ell^2}\right)+{\cal O}(\epsilon^6)\right)
\end{align}

\section{Perturbative Fields}
\label{app:fields}

In this Appendix I catalogue all of the gravitational and scalar fields for the perturbative boson stars and hairy black holes in space-time dimension $D=n+2$ for $n=3,5$.  The fields are labeled as $F_{n;p,q}$, where $p$ denotes the order in $\epsilon$ and $q$ denotes the order in $r_+$.  Note that $\mathrm{Li}_2[x]$ is the dilogarithm function while $P_\nu[x]$ is the Legendre function of the first kind.  

\subsection{n=3}

\begin{align}
\tilde{s}_3(x)&=22x^6+31x^4-250x^2+75\nonumber\\
\Pi^{out}_{3;1,2}&=\frac14 \left(-\ell^2 (\ell^2 + 11 r^2) + 21 r^2 (\ell^2 + r^2) \log\left[1 + \frac{\ell^2}{r^2}\right]\right)\nonumber\\
\Pi^{out}_{3;1,4}&=\frac{-\ell^2}{64(r^2+\ell^2)}\bigg(2r^{10}+(19+4\pi^2)r^8\ell^2+2(2163+4\pi^2r^6\ell^4+4(1156+\pi^2)r^4\ell^6\nonumber\\
&+600r^2\ell^8+5\ell^{10}\bigg)-\frac{1}{32}\bigg(24r^4\ell^4(r^2+\ell^2)\log^2\left[\frac{r}{\ell}\right]+2\ell^4(r^2+\ell^2)\log\left[\frac{r_+}{4\ell}\right]\bigg(8r^2\ell^2\nonumber\\
&+\ell^4+12r^4\log\left[1+\frac{\ell^2}{r^2}\right]\bigg)-\log\left[1+\frac{\ell^2}{r^2}\right]\bigg(r^{10}+9r^8\ell^2+3745r^6\ell^4+3267r^4\ell^6\nonumber\\
&-51r^2\ell^8-\ell^{10}+441r^4\ell^4(r^2+\ell^2)\log\left[1+\frac{\ell^2}{r^2}\right]\bigg)\bigg)-\frac{3r^4\ell^4}{8}(r^2+\ell^2)\mathrm{Li}_2\left[-\frac{r^2}{\ell^2}\right]\nonumber
\end{align}
\begin{align}
f_{3;2,0}&=\frac{5r^4+20r^2\ell^2+6\ell^4}{9}\nonumber\\
\Omega_{3;2,0}&=\frac{r^4+4r^2\ell^2+6\ell^4}{12}\nonumber\\
f^{in}_{3;2,2}&=\frac{2120+21\pi^2x^4P_{\frac12}\big[2x^2-1\big]\left(2\big(1-2x^2\big)P_{\frac12}\big[2x^2-1\big]+3P_{\frac32}\big[2x^2-1\big]\right)}{168x^3}\nonumber\\
g^{in}_{3;2,2}&=-\frac{145}{14}\left(1-\frac{\ell^4}{z^4}\right)\nonumber\\
\Omega^{in}_{3;2,2}&=\frac{190}{21}+\frac{25y^2}{7}\nonumber\\
f^{out}_{3;2,2}&=-\frac{7}{6}r^2(r^2 + \ell^2)(5r^4+20r^2\ell^2+6\ell^4)\log\left[1 + \frac{\ell^2}{r^2}\right] - \frac{1}{72} \bigg(191r^{8} + 535r^6\ell^2 +20r^4\ell^4\nonumber\\
&-84r^2\ell^6-48\ell^8\bigg)\nonumber\\
g^{out}_{3;2,2}&=(r^2 + \ell^2)\left(\frac{20}{3}(r^2+\ell^2)^5 - 7 \ell^8 (4 r^2 + \ell^2)\right)\log\left[1 + \frac{\ell^2}{r^2}\right] - \frac{\ell^2}{9} \bigg(60 r^{10} + 330r^8\ell^2\nonumber\\
&+740r^6\ell^4+855 r^4\ell^6+378r^2\ell^8+83\ell^{10}\bigg)\nonumber\\
\Omega^{out}_{3;2,2}&=(r^2 + \ell^2)\left(-4(r^2+\ell^2)^4 +\frac{7\ell^4}{8} (r^4+4r^2\ell^2+6\ell^4)\right)\log\left[1+\frac{\ell^2}{r^2}\right]+\frac{\ell^2}{96}\bigg(384r^{8}\nonumber\\
& + 1771r^6\ell^2+3139r^4\ell^4+2516r^2\ell^6+668\ell^{8}\bigg)\nonumber\\
\Pi_{3;3,0}&=\frac{900r^6+3935r^4\ell^2+5548r^2\ell^4+1540\ell^6}{2016}\nonumber\\
\Pi^{out}_{3;3,2}&=\frac{-\ell^2}{48384}\bigg(14196r^{14}+7098(4\pi^2+25)r^{12}\ell^2+14(10140\pi^2-6037)r^{10}\ell^4\nonumber\\
&+1183(240\pi^2-1667)r^{8}\ell^6
+65(4368\pi^2-55327)r^{6}\ell^8+5(28392 \pi^2-450949)r^{4}\ell^{10}\nonumber\\
&+3(9464\pi^2-180149)r^{2}\ell^{12}+9240\ell^{14}\bigg)
+\frac{r^3}{8064}(r^2 + \ell^2)\log\left[1 + \frac{\ell^2}{r^2}\right] \bigg(2366r^{12}\nonumber\\
& + 28392r^{10}\ell^2+ 34096r^8\ell^4 - 43516r^6\ell^6 -8889r^4\ell^8+145204r^2\ell^{10}+41208\ell^{12}\bigg)\nonumber\\
&-\frac{169}{96}r^2\ell^4(r^2 + \ell^2)^5\left(2\mathrm{Li}_2\left[-\frac{r^2}{\ell^2}\right]-\log\left[1+\frac{r^2}{\ell^2}\right]\log\left[\frac{\ell^2(r^2 + \ell^2)}{r^4}\right]\right)\nonumber
\end{align}
\begin{align}
 f_{3;4,0}&=\frac1{1270080}\bigg(514952r^{14}+4631027r^{12}\ell^2+18512283r^{10}\ell^4+40913902r^8\ell^6\nonumber\\
&+51954798r^6\ell^8+36154839r^4\ell^{10}+11249595r^2\ell^{12}+1315860\ell^{14}\bigg)\nonumber\\
g_{3;4,0}&=\frac1{1270080}\bigg(3541r^{16}+35410r^{14}\ell^2+149055r^{12}\ell^4+3052440r^{10}\ell^6+16099475r^8\ell^8\nonumber\\&+34403186r^6\ell^{10}+25939155r^4\ell^{12}+7971520r^2\ell^{14}+872290\ell^{16}\bigg)\nonumber\\
h_{3;4,0}&=\frac{1}{1270080}\bigg(3541r^{12}+31869r^{10}\ell^2+123066r^8\ell^4+260694r^6\ell^6+311661r^4\ell^8\nonumber\\
&+183645r^2\ell^{10}+22260\ell^{12}\bigg)\nonumber\\
\Omega_{3;4,0}&=\frac1{2540160}\bigg(167242r^{14}+1505178r^{12}\ell^2+6020712r^{10}\ell^4+14139048r^8\ell^6\nonumber\\&+20982192r^6\ell^8+19004760r^4\ell^{10}+8795055r^2\ell^{12}+1598455\ell^{14}\bigg)\nonumber\\
\Pi_{3;5,0}&=\frac1{853493760}\bigg(428716940r^{18}+4416801537r^{16}\ell^2+20395866890r^{14}\ell^4\nonumber\\&+55586393870r^{12}\ell^6+98320298706r^{10}\ell^8+115794392980r^8\ell^{10}+88872056182r^6\ell^{12}\nonumber\\
&+41756607180r^4\ell^{14}+10678880150r^2\ell^{16}+1128452101\ell^{18}\bigg)\nonumber
\end{align}

\subsection{n=5}

\begin{align}
\tilde{s}_5(x)&=44x^{12}+88x^{10}+113x^8-1210x^6-784x^4+490x^2+245\nonumber\\
\Pi^{out}_{5;1,4}&=\frac{1}{8}\left(\ell^2\big(100r^4+71r^2\ell^2-\ell^4\big)- 30 r^4 (\ell^2 + r^2) \log\left[1 + \frac{\ell^2}{r^2}\right]\right)\nonumber\\
\Pi^{out}_{5;1,6}&=\frac{\ell^{8}}{6}\bigg(3456r^4+2295r^2\ell^2-111\ell^4+(r^2+\ell^2)(45r^4+9r^2\ell^2+\ell^4)\frac{\Gamma^2\left[\frac{1}{4}\right]}{\Gamma^2\left[\frac{-1}{4}\right]}\bigg)\nonumber\\&
-r^6\ell^6(r^2+\ell^2)\left(261-10\frac{\Gamma^2\left[\frac{1}{4}\right]}{\Gamma^2\left[\frac{-1}{4}\right]}\right)\log\left[1+\frac{\ell^2}{r^2}\right]\nonumber\\
f_{5;2,0}&=\frac{7r^6+42r^4\ell^2+105r^2\ell^4+20\ell^6}{50}\nonumber
\end{align}
\begin{align}
\Omega_{5;2,0}&=\frac{r^6+6r^4\ell^2+15r^2\ell^4+20\ell^6}{60}\nonumber\\
f^{in}_{5;2,2}&=\frac{1}{2145\pi x^3}\bigg(19831\pi(1+x^4)+8580\Gamma^4\big[\frac{5}{4}\big]x^6P_{\frac14}\big[2x^4-1\big]\left(3\big(1-2x^4\big)P_{\frac14}\big[2x^4-1\big]\right.\nonumber\\
&\left.+5P_{\frac54}\big[2x^4-1\big]\right)\bigg)\nonumber\\
g^{in}_{5;2,2}&=-\int_1^{z/\ell}{\frac{14\big(2458 + 4916 x^2 + 3173 x^4\big)}{715x^7(1+x^2)^2}dx}\nonumber\\
\Omega^{in}_{5;2,2}&=\frac{67368}{715}-\frac{47537y^4}{715}\nonumber\\
f^{out}_{5;2,4}&=\frac{3}{20}r^4(r^2 + \ell^2)(7r^6+42r^4\ell^2+105r^2\ell^4+20\ell^6)\log\left[1 + \frac{\ell^2}{r^2}\right] - \frac{1}{1200} \bigg(3463r^{12}\nonumber\\&
+25501r^{10}\ell^2+80913r^8\ell^4+143675r^6\ell^6+91100r^4\ell^8+13140r^2\ell^{10}-360\ell^{12}\bigg)\nonumber\\
g^{out}_{5;2,4}&=-r^2(r^2 + \ell^2)\left(\frac{133}{5}(r^2+\ell^2)^7 - 3 \ell^{12} (6 r^2 + \ell^2)\right) \log\left[1 + \frac{\ell^2}{r^2}\right] +\frac{\ell^2}{300} \bigg(7980 r^{16} \nonumber\\
&+ 59850r^{14}\ell^2 +194180r^{12}\ell^4+354445 r^{10}\ell^6+395276r^8\ell^8+272118r^6\ell^{10}\nonumber\\
&+92268r^4\ell^{12}+7401r^2\ell^{14}-90\ell^{16}\bigg)\nonumber\\
\Omega^{out}_{5;2,4}&=(r^2 + \ell^2)\left(\frac{169}{6}(r^2+\ell^2)^6 -\frac{\ell^6}{8} (r^6+6r^4\ell^2+15r^2\ell^4+20\ell^6)\right)\log\left[1+\frac{\ell^2}{r^2}\right]\nonumber\\
&-\frac{\ell^2}{1440}\bigg(40560r^{14}+263640r^{12}\ell^2+722791r^{10}\ell^4+1074337r^{8}\ell^6+918573r^6\ell^{8}\nonumber\\
&+430519r^4\ell^{10}+81352r^2\ell^{12}-3780\ell^{14}\bigg)\nonumber\\
\Pi_{5;3,0}&=\frac{53970r^{10}+313026r^8\ell^2+773598r^6\ell^4+1035891r^4\ell^6+767826r^2\ell^8+152581\ell^{10}}{257400}\nonumber\\
f_{5;4,0}&=\frac1{618377760000}\bigg(35065241462r^{20}+455760809811r^{18}\ell^2+2733953554501r^{16}\ell^4\nonumber\\
&+10021950443882r^{14}\ell^6+24529389790620r^{12}\ell^8 +41236810544215r^{10}\ell^{10}\nonumber\\
&+47555951613885r^8\ell^{12}+36498131558460r^6\ell^{14}+17103932986140r^4\ell^{16}\nonumber\\
&+3682611572640r^2\ell^{18}+295557662400\ell^{20}\bigg)\nonumber
\end{align}
\begin{align}
\Pi^{out}_{5;3,4}&=\frac{\ell^2}{883396800}\bigg(464785464r^{22}+7204174692r^{20}\ell^2+309856976(15\pi^2+182)r^{18}\ell^4\nonumber\\&
+2(16267491240\pi^2+99389066437)r^{16}\ell^6+2(48802473720\pi^2+175884179161)r^{14}\ell^8\nonumber\\&
+88(1848578550\pi^2+3555421769)r^{12}\ell^{10}+3080(52816530\pi^2+36628159)r^{10}\ell^{12}\nonumber\\&
+4(24401236860\pi^2+2936476817)r^{8}\ell^{14}+(32534982480\pi^2+44985717097)r^{6}\ell^{16}\nonumber\\&
+429(10834160 \pi^2+89900771)r^{4}\ell^{18}+5998786365r^{2}\ell^{20}-65457249\ell^{22}\bigg)\nonumber\\&
-\frac{r^4}{1029600}(r^2 + \ell^2)\log\left[1 + \frac{\ell^2}{r^2}\right] \bigg(541708r^{18}+8125620r^{16}\ell^2+61754712r^{14}\ell^4\nonumber\\&+221505790r^{12}\ell^6
+ 421393950r^{10}\ell^8+447202548r^8\ell^{10} +255856550r^6\ell^{12}\nonumber\\& +66949977r^4\ell^{14}+6608370r^2\ell^{16}
-1853965\ell^{18}\bigg)+\frac{135427}{8580}r^4\ell^6(r^2 + \ell^2)^7\times\nonumber\\&
\times\bigg(2\mathrm{Li}_2\left[-\frac{r^2}{\ell^2}\right]-\log\left[1+\frac{r^2}{\ell^2}\right]\log\left[\frac{\ell^2(r^2 + \ell^2)}{r^4}\right]\bigg)\nonumber\\
g_{5;4,0}&=\frac1{618377760000}\bigg(87329195r^{22}+1222608730r^{20}\ell^2+7946956745r^{18}\ell^4\nonumber\\
&+31598878220r^{16}\ell^6+677658846865r^{14}\ell^8+4244040931130r^{12}\ell^{10}\nonumber\\
&+12564927910535r^{10}\ell^{12}+21539953150144r^8\ell^{14}+22932404913236r^6\ell^{16}\nonumber\\&+11564746522784r^4\ell^{18}+2586519623296r^2\ell^{20}+198795488864\ell^{22}\bigg)\nonumber\\
h_{5;4,0}&=\frac1{123675552000}\bigg(17465839r^{18}+227055907r^{16}\ell^2+1362335442r^{14}\ell^4\nonumber\\&+4978052794r^{12}\ell^6+12277654675r^{10}\ell^8+21295887327r^8\ell^{10}+25938182556r^6\ell^{12}\nonumber\\&+20983798836r^4\ell^{14}
+9390693312r^2\ell^{16}+769728960\ell^{18}\bigg)\nonumber\\
\Omega_{5;4,0}&=\frac1{46378332000}\bigg(327248543r^{20}+4254231059r^{18}\ell^2+25525386354r^{16}\ell^4\nonumber\\&
+93593083298r^{14}\ell^6+234379619045r^{12}\ell^8+420702504651r^{10}\ell^{10}\nonumber\\&
+547149789758r^8\ell^{12}+505325794688r^6\ell^{14}+311179210446r^4\ell^{16}\nonumber\\&
+106176159440r^2\ell^{18}+15058082990\ell^{20}\bigg)\nonumber
\end{align}
\begin{align}
\Pi_{5;5,0}&=\frac1{453635740958400000}\bigg(74253000574956420r^{26}+1032668691070620996r^{24}\ell^2\nonumber\\&+6684077146010747418r^{22}\ell^4+26698941169899723207r^{20}\ell^6\nonumber\\&
+73557029936994344181r^{18}\ell^8+147920037831416512914r^{16}\ell^{10}\nonumber\\&+223871564194330948248r^{14}\ell^{12}+258620332265962810395r^{12}\ell^{14}\nonumber\\&+228077768805842399885r^{10}\ell^{16}+151159034180233581556r^8\ell^{18}\nonumber\\&+72350175319616674844r^6\ell^{20}+23112600885726752792r^4\ell^{22}\nonumber\\&
+4271077136958547132r^2\ell^{24}+337633104499816268\ell^{26}\bigg)\nonumber
\end{align}



\bibliographystyle{plain}
\renewcommand{\bibname}{References}
\bibliography{ThesisBib}
\addcontentsline{toc}{chapter}{\textbf{References}}

\nocite{*}

\end{document}